\makeatletter 
\@ifundefined{@parse@version@dash}{%
\def\@parse@version#1{\@parse@version@0#1}
\def\@parse@version@#1/#2/#3#4#5\@nil{%
\@parse@version@dash#1-#2-#3#4\@nil}
\def\@parse@version@dash#1-#2-#3#4#5\@nil{%
  \if\relax#2\relax\else#1\fi#2#3#4 }
}{}
\makeatother
\documentclass[aps,twocolumn,superscriptaddress,floatfix,preprintnumbers,nofootinbib]{revtex4-2}
\usepackage[utf8]{inputenc}
\usepackage{amsmath}
\usepackage{amssymb}
\usepackage{stackrel}

\usepackage{enumerate}
\usepackage{enumitem}
\usepackage{amsmath}
\usepackage{tikz}
\usepackage{feynmf}
\usepackage{slashed}
\usepackage{braket}
\usepackage[toc,page]{appendix}
\usepackage{url}
\usepackage{natbib}
\usepackage{graphicx}
\usepackage[pdftex,bookmarks,linktocpage,pdfpagelabels,plainpages=false,hyperfigures,linkcolor=blue,citecolor=blue]{hyperref} 
\hypersetup{colorlinks=true}

\usepackage{mathrsfs,amssymb}  
\usepackage{cancel}
\usepackage[normalem]{ulem}
\usepackage{multirow}

\newcommand{\ourT}{\ensuremath{\top}}

\newcommand{\genericT}{\ensuremath{T}}
\newcommand{\met}{\slashed{E}_\genericT}
\newcommand{\mht}{\slashed{H}_\genericT}

\newcommand{\mpt}{\slashed{p}_\genericT}
\newcommand{\mptvec}{\slashed{\vec{p}}_\genericT}

\makeatletter
\newcommand{\fmslash}[2][0mu]{%
  \mathchoice
    {\fmsl@sh\displaystyle{#1}{#2}}%
    {\fmsl@sh\textstyle{#1}{#2}}%
    {\fmsl@sh\scriptstyle{#1}{#2}}%
    {\fmsl@sh\scriptscriptstyle{#1}{#2}}}
\newcommand{\fmsl@sh}[3]{%
  \m@th\ooalign{$\hfil#1\mkern#2/\hfil$\crcr$#1#3$}}
\makeatother

\newcommand{\mptvecperp}{{\vec{\fmslash P}_{T_\perp}}}
\newcommand{\mptvecpar}{{\vec{\fmslash P}_{T_\parallel}}}

\newcommand{\ourperp}{\ensuremath{\vee}}

\newcommand{\myT}{\ensuremath{\perp}}
\newcommand{\massless}{{\ensuremath{\circ}}}

\newcommand{\beq}{\begin{equation}}
\newcommand{\eeq}{\end{equation}}
\newcommand{\bea}{\begin{eqnarray}}
\newcommand{\eea}{\end{eqnarray}}

\graphicspath{{graphics/}}

\allowdisplaybreaks

\begin{document}

\preprint{MI-HET-780}
\preprint{FERMILAB-PUB-22-481-QIS}

\author{Roberto Franceschini}
\email{roberto.franceschini@uniroma3.it}
\affiliation{Universit\`{a} degli Studi Roma Tre and INFN Roma Tre, Via della Vasca Navale 84, I-00146 Roma, Italy}

\author{Doojin~Kim}
\email{doojin.kim@tamu.edu}
\affiliation{Mitchell Institute for Fundamental Physics and Astronomy,
Department of Physics and Astronomy, Texas A\&M University, College Station, TX 77843, USA}

\author{Kyoungchul Kong}
\email{kckong@ku.edu}
\affiliation{Department of Physics and Astronomy, University of Kansas, Lawrence, KS 66045, USA}

\author{Konstantin T. Matchev}
\email{matchev@ufl.edu}
\affiliation{Institute for Fundamental Theory, Physics Department, University of Florida, Gainesville, FL 32611, USA}

\author{Myeonghun Park}
\email{parc.seoultech@seoultech.ac.kr}
\affiliation{Faculty of Natural Sciences, Seoultech, 232 Gongneung-ro, Nowon-gu, Seoul, 01811, Korea}
\affiliation{School of Physics, KIAS, Seoul 02455, Korea}

\author{Prasanth Shyamsundar}
\email{prasanth@fnal.gov}
\affiliation{Fermilab Quantum Institute, Fermi National Accelerator Laboratory, Batavia, IL 60510, USA}

\title{Kinematic Variables and Feature Engineering for Particle Phenomenology}

\begin{abstract}
Kinematic variables have been playing an important role in collider phenomenology, as they expedite discoveries of new particles by separating signal events from unwanted background events and allow for measurements of particle properties such as masses, couplings, spins, etc. For the past 10 years, an enormous number of kinematic variables have been designed and proposed, primarily for the experiments at the Large Hadron Collider, allowing for a drastic reduction of high-dimensional experimental data to lower-dimensional observables, from which one can readily extract underlying features of phase space and develop better-optimized data-analysis strategies. We review these recent developments in the area of phase space kinematics, summarizing the new kinematic variables with important phenomenological implications and physics applications. We also review recently proposed analysis methods and techniques specifically designed to leverage the new kinematic variables. As machine learning is nowadays percolating through many fields of particle physics including collider phenomenology, we discuss the interconnection and mutual complementarity of kinematic variables and machine learning techniques. We finally discuss how the utilization of kinematic variables originally developed for colliders can be extended to other high-energy physics experiments including neutrino experiments. 
\end{abstract}

\maketitle

\tableofcontents

\section{Introduction}
\label{sec:intro}

The defining objective of particle physics is to understand the elementary constituents of our Universe and their interactions at the most fundamental level. Advancing our understanding of Nature at these smallest possible scales requires in turn extraordinarily large and complex particle physics experiments. For example, the Large Hadron Collider (LHC) at CERN is not only the largest man-made experiment on Earth, but also the most prolific producer of scientific data. The data delivery rate at its upcoming upgrade, the High-Luminosity LHC (HL-LHC), will increase 100-fold to about 1 exabyte per year, bringing quantitatively and qualitatively new challenges due to its event size, data volume, and complexity, thereby straining the available computational resources. New particle physics discoveries in this era of big data will only be possible with novel methods for data collection, processing, and analysis. 

\subsection{The curse of dimensionality and the zoo of kinematic variables}

Modern particle physics data is extremely high-dimensional --- typical events result in multiple ($\sim 1000$) particles in the final state. The dimensionality of the data will increase even further at the HL-LHC. Ideally, one would like to make use of the full information encoded in the raw experimental data, but this approach would run into serious challenges:
\begin{itemize}
    \item From a theorist's point of view, the ultimate goal is to understand the fundamental laws of Nature at the microscopic level. However, it is highly nontrivial to decipher the underlying physics and/or develop physical intuition by looking at the raw data.
    \item From a practitioner's point of view, working with the full (raw) dataset very quickly becomes computationally prohibitive as the dimensionality of the data increases \cite{Albertsson:2018maf}. 
\end{itemize}

\begin{figure}[t]
\begin{center}
\includegraphics[width=.45\textwidth]{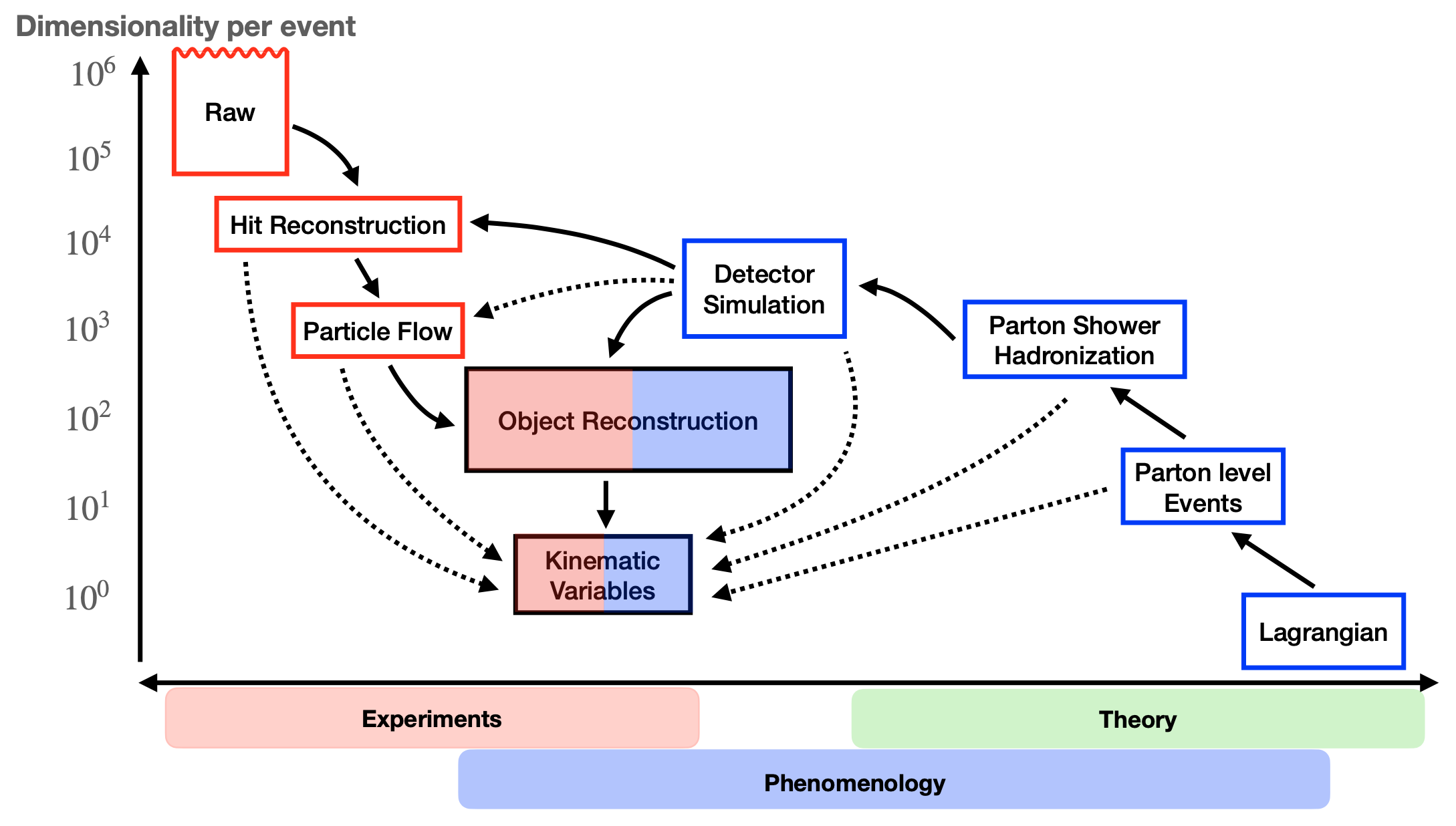}
\end{center}
\caption{Illustration of the dimensionality reduction in collider experiments and phenomenological studies. The common goal is to find the optimal low-dimensional kinematic observables. On the experimental side (red boxes on the left) this is accomplished by reconstructing the low-level detector data into progressively more physically motivated quantities. On the theory side (blue boxes on the right) the kinematic variables (or features in their distributions) are meant to reflect fundamental parameters in the theory Lagrangian. The solid (dotted) arrows indicate the typical flow (or simplifying shortcuts) in the high energy physics simulation chain \cite{Ask:2012sm}.} 
\label{fig:dim_reduction}
\end{figure}

Given the size and nature of the experimental dataset, modern particle physics analyses inevitably involve some kind of dimensionality reduction to fewer variables (features), which are suitably chosen to be optimal for the goal of the particular experiment. These higher-level variables are derived from the measured particle kinematic information, and therefore, are generically referred to as ``kinematic variables'', see Figure~\ref{fig:dim_reduction}. 
Naturally, there is no unique or ``best'' way to perform this dimensional reduction --- the perceived benefits of any given technique depend on a variety of factors, e.g., the experimental signature, the goal of the analysis, the control over the physics and instrumental backgrounds, and finally, one's judging criteria, which can be rather subjective to begin with. Moreover, if the final state contains invisible particles such as neutrinos and dark-matter candidates which appear as missing energy, their treatment opens the door for many new possibilities. This is why many different approaches have been tried, and as a result, a great number of kinematic variables have been proposed and investigated in the literature. Depending on the underlying event topology and the target study point, they may show different levels of performance and capability (i.e., no single variable exhibits absolute superiority to the others), hence, it is prudent to keep as many tools as possible in the analysis toolbox.

\subsection{Goal, scope and organization of this review}

This review is meant to provide a comprehensive guide to commonly used kinematic variables with a special focus on the recent developments within the last decade. Such a review is important and timely for the following reasons:
\begin{itemize}
    \item {\em A comprehensive list of kinematic variables.} Kinematic variables are routinely used in experiments to search for new signals, as well as to perform parameter measurements in observed processes. The use of the right kinematic variables can expedite the discovery of new physics, as well as increase the sensitivity to a given parameter. This review will provide a comprehensive menu from which practitioners can either pick existing kinematic variables which are the right ones for their task, or derive new kinematic variables following the methodology presented here.
    \item {\em Feature engineering.} Machine learning is now increasingly being used for data analysis in high energy physics. It is known that the performance (and the training efficiency) of the algorithms depends crucially on the parametrization of the input features. Using the right kinematic variables to describe the data would greatly enhance the performance of machine learning techniques in analyzing the data. Finding the right balance between attributes of the data that one wants to be sensitive to and those which are irrelevant for the question at hand, is an art. This review can thus be used either to optimize the input for various machine-learning algorithms and tasks or to properly interpret the output from the machine in terms of human-engineered kinematic quantities.
    \item {\em The need for an up-to-date review.} The last such review of comparable scope was written ten years ago~\cite{Barr:2010zj}. There also exist several sets of pedagogical lectures targeting newcomers in the field which focus on   standard material~\cite{Han:2005mu,Perelstein:2010hh,Schwartz:2017hep}. A few other, more limited in scope, reviews have appeared recently as well, e.g., focusing on energy peaks~\cite{Franceschini:2017dxe} or
    minimum invariant mass bounds~\cite{Barr:2011xt}.
\end{itemize}

The organization of the paper is as follows. Section~\ref{sec:preprocessing} provides the necessary background, motivation and context for the construction of kinematic variables. Our conventions and notation for the particle kinematics are then presented in Section~\ref{sec:notation}. Some basic kinematic observables and their use in a few benchmark processes from the Standard Model (SM) are reviewed in Section~\ref{sec:standard}. In Section~\ref{sec:inclusive} (Sections~\ref{sec:exclusive} through \ref{sec:otherexc}) we describe inclusive (exclusive) event variables used to characterize a {\it single event}. Variables and methods relying on {\em ensembles of events} are discussed in Section~\ref{sec:ensemblesn}. The interplay between the classic kinematic methods and the more recent machine-learning approaches is discussed in Section~\ref{sec:ML}. Section~\ref{sec:otherexps} contains examples of kinematic variables which are experiment-specific, while Section~\ref{sec;conclusion} is reserved for conclusions and outlook. Appendix~\ref{sec:tools} provides a guide to some commonly used tools and codes for kinematic variables.

\section{Kinematic variables run-through}
\label{sec:preprocessing}

\subsection{Pre-processing of the input data}
The primary objective of a particle physics experiment is to test a theory model, which is usually encoded in a Lagrangian in the quantum field theory. Then, as depicted in the rightmost side of Figure~\ref{fig:dim_reduction}, one can use this Lagrangian to predict the kinematic distributions of relevant quantities of interest. This is a relatively straightforward procedure, which takes advantage of established theoretical tools like the perturbative expansion in the quantum field theory. However, this can only be done at the parton level, i.e., in terms of the fundamental particles represented by the fields appearing in the Lagrangian. Therefore, a necessary step in any analysis is to measure the four-momenta of the fundamental particles emerging from the hard collision.

For leptons like electrons and muons, this is relatively easy, since the object measured in the detector represents the fundamental particle itself. For tau-leptons, the situation is a bit more complicated, since taus can only be identified through their hadronic decays, in which there is a neutrino gone missing. However, the biggest challenge is presented by colored partons (quarks and gluons), which are observed as streams of hadrons called jets. The parton showering and hadronization processes are described by taking limits of perturbative QCD and by phenomenological models implemented in the general purpose event generators. Jet reconstruction algorithms are then needed to cluster the particles observed in the detector into individual jets, and thus, to obtain the jet four-momenta, which can be related to the four-momenta of the underlying partons. At the end of the day, as a result of the so-described ``object reconstruction'' procedure (see Figure~\ref{fig:dim_reduction}), one ends up with a set of four-momenta for the relevant fundamental particles in the event.\footnote{Recently there have been suggestions to bypass the ``object reconstruction" stage altogether and directly leverage low-level data. Examples include end-to-end analyses \cite{Andrews:2018nwy,Andrews:2019faz}, the use of jet images \cite{Cogan:2014oua,Kagan:2020yrm}, etc.} These four-momenta serve as the basis for constructing the kinematic variables discussed in this review. Each kinematic variable is a certain mapping $f$ from the measured four-momenta to a single, typically scalar, quantity. However, there are some practical challenges in defining the proper mapping $f$, as discussed next. 
\subsection{Constructing kinematic variables and associated challenges}
\label{sec:challenges}

In the construction of any derived kinematic quantity, in general, one may encounter a number of practical problems which we briefly discuss below.

\smallskip

\noindent {\bf Particle ID and reconstruction.} Object reconstruction involves a set of criteria applied on the low-level data, for example, the presence or absence of a track, the ratio of the energy deposits in the electromagnetic and hadronic calorimeter, isolation requirements, etc. In principle, particle identification and reconstruction are never perfect --- sometimes the ``wrong" types of particles may pass the requirements, leading to fake leptons, fake photons, and so on.
This is a potential problem in the construction of exclusive kinematic variables, which assume a certain event topology and therefore are defined in terms of the momenta of the correspondingly identified objects.

\smallskip

\noindent {\bf Combinatorial problem.} It arises whenever the final state contains several reconstructed objects of the same type. The association of reconstructed objects at the detector level to their parton-level counterparts is not unique, and one has to deal with the resulting combinatorial ambiguity. The problem is exacerbated by the fact that several types of partons, namely the light quarks and the gluons, yield jets which appear very similar in the detector and can only be discriminated on a statistical basis \cite{CMS:2013kfa,ATLAS:2014vax,Komiske:2016rsd}.
In most practical applications, the combinatorial problem manifests itself as a partitioning ambiguity whenever we try to select the decay products of a common parent particle. For example, in the case of pair-production of two parent particles, the reconstructed objects need to be separated into two groups, e.g., with the hemisphere method~\cite{Ball:2007zza,Matsumoto:2006ws}. 
References~\cite{Lester:2007fq,Alwall:2009zu} extended this idea to account for jets from initial state radiation, which are considered as a separate category. Other techniques to mitigate the combinatorial problem include event-mixing \cite{Albrow:1976jm}, mixed event subtraction \cite{Hinchliffe:1996iu,Agashe:2015wwa}, the use of ranked variables \cite{Kim:2015bnd},  recursive Jigsaw reconstruction method~\cite{Jackson:2017gcy}, etc. 
Since different partitions of the final state objects typically result in different values for the kinematic variables, one could use this to select the correct partition. Specific applications of this idea to the dilepton $t\bar{t}$ event topology, using the $M_{T2}$ and the constrained $M_2$ variables were considered in Refs.~\cite{Baringer:2011nh} and~\cite{Debnath:2017ktz}, respectively.

\smallskip

\noindent {\bf Imperfect detectors.} The observed experimental objects and their kinematics can be different from the actual event due to imperfect detectors. Similarly, the observed objects can differ from the simulated Monte Carlo truth, which necessitates the detector simulation stage in the Monte Carlo chain depicted in Figure~\ref{fig:dim_reduction}. On the one hand, the measured energies, momenta, and timing are in general smeared from their parton-level values. While this is not necessarily a roadblock for the calculation of the kinematic variables per se, it should be kept in mind when interpreting the results. The more serious problem, already mentioned above, is the misidentification of particles  --- e.g., imperfect $b$-tagging would reintroduce the combinatorial problem of selecting the correct $b$-jet among the many jet candidates in an event. Finally, an important variable, used either by itself, or in the construction of many kinematic variables, is the missing transverse momentum, which is defined as the transverse recoil against all visible objects in the event, and is therefore susceptible to their mismeasurement.

\smallskip

\noindent {\bf Unknown new physics parameters.} For new physics signals, one does not know a priori the values of the new model parameters, e.g., the mass of a dark-matter candidate. In such cases, the definitions of the kinematic variables often involve a test value for the corresponding parameter, which needs to be chosen judiciously. In what follows, we shall use a tilde to denote such trial parameter values, e.g., $\tilde m$ for the mass of a dark-matter candidate.

\smallskip

\noindent {\bf Multiple solutions.} Whenever the kinematic variables are defined as solutions to non-linear constraints, there may appear multiple solutions (e.g., the $z$-momentum of the neutrino in a semi-leptonic $t\bar{t}$ event), and one has to design a suitable procedure to arrive at a unique answer.

\smallskip

Depending on the case at hand, there are different approaches to tackling these problems, some of which will be illustrated in the subsequent sections. 

\subsection{Typical uses and applications}

In analogy to the ``no free lunch theorem" in machine learning, no single kinematic variable is optimal for all conceivable tasks in particle phenomenology. Even if we fix the task, the optimal variable can change with time, e.g., depending on the running conditions of the experiments or the evolution in our theoretical understanding of the background processes. This is why a great number of kinematic variables have been considered in the recent literature, with a wide range of applications, for example:
\begin{itemize}
    \item Cleverly chosen kinematic variables are often used for signal versus background discrimination in new physics searches. The choice of variable(s) is tied up to the hypothesized event topology (typically in the form of a simplified model). The ideal variable would capture the salient features of the process at hand and would not be too sensitive to the full details of the underlying new physics model.
    \item Kinematic variables are key inputs to modern multivariate analyses, including machine learning approaches (see Section~\ref{sec:ML} below).
    \item Known kinematic variables can be used to define new, higher-level, kinematic variables, e.g., by using the existing correlations between different variables~\cite{Allanach:2000kt}, or by incorporating them into the algorithmic definition~\cite{Lester:1999tx}.  
    \item Kinematic variables can be used to identify events with special kinematics. For example, one can cut on the value of the $M_{T2}$ variable to select a sample of events in which the true momenta of the invisible particles are fully determined~\cite{Cho:2008tj,Kim:2017awi}.
    \item Certain kinematic variables can be used to test and validate the results from alternative machine learning approaches \cite{Kim:2021pcz}.
    \item The distributions of some kinematic variables exhibit features (bumps, edges, kinks, cusps, etc.) and/or shapes which can be directly correlated with fundamental parameters in the theory Lagrangian~\cite{Hinchliffe:1996iu,Cho:2007qv,Burns:2008cp,Han:2009ss}.
\end{itemize}

\section{Conventions and notation}
\label{sec:notation}

\begin{figure}[t]
\begin{center}
\includegraphics[width=.45\textwidth]{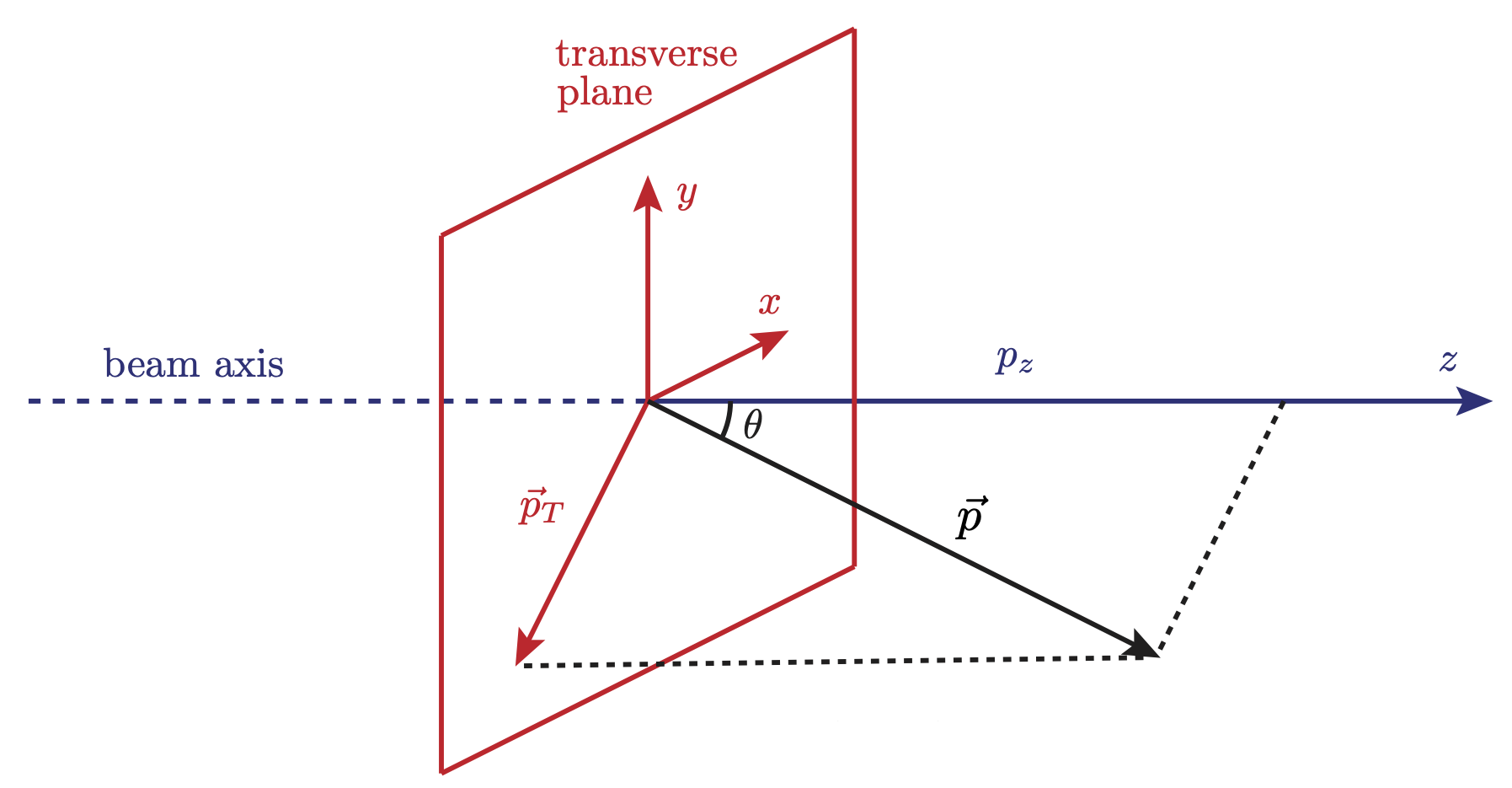}
\end{center}
\caption{The standard geometry of a collider experiment. The $z$ axis (in blue) is oriented along the beam, while the $x$ and $y$ axes (in red) define the transverse plane.  Any 3-dimensional vector~$\vec P$ can be uniquely decomposed into a longitudinal component $p_z$ and a transverse component~$\vec p_T$. Taken from Ref. \cite{Barr:2011xt}.  
}
\label{fig:convention}
\end{figure}

Collider experiments usually employ a Cartesian coordinate system in which the $z$-axis is aligned with the beam direction, while the $x$ and $y$ axes define the transverse plane orthogonal to the beam (see Figure~\ref{fig:convention}). For example, in this system a particle's three-momentum $\vec{p}\equiv (p_x,p_y,p_z)$ is decomposed into a longitudinal component $p_z$ along the $z$ axis and a transverse component $\vec{p}_T\equiv (p_x,p_y)$ within the transverse plane. As shown in Figure~\ref{fig:convention}, some of these Cartesian components can be traded for the magnitude of the transverse momentum
\beq
p_T \equiv \sqrt{p_x^2+p_y^2},
\eeq
the azimuthal angle $\varphi$ defined as
\beq
\varphi \equiv \tan^{-1} \left ( \frac{p_y}{p_x} \right ) \in [0, 2\pi),
\eeq
and/or the polar angle $\theta$
\beq
\theta \equiv \tan^{-1} \left( \frac{p_T}{p_z}\right) \in [0, \pi].
\eeq

The energy $E$ and three-momentum $\vec{p}$ of a particle form a four-vector $p^\mu=(E,\vec{p}\, )$ whose 1+3 dimensional components will be denoted with mid-alphabet Greek indices. The invariant mass $m$ is then defined as 
\beq
m\equiv \sqrt{p^\mu p_\mu} = \sqrt{E^2-\vec{p\,}^2}.
\label{eq:3dmassdef}
\eeq
By analogy, the transverse energy $E_T\equiv \sqrt{m^2+p_T^2}$ and the transverse momentum $\vec{p}_T$ form a 1+2 dimensional vector $p^\alpha =(E_T,\vec{p}_T)$, whose components will be denoted with Greek letters from the beginning of the alphabet. 

The energy $E$ and the longitudinal momentum component $p_z$ can be used to define the rapidity 

\beq
y \equiv \frac{1}{2} \ln \left (\frac{E + p_z}{E - p_z} \right ),
\label{eq:rapidity}
\eeq
which in the case of massless particles reduces to the pseudo-rapidity
\beq
\eta \equiv  - \ln \left [ \tan \left (  \frac{\theta}{2} \right ) \right ] .
\label{eq:pseudorapidity}
\eeq
In what follows, we shall use the letter $p$ to denote momenta of visible particles seen in the detector, while the letter $q$ will be reserved for the momenta of invisible particles (dark-matter candidates, neutrinos, or other very long-lived weakly interacting particles).

In a collider experiment, the transverse momentum of the initial state is zero, which places a constraint on the final state transverse momenta:
\beq
\sum_a \vec{p}_{aT} + \sum_b \vec{q}_{bT} = 0.
\eeq
The measured total missing transverse momentum $\mptvec$ is therefore given by
\beq
\mptvec \equiv \sum_b \vec{q}_{bT} = - \sum_a \vec{p}_{aT}. 
\eeq
See also Section~\ref{sec:misse} for more detailed discussion.

At lepton colliders, the center-of-mass energy $\sqrt{s}$ is fixed and the longitudinal momentum of the initial state is also fixed (often zero), while at hadron colliders, the parton-level center-of-mass energy $\sqrt{\hat s}$ varies from one event to the next, and the longitudinal momentum of the initial state is a priori unknown. This motivates the use of kinematic variables like (\ref{eq:rapidity}) and (\ref{eq:pseudorapidity}) which have convenient transformation properties under longitudinal Lorentz boosts (along the $z$-axis). 

\begin{figure}[t]
\begin{center}
\includegraphics[width=.45\textwidth]{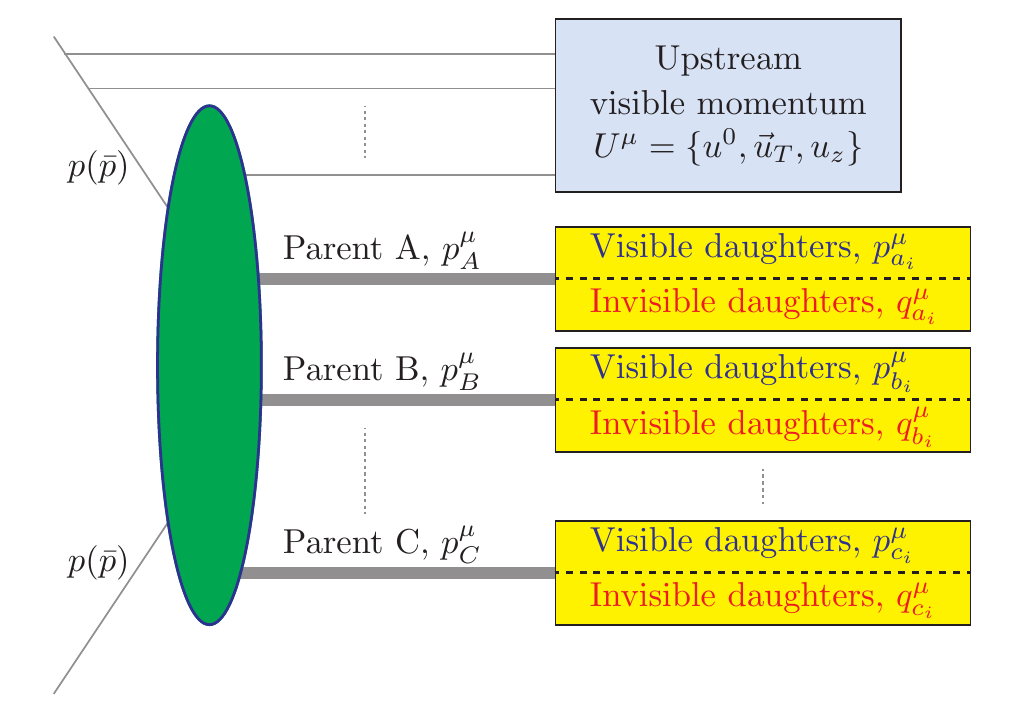}
\end{center}
\caption{The generic event topology of a collider event. The result of the initial collision (for definiteness, we show the case of hadron colliders where the beams consist of protons $p$ or antiprotons $\bar p$) is a set of final state particles which can be grouped into collections $A,B,C,\cdots$, each containing sets of visible and invisible daughters. Adapted from Ref.~\cite{Barr:2011xt}.  \label{fig:convention2}}
\end{figure}

Individual final state particles will be labelled with $a,b,c,\cdots$. Collections of such particles (which, for example, are hypothesized to have a common origin) are labelled with $A,B,C,\cdots$ (see Figure~\ref{fig:convention2}). For example, let $A=\{a_1,a_2,a_3,\cdots\}$ be a collection of final state particles. The four-momentum of the whole collection will be $p_{A}^\mu$, while the four-momenta of the individual particles will be denoted with $p^\mu_{a_i}$ or $q^\mu_{a_i}$, respectively, depending on whether the particle is visible or invisible. We shall use lowercase $m$ for the masses of final state particles and uppercase $M$ for kinematic mass variables, which typically are related to the masses of parent collections $A, B, C, \cdots$. Taking jets and leptons to be massless ($m=0$) is usually a good approximation, but for $W$, $Z$, $t$, and dark-matter candidates, we shall keep the explicit dependence on $m$. In the case of invisible final state particles, as mentioned earlier, it is often useful to treat their mass as a test parameter (denoted with a tilde, $\tilde m_\chi$), regardless of whether the true mass is known or not.

As shown in Figure~\ref{fig:convention2}, we shall use $U$ to denote the collection of particles which are not assigned to any other groups. In practice, those arise from initial state radiation (ISR) or from decays upstream.\footnote{As explained later in Section~\ref{sec:mt2variables}, ungrouped particles downstream can be effectively eliminated from the discussion by introducing the intermediate resonances as effective invisible particles.}

\section{Standard kinematic information \label{sec:standard}}

\subsection{Simple kinematic observables}
The standard kinematic information is what is directly measured by detectors for individual particles reconstructed by Particle Flow (see the left side of Figure~\ref{fig:dim_reduction}).
The four-momenta of the particles can be represented in Cartesian coordinates $(E, p_x,p_y,p_z)$, or, more commonly, in cylindrical coordinates $(E, p_T, \varphi, p_z)$ where $p_z$ can also be traded for the pseudo-rapidity which preserves relative distance under longitudinal boosts.
This kinematic information can then be compared to the expected distributions from a given theory model, which are typically done at the parton level (see the right side of  Figure~\ref{fig:dim_reduction}). Therefore, one would like to match the particles observed in the detector to the fundamental (parton-level) particles in the SM:
\begin{itemize}
\item For non-hadronic particles which are stable on the detector scale (electrons, muons, and photons), this correspondence is direct. 
\item The neutrinos, on the other hand, are invisible and not reconstructed individually. Nevertheless, the sum of their momenta can be inferred from the imbalance between the four-momentum of the initial and final state in the event (see Section~\ref{sec:misse}).
\item The case of hadronic particles is much more complicated, due to confinement --- the quarks and gluons at the parton level appear as collections of hadrons (``jets''), which necessitates jet reconstruction algorithms to recover the parton level information. The situation is even more complicated due to the presence of initial and final state radiation, which results in additional jets which further muddle up the picture. In a typical jet reconstruction algorithm, particles are grouped based on their relative distance in some suitably chosen metric, e.g.,  $\Delta R = \sqrt{\Delta \eta^2 + \Delta \varphi ^2}$ in the $(\eta, \varphi)$ space, where $\Delta \eta$ and $\Delta \varphi$ are the differences in the pseudo-rapidities and the azimuthal angles of the two objects, respectively.
\item The heavy particles in the SM ($W$, $Z$, Higgs, and top) are then reconstructed probabilistically by grouping their decay products as illustrated in Figure~\ref{fig:convention2} and demanding that the invariant mass of the respective collection of decay products is consistent with the mass of the parent particle.
\end{itemize}
Additional variables which could be used to cut on (select events) are the number of reconstructed objects from each type: $N_j/N_b/N_\ell/N_e/N_\mu/N_\tau/N_\gamma$. 

In a traditional cut-and-count analysis, one would i) narrow down the number of variables to consider (dimensionality reduction); ii) place cuts on them to define a signal region, and iii) perform a counting experiment in the signal region. This dimensionality reduction, however, necessarily leads to some information loss. The goal of the experimenter is to utilize kinematic variables which minimize the information loss. In practice, the following two approaches (or a combination thereof) have been used:

\begin{itemize}
\item Make direct use of some of the simple kinematic variables described above, e.g., $p_T$, pseudo-rapidity, $\Delta R$, invariant mass of a collection of particles, number of reconstructed objects of a given type, etc. One could even imagine using the full kinematic information from the event as an input to a Machine Learning (ML) algorithm like a Neural Network (NN) classifier (see Section~\ref{sec:ML}).
\item Perform the dimensionality reduction in a more optimal way, by forming suitable high-level kinematic variables, which are functions of the simple observables, and retain as much of the relevant information as possible. The main purpose of the present article is to review precisely these types of observables.
\end{itemize}

The interplay between those two approaches illustrates the tension between optimality and generalizability. The simple kinematic variables are robust and universally applicable (model-independent), but are not as sensitive. The high-level variables bring about higher sensitivity and physics performance, but are not easily generalizable to other signal processes. With either of the two approaches, one must connect the kinematic measurements to the parton-level kinematics. This ``unfolding" needs to overcome the two classes of challenges discussed in the next two subsections.

\subsection{Experimental uncertainties} 
Realistic measurements of kinematic variables are affected by various experimental uncertainties. First and foremost, the low-level measurements are subject to intrinsic uncertainties, e.g., missing tracker hits, calorimeter activity below the detectable threshold, and instrumental noise, etc.
In addition, when high-level objects are reconstructed (e.g., a jet of particles), the measurements are further affected by uncertainties arising from the definition of the high-level object. 

The energy resolution $\Delta E$ of a calorimeter is typically parametrized by a noise ($N$), a stochastic ($S$), and a constant ($C$) terms
\begin{align}
\frac{\Delta E}{E} =  \sqrt{
\bigg( \frac{N}{E} \bigg)^2 +\bigg( \frac{S}{\sqrt{E}}\bigg)^2  +C^2~
} \, ,
\end{align}
where the constants $N$, $S$, and $C$ are specific to a given experiment and calorimeter type \cite{Han:2005mu,deFavereau:2013fsa}. The momentum resolution $\Delta p_T$ based on a curvature measurement can be generically expressed as \cite{Han:2005mu,deFavereau:2013fsa}
\begin{equation}
    \frac{\Delta p_T}{p_T} = a \, p_T \oplus b \,,
\end{equation}
where $a$ and $b$ are resolution parameters specific to the detector of interest. 

The experimental environment brings additional challenges in the measurements of kinematic quantities. For example, when the average number of interactions per bunch crossing significantly exceeds 1, a number of soft (minimum bias) events accompany the hard scattering event, confusing its interpretation and biasing the kinematic measurements. 
Such pileup effects may be mitigated by installing new precision timing detectors~\cite{CERN-LHCC-2017-027} or by analysis techniques using substructure~\cite{Kogler:2018hem,Soyez:2018opl,ATLAS:2012am,Bertolini:2014bba} or machine learning~\cite{Komiske:2017ubm,ArjonaMartinez:2018eah}. 

These effects can be controlled and improved by exploiting the data itself. 
Extensive review of the progress in understanding the experimental systematics is beyond the scope of this review. 
For our purposes, the effect of the detector resolution is to smear the sharp kinematic features which are expected in the ideal case with perfect resolution. For example, the extraction of kinematic endpoints will have to be done by modelling the shape of the distribution in the vicinity of the endpoint, taking into account the detector resolution.
This highlights the importance of designing the right kinematic variables which are as robust as possible to all these experimental effects.   

\subsection{Theoretical uncertainties}

Establishing the usefulness of a kinematic variable, e.g., in measuring a parameter of the fundamental Lagrangian, requires extensive calculations of i) the theoretical predictions for the observable under study; ii) the sensitivity of the designed kinematic variable to the quantity that we want to extract; iii) a number of auxiliary quantities that need to be controlled in experiments. 
It is crucial to control all the details, and especially approximations, which characterize these theoretical computations. 

\smallskip

\noindent {\bf Higher orders in perturbation theory.}
The vast majority of calculations, especially automated ones, are done at a fixed-order in perturbation theory. The first category of uncertainties arises due to missing next-to-leading order (NLO) contributions. Corrections of this sort can arise from QCD or electroweak interactions or both. The impact of missing higher orders is typically evaluated by variations of scales and other possible unphysical parameters that are introduced just for computational purposes and should have zero impact for an all-order calculation.  

\smallskip 

\noindent {\bf Fragmentation and hadronization modelling.} Theoretical computations in the perturbation theory are done at the parton level and describe processes limited to a small total number of particles. Often this number is completely fixed, or it can be fuzzily defined if the calculation is carried out beyond the leading order (LO) in the perturbation theory and virtual and real corrections are included.
The hadronization of colored partons is described by phenomenological models, which introduce another category of theoretical uncertainties. They can be estimated by i) comparing the results from different event generators, ii) varying the underlying model parameters within acceptable ranges, etc.

\smallskip

\noindent {\bf Parton distribution functions.} At hadron colliders, parton-level calculations need to be convoluted with the parton distribution functions (PDFs) which contain a lot of uncertain parameters. In order to propagate the PDF uncertainty to some kinematic variable, the latter must be evaluated for each member of the PDF set~\cite{Butterworth:2015oua}.

\smallskip

\noindent {\bf Narrow-width approximation.}
Another commonly used approximation relies on the fact that in the limit of a narrow particle width, the Breit-Wigner distribution approaches a Dirac $\delta$ function. This narrow-width approximation simplifies the treatment of multi-particle final states by iteratively factorizing the computation into the production of parent particles and their subsequent decay. In this approximation the parent particles are exactly on their mass shell and their quantum numbers, including polarization, are in well defined quantum mechanical pure states. In reality, due to the unstable nature of the parent particles, their momenta should be smeared over a region close to their mass shell and furthermore, their polarization should be treated as a density matrix with fully quantum mechanical interference properly taken into account. Depending on the kinematic variable under consideration, the offshellness or polarization effects may play important roles.

\smallskip

\noindent {\bf Finite Monte Carlo statistics.}
Yet another source of theoretical uncertainty is due to the finiteness of the simulated Monte Carlo (MC) samples for the relevant theory models under consideration. It is important to keep this MC-statistical uncertainty under control, so that it does not bottleneck the overall sensitivity of the experiment. Fortunately, the MC-statistical uncertainty can be made arbitrarily small by increasing the simulation statistics. However, there are often limitations on the amount of computational resources available, which in turn limits the number of events that can be produced. In this context, keeping in mind the increased computational demands at future colliders, it is important to i) speed up the MC production pipelines and ii) achieve a better bang for the buck in terms of sensitivity reached per event simulated. The latter can be accomplished by preferentially producing events with high utility to the experiment (with appropriate biasing techniques) \cite{Valassi:2020ueh,Matchev:2020jqz}.

\section{Real world examples: top and \texorpdfstring{$W$}{W} physics}
In this section we shall review the historical developments with pointers to the subsequent sections where the kinematic variables are introduced and explained in more detail.

Given the above described uncertainties from experiment and from theory, the design of suitable kinematic variables is a key part of the extraction of useful information out of experiments. Kinematic variables need to be designed taking into account the specific goals of each experiment with the aim to exploit the strengths and to minimize the weaknesses of the data obtained in experiments. A representative situation may be find in measurements of SM properties concerning the top quark and the $W$ boson. These SM particles are relatively well known and it is well established that they decay in several possible decay modes. Some of these include hadronic jets and enjoy the largest rate, but also suffer larger experimental uncertainties due to mismeasurement of jet properties. Alternative modes not containing jets contain well measurable charged leptons, but at the same time are less copious. Decay modes with charged leptons also bring along neutrinos, which cannot be measured in high energy experiments at colliders and make complicated any global reconstruction of the kinematics of a single event.  All these considerations must be weighed carefully in the design of a kinematic observables. The optimal solution is typically ``evolving with time'', as the experiments accumulate more data and they acquire more control on the instrumentation thanks to the gained experience.

It is remarkable to look for instance at the evolution of the top quark mass measurement, which could be measured with very little data, one putative $t\bar{t}$ event at Run I Tevatron~\cite{PhysRevD.45.1531}, using very heavily the properties of the top quark predicted in the SM. Modern measurements instead tend to rely as little as possible on theoretical input, aiming at finding measurement that can withstand heavy degree of ``modeling uncertainty''. The difference has to do with the accuracy sought in these measurements. Early measurements aimed at extracting the most information of the data and, given the relatively low precision of the measurement, could safely ignore a large number of issues which are instead very important today for precision measurement. Indeed, measurement from LHC aim at uncertainties of the order of $\Lambda_{QCD}$, where a number of theoretical issue emerge most prominently. Measurements that will be carried out at the HL-LHC will face similar theoretical issues, but the detectors will be improved compared to the current ones, hence different kinematic variables will be the best suited for the job. 

The importance of theory aspects in modern measurements about the $W$ boson and the top quark make them an ideal place to illustrate the design of kinematic variables. In addition, the top quark and $W$ boson provide useful laboratories to test  various ideas motivated by beyond-the-Standard-Model (BSM) searches and possible measurements of new physics states. All the more reasons for which we will use the top quark and the $W$ boson  to illustrate many of the methods which we will discuss. 

Furthermore, as new physics has not been found at the LHC so far, concrete experience has been accumulated in Run1 and Run2 only about the measurement of SM particles masses. In this context the measurement of the masses of $W$ boson and top quark masses have played both the role of a playground for new ideas to be used in future new physics measurements and a test-bench in which sharpen our more traditional variables and better understand the delicate theory aspects that enter in their measurements. 

Given the strong motivations for the precision measurement of $m_t$ and $m_W$ a huge theoretical and experimental effort has taken place in recent years to put these measurements under control. In fact, current and future accumulated data at the LHC {\it in principle} allows to extract these masses at a extraordinary precision level~\cite{CMS:2013wfa}, but we are currently unable to exploit this huge data set because of systematic uncertainties in measurements and theoretical uncertainties in the computations needed to even define properly the observables used to extract  $m_t$ and $m_W$. The target for these measurement is to attain a measurement at the $10^{-3}$ level and, most importantly, to obtain such level of accuracy through observables and kinematic quantities that are as robust as possible to possible mismodeling of detector effects, not sufficiently accurate theoretical calculations, and other sources of systematic errors.

In the following we will briefly review the challenges posed in the measurements of $m_t$ and $m_W$. We will also review the kinematical quantities proposed to extract these masses in a most reliable and precise way. In Section~\ref{mtsection} we deal with $m_t$~\cite{Frixione:2014ala,FerrarioRavasio:2019vmq,FerrarioRavasio:2019glk,Corcella:2017rpt,ATLAS:2019ezb,Aaltonen:2009zi}, in Section~\ref{mwsection} we deal with $m_W$. A discussion on the mass of the Higgs boson is deferred to a later  Section~\ref{sec:mgammagamma}, as the measurement itself is rather straightforward in the 4 lepton and two photon channels, but it requires some care in dealing with interference effects specific to that case.

\subsection{\texorpdfstring{$m_t$}{mt} \label{mtsection}}

\begin{table}[t]
    \centering
    \begin{tabular}{c|c|c}
    \hline \hline
    Channel & Kinematic variables & References\\
    \hline
    $\ell\ell\oplus\ell j$   &  $L_{x,y}$ & \cite{Khachatryan:2016wqo,Hill:2005zy,CMS-PAS-TOP-12-030} \\  $e\mu$ & ``basic'' & \cite{Frixione:2014ala,Czakon:2020qbd,ATLAS:2017kdr} \\ 
       $e\mu$ & $m_{b,\ell}$  & \cite{CMS-PAS-TOP-14-014} \\
      -  & $m_{J/\psi\ell}$ &   \cite{Kharchilava:1999yj,Czakon:2021ohs} \\
     $\ell j$  &  $s_{ttj}$ & \cite{Alioli:2013mxa,Aad:2019mkw} \\
      $e\mu$    & $E_b$ &  \cite{Agashe:2012bn,CMS-PAS-TOP-15-002} \\
    \hline \hline
    \end{tabular}
    \caption{A summary of top quark mass measurements using various kinematic variables.    \label{tab:summary_topmass}}

\end{table}
The measurement of the top quark mass is the subject of several experimental works at the LHC and at the TeVatron. Specialized reviews exist \cite{Corcella:2019tgt,Hoang:2020iah}, so we will not try not to be comprehensive, but rather we want to highlight the diversity of efforts put in place to attack this difficult problem with several possible complementary strategies. 

Methods used for $ t\bar{t}$ pair production at LHC or TeVatron experiments are reported in Table~\ref{tab:summary_topmass} together with the decay  in which they are used.  

The simplest idea, from a point of view of the kinematic variable used, is the measurement of the invariant mass of all the decay products of the top quark. This method is conceptually quite simple and lies at the heart of the most precise results currently available.  Still, it suffers several effects that are particularly difficult to estimate and that we review in the following. 

First of all for a full reconstruction of the top decay products the most straightforwards channel would be fully hadronic, so that all the hard decay products of the top quark can be measured, or, in other words, there are no invisible particles such as neutrinos stemming directly from the decay and carrying away unknown amounts of energy. The measurement of hadrons, unfortunately, is quite imprecise.  In fact hadrons are usually dealt with in jets, which offer the possibility to relate the hadron-level measurement to perturbative QCD calculations with few particles. In these measurements one faces the difficulty to  track a large number of particles, some of which are not even energetic enough to be recorded by the detectors. In additions there is an inherent problem in the matching of neutral and charged hadrons, that are measured by different sub-detectors. All in all, hadronic measurements have serious problem to be very precise (see Ref.~\cite{1702.07546v1} for a recent result) and the best measurements of the top quark mass are presently obtained from the semi-leptonic channel. In this channel it is possible to be somewhat less sensitive to the imprecise measurement of  jets, as one can attempt to measure the invariant mass of the leptonically decaying top quark. Here the challenge lies in indirectly reconstructing the momentum of the neutrino from momentum conservation. For this reconstruction it is necessary to use all the other measured particles in the event, including hadronic jets, hence there is still a dependence on the quality of the measurement of hadronic jets which poses a challenge. Methods to ameliorate the knowledge of jets  in this measurement try to use the knowledge of the $W$ boson mass to put constraints on the jet reconstruction~\cite{CMS-PAS-TOP-14-001,Chatrchyan:2012kl} as to obtain a measurement of the top quark mass together with a dedicated calibration of the jet energy. 

In addition to these experimental issues, the definition of the top quark mass as the peak of an invariant mass has proven to be difficult to interpret on theoretical grounds~\cite{Hoang:2014la,2004.12915v1,Nason:2017aa,Beneke:2016lr}. In fact the top quark, being colored, cannot exist as a long-distance object. It has to turn into a color singlet object either forming hadrons of its own flavor or thanks to the hadronization of its decay products. The theoretical definition of a mass for the top quark that can be used beyond the LO of perturbation theory, a very necessary requirement when we aim for 1 GeV or less uncertainty for this measurement, has required quite a review of the whole strategy to measure this quantity. Indeed the extraction of the top quark mass from templates of theoretical predictions based on detailed event simulation from fixed (often leading) order approximations, possibly supplemented with leading logarithm parton showers, is questioned when precisions around 1~GeV are claimed. 
Efforts are in place to obtain more precise theoretical template for this type of method, see e.g \cite{Ferrario-Ravasio:2019ab,Ferrario-Ravasio:2019aa,Ferrario-Ravasio:2018aa,Jezo:2016oj}. In any case, being these calculations at the edge of what is currently computable,   there is much need to validate any of the measurements that they will enable. 

For this validation it is key to find new independent methods, which may suffer less the  theory uncertainty in the definition of the top quark mass itself, as well as suffer different kind of  experimental uncertainty. This need has spurred a large activity in the proposal of new mass measurements for the top quark mass. 

One method proposed to measure the top quark mass has to do with a strict inequality for the invariant mass of a sub-system of the decay products, and in particular, considering the bottom jet and the charged lepton from the top decay one can exploit the relation
\begin{equation} 
\label{eq:bl-endpoint} m_{b\ell}\leq m_{t}\,.
\end{equation}
The measurement of the end-point, or the shape around the end-point, of the bottom-lepton invariant mass distribution has lead to new determinations of the top quark mass~\cite{CMS-PAS-TOP-11-027}, which probe the uncertainty due to jet energy measurements in a different way than other methods, as the jets involved are mostly $b$-jets. In addition this method has a sensitivity to off-shell effects as the relation eq.(\ref{eq:bl-endpoint}) assumes perfectly on-shell top quarks. Therefore this method can be used as diagnostic for the importance of off-shell effects in the measurement. 

As leptons from the top quark decay are arising from a color-singlet $W$ boson, it has been proposed to use kinematic variables based solely on leptons to measure the top quark mass~\cite{Frixione:2014jk}. The proposed kinematic variables, e.g., 
\beq
m_{\ell^{+}\ell^{-}}=\sqrt{(p_{\ell^+}+p_{\ell^-})^2}\,,
\eeq
the invariant mass of two leptons from the fully leptonic top decay, being based on inclusive definitions top-like events with leptons, have the merit to not require any top quark explicit reconstruction, hence potentially freeing the measurement from the burden of specifying ``what is a top quark''. This potentially alleviates the issues from the definition of the top quark mass as it essentially treats the top quark mass as a parameter of the Lagrangian, i.e., a couplings, which impacts measurable kinematic observables. Another important aspect of these inclusive leptonic measurements has to do with QCD effects. In fact the importance of QCD corrections and hadronization physics in these leptonic observables is expected to be reduced as they only feel jets physics from recoiling against hadrons and other similar inevitable interrelations from particles belonging to the same process. At the same time, the leptons being daughter particles from the $W$ boson, do not directly feel the top quark mass. This reduction of sensitivity to both the interesting parameters (the top quark mass) and the uncertainties that plague other methods, require a quantitative evaluation of the concrete merit of these observables.  Concrete studies~\cite{Frixione:2014jk} revealed that a detailed theoretical description of the hard-scattering and of the parton shower is needed to obtain reliable measurement at the GeV uncertainty scale. 

Based on purely leptonic measurements it as been proposed to correlate the top quark mass to a suitably defined integral of the energy distribution of leptons~\cite{Kawabataa:2014osa}. The quantity of interest is an integral of the energy distribution  times a special weight function $w$, which is derived from kinematic properties of the top quark decay in perturbation theory
\beq
I(m)=\int dE_{\ell} \frac{d\Gamma (m)}{dE_{\ell}} w(E_{\ell})\,,
\eeq
such that when the integral is computed for the value $\hat{m}$ realized in data one expects $I(m=\hat{m})=0$.

The idea of using only leptons to construct an observable sensitive to the top quark mass has also been explored in the context of pairs of leptons arising from the same top quark, e.g., one from the leptonic decay of the $W$ boson and the muon originating from the semi-leptonic decay of the $b$-quark-initiated hadrons.  This type of measurement~\cite{Aaltonen:2009zl,ATLAS:2019ezb} is called ``soft-leptons'' as it uses a (non-prompt, soft) muon from $B$-hadron decays that appear in the top decay, which are softer than those from $W$ boson decays. The computation of templates for 
\beq
m_{\ell\mu}=\sqrt{\left(p_\ell +p_{\mu}\right)^2} \,
\eeq
 relies  on the  hadronic physics of $B$ hadrons and their semi-leptonic decays, thus this method is important as it exposes hadronization effects. Variations of this idea are considered: for example, it has been proposed to use 
 \beq
 m_{3\ell}=\sqrt{\left(\sum p_\ell  \right)^2}
 \eeq 
 formed by three leptons  from the same top quark, following  an early proposal to use rare $B\to J/\psi+X$ decays, which can be tagged \cite{Kharchilava:2000yk} in clean leptonic modes of the $J/\psi$. 

Kinematic variables have been studied~\cite{Corcella:2017aa} with the goal of testing  our understanding of hadronization in top quark events, as to aid sharpening results from the method of soft-leptons mentioned in the previous paragraph, as well as other methods based on hadrons, which we will discuss below. The exploration of  Ref.~\cite{Corcella:2017aa} reveal that a thorough understanding of QCD up to minute effects in the description of radiation and hadronization is in general necessary to warrant sub-GeV precision in the top quark mass extraction. Keeping in mind this ambitious goal for hadron-based measurements, traditional variables are considered in experiments to calibrate the tiny but relevant QCD effects that one faces at sub-GeV precision, see e.g., \cite{ATLAS-CONF-2020-050,2108.11650v1,ATLAS-Collaboration:2015ng}. In additions, new kinematic variables have been proposed in Ref.~\cite{Corcella:2017aa} to provide a calibration on data of these minute QCD effects and put these effects under control using data.  

A top quark mass measurement has also been proposed using only the measured energy of $b$ quarks (see also Section~\ref{sec:energy}). This method is purely based on the energy spectrum of the $b$-jet.
Similarly to some of the proposals based on leptons, this method does not require any definition of reconstructed top quark. In addition, the position of the peak of the distribution is predicted to be insensitive to the production mechanism of the top quarks as long as the sample of measured $b$ jets arises from an equal mixture of left-handed and right-handed top quarks (i.e., unpolarized)~\cite{Agashe:2013sw}. The observable is simple enough that it can reliably be computed in perturbation theory, so far up to NLO in QCD both at the jet level and at the hadronic level~\cite{Agashe:2016xq}. Uncertainties from jet energy measurements and hadronization uncertainties are the most important ones in application of this method at the jet level and hadron level method, respectively. 

As there is a certain abundance of methods based on jets or hadron energies, alternative methods have been proposed, as they may help to get a truly independent determination of the top quark mass. One proposal that goes away from energetics has been put forward in Ref.~\cite{Hill:2005dq}. The idea is to measure $B$ hadron flight lengths in the detector, relying on the fact that the hadron decay is controlled by its proper lifetime and its boost, the latter being larger when the the $B$ arises as a decay product of a heavier particle. From the experimental point of view this method has the advantage to use length measurements, that are very precise, thanks to tracking, and not at all affected by jet energetics, nor the definition of jets. So far this method has been implemented in experiments only measuring the transverse decay length $L_{xy}$ flown in the plane orthogonal to the beam axis
\begin{equation}
L_{xy} = \sqrt{L_x^2+L_y^2} \,.
\end{equation}
The measurement~\cite{CMS-PAS-TOP-12-030} has proven to be quite sensitive to hadronization effects, which is expected as the nature of the $B$ hadrons impacts the measurement via their proper lifetime and boost. A large sensitivity to the top quark production mechanism has also been remarked in this measurement. This is in part expected as a production mechanism characterized by larger top quark boost can mimic larger boost of the $B$ hadron, as the length flown by the $B$ hadron is sensitive to the top quark total energy, without distinction if  it is from mass or from momentum. Nevertheless, reduced dependence of the production mechanism can be gained by a     study more focused on properties that are stable upon changes of the production mechanism, e.g., the peak of the $B$ hadron boost distribution~\cite{peaklength} that is in a one-to-one relation with the $b$ energy peak discussed above.

Other mass measurement methods have to do with threshold effects, which manage to exploit basic kinematic inequalities in the context of $pp$ collisions in which some quantities are not readily accessible or controllable. One key observation is that the production rate of massive particle is very sensitive to the energy that one has at disposal to form these particles, e.g., the formation of a pair of massive particles is very suppressed when the available center-of-mass energy is below twice the mass of the particle. The rate quickly rises once the center of mass energy that goes in the process to produce the pair of particles passes the threshold of twice the mass of heavy particle, and then for much larger center of mass energy compared to the heavy particle mass the cross-section follows usual geometrical scaling. With this idea in mind it has been proposed to study $ttj$ events at the LHC and to use the hardness of the top to control the total invariant mass that enters the actual partonic process giving rise to $gg\to t\bar{t}j$ or $q\bar{q}\to t\bar{t} j$. Exploiting the dependence of the rate on the hardness of the jet, or using a more comprehensive measure of the partonic center of mass energy, such as
\begin{equation}
\sqrt{s_{t\bar{t}j}}=\sqrt{(p_t+p_{\bar{t}}+p_j)^2}\,,
\end{equation}
in Ref.~\cite{Alioli:2013mxa,Aad:2019mkw} a method has been proposed using templates computed in at NLO in perturbation theory, including matching to parton shower. This method, as other that do not require to reconstruct an object called ``top quark'' we saw above, lends itself to an interpretation of the measurement as the dependence of a suitable observables on a Lagrangian parameter, hence it is considered to give theoretically cleaner results compared to invariant mass peak we saw at the beginning of this section.  

Related to the threshold of $t\bar{t}$ production a proposal has been put forwards to identify the  top quark mass from bound state effects in the diphoton mass spectrum~\cite{Kawabata:2016aya}. This approach would benefit of a clean definition of the top quark mass definition in QFT relevant for this phenomenon. 

\begin{figure}[t]
\begin{center}
\includegraphics[width=.45\textwidth]{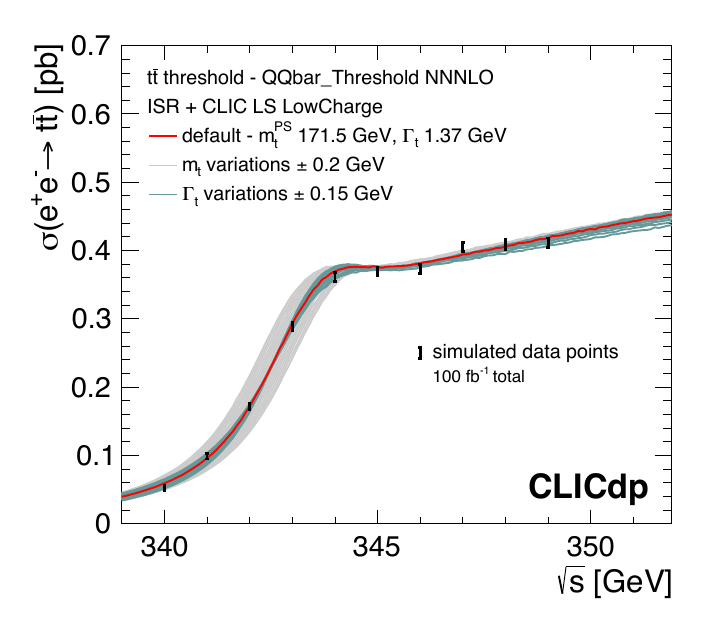}
\end{center}
\caption{Top quark mass from high precision rate calculation at $e^+e^-$ collider \cite{Abramowicz:2018aa}.}
\label{fig:threshold}
\end{figure}

The need for the evaluation of theoretical uncertainties and precision control of detector effects is even more marked in the context of measurements to be carried out at future colliders. That is the case, for instance, of the measurement of the top quark using the dependence of the production cross-section on the center of mass energy~\cite{Seidel:1498599,Seidel:2013sqa,Maier:2019vll,Nowak:2021xmp}, i.e., a fit of the measurements with precise theory predictions upon variations of $m_t$ and other relevant quantities (e.g., $\Gamma_t$ and $m_t$ as shown in Figure~
\ref{fig:threshold}). In this context is it is of utmost importance to compute rates taking into account \cite{Beneke:2016kkb,Beneke:2015kwa,Beneke:2013kia,Beneke:2010mp,Hoang:1999zc,Hoang:1998xf,Strassler:1990nw} bound state dynamics, off-shell effects, non-relativistic corrections, electroweak effects, soft corrections that may need resummation, and transfer factors that account for what fraction of the total cross-section end up in the detector acceptance  \cite{Nejad:2016bci,Bach:2017ggt,Hoang:2013uda}. Especially for the matching between the measured fiducial cross-sections and the theoretically cleaner total ones, it will be key to exploit suitably defined kinematic variables that can serve as diagnostic of the theoretical computations. Furthermore, methods (e.g.~\cite{Boronat:2019cgt}) applicable at center of mass energies slightly larger than the threshold (if attainable by the machine that will perform the threshold scan), will be of key importance to validate the very precise measurement from the threshold scan.

\subsection{\texorpdfstring{$m_W$}{mW}   \label{mwsection}}

\begin{table}
\begin{centering}
\begin{tabular}{c|c}
 \hline \hline 
 & References \tabularnewline

\hline 
$p_{T,\ell}$ & \cite{ATLAS-Collaboration:2017ac}\tabularnewline
$m_{T}$ & \cite{PhysRevLett.50.1738}\tabularnewline
derivatives of the energy distribution & \cite{Bianchini:2019aa}\tabularnewline
singularity variables & \cite{De-Rujula:2011lr}\tabularnewline
\hline \hline 
\end{tabular}
\par\end{centering}
\caption{\label{tab:mw}Summary of methods proposed for the measurement of the $W$ boson
mass.}
\end{table}

The measurement of the $W$ boson mass is a simple example to showcase
the importance of employing smart kinematic variables. This measurement is also a good example of the role of theory in performing precise measurements and scrutinizing possible sources of uncertainties. This measurement has been performed so far at $e^{+}e^{-}$ colliders~\cite{The-ALEPH-Collaboration:2013kq}
and hadron colliders~\cite{1907.02029v1,ATL-PHYS-PROC-2017-051,ATLAS-Collaboration:2017ac,1307.7627v2}.
Future prospects for LHC and circular $e^{+}e^{-}$ colliders are
discussed in \cite{ATL-PHYS-PUB-2018-026,2107.04444v1}. As the $W$ boson mass is one of the possible input parameters to define the SM, this measurement has foundational importance for precision tests of the SM.

The target is to reach a total uncertainty of order 10 MeV, that is about $10^{-4}$ relative accuracy, which would allow to obtain a comparable precision to that of the indirect determination of $m_{W}$ from the SM electroweak fit~\cite{Baak:2014oq}. Given this ambitious target a great part of the discussion on how to measure this mass
has to do with the reduction and modeling of both experimental and
theoretical uncertainties. Kinematic variables have played an important
role in devising measurements robust to these uncertainties and will
continue to provide useful insights to steer the effort towards a
precision measurement of the $W$ boson mass. A summary of the techniques available for this measurement is presented in Table~\ref{tab:mw}.

Presently employed methods use the spectrum of transverse momentum
of the charged leptons (see e.g., \cite{ATLAS-Collaboration:2017ac}) and the transverse mass~\cite{PhysRevLett.50.1738}
in $W$ leptonic decays. Already considering these two quite simple variables it is possible to see how the evolving performances of the experiments and the depth of the theoretical interpretation of the measurement forces a continuous evolution of the kinematic variables best suited for the job. Indeed the transverse mass is an early example of
kinematic variable, which has played a very important role in early
determinations of $m_{W}$ thanks to its robustness against PDF uncertainties
at hadron colliders. In recent years, as the precision target has shifted toward ever smaller uncertainties, the transverse mass hit a bottleneck arising from the necessity of using
missing transverse momentum, hence in modern measurements of $m_{W}$ it needs to be complemented by
other observables. 

At first sight one might think that in a simple process such as $pp\to W\to\ell\nu$ there is very limited
choice for observables alternative to $m_{T}$, hence the job of designing new kinematic variables might be deemed trivial by a too lighthearted judgment. On the contrary quite a number of alternative approaches have been proposed, starting from strategies on how to combine the simplest pieces of information from $m_{T}$ and lepton $p_{T}$ distributions, for example, by using the combined information from these two variables the latest measurements (e.g.~\cite{ATLAS-Collaboration:2017ac}). Putting aside for a moment the great improvement on the determination of the proton PDF, the combination of these two methods has been greatly beneficial.  In fact the  the bottleneck of $m_{T}$ due to invisible moments involved can be surpassed thanks to the extra sensitivity to $m_{W}$ from $p_{T}$ and the PDF sensitivity of $p_{T}$ distributions can be tamed looking at more stable $m_{T}$ features.

In addition to targeted kinematic variables design, a great amount of further theory inputs ameliorated the robustness of this mass measurement in recent years.
In fact, at the precision we aim to carry out the mass measurement, we need
to keep under excellent control not only the effect of PDF uncertainties but also their 
related correlations \cite{1501.05587v2,1104.2056v1,1910.04726v2,1508.06954v2},
as well as high-order QCD and EW corrections (see e.g., \cite{2103.02671v1}
and references therein) which can bias the measurement. 

Beyond the simple variables $p_{T}$ and $m_{T}$ other ideas have been explored in the literature. 
The utility of singularity conditions and singularity variables has
been explored e.g., in Refs.~\cite{De-Rujula:2011lr,Kim:2009si}.
The underlying idea is to formulate a kinematic variable that maximizes
the amount of information on $m_{W}$ that can be extracted from events
at hadron colliders and that helps to focus the information in particular
regions of the phase-space of visible particles. In this approach a certain amount
of knowledge of partially unknown longitudinal momenta is still necessary. Therefore 
PDF are still a necessary input. Still, the concentration of the information on $m_{W}$ in special features of the distributions, such as singular points, can help to test the measurement carried out with the standard methods. We are not aware of experimental
studies using these type of variables nor of theory studies seeking
to quantify their robustness beyond the LO picture on which the variables
are built.

A different approach has been proposed focusing on just the observable
momentum of the charged lepton. Using the fact that at LO in perturbation
theory the decay of a spin-1 into a pair of spin-half particles can
contain only few spherical harmonics, Ref.~\cite{Bianchini:2019iey}
has proposed to use the energy distribution of the leptons from the
$W$ boson decay in way similar to Ref.~\cite{Agashe:2013sw,Agashe:2016xq}.
For the specific case of the decay of a spin-1 into a pair of spin-half
particles Ref.~\cite{Bianchini:2019iey} has identified possible features
in the first and second derivative of the energy distribution, which
can provide further information on the mass of the $W$ boson, including
in cases in which the peak of the energy distribution does not strictly
speaking enjoy the properties exploited in Ref.~\cite{Agashe:2013sw}.

All in all there is a variety of methods that can be exploited to measure $m_{W}$ at hadron colliders thanks to careful design of kinematic variables. These methods leverage different strengths of the measurement and try to minimize the exposure to the theoretical and experimental weaknesses in different ways. The combination of the information that can be attained by this variety of methods will help up gain confidence in the results of such a challenging measurement.

\section{Inclusive event variables}
\label{sec:inclusive}

\begin{figure}[t]
\begin{center}
\includegraphics[width=.4\textwidth]{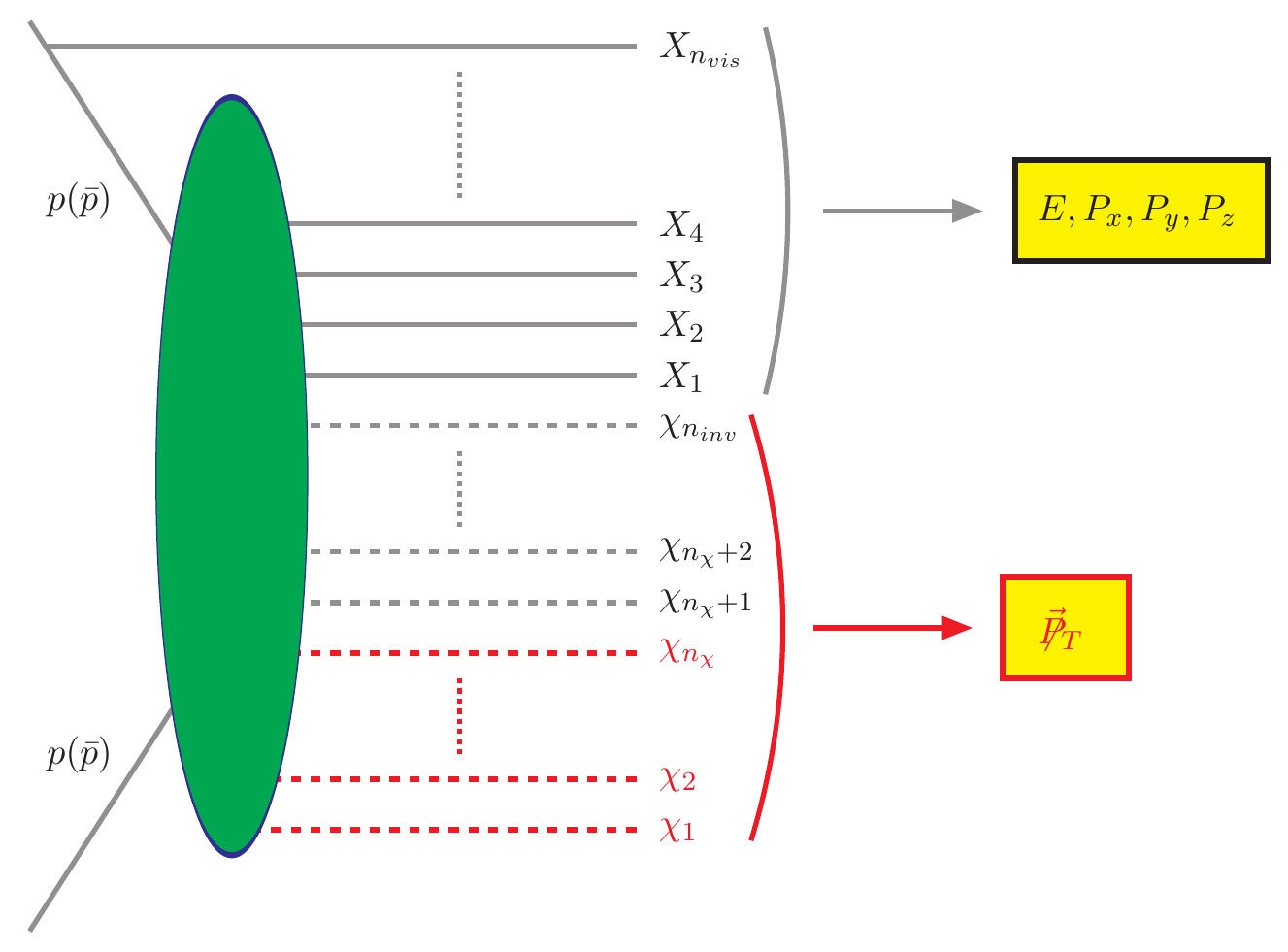}
\end{center}
\caption{The generic event topology illustrating the use of inclusive event variables from Section~\ref{sec:inclusive}.
Taken from Ref.~\cite{Konar:2008ei}.
}
\label{fig:generic_event_topology}
\end{figure}

In this section we shall focus on {\it inclusive} kinematic variables. They are robust and model-independent since one does not make any assumptions about the underlying event topology. The downside is that they are not as sensitive to specific signals as their {\it exclusive} cousins discussed later in Sections~\ref{sec:exclusive}-\ref{sec:otherexc}, which are intentionally designed to look for such signals. Nevertheless, due to their simplicity, inclusive variables have proven to be valuable and have found wide usage at both the trigger and the analysis level. 

Inclusive event variables are applicable to a generic event topology shown in Figure~\ref{fig:generic_event_topology}. Unlike the case in Figure~\ref{fig:convention2}, here we make no assumptions about the underlying process, hence there is no partitioning of the final state particles other than dividing them into visible (solid lines) and invisible (dashed lines). Black solid lines correspond to SM particles which are visible in the detector, e.g.~jets, electrons, muons and photons. The SM particles may originate either from initial state radiation, or from the hard scattering and subsequent cascade decays (indicated with the green-shaded ellipse). Dashed lines denote neutral stable particles which are invisible in the detector. In general, the set of invisible particles consists of some number of SM neutrinos (denoted with the black dashed lines), as well as some number of BSM particles (indicated with the red dashed lines) which could be dark matter candidates. The identities and the masses of the BSM invisible particles do not necessarily have to be all the same, allowing for the simultaneous production of several {\em different} species of dark  matter particles. A few global event variables describing the visible particles are: the total energy $E$, the transverse components $P_x$ and $P_y$ and the longitudinal component $P_z$ of the total visible momentum $\vec{p}$. The only experimentally available information regarding the invisible particles is the missing transverse momentum $\mptvec$.

\subsection{Event-shape-type variables \label{sec:eventshape}}

In this section we review some classic event shape variables summarized in Table~\ref{tab:eventshapevariables}. Other modern approaches involving jet substructure variables will be reviewed in a sister white paper submitted to Snowmass. The basic idea of the event shape variables is to give more information than just the cross section by defining the ``shape'' of an event (pencil-like, planar, spherical etc.) \cite{Banfi:2010xy}. Event shape variables describe the patterns and correlations of energy flow resulting from the particle collisions. 

\begin{table*}[t]
\centering
\begin{tabular}{c | c | c c c | c}
\hline \hline
\multirow{2}{*}{Observable}  & \multirow{2}{*}{Definition} & \multicolumn{3}{|c|}{Typical values for} & \multirow{2}{*}{References}  \\
 & & Pen. & Copl. & Iso. &  \\
\hline
 \multirow{2}{*}{Sphericity}   & $S=\frac{3}{2}\left(\lambda_2+\lambda_3 \right)$, $\lambda_i (\lambda_1 \geq \lambda_2 \geq \lambda_3)$, & \multirow{2}{*}{0} & \multirow{2}{*}{$\leq \frac{3}{4}$} & \multirow{2}{*}{$\leq 1$} & \multirow{2}{*}{\cite{Bjorken:1969wi}} \\
    & eigenvalues of $M_{i j} = \frac{\sum_{a=1}^{n_j} p_{a,i} p_{a,j}}{\sum_{a=1}^{n_j} |\vec p_a|^2}$ with $i,j\in \{x,y,z \}$  & & & & \\
 Transverse sphericity   & $S_T=\frac{2\lambda_2}{\lambda_1+\lambda_2}$ & & & & \cite{Bjorken:1969wi} \\
  Aplanarity   & $A=\frac{3}{2}\lambda_3$ & 0 & 0 & $\leq \frac{1}{2}$ & \cite{Bjorken:1969wi} \\
  Planarity   & $P=\lambda_2-\lambda_3$ & & & & \cite{Bjorken:1969wi} \\
  (Transverse) spherocity   & $S_0=\frac{\pi^2}{4}\stackrel[{\hat n}]{}{\min} \left( \frac{\sum_a |\vec{p}_{a,T} \times \hat{n}| }{\sum_a \vec{p}_{a,T}} \right)^2$ & 0 & 0 & $\leq 1$ & \cite{Banfi:2010xy} \\
  \hline
  Thrust   & $T = \stackrel[{\hat n}]{}{\max} \left( \frac{\sum_a \left | \vec p_a \cdot \hat n \right | }{ \sum_a \left | \vec p_a \right |} \right) $  & 1 & $\geq \frac{2}{3}$ & $\geq\frac{1}{2}$ & \cite{Brandt:1964sa,Farhi:1977sg} \\
  Thrust major   & $T_{\rm major} = \stackrel[{\hat n_{\rm ma} \perp \hat n_T}]{}{\max}  \left( \frac{\sum_a  | \vec p_a \cdot \hat n_{\rm ma} | }{ \sum_a  | \vec p_a |} \right )$ & 0 & $\leq \frac{1}{3}$ & $\leq \frac{1}{\sqrt{2}}$ & \cite{Brandt:1964sa,Farhi:1977sg} \\
  Thrust minor   & $T_{\rm minor} = \frac{\sum_a \left | \vec p_a \cdot \hat n_{\rm mi} \right | }{ \sum_a \left | \vec p_a \right |}$ with $\hat n_{\rm mi} = \hat n_T \times \hat n_{\rm ma}$  & 0 & 0 & $\leq \frac{1}{2}$ & \cite{Brandt:1964sa,Farhi:1977sg} \\
  Oblateness   & $\mathcal{O}=T_{\rm major}-T_{\rm minor}$ & 0 & $\leq \frac{1}{3}$ & 0 & \\
  \hline
  \multirow{2}{*}{Normalized hemisphere mass} & $M_{1(2)}^2 = \frac{1}{E_{\rm CM}^2} \left ( \sum_{a \in H_{1(2)}} p_a \right )^2$ with $H_{1(2)}$ being   & & & & \\
   & hemispheres divided by the plane normal to $\hat{n}_T$ & & & & \cite{Banfi:2010xy} \\
   Heavy jet mass & $M_{H}^2 = \max\left(M_1^2, M_2^2 \right)$ & 0 & $\leq \frac{1}{3}$ & $\leq \frac{1}{2}$& \cite{Banfi:2010xy}\\
   Light jet mass & $M_{L}^2 = \min \left(M_1^2, M_2^2 \right)$ & & & & \cite{Banfi:2010xy}\\
   Jet mass difference & $M_{D}^2 = \left |M_1^2 - M_2^2 \right |$ & 0 & $\leq \frac{1}{3}$ & 0 & \cite{Banfi:2010xy}\\
       Jet broadening   & $B_{1(2)} = \frac{\sum_{a \in H_{1(2)}} | \vec p_a \times \hat n_T|}{ 2\sum_b |\vec p_b|^2}$ & & & & \cite{Banfi:2010xy}\\
    Wide/narrow, total broadening & $B_{W/N}=\max/\min(B_1,B_2)$, $B_T=B_W+B_N$ & & & &\cite{Banfi:2010xy} \\
  \hline
  Fox-Wolfram moments   & $H_\ell = \sum_{i,j} \frac{|\vec p_i| |\vec p_j|}{E^2} P_{\ell} (\cos\theta_{ij})$ & & & & \cite{Fox:1978vw} \\
  $N$-jettiness & $\tau_N=\frac{2}{Q^2}\sum_k \min\{q_a\cdot p_k,q_b\cdot p_k, q_1\cdot p_k, \cdots, q_N\cdot p_k\}$ & & & & \cite{Stewart:2010tn} \\
  $N$-subjettiness & $\tau_N =\frac{1}{\sum_k p_{T,k}R_0} \sum_k p_{T,k}\min \left\{\Delta R_{1k}, \Delta R_{2k}, \cdots, \Delta R_{Nk} \right\}$ & & & & \cite{Thaler:2010tr} \\
  Energy-energy correlation   & $EEC(\chi) = \frac{1}{\sigma} \frac{d\Sigma}{d\cos\chi} = \sum_{i,j} \int \frac{E_i E_j}{Q^2} \delta( \hat p_i \cdot \hat p_j - \cos\chi) d\sigma$ & & & & \cite{Basham:1978bw,Basham:1978zq} \\
 \hline \hline
\end{tabular}
\caption{\label{tab:eventshapevariables} A summary of event shape variables in their definition, typical values, and associated references. Inspired by Fabio Maltoni's lecture \cite{ShapeVariableTable:Maltoni} given at the 2013 CERN - Latin-American School of High Energy Physics \cite{Mulders:1484921}.}
\end{table*}

A very common observable is the thrust, which is defined as
\begin{eqnarray}
T &=& \max_{ \vec n} \left ( \frac{\sum\limits_i \Big | \vec p_i \cdot \vec n \Big | }{ \sum\limits_i \Big | \vec p_i \Big |} \right ) \, .
\label{eq:thrust}
\end{eqnarray}
Here the so-called thrust axis $\vec n_T$ is defined in terms of the unit vector $\vec{n}$ which maximises $T$. This definition implies that for $T = 1$ the event is perfectly back-to-back, while for $T = 1/2$ the event is spherically symmetric. The unit vector which maximises the thrust in the plane perpendicular to $ \vec{n}_T $ is called the ``thrust major'' direction, and the vector perpendicular to both the thrust and the thrust major is called the thrust minor direction. The thrust major and the thrust minor variables are defined as 
\begin{eqnarray}
T_{\rm major} &=& \max_{\vec n_{\rm ma} \perp \vec n_T} \left ( \frac{\sum\limits_i \Big | \vec p_i \cdot \vec n_{\rm ma} \Big | }{ \sum_i \Big | \vec p_i \Big |} \right ) \, , \\
T_{\rm minor} &=& \frac{\sum\limits_i \Big | \vec p_i \cdot \vec n_{\rm mi} \Big | }{ \sum_i \Big | \vec p_i \Big |} \, , 
\end{eqnarray}
where $\vec n_{\rm mi} = \vec n_T \times \vec n_{\rm ma}$. 
The oblateness $O$ is defined as the difference between the thrust major and thrust minor, $O = T_{\rm major} - T_{\rm minor}$. 
Transverse thrust and its minor component are defined similarly but using transverse momenta ($\vec p_{T,i}$ instead of $\vec p_i$) of particles in the events.

The sphericity ($S$), transverse sphericity ($S_T$), aplanarity ($A$) and planarity ($P$) provide additional global information about the full momentum tensor, $M$, of the event via its eigenvalues:
\begin{eqnarray}
    M_{i j} &=& \frac{\sum\limits_{a=1}^{n_j} p_{i a} p_{j a}}{\sum\limits_{a=1}^{n_j} |\vec p_a|^2} \, , 
\end{eqnarray}
where $i,j$ are the spatial indices and the sum runs over all particles (or in some applications, over the reconstructed jets). The ordered eigenvalues $\lambda_i$ ($\lambda_1 > \lambda_2 > \lambda_3$) with the normalization condition $\sum_i \lambda_i = 1$ define the sphericity, transverse sphericity, aplanarity, and planarity as follows:
\begin{eqnarray}
S   &=& \frac{3}{2} \Big ( \lambda_2 + \lambda_3 \Big )\, , \\
S_T &=& \frac{2 \lambda_2}{\lambda_1 + \lambda_2} \, ,\\
A   &=& \frac{3}{2}  \lambda_3 \, , \\
P   &=& \frac{2}{3}\Big (S - 2A  \Big ) =  \lambda_2 - \lambda_3 \, .
\end{eqnarray}
The sphericity axis is defined along the direction of the eigenvector of $\lambda_1$ and the semi-major axis is along the eigenvector for $\lambda_2$. 
The sphericity and transverse sphericity measure the total transverse momentum with respect to the sphericity axis defined by the four-momenta in the event. In other words, the sphericity of an event is a measure of how close the spread of energy in the event is to a sphere in shape. The allowed range for $S$ is $0 \leq  S \le 1$. The transverse sphericity is defined by the two largest eigenvalues, and the allowed range is again $0 \leq S_T < 1$.
Aplanarity measures the amount of transverse momentum out of the plane formed by the two leading jets. The allowed range for $A$ is $0 \leq A < 1/2$. The planarity is a linear combination of the second and third eigenvalue of the quadratic momentum tensor. 

A plane through the origin whose normal vector is the thrust vector ($\vec n_T$) divides an event into two hemispheres, $H_1$ and $H_2$. The corresponding normalized hemisphere invariant masses are defined as
\begin{eqnarray}
M_i^2 = \frac{1}{E_{CM}^2} \left ( \sum_{a \in H_i} p_a \right )^2 \, , ~~~~i=1,2 \, ,
\end{eqnarray}
where $p_a$ is the four-momentum of the $a$-th jet. 
The larger of the two is called the heavy jet mass $M_H$ and the smaller is called the light jet mass $M_L$,
\begin{eqnarray}
M_H &=& \max ( M_1^2, M_2^2 ) \, , \\
M_L &=& \min ( M_1^2, M_2^2 ) \, .
\end{eqnarray}
The difference between the two is called the jet mass difference $M_D = M_H - M_L$. 

A measure of the broadening of particles in the transverse momentum with respect to the thrust axis $\vec n_T$ is calculated as follows
\begin{equation}
    B_i = \frac{\sum\limits_{a \in H_i} | \vec p_a \times \vec n_T|}{ 2\sum\limits_b |\vec p_b|^2} \, ~~~i=1,2 \, ,
\end{equation}
where $b$ runs over all particles and $a$ runs over particles in one of the two hemispheres. The larger of the two hemisphere broadenings is called the wide jet broadening [$B_W = \max(B_1, B_2)$], while the smaller is called the narrow jet broadening [$B_N=\min (B_1, B_2)$]. The total jet broadening is the sum of the two, $B_T = B_W + B_N$. 

The $C$-parameter
\begin{equation}
C = 3 (\lambda_1 \lambda_2 + \lambda_2 \lambda_3 + \lambda_3 \lambda_1) \, ,
\end{equation} 
is derived from the  eigenvalues ($\lambda_i$) of the linearized momentum tensor $\Theta_{ij}$,
\begin{equation}
    \Theta_{ij} = \frac{1}{\sum_a |\vec p_a|} \sum_b \frac{p_{ib} p_{j b}}{|\vec p_b|}\, , ~~~i,j=1,2,3 \, .
\end{equation}
Many of these shapes variables are used to analyze data at both lepton colliders \cite{Ford:2004dp,DELPHI:2003yqh} and hadron colliders \cite{Weber:2011kor,Weber:2009bhh,ATLAS:2020vup,Banfi:2010zz,Lenz:2017lqo}. 

The Fox-Wolfram moments \cite{Fox:1978vw,Fox:1978vu} are defined as 
\begin{equation}
    H_\ell = \sum_{i,j} \frac{|\vec p_i| |\vec p_j|}{E_{\rm total}^2} P_{\ell} (\cos\theta_{ij}) \, ,
\end{equation}
where $\theta_{ij}$ is the opening angle between energy clusters $i$ and $j$, $E_{\rm total}$ is the total energy of the clusters (in the event center-of-mass frame), $P_{\ell}(x)$ is the Legendre polynomial. 
For an event which has the structure of two back-to-back jets in the center-of-mass frame, $H_0=0$, $H_{\ell}\approx1$ for even $\ell$, and $H_{\ell}\approx 0$ for odd $\ell$. Often the ratio between the Fox-Wolfram moments could be a useful discriminating variable against backgrounds ---
see Refs. \cite{Bernaciak:2012nh,Chen:2011ah,Englert:2012ct} for application of the Fox-Wolfram moments in Higgs physics and in jet-substructure.

The transverse spherocity \cite{Banfi:2010xy} is defined as
\begin{equation}
    S_\perp = \frac{\pi^2}{4} \min_{\vec n_T} \left ( \frac{\sum\limits_i | \vec p_{i T} \times \vec n_T |}{\sum\limits_i p_{i T}} \right )^2 \, ,
\end{equation}
where the minimization is performed over all possible unit transverse vectors $\vec n_T = (n_x, n_y, 0)$ [not to be confused with the thrust axis defined in (\ref{eq:thrust})]. This variable ranges from 0 for pencil-like events, to a maximum of 1 for circularly symmetric events. 

The centrality
\begin{equation}
    C = \frac{\sum |\vec p_{vis,i}|}{\sum E_{vis, i}} 
\end{equation}
is a measure of how much of the event is contained within the central part of the detector.

The energy-energy correlation ($EEC$) function \cite{Basham:1978bw,Basham:1978zq} is defined as \begin{equation}
    EEC(\chi) = 
    \frac{d\Sigma} {d\cos\chi} = \sum_{i,j} \int \frac{E_i E_j}{E_{\rm total}^2} \delta( \hat p_i \cdot \hat p_j - \cos\chi) d\Phi \, ,
\end{equation} 
where $i,j$ run over all final state particles, which have four-momenta $p^\mu_i = (E_i, \vec p_i)$ and $p^\mu_j = (E_j, \vec p_j)$, $E_{\rm total}$ is the total energy of the system in the center-of-mass frame and $d\Phi$ is the phase space measure \cite{Dixon:2018qgp}. The unit vectors $\hat p_i$ and $\hat p_j$ point along the spatial components of $p_i$ and $p_j$, respectively. $EEC$ measures the differential angular distribution of particles that flow through two cells in the calorimeter separated by an angle $\chi \in (0, \pi)$ and is defined as an energy-weighted cross section corresponding to the process of interest. 

Another example of a simple shape variable is $y_{23}$, a measure of the third-jet $p_T$ relative to the sum of the transverse momenta of the two leading jets in a multi-jet event, which is defined as \cite{Akrawy:1989rg,Catani:1991hj}:
\begin{equation}
    y_{23} = \frac{p_{T,j_3}^2}{(p_{T,j_1} + p_{T, j_2})^2} \, ,
\end{equation}
where $p_{T,j_1}$, $p_{T,j_2}$ and $p_{T,j_3}$ represent the leading, subleading, and third-leading jet in the event, respectively. The allowed range for $y_{23}$ is $0 \leq y_{23} < 1/4$. 

There are many other event shape variables not discussed in this review. We refer to Ref. \cite{Moult:2016cvt} for new insight in the energy correlation functions, Refs. \cite{Stewart:2010tn,Thaler:2010tr} for N-jettiness, Refs. \cite{Cesarotti:2020hwb,Cesarotti:2020ngq} for event isotropy using the energy mover’s distance (EMD) and Ref. \cite{Banfi:2010xy} for other interesting event shape variables. 

\subsection{Missing momentum \label{sec:misse} }

Missing energy (missing momentum) refer to the amount of energy (momentum) that is not measured or detected in a particle detector, but can be inferred from the laws of energy-momentum conservation. In hadron colliders, the initial momenta of the colliding partons along the beam axis are unknown, so the missing energy and the missing total momentum cannot be determined. However, the total momentum of initial particles in the plane orthogonal to the beam is zero, and therefore, any net visible momentum in the transverse direction is indicative of missing transverse momentum, $\mptvec$. 

Missing (transverse) momentum arises whenever the final state includes particles that do not interact with the electromagnetic or strong forces, and therefore escape the detector. A typical example in the SM is neutrino production. More importantly, dark-matter candidates in BSM models are also invisible in the detector, making the $\mptvec$ signature a smoking gun for the existence of non-gravitationally interacting dark matter. Therefore, an extensive range of dark-matter searches have been performed in collider experiments, centered around the missing transverse momentum signature: for example, $\mptvec$ plus mono-jet~\cite{CMS:2017jdm,ATLAS:2017bfj}, mono-photon~\cite{ATLAS:2017nga,CMS:2017qyo}, mono-$Z/W$~\cite{CMS:2017ret,CMS:2017jdm,ATLAS:2017nyv,ATLAS:2018nda,CMS:2020ulv}, and mono-higgs~\cite{ATLAS:2017pzz,ATLAS:2017uis,CMS:2019ykj}. 

The missing transverse momentum, $\mptvec$, of the hard scattering interaction is defined as the negative vectorial sum of the transverse momenta of the set of reconstructed objects including hard  and soft objects~\cite{CMS:2011bgj,ATLAS:2018txj}:
\begin{equation}
\mptvec
=- \sum_{i \in {\rm hard~objects}} \vec p_{T,i} - \sum_{j \in {\rm soft~objects}} \vec p_{T, j}\,, \label{eq:met1}
\end{equation}
whose magnitude and angle on the transverse plane are respectively defined as 
\bea
\met &=& E_T^{\rm miss}  = \mpt  
=\sqrt{\slashed{p}_{Tx}^2+\slashed{p}_{Ty}^2}  \label{eq:met2} \\
\varphi &=& \tan^{-1} \left (  \frac{\slashed{p}_{Ty}}{\slashed{p}_{Tx}} \right ). \label{eq:met3}
\eea 
As indicated in (\ref{eq:met2}), it has become a custom to refer to the magnitude of the missing transverse momentum as the ``missing transverse energy", or MET for short.
Here the hard objects consist of selected $e^\pm$, $\mu^\pm$, and accepted $\gamma$, $\tau^\pm$, and jets, while the soft objects are not associated with any of the aforementioned hard objects but identified as the unused tracks from the primary vertex~\cite{ATLAS:2018txj}. In order to  reduce effects from pile-up, in ATLAS~\cite{ATLAS:2018txj}, these tracks are required to have $p_{T} > 0.4$ GeV,  $|\eta|<2.5$ and transverse (longitudinal) impact parameter $|d_0| <1.5\,\textrm{mm}\,(|z_0 \sin\theta| < 1.5\,\textrm{mm})$.
The scalar sum of all transverse visible momenta is defined as
\beq
H_T = \sum_{i \in {\rm hard~objects}} p_{T,i} + \sum_{j \in {\rm soft~objects}} p_{T, j}\,.\label{eq:met4}
\eeq
The quantities defined in Eqs.~\eqref{eq:met1} through \eqref{eq:met4} are often used to estimate the hardness of the hard scattering event in the transverse plane, and thus provide a measure for the event activity in physics analyses.

\subsection{Variables sensitive to the overall energy scale\label{sec:energyscale}}
In the case of fully visible final states, the total invariant mass in the event provides an estimate of the energy scale $\sqrt{\hat s}$ of the hard scattering, where $\hat s$ is the parton-level Mandelstam variable. However, if the final state includes invisible particles as in Figure~\ref{fig:generic_event_topology}, the task becomes more challenging, which has motivated the introduction of several inclusive variables for this purpose.  

One class of such variables were originally  explored in the context of supersymmetry, where strong production of gluinos and/or squarks results in a multijet plus $\met$ signature. 
Several versions of an ``effective scale" variable $M_{\rm eff}$ for that case have been used throughout the literature \cite{Hinchliffe:1996iu,Tovey:2000wk}; they are closely related to (\ref{eq:met4}) and differ by i) the number of jets $N_{j}$ included in the sum - typical choices for $N_{j}$ are either 4 or ``all", and ii) whether the value of the $\met$ is added as well or not: 
\bea
M_{\rm eff} (N_{j}, I)&&= \sum_i^{N_{j}} p_{T,i}+ I*\met,
\label{eq:Meffdef}
\eea
where $I\in \{0,1\}$ parametrizes the binary choice for including the $\met$ or not. The main advantage of the effective mass variable (\ref{eq:Meffdef}), which led to its widespread usage in the LHC community, is its simplicity. However, it also has drawbacks --- for example, it misses the potential dependence on the masses of any invisible particles. Being empirically derived, it is not on a firm theoretical footing, which explains the large number of different $M_{\rm eff}$ variants being used.

\begin{figure}[t]
\begin{center}
\includegraphics[width=.23\textwidth]{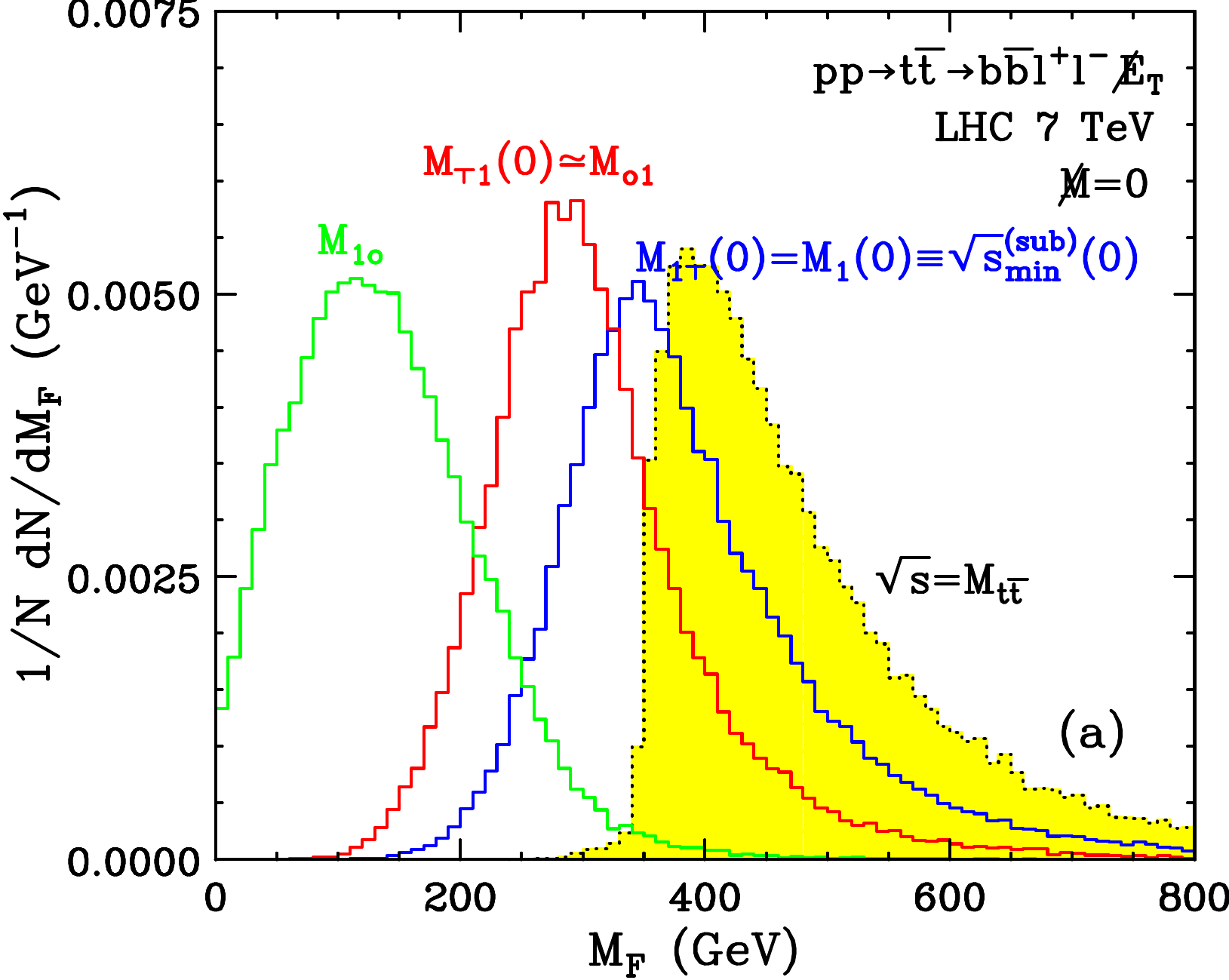} 
\includegraphics[width=.23\textwidth]{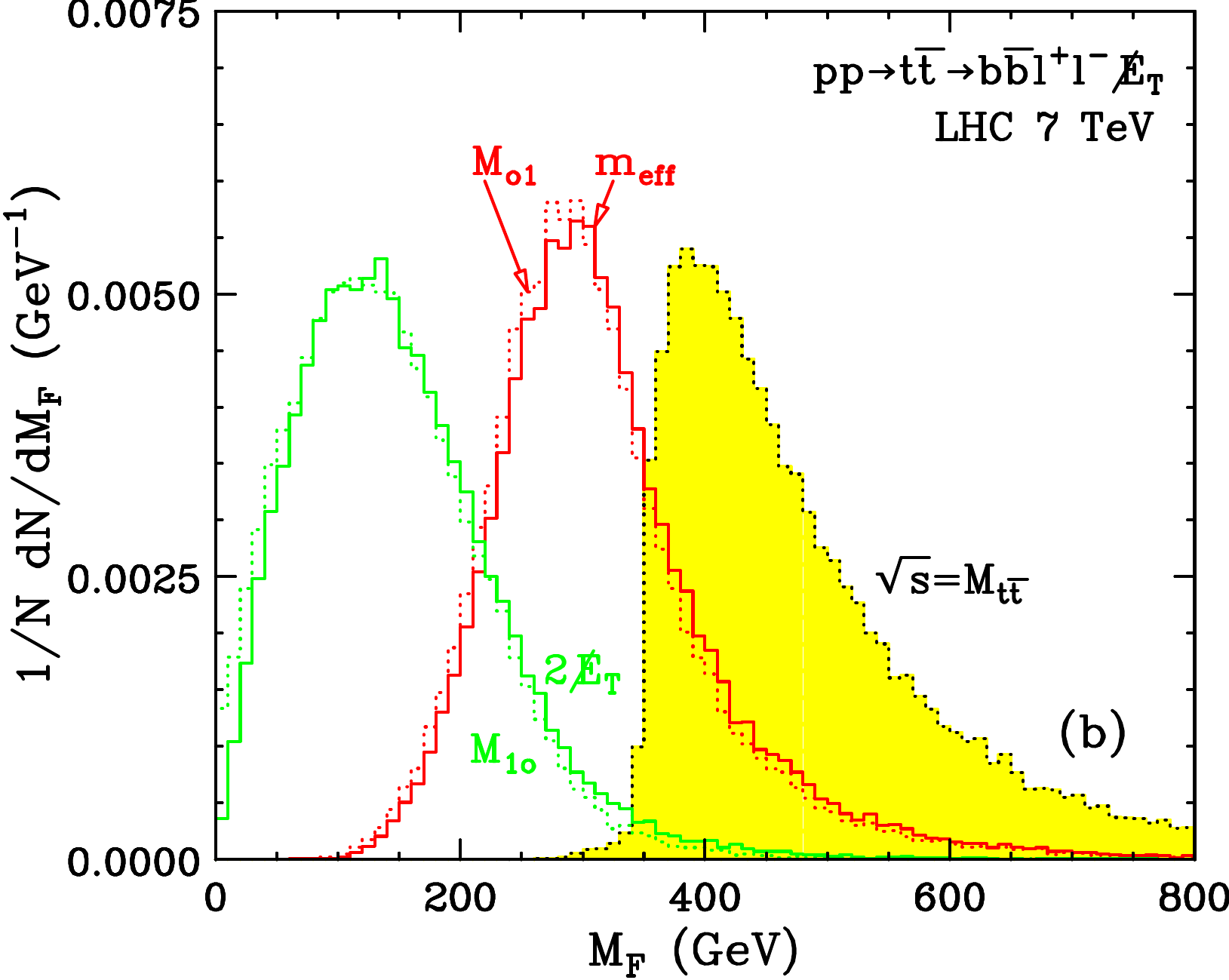}
\end{center}
\caption{Unit-normalized distributions of various inclusive event variables ($\met$, $M_{\rm eff}$, $H_T$, $M_1$, $\sqrt{\hat{s}}_{\rm min}$) for top quark pair production ($pp \to t \bar t \to b \bar b \ell^+ \ell^- \nu \bar\nu$). The yellow-shaded histogram shows the true $\sqrt{\hat s}$ distribution in the sample. Taken from Ref. \cite{Barr:2011xt}.}
\label{fig:inclusive}
\end{figure}

An alternative approach, advocated in Ref.~\cite{Konar:2008ei}, was to enforce the missing energy constraint in Eq.~(\ref{eq:met1}) and then utilize a minimum energy principle to fix the momenta of the invisible particles and thus arrive at a more precise estimate of $\sqrt{\hat s}$, 
\bea
\sqrt{\hat s}_{\min}(M_{\rm inv}) &&= \sqrt{E^2 -P_z^2}+\sqrt{\met^2+M_{\rm inv}^2}, \label{eq:smin}
\eea
where $E$ and $P_z$ are the total visible energy and the total longitudinal visible momentum in the event, respectively. The hypothesized parameter $M_{\rm inv}$ is the total mass of all invisible particles in the event. By construction, $\sqrt{\hat s}_{\min}$ is the minimum possible center-of-mass energy (for a given value of $M_{\rm inv}$) which is consistent with the measured values of the total energy $E$ and the total visible momentum $\vec{P}$ and thus has a well-defined physical meaning. However, when applied to the full event, $\sqrt{\hat s}_{\min}$ receives large contributions from the intense QCD radiation in the forward direction, which disrupt the connection to the underlying new physics parameters \cite{Papaefstathiou:2009hp}. This motivated ``subsystem" variants of $\sqrt{\hat s}_{\min}$ where one focuses on the central region, with measured total energy $E_{(sub)}<E$ and total longitudinal momentum $P_{z(sub)}$, away from the dangers of the forward QCD radiation \cite{Konar:2010ma,Robens:2011zm}:
\bea
\sqrt{\hat s}_{\min}^{(sub)}(M_{\rm inv}) &=& \Biggl\{ \Biggl(\sqrt{E_{(sub)}^2 -P_{z(sub)}^2} \label{eq:sminsub} \\
 &&+ \sqrt{\met^2+M_{\rm inv}^2}\Biggr)^2 -\vec u_{T}^2 \Biggr\}^{\frac{1}{2}}, 
 \nonumber
\eea
where $\vec{u}_{T} = \vec{P}_T - \vec{P}_{T(sub)}$ is the upstream transverse momentum due to QCD radiation and/or visible particle decays outside the subsystem (see Figure~\ref{fig:convention2}). The $s_{\rm min}$ variables have been further extended including additional constraints during the minimization \cite{Swain:2014dha} and such constrained $s_{\min}$ variables have been applied to physics processes like $h \to \tau\tau$ \cite{Bhardwaj:2016lcu,Konar:2016wbh}. A sample menagerie of inclusive event variables is shown in Figure~\ref{fig:inclusive} for the case of dilepton top quark pair production ($pp \to t \bar t \to \ell^+ \ell^- \nu \bar\nu$) \cite{Barr:2011xt}.

\section{Exclusive event variables: invariant mass \label{sec:exclusive}
}
In the following three sections we discuss kinematic variables that can be constructed and evaluated by processing information restricted to/associated with a particular set of final-state visible particles in an event by a suitable partitioning as illustrated in Figure~\ref{fig:convention2}. The current section will be devoted to invariant mass variables, which can be reconstructed from collections of visible particles only (Section~\ref{sec:mass}) or from semi-invisible collections of particles (Section~\ref{sec:mt2variables}).  

Mass variables have played a major role not only for measuring the masses of new particles but for discovering new physics in resonance-type searches. 
Techniques utilizing mass variables received a major boost in the LHC era, and have been actively and extensively investigated for LHC phenomenology. Examples range from traditional (1+3)-dimensional invariant masses and (1+2)-dimensional transverse masses to the stransverse mass~\cite{Lester:1999tx} and its variations, $M_2$~\cite{Cho:2014naa}, the razor~\cite{Rogan:2010kb}, $\Delta_4$~\cite{Byers:1964ryc}, etc. In the following we discuss the main ideas and mathematical understanding of these variables, their collider implications, and typical applications.   

\subsection{Mass variables of collections of visible particles \label{sec:mass}}
In this subsection we review mass variables\footnote{We remind the reader that in our convention the masses of individual particles are denoted with lowercase $m$, while any mass of a collection of particles is denoted with a capital $M$, see Section~\ref{sec:notation}.} which do not make use of the measured $\met$. The standard example is the invariant mass of a set of visible particles, $M_{\rm vis}$,
\begin{equation}
    M_{\rm vis}^2 = \left(\sum_i p_i \right)^2 = \left(\sum_i E_i  \right)^2- \left(\sum_i \vec{p}_i  \right)^2\,,    
\label{eq:Mvisdef}
\end{equation}
where $i$ runs over the visible particles of interest.
Since it is a Lorentz-invariant quantity by definition, its physical implications can be understood consistently irrespective of the frame in which one performs measurements or analyses. This is why (\ref{eq:Mvisdef}) is routinely being used in a wide range of high energy experiments including accelerator-based ones.

The simplest (but sufficiently nontrivial) application is a heavy resonance, $A_1$, decaying to a pair of visible particles $a_1$ and $a_0$, i.e., $A_1\to a_1 a_0$. 
The energy-momentum conservation, i.e., $p_{A_1}=p_{a_1}+p_{a_0}$, implies that the resonance mass can be reconstructed from the four-momenta of the visible decay products:
\begin{equation}
    M_{A_1}^2=p_{A_1}^2 = (E_{a_1}+E_{a_0})^2-(\vec{p}_{a_1}+\vec{p}_{a_0})^2\,.
\end{equation}
Mathematically, the distribution of $M_{A_1}$ is $\delta$-function-like at the true mass $m_{A_1}$ of $A_1$. However, the virtuality of the unstable $A_1$ forces events to spread and populate the region around $m_{A_1}$ according to the Breit–Wigner distribution in $M_{A_1}$ as follows
\begin{eqnarray}
    \frac{dN}{dM_{A_1}} \propto \frac{1}{(M_{A_1}^2-m_{A_1}^2)^2 + m_{A_1}^2 \Gamma_{A_1}^2}\,,
\label{eq:BWdistribution}    
\end{eqnarray}
where $\Gamma_{A_1}$ is identified as the decay width of $A_1$. As a consequence, the $M_{A_1}$ distribution allows for a simultaneous determination of $m_{A_1}$ and $\Gamma_{A_1}$. 
Since most events lie within a few $\Gamma_{A_1}$ from $m_{A_1}$, by restricting to a narrow invariant mass window around $m_{A_1}$, one can efficiently isolate the resonance events from unwanted background events. Due to this great background-rejection capability, the invariant mass variable (\ref{eq:Mvisdef}) has played a crucial role in the discovery of many particles including the $Z$ gauge boson~\cite{UA1:1983mne,UA2:1983mlz}, hadrons such as $J/\psi$~\cite{E598:1974sol,SLAC-SP-017:1974ind} and $\Upsilon$~\cite{Herb:1977ek}, and the SM higgs~\cite{ATLAS:2012yve,CMS:2012qbp}.

Once some of the decay products are invisible in the detector, the resonance feature is no longer available. Nevertheless, the invariant mass of the remaining visible decay products still provides useful information about the underlying dynamics, and its features have been thoroughly investigated. To have a nontrivial invariant mass variable, at least two visible final-state particles are required on top of the invisible particle(s). The most renowned example is the leptonic decay of a top quark, i.e., $t\to bW,~W\to \ell \nu_\ell$, giving rise to the invariant mass  $m_{b\ell}$ formed by the bottom quark and the lepton
\begin{equation}
m_{b\ell}=(p_b+p_\ell)^2\,.
\end{equation}
When it comes to models of BSM, there exist many such processes in connection with dark-matter candidates, for example, the decay of a supersymmetric lepton to a pair of leptons and the lightest neutralino (an invisible dark-matter candidate) via a heavier neutralino intermediary state. 

\begin{figure}[t]
    \centering
    \includegraphics[width=0.48\textwidth]{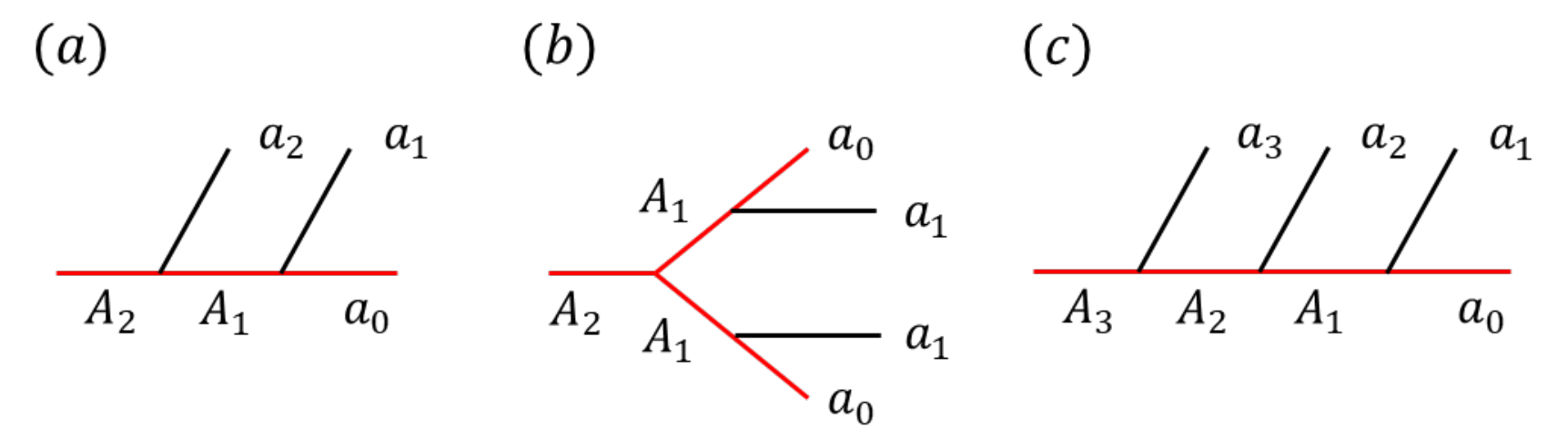}
    \caption{($a$) Two-step two-body cascade decay topology, ($b$) Antler decay topology, ($c$) Three-step two-body cascade decay topology.}
    \label{fig:invmassdiagrams}
\end{figure}

Let us work out the generic two-step two-body cascade decay case, $A_2 \to a_2 A_1,~A_1 \to a_1 a_0$, and assume that $a_2$ and $a_1$ are visible and massless while $a_0$ is invisible [see Figure~\ref{fig:invmassdiagrams}($a$)]. For simplicity, we further assume that all particles are spinless or produced in an unpolarized fashion, and focus on decay kinematics purely governed by phase space. 
Since $a_1$ and $a_2$ were assumed to be massless, the invariant mass squared $M_{a_2a_1}^2$ is simply given by 
\begin{equation}
    M_{a_2a_1}^2 = 2E_{a_2}E_{a_1}\left(1-\cos\theta_{a_2a_1}\right),
\label{eq:M2body}\end{equation}  
where $\theta_{a_2a_1}$ is the angle between $\vec{p}_{a_2}$ and $\vec{p}_{a_1}$.

Using Lorentz-invariance, one can evaluate this quantity in a convenient frame. In the $A_1$ rest frame, the energies of $a_2$ and $a_1$, $E_{a_2}^{*}$ and $E_{a_1}^{*}$ are 
\begin{eqnarray}
    E_{a_2}^{*}&=& \frac{m_{A_2}^2-m_{A_1}^2}{2m_{A_1}}\,, \\
    E_{a_1}^{*}&=& \frac{m_{A_1}^2 -m_{a_0}^2}{2m_{A_1}}\,.
    \label{eq:E1E2A1rest}
\end{eqnarray}
Note that in this frame the distribution of  $\cos\theta_{a_2a_1}^{*}$ becomes flat. Then, using Eqs.~(\ref{eq:M2body}-\ref{eq:E1E2A1rest}), we can derive the unit-normalized distribution of $M_{a_2a_1}$, 
$\frac{1}{\Gamma}\frac{d\Gamma}{d M_{a_2a_1}}$, 
as follows:
\begin{eqnarray}
    \frac{1}{\Gamma}\frac{d\Gamma}{d M_{a_2a_1}}&=&\frac{M_{a_2a_1}}{2E_{a_2}^{*}E_{a_1}^{*}}\Theta(M_{a_2a_1}^{\max}-M_{a_2a_1}) \nonumber \\
    &=& \frac{2M_{a_2a_1}}{M_{a_2a_1}^{\max}} \Theta(M_{a_2a_1}^{\max}-M_{a_2a_1})\,, \label{eq:invmassdist}
\end{eqnarray}
where $M_{a_2a_1}^{\max}$ denotes the maximum value of $M_{a_2a_1}$ arising at $\cos\theta_{a_2a_1}^{*}=-1$, i.e., when $a_2$ and $a_1$ move in the back-to-back direction. It is a function of the three input mass parameters:
\begin{eqnarray}
    \left(M_{a_2a_1}^{\max}\right)^2=m_{A_2}^2(1-R_{12})(1 -R_{01}),
    \label{eq:endpoint1}
\end{eqnarray}
where we introduce a mass ratio symbol for purposes of later convenience\footnote{Note that $R_{01} \equiv m_{a_0}^2/m_{A_1}^2$.} 
\begin{equation}
R_{ij}\equiv m_{A_i}^2/m_{A_j}^2.
\end{equation}
As suggested by Eq.~\eqref{eq:invmassdist}, the $M_{a_2a_1}$ distribution increases linearly and sharply falls off at the kinematic endpoint defined in Eq.~\eqref{eq:endpoint1}. Therefore, the $M_{a_2a_1}$ invariant mass variable can be used as a kinematic cut to define the signal-rich region, and the measurement of the kinematic endpoint provides a relation among the three underlying mass parameters. Numerous experimental and phenomenological studies have adopted this variable for various physics applications. Examples include the top quark mass measurement~\cite{CMS-PAS-TOP-14-014}, as well as new particle searches and mass determinations in the context of supersymmetry, extra dimensions, and other BSM exotica.   

The shape described in Eq.~\eqref{eq:invmassdist} is valid as far as $A_1$ is either scalar or unpolarized and is produced on mass-shell with a negligible particle width. A nontrivial matrix element reshuffles and reweighs the relevant phase-space density, resulting in a shape distortion while keeping the endpoint unchanged. Indeed, many new physics models conceive the same experimental signatures, potentially along with the same decay topology \cite{Cheng:2002ab}. It has been realized that the shape analysis can be an important tool to understand the underlying dynamics~\cite{Barr:2004ze,Smillie:2005ar,Datta:2005zs,Athanasiou:2006ef,Alves:2006df,Wang:2006hk,Burns:2008cp}. 
Different spin correlations between the visible particles result from different spin assignments of $A_2$, $A_1$, and $a_0$, giving rise to different shapes of the $M_{a_2 a_1}$ distributions. For example, supersymmetric models and extra-dimensional models often give rise to an identical set of final-state visible particles under the same event topology; the shape analysis allows to discriminate the underlying scenarios~\cite{Barr:2004ze,Smillie:2005ar,Datta:2005zs,Athanasiou:2006ef,Alves:2006df,Wang:2006hk,Burns:2008cp,Kilic:2007zk,Csaki:2007xm}. A departure from Eq.~\eqref{eq:invmassdist} may also arise even in the absence of non-trivial spin correlations. It has been demonstrated that the non-negligible particle width of the intermediary particle $A_1$ encoded in its propagator can affect the shape, resulting in the extension of the $M_{a_2 a_1}$ distribution beyond its nominal endpoint in Eq.~\eqref{eq:endpoint1}~\cite{Grossman:2011nh}. The study of this sort has been generalized in a more systematic manner to the case where not only $A_1$ but $A_2$ and $a_0$ also have non-negligible particle widths~\cite{Kim:2017qdi}. Again the distributions are extended beyond the nominal endpoint, and, depending on the underlying mass spectrum, this endpoint ``violation'' effect can be appreciable for $\Gamma/m$ as low as 1\%, even in the presence of detector smearing~\cite{Kim:2017qdi}. In particular, this effect allows to test the nature of the invisible $a_0$, which is typically assumed to be a stable dark-matter candidate. However, it is also possible that it has a non-zero width due to its invisible decays to lighter dark-sector states. Therefore, this kind of shape analysis could discriminate between a true dark-matter candidate or an unstable (invisibly-decaying) dark-sector state~\cite{Kim:2017qdi}. 

The shape in Eq.~\eqref{eq:invmassdist} may differ from the expectation if the underlying physics does not obey the assumed two-step two-body cascade decay topology. For example, the intermediary state $A_1$ could be highly off-shell, or more invisible particles could be involved in the process in addition to $a_0$, or $A_2$ may decay to a pair of $A_1$'s, each of which decays to $a_0$ and $a_1$ [i.e., the so-called ``antler'' topology~\cite{Han:2009ss,Han:2012nm,Edelhauser:2012xb}, $A_2\to A_1 A_1 \to a_1 a_0 a_1 a_0$ depicted in Figure~\ref{fig:invmassdiagrams}($b$)]. The $M_{a_1a_1}$ distribution resulting from the antler event topology is given by 
\begin{equation}
    \frac{1}{\Gamma}\frac{d\Gamma}{d M_{a_1a_1}}\propto\left\{
    \begin{array}{l l}
   2 \eta M_{a_1a_1}    &  0< \frac{M_{a_1 a_1}}{M_{a_1 a_1}^{\max}}<e^{-2\eta} \\ 
      M_{a_1 a_1}\ln\frac{M_{a_1 a_1}^{\max}}{M_{a_1 a_1}}   & e^{-2\eta}< \frac{M_{a_1 a_1}}{M_{a_1 a_1}^{\max}}<1
    \end{array} \right., \label{eq:endpoint2}
\end{equation}
where $\cosh \eta=\frac{\sqrt{R_{12}}}{2}$ and the endpoint $M_{a_1 a_1}^{\max}$ is given by
\begin{equation}
    \left(M_{a_1 a_1}^{\max}\right)^2=m_{A_1}^2e^{2\eta}(1-R_{01})^2.
\end{equation}
It is noteworthy that the distribution shows a derivative discontinuity, i.e., cusp, at $M_{a_1a_1}=e^{-2\eta}M_{a_1a_1}^{\max}$~\cite{Han:2009ss}.
This is a kinematic feature unaffected by the underlying dynamics; the cusp feature remains intact in the presence of non-trivial spin correlations~\cite{Edelhauser:2012xb}.

In general, different event topologies resulting in only two visible particles $a_1$ and $a_2$ will give different $M_{a_2 a_1}$ distributions, and the phase-space shape information allows one to distinguish the underlying physics without any prior assumptions on the process and its detailed dynamics~\cite{Cho:2012er}. As an application, one can infer the number of invisible or dark-matter particles from the shape analysis and check whether or not the associated dark-matter stabilization symmetry is a $Z_2$ parity~\cite{Agashe:2010gt}.
As another application, it has been demonstrated that these various $M_{a_2 a_1}$ distributions can mimic resonance-induced distributions with a broad width especially at earlier stages of experiments, in the context of the 750 GeV diphoton excess~\cite{Cho:2015nxy}. 

These considerations have been extended to other event topologies resulting in a larger number of visible particles, e.g., three-step cascade decays~\cite{Allanach:2000kt,Lester:2001zx,Gjelsten:2004ki,Gjelsten:2005aw,Miller:2005zp,Costanzo:2009mq,Burns:2009zi,Matchev:2009iw,Agashe:2010gt,Kim:2015bnd,Matchev:2019sqa} and $\geq$3-body invariant mass variables~\cite{Gjelsten:2005aw,Alves:2006df,Bisset:2008hm,Kim:2015bnd}.  Let us illustrate the three-step two-body cascade decay case, $A_3 \to a_3 A_2$, $A_2 \to a_2 A_1$, $A_1 \to a_1 a_0$, and assume again that $a_3$, $a_2$ and $a_1$ are visible and massless, while $a_0$ is invisible and potentially massive [see Figure~\ref{fig:invmassdiagrams}($c$)]. The shapes and endpoints of the $M_{a_3 a_2}$ and $M_{a_2 a_1}$ invariant mass distributions follow Eqs.~\eqref{eq:invmassdist} and \eqref{eq:endpoint1} with mass parameters appropriately replaced. 
The unit-normalized $M_{a_3 a_1}$ distribution is~\cite{Miller:2005zp} 
\begin{equation}
   \frac{1}{\Gamma}\frac{d\Gamma}{d M_{a_3a_1}}=\left\{
    \begin{array}{l l}
    \frac{2M_{a_3 a_1}\ln R_{21}}{(M_{a_3 a_1}^{\max})^2 (1-R_{12})}    &  0< \frac{M_{a_3 a_1}}{M_{a_3 a_1}^{\max}}<\sqrt{R_{12}} \\ 
      \frac{2M_{a_3 a_1}\ln\frac{(M_{a_3 a_1}^{\max})^2}{M_{a_3 a_1}^2}}{(M_{a_3 a_1}^{\max})^2 (1-R_{12})}   & \sqrt{R_{12}}< \frac{M_{a_3 a_1}}{M_{a_3 a_1}^{\max}}<1
    \end{array} \right., \label{eq:endpoint3}
\end{equation}
where the endpoint $M_{a_3 a_1}^{\max}$ is given by
\begin{equation}
    \left(M_{a_3 a_1}^{\max}\right)^2=m_{A_3}^2(1-R_{23})(1-R_{01}).
\end{equation}
Note that the distribution again features a cusp at $M_{a_3 a_1}=M_{a_3 a_1}^{\max}\sqrt{R_{12}}$, as in the ``antler'' topology. The existence of the cuspy structure can be utilized in a model-independent fashion to distinguish $Z_3$-stabilized dark-matter models 
from $Z_2$-stabilized dark-matter models~\cite{Agashe:2010gt}, as it is unaffected by the underlying model details.  

As discussed in Section~\ref{sec:challenges}, it is often rather difficult to uniquely identify $a_1$, $a_2$ and $a_3$ on an event-per-event basis. For example, (light) quarks and gluons are never observed as isolated objects but as a clustering of hadronic objects, and it is very difficult to tell them apart on an individual case basis. For charged leptons, their electric charges are identified, but the order of appearance is not resolvable unless they come from the decay of long-lived particles and timing is appropriately measured. Therefore, the two-body invariant mass variables in the process accompanying $\geq 3$ visible particles are generally plagued by this combinatorial ambiguity.

The classic example from supersymmetry, which is often referred to as the ``$q\ell\ell$ chain'', is the decay of a supersymmetric quark $\tilde{q}$ to a quark, an opposite-sign lepton pair, and a lightest neutralino $\tilde{\chi}_1^0$ through two intermediary states: a heavier neutralino $\tilde{\chi}_2^0$ and a supersymmetric lepton $\tilde{\ell}$, i.e., $\tilde{q}\to q \tilde{\chi}_2^0, \tilde{\chi}_2^0 \to \ell^\pm_n \tilde{\ell}^\mp, \tilde{\ell}^\mp \to \ell^\mp_f \tilde{\chi}_1^0$. Here $\ell_{n(f)}$ denotes the final-state lepton closer (further) to the quark. 
Due to the combinatorial ambiguity associated with the leptons, $M_{q\ell_n}$ and $M_{q\ell_f}$ are not experimentally measurable quantities.
One possible trial would be to put them together into a single combined distribution $M_{q\ell} \equiv M_{q\ell_{n}}\cup M_{q\ell_{f}}$~\cite{Matchev:2009iw}. The larger endpoint, i.e., $\max\left(M_{q\ell_{n}}^{\max}, M_{q\ell_{f}}^{\max}\right)$, is measurable, whereas the smaller one may be buried in the middle of the $M_{q\ell}$ distribution. 
Nevertheless, this approach may be advantageous in the sense that the associated mass inversion formulas have a two-fold rather than a three-fold ambiguity~\cite{Matchev:2009iw}.
On the other hand, the two ordered invariant masses, $M_{q\ell}^>\equiv \max\left(M_{q\ell_{n}}, M_{q\ell_{f}}\right)$ and $M_{q\ell}^<\equiv \min\left(M_{q\ell_{n}}, M_{q\ell_{f}}\right)$, are experimentally measurable, and their respective endpoints can provide two independent mass relations depending on the underlying particle mass hierarchy. These ordered invariant masses have been extensively studied in the context of the  supersymmetric $q\ell\ell$ chain~\cite{Allanach:2000kt,Lester:2001zx,Gjelsten:2004ki,Gjelsten:2005aw,Miller:2005zp,Burns:2009zi,Matchev:2009iw}, and in a seesaw scenario~\cite{Dev:2015kca}. 

The idea of ordering the invariant masses was later generalized to the case where all visible particles in the final state are completely indistinguishable~\cite{Kim:2015bnd}. For a chosen fixed number of final state particles sampled from the final state, all possible invariant mass combinations are formed and then ranked, and the corresponding distributions are then inspected for the appearance of any upper kinematic endpoints. This systematic approach allows to access particular phase-space configurations through the respective kinematic endpoints and thus obtain independent mass relations which would be unavailable with the standard unranked invariant mass combinations. 

\subsection{Mass variables of semi-invisible collections of particles  \label{sec:mt2variables}}
In this section we discuss invariant mass-type variables whose definition takes advantage of the knowledge of the missing transverse momentum $\mptvec$ in the event. They usually target parent particles, whose decay products may include invisible particles.

The first kinematic variable we introduce is the transverse mass $M_T$ which is applicable to the case when a parent particle of mass $m_P$ decays semi-invisibly to a collection of visible daughter particles with total momentum $p^\mu = (E_{{\rm vis}}, \vec p_{{\rm vis},T}, p_{{\rm vis},z})$ plus an invisible daughter particle with total momentum $q^\mu = (E_{\rm inv}, \vec p_{{\rm inv},T}, q_{{\rm inv},z})$. Two examples of such semi-invisible decays are shown in Figures~\ref{fig:invmassdiagrams}($a$) and \ref{fig:invmassdiagrams}($c$). 
The momentum of the decaying parent particle is therefore $p^\mu + q^\mu$ and the reconstructed parent invariant mass $M_P$ is 
\begin{eqnarray}
M_P^2 &=& \big ( p + q \big )^2  
= m_{{\rm vis}}^2 + m_{\rm inv}^2 
\label{eq:MPdef}
\\
&& + 2 \big ( E_{{\rm vis}, T} E_{{\rm inv}, T} \cosh(\Delta \eta) - \vec p_{{\rm vis}, T} \cdot \vec q_{{\rm inv}, T} \big ) \, , \nonumber 
\end{eqnarray}
where the invariant mass $m_{\rm vis}$ of the visible sector is defined by $m_{\rm vis}^2 = p_\mu p^\mu$,  the mass $m_{\rm inv}$ of the invisible daughter is given by $m_{inv}^2 = q_\mu q^\mu$, and $\Delta \eta = \eta_{\rm vis} - \eta_{\rm inv}$ is the pseudo-rapidity difference between $p^\mu$ and $q^\mu$. The corresponding transverse energies are given by
\begin{eqnarray}
E_{{\rm vis}, T} &=& \sqrt{m_{\rm vis}^2 + \vec p^{\,2}_{{\rm vis}, T}} \, , \\ E_{{\rm inv}, T} &=& \sqrt{m_{\rm inv}^2 + \vec p^{\, 2}_{{\rm inv}, T}} \, .
\end{eqnarray}
If both $p^\mu$ and $q^\mu$ were observable, a particle of mass $m_P$ would appear as a resonance peak at $M_P=m_P$ in the invariant mass distribution $dN/dM_P$ [see Eq.~\eqref{eq:BWdistribution}]. However, if the daughter particle with momentum $q$ escapes the detector, one can only infer $\vec q_{{\rm inv},T}$ from the momentum conservation on the transverse plane as 
\begin{equation}
\vec q_{{\rm inv},T} = \mptvec \, .
\label{eq:qinvFormula}
\end{equation}
The longitudinal component $q_{{\rm inv},z}$ remains unknown and most of the discussion in the literature on kinematic variables has centered around the question how to deal with such missing information not just in this simple example, but in more general cases as well \cite{Barr:2010zj,Barr:2011xt}.

\begin{figure}[t]
\centering
\includegraphics[width=8.8cm]{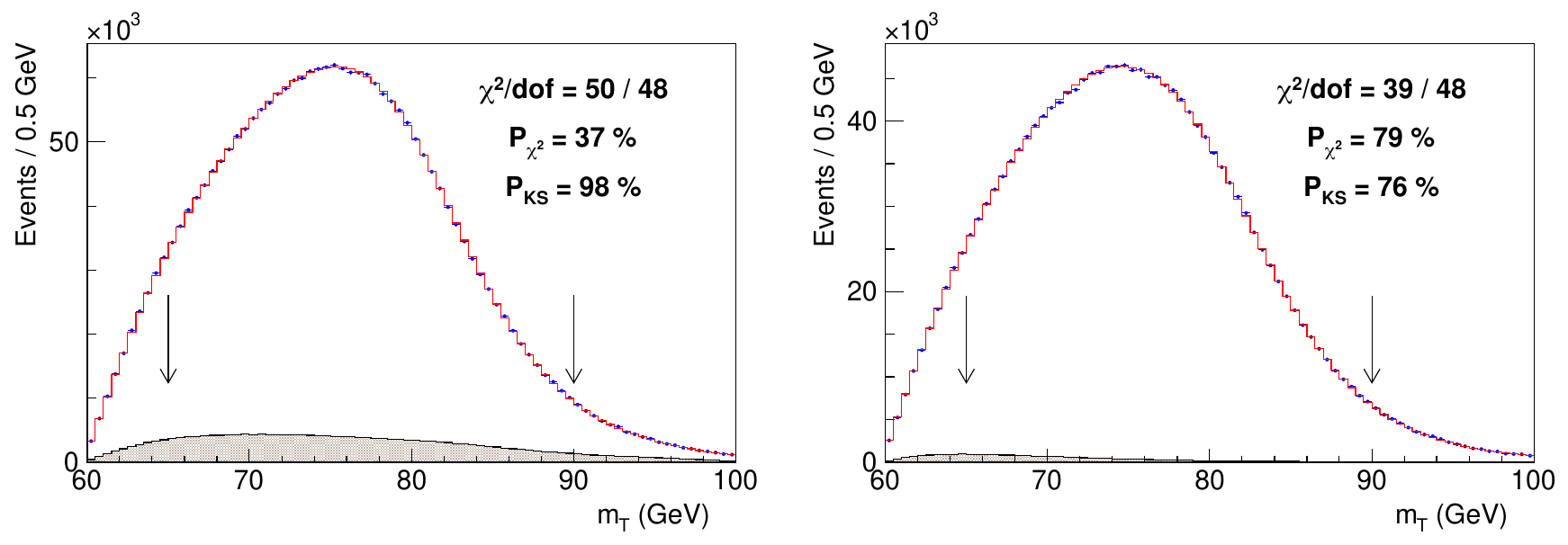}
\caption{\label{fig:mtw} Distribution of $M_T$ for $W$ boson to $\mu \nu_\mu$ (left) and $e\nu_e$ (right) final states for $m_{\rm inv}=0$. Simulation and data points are shown as histogram and data points, respectively. Taken from Ref. \cite{CDF:2022hxs}.}
\end{figure}

One very general approach is to obtain a variable which provides an event-wise {\em lower} bound on the parent mass $m_P$ \cite{Barr:2011xt}. For this purpose, one considers all possible values of the unknown invisible momentum components (in this simple case $q_{{\rm inv},z}$) and picks the smallest resulting value of the reconstructed parent mass (\ref{eq:MPdef}).
By minimizing Eq. (\ref{eq:MPdef}) over $q_{{\rm inv},z}$ [or simply by noticing that $\cosh(\Delta \eta) \geq 1$] one obtains the so-called transverse mass
\begin{eqnarray}
&& M^2_{TP}(\vec q_{{\rm inv}, T},m_{\rm inv}) = m_{\rm vis}^2 + m_{\rm inv}^2  \nonumber \\
 && \qquad\qquad + 2 \big ( E_{{\rm vis}, T} E_{{\rm inv}, T}  - \vec p_{{\rm vis}, T} \cdot \vec q_{{\rm inv}, T} \big ) ,
\label{eq:MTPdef} 
\end{eqnarray}
where $\vec q_{{\rm inv}, T}$ is given by (\ref{eq:qinvFormula}). 
By construction (since the minimization over $q_{{\rm inv},z}$ would inevitably include its true value in the event), the transverse mass satisfies the inequality 
\begin{equation}
M_{T P} \leq m_P.
\label{MTinequality}
\end{equation} 
The equality holds for events with $\Delta \eta = 0$. 
The most famous example for the use of the transverse mass is the discovery of the $W$ boson \cite{UA1:1983crd,UA2:1983tsx}. A recent measurement of the $W$ mass done by the CDF Collaboration (see Figure~\ref{fig:mtw}), which shows some tension with the SM and with previous measurements, is also based on the transverse mass \cite{CDF:2022hxs}, since $M_{T P}$ is less sensitive to the modelling of the $W$-$p_{T}$ spectrum. The distributions in Figure~\ref{fig:mtw} exhibit an upper kinematic endpoint, which, however, is smeared beyond the naive theoretical prediction (\ref{MTinequality}) due to the finite $W$ width and the detector resolution. We note that such use of the transverse mass relies on the assumption that there is only one missing particle in each event, so that (\ref{eq:qinvFormula}) can be used to find its transverse momentum $\vec q_{{\rm inv}, T}$. Next we shall discuss the case of multiple invisible particles in the same event.

\begin{figure}[t]
\centering
\includegraphics[width=8.5cm]{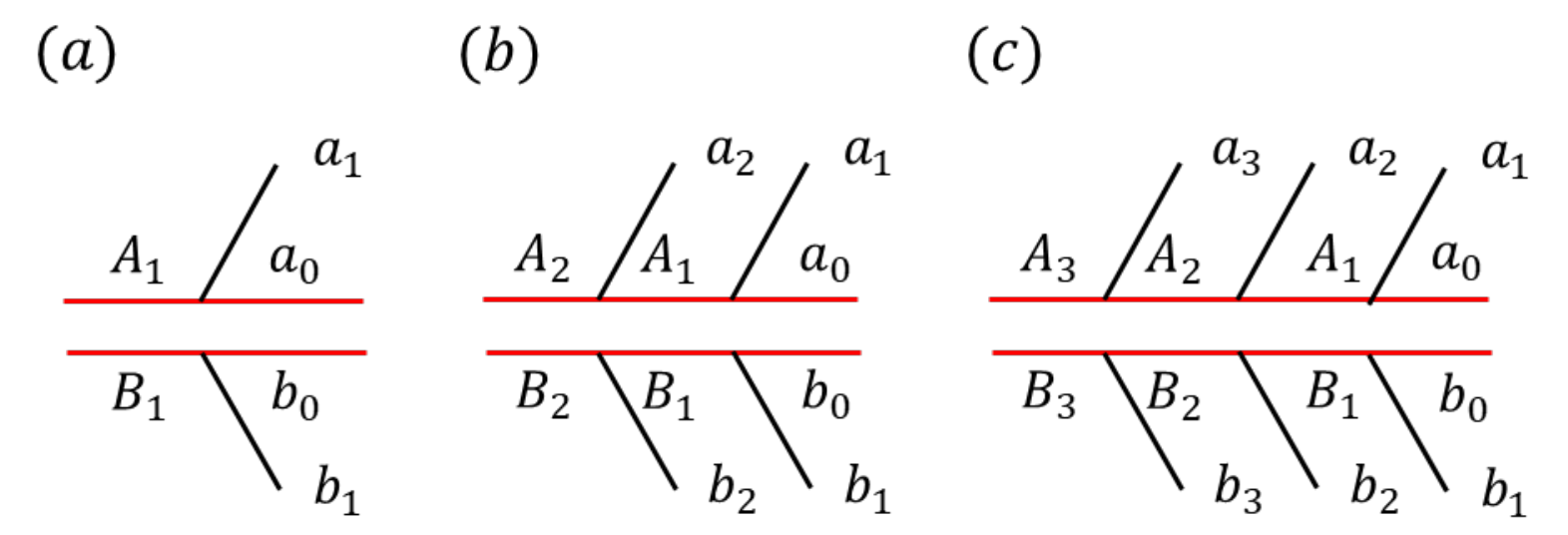}
\caption{\label{fig:mt2diagram}
Event topologies of pair-produced heavy resonances each of which undergoes ($a$) a one-step two-body decay, ($b$) a two-step two-body decay, and ($c$) a three-step two-body decay.}
\end{figure}

A well-motivated class of new physics models which generically predict a $\met$ signature, are models with dark-matter candidates. In such models, the lifetime of the dark-matter particle is typically protected by an exact discrete symmetry, which implies that the collider signals will involve not one, but {\it two} decay chains, each terminating in a dark-matter particle invisible in the detector. A few simple examples of such event topologies are shown in Figure~\ref{fig:mt2diagram}.

Let us start with the simplest case of a single two-body decay on each side of the event as in Figure~\ref{fig:mt2diagram}($a$) which was the inspiration for inventing the famous Oxbridge $M_{T2}$ variable~\cite{Lester:1999tx,Barr:2003rg}. Now we can form two transverse parent masses: $M_{TP_1}(\vec{q}_{1T},m_{a_0})$ for the first parent particle $P_1=A_1$, which depends on the transverse momentum $\vec{q}_{1T}$ and the mass $m_{a_0}$ of the invisible particle $a_0$; and $M_{TP_2}(\vec{q}_{2T},m_{b_0})$ for the second parent particle $P_2=B_1$, which depends on the transverse momentum $\vec{q}_{2T}$ and the mass $m_{b_0}$ of the invisible particle $b_0$. For simplicity, in what follows we shall assume symmetric event topologies, in which the parents $A_1$ and $B_1$, as well as the daughters $a_0$ and $b_0$ are the same (this assumption can be easily avoided, see Refs.~\cite{Barr:2009jv,Konar:2009qr}).  
In that case, the kinematics is governed by a single parent mass $m_P=m_{A_1}=m_{B_1}$ and a single daughter mass $m_0=m_{a_0}=m_{b_0}$.  If $\vec{q}_{1T}$ and $\vec{q}_{2T}$ were separately known, we would be assured that
\begin{equation}
\max [ M_{TP_1}(\vec{q}_{1T},m_{0}), M_{TP_2}(\vec{q}_{2T},m_{0}) ] 
\leq m_P.  
\end{equation}
However, $\vec{q}_{1T}$ and $\vec{q}_{2T}$ are not uniquely fixed by the $\mptvec$ constraint, as they are related by
\begin{equation}
\vec{q}_{1T} + \vec{q}_{2T} = \mptvec\,,
\label{eq:q1Tq2TMET}
\end{equation}
and the best we can do is to perform a minimization over all possible partitions of the $\mptvec$ into $\vec{q}_{1T}$ and $\vec{q}_{2T}$. This naturally leads to the definition of the Oxbridge $M_{T2}$ variable as \cite{Lester:1999tx,Barr:2003rg}
\bea
M_{T2} (\tilde m) &\equiv& \min_{\vec{q}_{1T},\vec{q}_{2T}}
\hspace*{-0.1cm}
\left\{\max\left[M_{TP_1}(\vec{q}_{1T},\tilde m),\,M_{TP_2} (\vec{q}_{2T},\tilde m)\right] \right\} ,\nonumber\\
\mptvec  &=& \vec{q}_{1T}+\vec{q}_{2T} \;, \label{eq:mt2def}
\eea 
where the {\it a priori} unknown daughter mass $m_0$ has been replaced with a test mass parameter $\tilde m$.
This construction guarantees on an event-by-event basis that
\begin{equation}
m_0 \leq M_{T2}(\tilde m = m_{0} ) \leq m_P \, .
\end{equation}
This fact can be used by constructing the $M_{T2}$ distribution, reading off its upper kinematic endpoint $M_{T2}^{\rm max}$ and interpreting it as
\begin{equation}
M_{T2}^{\rm max}(\tilde m = m_0) = m_P.
\label{eq:MT2max}
\end{equation}

The $M_{T2}$ concept can be readily applied to the more complex event topologies in Figures~\ref{fig:mt2diagram}($b$) and~\ref{fig:mt2diagram}($c$), where one has several choices of designating parent and daughter particles, leading to a menagerie of different ``subsystem" $M_{T2}$ variables \cite{Kawagoe:2004rz,Burns:2008va}. 
In general, the minimization in Eq.~(\ref{eq:mt2def}) has to be done numerically (see Table \ref{tab:codes}). However, for certain special cases, analytical solutions have been derived \cite{Barr:2003rg,Lester:2011nj,Lally:2012uj,Lester:2007fq,Cho:2007qv,Cho:2007dh}. For example, when the minimization results in the case of $M_{TP_1} = M_{TP_2}$, which is known as the balanced solution $M_{T2}^{B}$, the analytic expression for the symmetric $M_{T2}$ variable is given by \cite{Cho:2007qv,Lester:2007fq} 
\begin{eqnarray}  
&&\Big[M_{T2}^{B}{(\tilde{m})}\Big]^{2}
= \tilde{m}^2 + A_T \label{eq:mt2oldB}\\
&&\hspace*{0.5cm}+ 
          \sqrt{ \left ( 1 + \frac{4 \tilde{m}^2}{2A_T-m_1^2-m_2^2} \right ) 
             \left ( A_T^2 - m_1^2 ~m_2^2  \right ) } \,,\nonumber
\end{eqnarray}
where $m_{i}$, $\vec{p}_{iT}$, and $E_{iT}$ are respectively the mass, the transverse momentum, and the transverse energy of the visible particles in the $i$th decay chain ($i=1,2$) and $A_T$ is a convenient shorthand notation introduced in \cite{Cho:2007dh} 
\begin{equation}
A_T = E_{1T} E_{2T} + \vec{p}_{1T}\cdot\vec{p}_{2T} \, .
\label{ATdef}
\end{equation}

\begin{figure}[t!]
\centering
\includegraphics[width=0.23\textwidth]{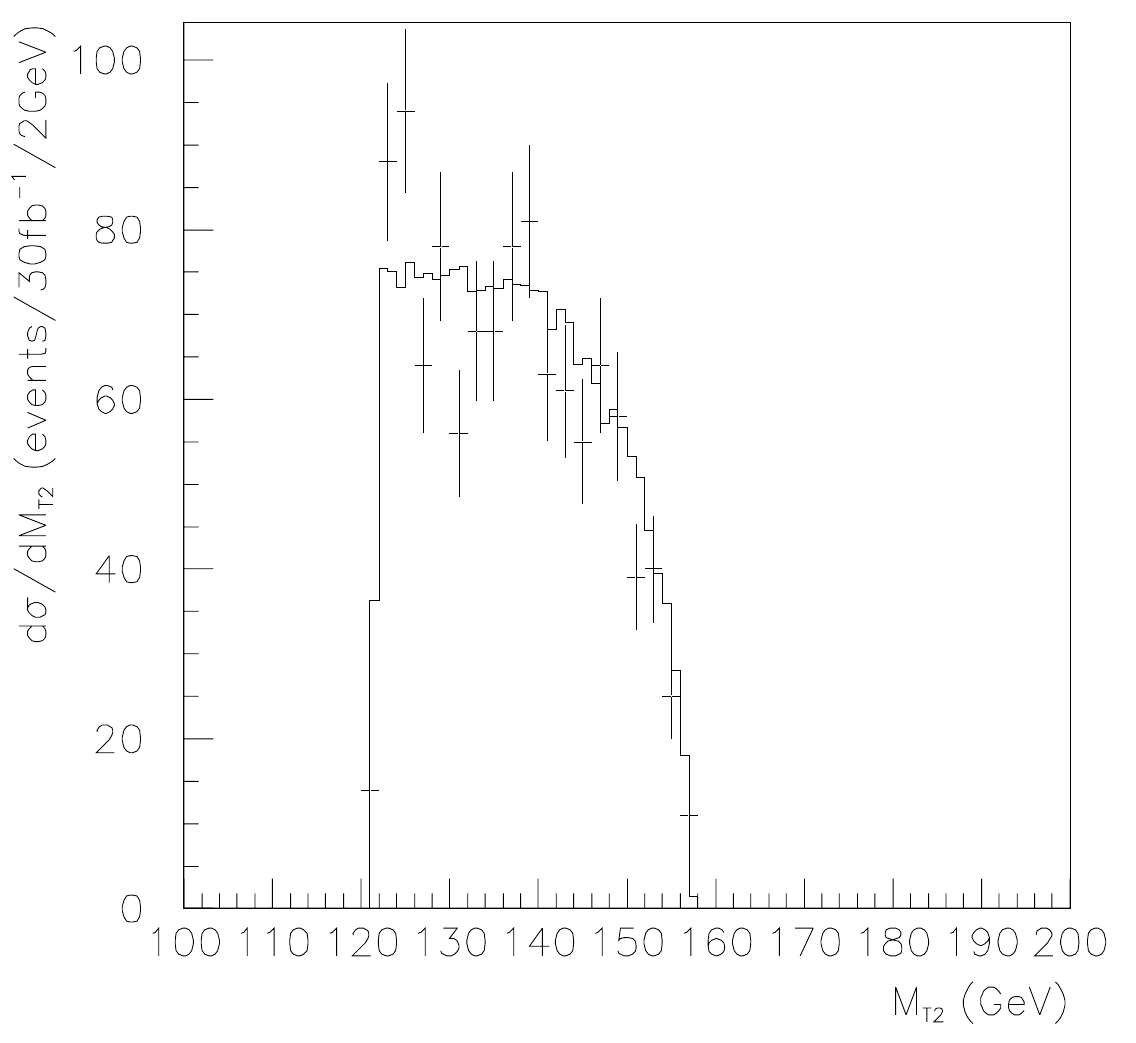}
\includegraphics[width=0.227\textwidth]{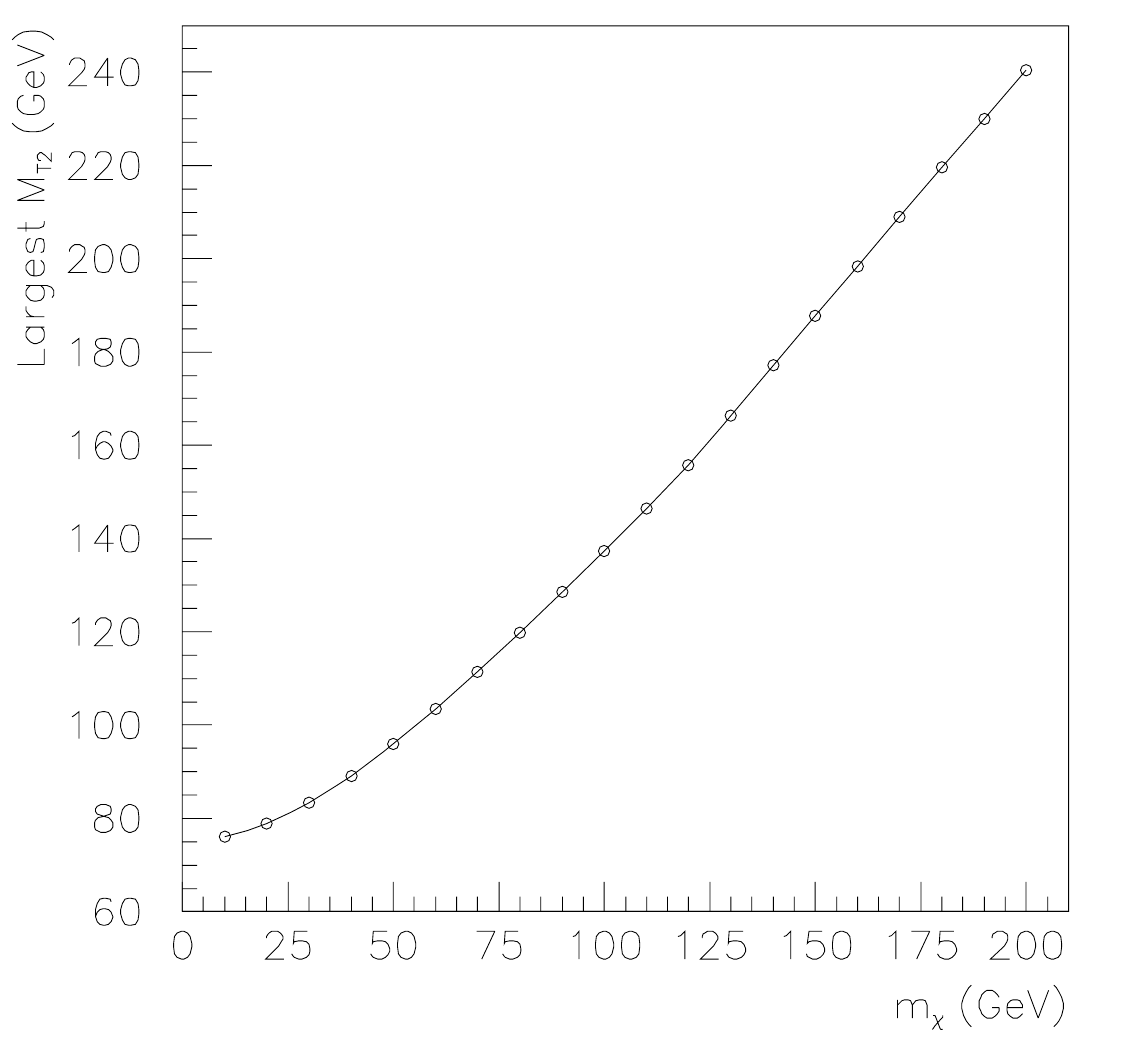} 
\caption{\label{fig:mt2slepton} Left: $M_{T2}$ distribution for slepton production $p p \to \tilde \ell \tilde\ell^\ast \to \ell^+\ell^- + \mptvec$ at the LHC, assuming the actual value for the test mass $\tilde m$. Slepton and neutralino masses are set to $m_{\tilde \ell} = 157.1$ GeV and $m_{\tilde \chi_1^0}=121.5$ GeV, respectively. Right: Values of $m_{\tilde\ell}$ as a function of the test mass. Taken from Ref. \cite{Lester:1999tx}.}
\end{figure}

\begin{figure}[t]
    \centering
    \includegraphics[width=6cm]{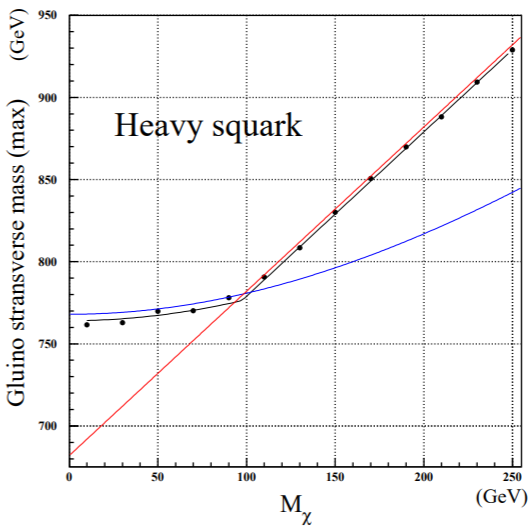}
    \caption{$M_{T2}^{\rm max}$ as a function of the test mass for gluino pair production at the LHC $p p \to \tilde g \tilde g \to j j j j + \mptvec$. Taken from Ref. \cite{Cho:2007qv}. }
    \label{fig:mt2kink}
\end{figure}

A sample distribution of $M_{T2}$ \cite{Lester:1999tx} for slepton production $p p \to \tilde \ell \tilde\ell^\ast \to \ell^+\ell^- + \mptvec$ at the LHC is shown in the left panel of Figure~\ref{fig:mt2slepton} for slepton mass $m_{\tilde \ell} = 157.1$ GeV and neutralino mass $m_{\tilde \chi_1^0}=121.5$ GeV. The test mass $\tilde m$ is taken to be equal to the true mass of the missing particle, $\tilde m = m_{\tilde \chi_1^0}$. The distribution clearly shows the expected endpoint (\ref{eq:MT2max}) at $M_{T2}^{\rm max} = m_{\tilde \ell} = 157.1$ GeV.
The right panel of Figure~\ref{fig:mt2slepton} shows the measured values of $M_{T2}^{\rm max}$ as a function of the test mass. Interpreting the endpoint ($M_{T2}^{\rm max}$) of $M_{T2}$ as the mass of the decaying particle according to (\ref{eq:MT2max}), Figure~\ref{fig:mt2slepton} reduces the two dimensional mass parameter space ($m_{\tilde \chi_1^0}$, $m_{\tilde \ell}$) to one dimension. 
One additional independent measurement would then be able to fix both the masses of the parent and the  daughter particles.
In fact, the $M_{T2}$ itself could provide such measurement via a kink structure, which may arise for a number of reasons, e.g., due to non-trivial invariant mass in the visible sector~\cite{Cho:2007qv,Cho:2007dh}, due to initial state radiation \cite{Gripaios:2007is,Barr:2007hy}, or due to upstream momentum from decays further up the chain \cite{Burns:2008va}. 
Figure~\ref{fig:mt2kink} shows an example of such a kink structure which appears in the gluino pair production at the LHC $p p \to \tilde g \tilde g \to j j j j + \mptvec$ \cite{Cho:2007qv}. 
The black dots are data points generated via simulation and the blue and red curve represent the best fit for $\tilde m < m_{\tilde \chi_1^0}$ and $\tilde m > m_{\tilde \chi_1^0}$, respectively. Their intersection corresponds to the true mass input, ($m_{\tilde \chi_1^0}$, $m_{\tilde \ell}$) = (780.3, 97.9) GeV. 
Another example of an $M_{T2}$ kink structure will be discussed in Figure~\ref{fig:ISRkink} below. 

\begin{figure}[t]
\centering
\includegraphics[width=8.8cm]{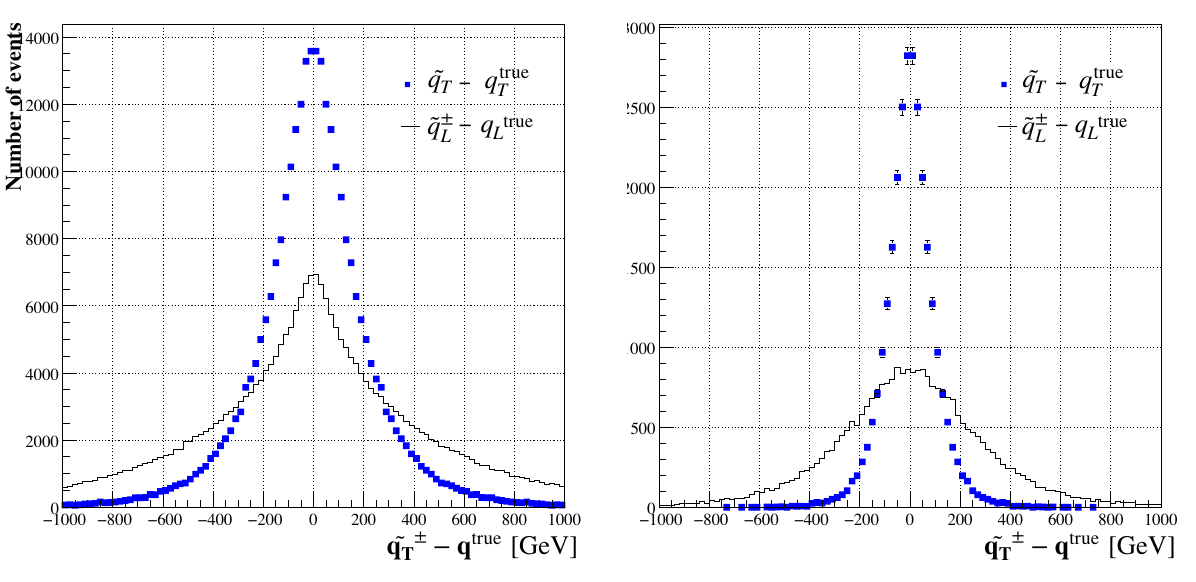}
\caption{\label{fig:maos}
The distributions of $\tilde{q}^{\pm} - q^{\rm true}$ for the full event set (left), and for the top 10\% events closest to the $M_{T2}$ endpoint (\ref{eq:MT2max}) (right). Here the MAOS momenta were constructed with $\tilde m=0$. Taken from Ref. \cite{Cho:2008tj}.}
\end{figure}

An interesting and important observation is that the result from the minimization involved in the $M_{T2}$ definition provides an ansatz $\tilde {\mathbf{q}}_{iT}$ for the transverse momentum of each missing particle \cite{Cho:2008tj}. The accuracy of this approximation improves in the vicinity of the upper kinematic endpoint (\ref{eq:MT2max}) of the $M_{T2}$ distribution, i.e., when $M_{T2} \approx m_{P}$. Armed with the ansatz for the transverse invisible momenta, one can use on-shell conditions to reconstruct the longitudinal momenta of the missing particles. In other words, fixing $\vec q_{iT}=\tilde{\mathbf{q}}_{i T}$, the longitudinal momenta of the missing particles can be determined as 
\begin{eqnarray}
\label{twofold} 
\tilde {q}_{iL}^\pm(\tilde m) &=& 
\frac{1}{(E_{iT})^2}\Big [ \,  p_{iL}A_i \\
&& \hspace*{0.3cm}  \pm
\sqrt{p_{iL}^2+(E_{iT})^2}\sqrt{A_i^2-(E_{iT}E^\chi_{iT})^2} \, \Big ] \, , \nonumber
\end{eqnarray}
where
$E_{iT}=\sqrt{m_{i}^2+|\mathbf{p}_{iT}|^2}$, $E^\chi_{iT}=\sqrt{\tilde m^2+|\tilde {\mathbf{q}}_{iT}|^2}$, and $A_i=\frac{1}{2}\left\{[M_{T2}^{\rm max}(\tilde m)]^2 - \tilde m^2 - m_{i}^2\right\}+\vec {p}_{iT}\cdot \tilde{\mathbf{q}}_{iT}$ for $\vec q_{iT} = \tilde{\mathbf{q}}_{iT}$.
This method of finding the momenta of missing particles is known as $M_{T2}$-Assisted On-Shell (MAOS) reconstruction \cite{Cho:2008tj}.
Figure \ref{fig:maos} illustrates the accuracy of the MAOS reconstruction for gluino pair production at the LHC, $ p p \to \tilde g \tilde g \to (jj\tilde \chi_1^0)(jj \tilde \chi_1^0)$ with $m_{\tilde \chi_1^0} = 122$ GeV, and $m_{\tilde g}=779$ TeV.
The MAOS reconstruction has been used for many collider studies including Higgs boson searches~\cite{Choi:2009hn,Choi:2010dw,Barr:2011he}, heavy resonance searches \cite{Park:2011uz}, spin measurement~\cite{Guadagnoli:2013xia} etc.

\begin{figure}[t]
\begin{center}
\includegraphics[width=.4\textwidth]{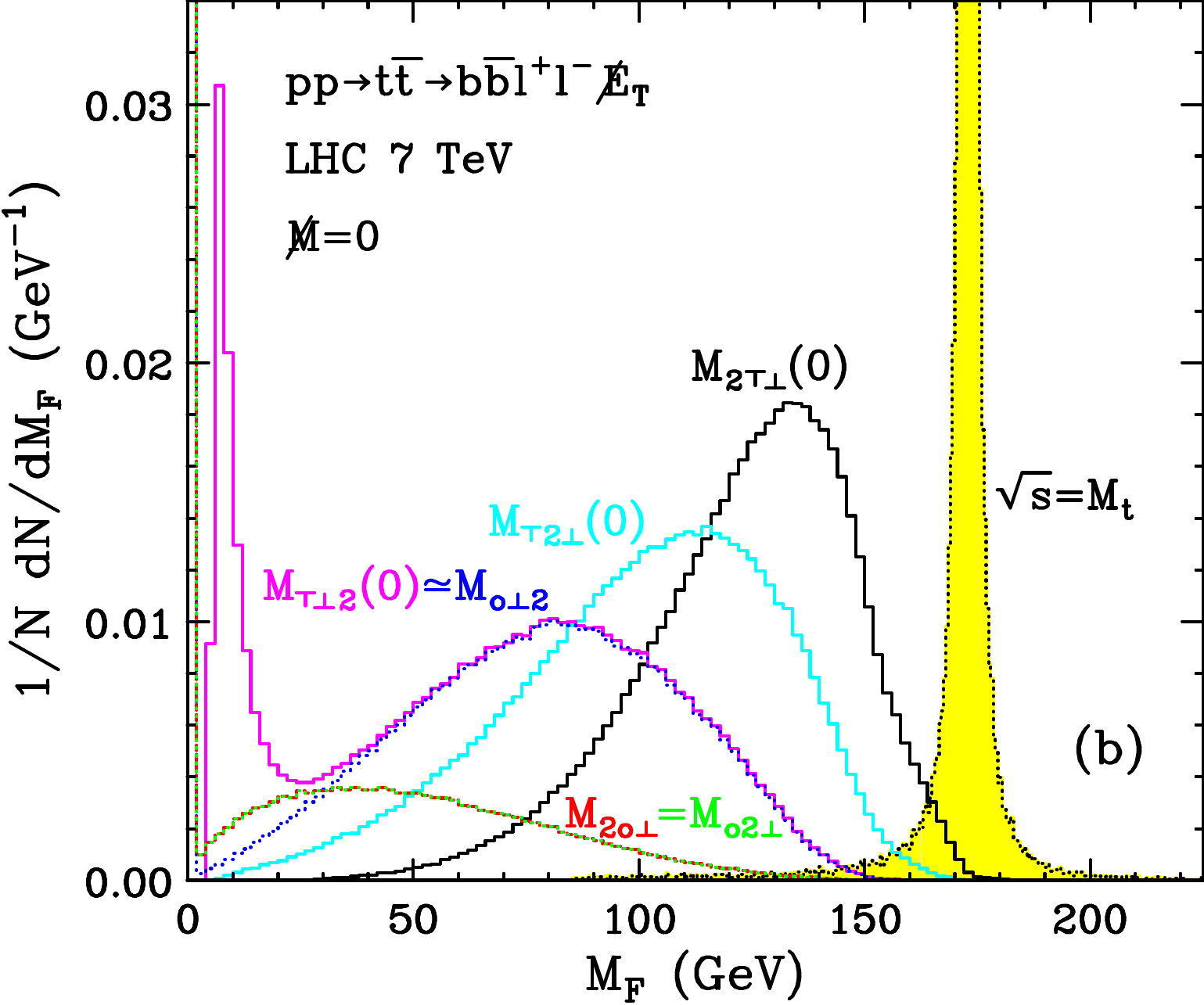}
\end{center}
\caption{Distributions of various doubly projected variables in the case of dilepton top quark production:
$M_{2 \ourT \myT}$ (black), 
$M_{\ourT 2  \myT}$ (cyan),
$M_{\ourT \myT 2 }$ (magenta),
$M_{2 \massless \myT}$ (red),
$M_{\massless 2  \myT}$ (green) 
and $M_{\massless \myT 2}$ (blue).
The yellow shaded distribution gives the average top quark mass in the event. 
The different types of projections (mass-preserving `$\myT$', speed-preserving `$\ourperp$' and massless `$\massless$') are defined in Ref.~\cite{Barr:2011xt}.
Taken from Ref.~\cite{Barr:2011xt}.}
\label{fig:doubly}
\end{figure}

In the presence of upstream (ISR) transverse momentum $\vec{P}_T$ and for arbitrary configurations of the visible transverse momenta $\vec{p}_{1T}$ and $\vec{p}_{2T}$, a general analytical formula for the calculation of $M_{T2}$ is still lacking. References~\cite{Konar:2009wn,Matchev:2009ad} discussed an interesting way of removing the effect of the ISR and retrieving an analytic solution. The basic idea is to decompose the transverse momenta $\vec{p}_{1T}$ and $\vec{p}_{2T}$ further onto the direction ($T_\parallel$) defined by the $\vec{P}_T$ vector and the direction ($T_\perp$) orthogonal to it:
\begin{eqnarray}
\vec p_{iT_\parallel} &\equiv& \frac{1}{P_T^2}\left(\vec{p}_{iT}\cdot\vec{P}_T\right)\vec{P}_T ,
\label{pitpar} \\
\vec p_{iT_\perp} &\equiv& \vec{p}_{iT}-\vec p_{iT_\parallel}
= \frac{1}{P_T^2} \vec{P}_T \times \left(\vec{p}_{iT}\times \vec{P}_T\right),
\label{pitperp}
\end{eqnarray}
and similarly for the transverse momenta $\vec{q}_{1T}$ and $\vec{q}_{2T}$ of the daughters and for $\mptvec$. Now consider the corresponding 1D decompositions of the transverse parent masses 
\begin{eqnarray}
M_{T_\parallel P_i}^{2} &\equiv&
m_i^2 + \tilde m^2 + 2\left(E_{iT_\parallel} E_{iT_\parallel}^\chi
- \vec{p}_{iT_\parallel}\cdot\vec{q}_{iT_\parallel}\right), \nonumber
\label{MTparallel}
\\
M_{T_\perp P_i}^{2}  &\equiv& 
m_i^2 + \tilde m^2 + 2\left(E_{iT_\perp} E_{iT_\perp}^\chi
- \vec{p}_{iT_\perp}\cdot\vec{q}_{iT_\perp} \right),\nonumber
\label{MTperp}
\end{eqnarray}
in terms of the 1D projected analogues of the transverse energy
\begin{eqnarray}
E_{iT_\parallel} &\equiv& \sqrt{m_i^2+|\vec p_{iT_\parallel}|^2}, \quad
E_{iT_\perp}     \equiv \sqrt{m_i^2+|\vec p_{iT_\perp}|^2}, \nonumber \\
E_{iT_\parallel}^\chi &\equiv& \sqrt{\tilde m^2+|\vec{q}_{iT_\parallel}|^2 }, \quad
E_{iT_\perp}^\chi \equiv \sqrt{\tilde m^2+|\vec{q}_{iT_\perp}|^2 }.
\nonumber
\end{eqnarray}
Now we define 1D $M_{T2}$ decompositions
in complete analogy with the standard $M_{T2}$ definition 
(\ref{eq:mt2def}):
\begin{eqnarray}
M_{T2_\parallel} 
&\equiv& 
\min_{\vec{q}_{1T_\parallel} + \vec{q}_{2T_\parallel} = \mptvecpar }
\left\{\max\left [ M_{T_\parallel P_1},M_{T_\parallel P_2}\right ] \right\} ,
\label{eq:mt2parallel} \\
M_{T2_\perp}     
&\equiv& 
\min_{\vec q_{1T_\perp} + \vec q_{2T_\perp} = \mptvecperp}
\left\{\max\left [ M_{T_\perp P_1},M_{T_\perp P_2}\right ]  \right\}.
\label{eq:mt2perp}
\end{eqnarray}
By construction, $M_{T2\perp}$ does not suffer from ISR effects, since it concerns the direction orthogonal to the ISR. Therefore one can use the existing formula \eqref{eq:mt2oldB} to compute $M_{T2_\perp}$ analytically. 
Some examples of the doubly projected variables are shown in Figure \ref{fig:doubly} taking dilepton top quark production as an example. We refer to Ref. \cite{Barr:2011xt} for more details on the different types of projections (mass-preserving `$\myT$', speed-preserving `$\ourperp$' and massless `$\massless$' projections) and the order of projection and agglomeration of visible particles.  

$M_{T2}$ has various applications in collider physics and has been further developed for more complicated topologies. Examples include $M_{TGen}$ (avoiding the combinatorics problem by iterating over all possible partitions of the visible set of final state particles) \cite{Barr:2003rg,Lester:2007fq}
asymmetric $M_{T2}$ (associated production and non-identical pair decays) \cite{Barr:2009jv,Konar:2009qr,Agashe:2010tu}, generalization to the case with multiple invisible particles \cite{Mahbubani:2012kx}, application to dark-matter stabilization symmetries \cite{Agashe:2010tu,Kim:2016ixu}, 
CDF top quark mass measurement using $M_{T2}$ \cite{CDF:2009zjw}, CMS top quark mass measurement using $M_{b\ell}$, $M_{T2}$, and $M_{T2,\perp}$/MAOS~\cite{CMS:2012eya, CMS:2016kgk}, and ISR tagging \cite{Alwall:2009zu,Kim:2015uea} etc. Since the analytic expression for the general case is unknown, one must use a code to compute $M_{T2}$ numerically \cite{Cho:2015laa,Park:2020bsu}. Special algorithms have been suggested for faster and more accurate calculation \cite{Cheng:2008hk,Lester:2014yga}. Some of these codes are summarized in Table \ref{tab:codes}. 

Another mass-constraining variable is the $M_2$ variable \cite{Barr:2011xt,Cho:2014yma,Cho:2015laa,Cho:2014naa}, which is the (3+1)-dimensional version of Eq.~(\ref{eq:mt2def}): 
\bea
M_{2} (\tilde m) &\equiv& \min_{\vec{q}_{1},\vec{q}_{2}}\left\{\max\left[M_{P_1}(\vec{q}_{1},\tilde m),\;M_{P_2} (\vec{q}_{2},\tilde m)\right] \right\} \, ,\nonumber\\
\vec{q}_{1T}+\vec{q}_{2T} &=& \mptvec  \;, \label{eq:m2def}
\eea 
where we use the {\em actual} parent masses $M_{P_i}$ from (\ref{eq:MPdef}) instead of their transverse masses $M_{T P_i}$ from (\ref{eq:MTPdef}).
Note that the minimization is now performed over the 3-component momentum vectors $\vec{q}_{1}$ and $\vec{q}_{2}$ \cite{Barr:2011xt,Cho:2014yma,Cho:2015laa,Cho:2014naa}.
In fact, at this point the two definitions (\ref{eq:mt2def}) and (\ref{eq:m2def}) are equivalent, in the sense that the resulting two variables, $M_{T2}$ and $M_2$, will have the same numerical value \cite{Ross:2007rm,Barr:2011xt,Cho:2014naa}. 

However, $M_2$ begins to differ from $M_{T2}$ when applying additional kinematic constraints 
beyond the missing transverse momentum condition $\vec{q}_{1T}+\vec{q}_{2T} = \mpt $. Then the $M_2$ variable can be further refined and one can obtain non-trivial variants as shown below \cite{Cho:2014naa}:
\bea
M_{2CX} (\tilde m) &\equiv& \min_{\vec{q}_{1},\vec{q}_{2}}\left\{\max\left[M_{P_1}(\vec{q}_{1},\tilde m),\;M_{P_2} (\vec{q}_{2},\tilde m)\right] \right\},  \nonumber
\\
\vec{q}_{1T}+\vec{q}_{2T} &=& \mpt  \label{eq:m2CXdef}  \\
M_{P_1}&=& M_{P_2} \nonumber 
\eea 
\bea
M_{2XC}(\tilde m) &\equiv& \min_{\vec{q}_{1},\vec{q}_{2}}\left\{\max\left[M_{P_1}(\vec{q}_{1},\tilde m),\;M_{P_2} (\vec{q}_{2},\tilde m)\right] \right\},  \nonumber\\
\vec{q}_{1T}+\vec{q}_{2T} &=& \mpt   \label{eq:m2XCdef} \\
M_{R_1}^2&=& M_{R_2}^2 \nonumber 
\eea 
\bea
M_{2CC}(\tilde m) &\equiv& \min_{\vec{q}_{1},\vec{q}_{2}}\left\{\max\left[M_{P_1}(\vec{q}_{1},\tilde m),\;M_{P_2} (\vec{q}_{2},\tilde m)\right] \right\}.\nonumber
\\
\vec{q}_{1T}+\vec{q}_{2T} &=& \mpt  \label{eq:m2CCdef}  \\
M_{P_1}&=& M_{P_2} \nonumber  \\
M_{R_1}^2&=& M_{R_2}^2 \nonumber 
\eea 
Here $M_{P_i}$ ($M_{R_i}$) is the mass of the parent (relative) particle in the $i$th decay chain and a subscript ``$C$'' 
indicates that an equal mass constraint is applied for the two parents (when ``$C$'' is in the first position) or for the relatives
(when ``$C$'' is in the second position). A subscript ``$X$'' simply means that no such constraint is applied.
In any given subsystem, 
these variables are related as follows \cite{Cho:2014naa}
\bea
M_{T2} = M_{2CX} \leq M_{2XC} \leq M_{2CC}. 
\label{eq:hierarchy}
\eea

Besides constraints enforcing mass equality between two different particles, we can also enforce the measured values of some masses. For example, in the $t \bar t$ event topology, we could use the experimentally measured $W$-boson mass, $m_W$, and introduce the following further constrained variable: 
\bea
M_{2CW}^{(b\ell)} (\tilde m =0) &\equiv& \min_{\vec{q}_{1},\vec{q}_{2}}\left\{\max\left[M_{t_1}(\vec{q}_{1},\tilde m),\;M_{t_2} (\vec{q}_{2},\tilde m)\right] \right\}  . \nonumber
\\
\vec{q}_{1T}+\vec{q}_{2T} &=& \mptvec  \label{eq:m2CWdef} \\
M_{t_1}&=& M_{t_2} \nonumber  \\
M_{W_1}&=& M_{W_2} = m_W \nonumber 
\eea 

Similarly, using the measured mass $m_t$ of the top quark, we can define a new variable in the $(\ell)$ subsystem:
\bea
M_{2Ct}^{(\ell)} (\tilde m =0) &\equiv& \min_{\vec{q}_{1},\vec{q}_{2}}\left\{\max\left[M_{W_1}(\vec{q}_{1},\tilde m),\;M_{W_2} (\vec{q}_{2},\tilde m)\right] \right\}.\nonumber
\\
\vec{q}_{1T}+\vec{q}_{2T} &=& \mptvec \label{eq:m2Ctdef}   \\
M_{W_1}&=& M_{W_2} \nonumber  \\
M_{t_1}&=& M_{t_2} = m_t \nonumber 
\eea 

\begin{figure}[t]
\centering
\includegraphics[width=4.2cm]{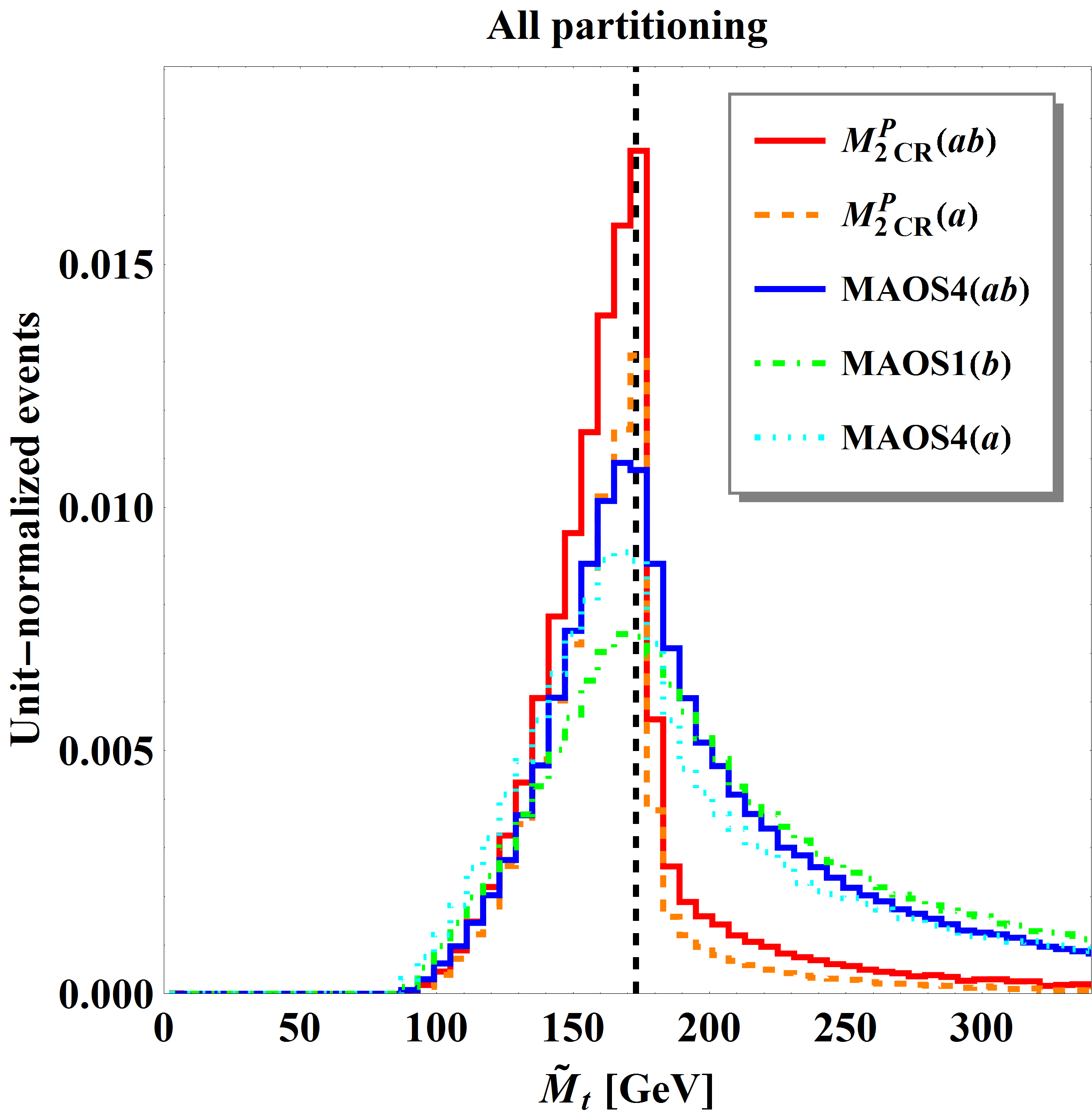} 
\hspace*{-0.2cm}
\includegraphics[width=4.2cm]{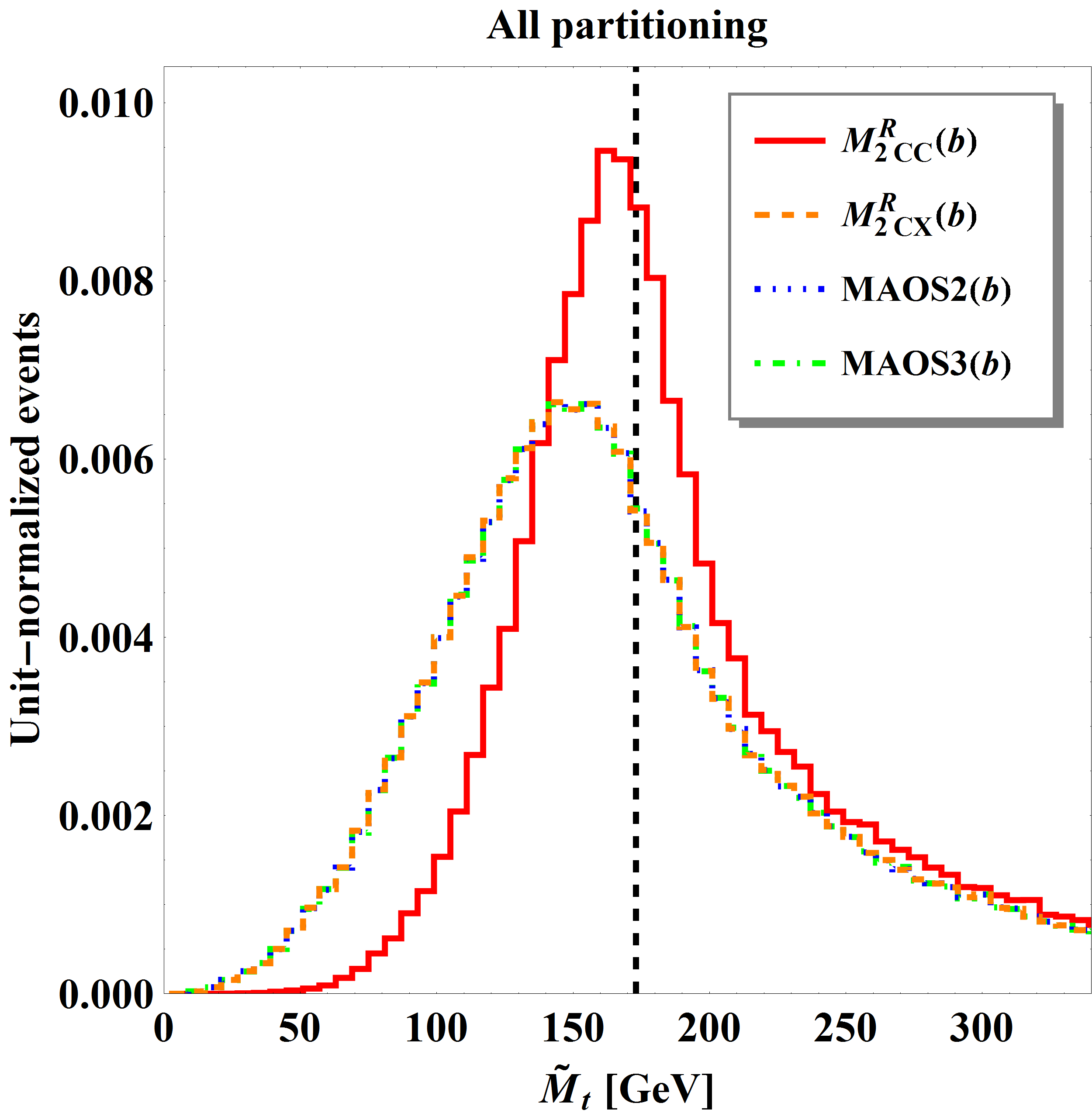}
\caption{\label{fig:reco_top} Comparison of the MAOS and $M_2$-assisted methods for top mass reconstruction. The left panel shows distributions of the reconstructed top mass $\tilde M_t$ with methods which use two mass inputs (the $W$-boson mass and the neutrino mass): the three MAOS methods, MAOS4($ab$) (blue solid line), MAOS1($b$) (green dot-dashed line), and MAOS4($a$) (cyan dotted line),
and the two $M_2$-based methods, $M_{2CR}(ab)$ (red solid line) and $M_{2CR}(a)$ (orange dashed line).  
The right panel shows distributions of the reconstructed top mass $\tilde M_t$ with methods which use a single mass input (the neutrino mass): MAOS2($b$) (blue dotted line) and MAOS3($b$) (green dot-dashed line) and $M_{2CX}(b)$ (orange dashed line) and $M_{2CC}(b)$ (red solid line). Taken from Ref. \cite{Kim:2017awi}.}
\end{figure}

\begin{figure}[tbp]
\centering
\includegraphics[width=4.2cm]{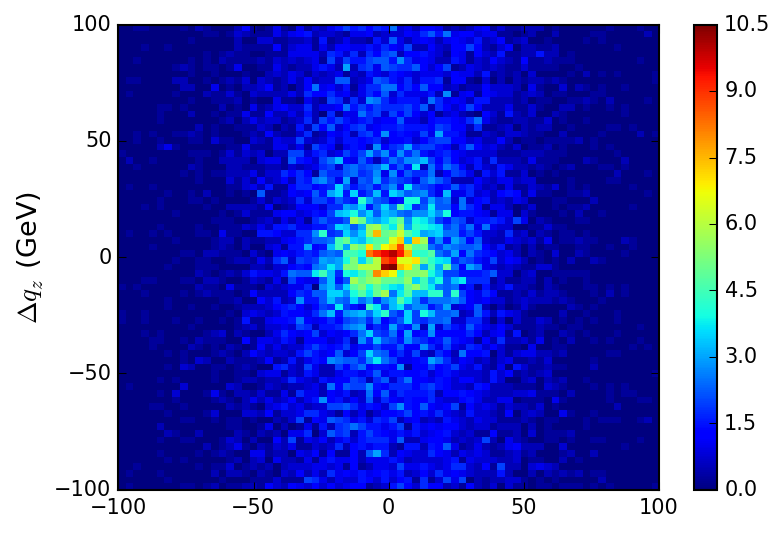} 
\hspace*{-0.2cm}
\includegraphics[width=4.2cm]{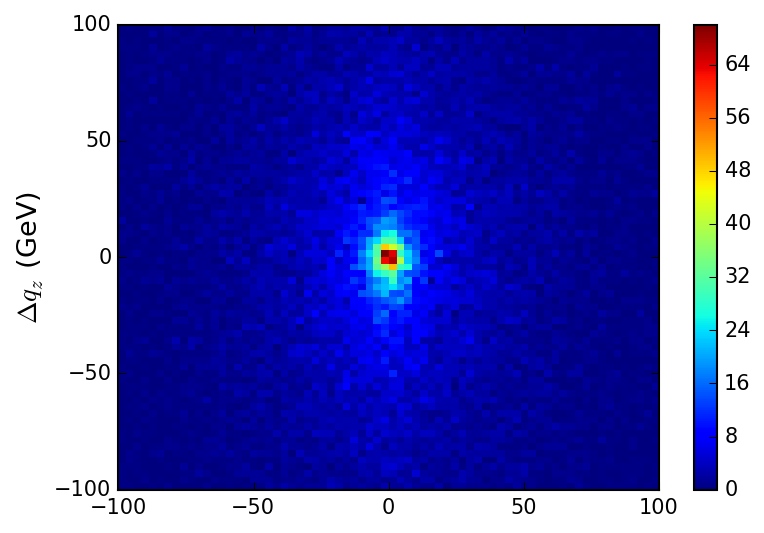} 
\caption{\label{fig:momentum2dmass}Correlations between $\Delta q_z$ and $\Delta q_x$ for MAOS1($b\ell$;$m_t$) (left) and $M_{2Ct}^{(\ell)}$ (right). Taken from Ref. \cite{Debnath:2017ktz}.}
\end{figure}

Just like the minimization in the $M_{T2}$ calculation allowed for the MAOS reconstruction of invisible momenta, the minimization in the $M_2$ computation provides a flexible and convenient reconstruction of the full missing momenta ($M_2$-assisted onshell reconstruction). Figure~\ref{fig:reco_top} compares the reconstructed top mass using a variety of MAOS and $M_2$ reconstruction schemes. In general, the momentum ansatz obtained from $M_2$ allows a sharper distribution with a shorter tail. This is due to better precision in the missing momentum reconstruction, as illustrated in Figure~\ref{fig:momentum2dmass}. In addition to invisible momentum reconstruction \cite{Kim:2017awi}, 
the $M_2$ has been used in various collider analyses including 
applications to Higgs mass measurement \cite{Konar:2016wbh}, new particle mass measurements \cite{Lim:2016ymd,Kim:2017awi,Baer:2016wkz}, distinguishing symmetric/asymmetric events \cite{Lim:2016ymd}, resolving combinatorial ambiguities \cite{Debnath:2017ktz,Alhazmi:2022qbf}, measurement of the Top-Higgs Yukawa CP structure \cite{Goncalves:2018agy,Goncalves:2021dcu}, applications to compressed stop search \cite{Cho:2014yma}, $Z'$ search \cite{Kim:2014ana}, and application in the antler-topology \cite{Konar:2015hea}, etc  \cite{Konar:2015hea,Konar:2016wbh,Swain:2014dha,Konar:2017kgm}.


There are a few other variants of the transverse mass. The cotransverse mass $M_{C}$ and the contransverse mass $M_{CT}$ are defined as \cite{Tovey:2008ui}
\begin{eqnarray}
M_C^2 &=& \Big ( E_{1} + E_{2} \Big )^2 - \Big ( \vec p_{1} - \vec p_{2} \Big )^2 \nonumber \\
&=& m_1^2 + m_2^2 + 2 ( E_{1} E_{2} + \vec p_{1} \cdot \vec p_{2}) \\
M_{CT}^2 &=& \Big ( E_{1T} + E_{2T} \Big )^2 - \Big ( \vec p_{1T} - \vec p_{2T} \Big )^2 \nonumber  \\
&=& m_1^2 + m_2^2 + 2 ( E_{1T} E_{2T} + \vec p_{1T} \cdot \vec p_{2T}) \, ,
\end{eqnarray}
where $E_i$ and $\vec p_i$ are the visible energy and three-momentum in the $i$th branch, and $E_{iT}$ and $\vec p_{iT}$ are the corresponding transverse energy and transverse momentum. They satisfy $M_C \geq M_{CT}$, just like $M  \geq M_T$. An interesting property of the $M_C$ variable is that it is invariant under the back-to-back boost of the two visible systems.

Similar to the stransverse mass $M_{T2}$, the constransverse mass variable $M_{CT2}$ is defined as \cite{Cho:2009ve,Cho:2010vz,Barr:2010ii}
\begin{eqnarray}
M_{CT2} = \min_{\vec{q}_{1T}+\vec{q}_{2T} = \mpt} \left [ \max \left \{ M_{CT}^{(1)}, M_{CT}^{(2)} \right \}  \right ]\, ,
\end{eqnarray}
where each $M_{CT}^{(i)}$ ($i=1,2$) is applied to the semi-invisible decay of parent particle $P_i$.

Another mass variable, $M_{2C}$, is defined as the minimum four-dimensional mass 
\begin{eqnarray}
M_{2C}^2 = \min_{q_1, q_2} \Big ( p_1 + q_1 \Big)^2 \, ,
\end{eqnarray}
under the following constraints
\begin{eqnarray}
\big ( p_1 + q_1 \big)^2 &=&\big ( p_2 + q_2 \big)^2 \\
q_1^2 &=& q_2^2 \\
\vec{q}_{1T} + \vec{q}_{2T} &=& \mptvec \\
\sqrt{ \big ( p_1 + q_1 \big)^2} - \sqrt{ \big ( q_1 \big)^2} &=& m_{P}-m_{0} \, ,
\end{eqnarray}
where the parent-daughter mass difference in the last constraint is assumed to be known from a preliminary measurement of an invariant mass endpoint (\ref{eq:endpoint1}) \cite{Ross:2007rm}. The $M_{2C}$ mass variable is bounded by the parent mass,  
$M_{2C}\leq m_{P}$.

Another set of kinematic variables extensively used by the CMS Collaboration in its searches for supersymmetry is the razor kinematic variables \cite{CMS:2018rst,CMS:2015adc,CMS:2012yua}. They are known to be sensitive to large mass differences between the parent particle and the invisible particles at the end of a decay chain. The razor variables are defined as \cite{Rogan:2010kb}
\begin{eqnarray}
M_R^2 &=& ( E_1 + E_2)^2 - ( q_{1z} + q_{2z})^2 \, , \\
( M_T^R )^2 &=& \frac{1}{2} \left [ \met (q_{1T} + q_{2T}) - \mptvec \cdot (
\vec q_{1T} + \vec q_{2T}) \right ] \, , \\
R^2 &=& \left ( \frac{M_T^R}{M_R} \right )^2 \, .
\end{eqnarray}
For QCD multijet background events, the distributions in both $M_R$ and $R^2$ fall exponentially, while for signal events they peak at finite values. 

\subsection{Singularity Variables 
\label{sec:singularity}}

\begin{figure}[t]
\begin{center}
\includegraphics[width=.45\textwidth]{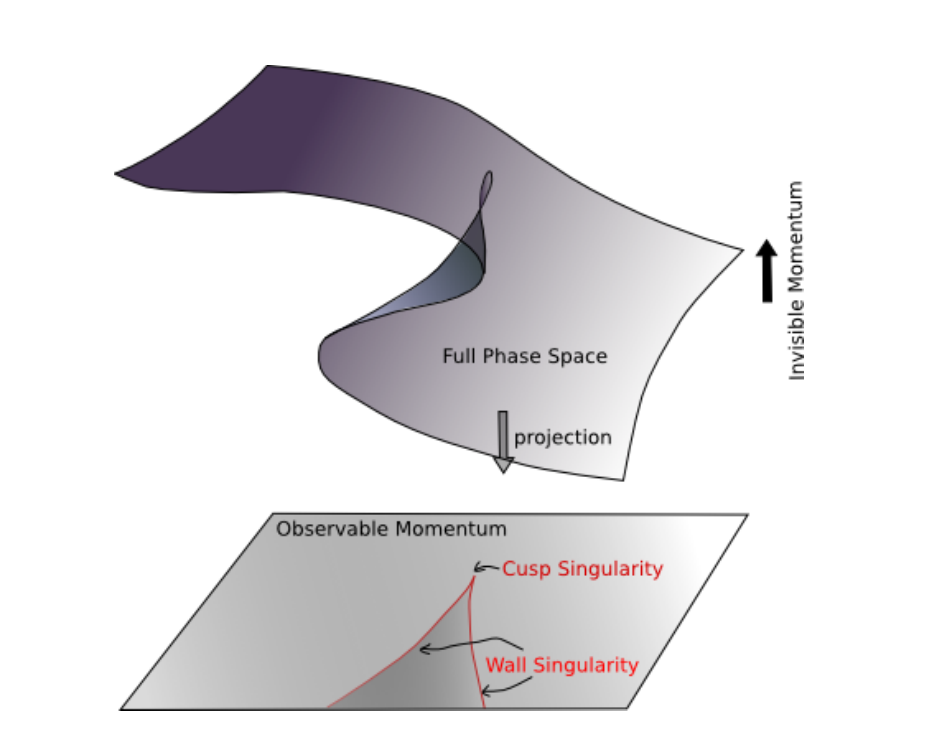}
\end{center}
\caption{A schematic depiction of the projection of the full phase space onto the space of observable momenta illustrating how folds in the allowed full phase space result in wall singularities in the observable space.  Taken from Ref.~\cite{Kim:2009si}.}
\label{fig:projection}
\end{figure}

The geometrical features of the high-dimensional phase space available to a given event topology are largely washed out when projecting to a single one-dimensional event observable. The so-called {\it singularity variables} \cite{Kim:2009si,Rujula:2011qn,DeRujula:2012ns,Matchev:2019bon}, however, provide an intuitive way to retain high-dimensional features as singularities in the corresponding one-dimensional kinematic distributions. The origin of such singularities is very well understood---similar to the phenomenon of caustics in optics, they are formed at points where the projection of the allowed phase space onto the observable space gets folded; see Figure~\ref{fig:projection} for a cartoon illustration. 

As previously discussed, such projections are inevitable in the presence of invisible momenta in the event, $\{q\}$, since event observables must be constructed out of visible momenta $\{p\}$ only. The geometrical structure in the allowed higher-dimensional phase space is imposed by a certain set of kinematic conditions (constraints) 
\begin{equation}
f_\alpha (\{p\}, \{q\}) = 0,~\qquad \alpha=1,2,\dots
\end{equation}
which arise in the assumed event topology. Mathematically the singularity condition is then expressed as the reduction in the rank of the Jacobian matrix of the coordinate transformation from the relevant set of kinematic constraints $f_\alpha$ to $\{q\}$, which is why such singularities are sometimes known as Jacobian peaks. Explicitly,
\begin{equation}
\det \left(\frac{\partial f_\alpha}{\partial q^\mu}\right) = 0.
\label{eq:singularity_condition}
\end{equation}

Given an event topology, the general procedure for deriving the singularity coordinate from Eq.~\eqref{eq:singularity_condition} was discussed and illustrated in Ref.~\cite{Matchev:2019bon} for the case of square Jacobian matrices and in Ref.~\cite{Kim:2009si} for the general case. Among the set of singularity variables one finds well-known examples like the transverse invariant mass $M_T$ discussed in Section~\ref{sec:mt2variables} \cite{Rujula:2011qn,DeRujula:2012ns,Matchev:2019bon}, the invariant mass $M_{a_2a_1}$ of the visible decay products in the two-step two-body cascade decay of Figure~\ref{fig:invmassdiagrams}($a$) in three \cite{Kim:2009si} or two \cite{Matchev:2019bon} spatial dimensions, and the Cambridge $M_{T2}$ variable from Section~\ref{sec:mt2variables}  \cite{Park:2021lwa}. However, there are also more recently discovered singularity variables like the $\Delta_4$ variable \cite{Agrawal:2013uka,Debnath:2018azt}, which is applicable to the three-step two-body decay chain in Figure~\ref{fig:invmassdiagrams}(c), the $\Delta_{\rm antler}$ variable \cite{Matchev:2019bon,Park:2020rol} relevant to the event topology of Figure~\ref{fig:invmassdiagrams}(b), and the $\Delta_{t\bar{t}}$ variable \cite{Kim:2019prx,Matchev:2019bon} relevant for the $t\bar{t}$ dilepton topology of Figure~\ref{fig:mt2diagram}(b). The  singularity variables are excellent analysis tools and can serve a dual purpose: in the age of discovery, they can be used to target signal-rich regions of phase space, and post-discovery, they form the basis for the focus point method for mass measurements discussed below in Section~\ref{sec:focus}.

\section{Exclusive event variables: energy, time, distance}
\label{sec:exclusive2}

In this section, we discuss a few basic quantities that are directly available in experiments and their non-trivial utilization. We begin with the energy variable followed by the timing and distance variables. 

\subsection{Energy peak  \label{sec:energy} }

While energy is not a Lorentz-invariant quantity, the peak position in the energy distribution of a visible particle coming from a two-body decay of the heavier resonance/parent particle carries a boost-distribution-invariant property. 
Suppose that particle $A_1$ is a scalar or produced in an unpolarized way and decays into a massless visible particle $a_1$ and another paricle $a_0$ which may be visible or invisible. In the rest frame of $A_1$, the energy of $a_1$ $E_{a_1}^*$ is simply given by 
\begin{equation}
    E_{a_1}^* = \frac{m_{A_1}^2-m_{a_0}^2}{2m_{A_1}}\,. \label{eq:epeak}
\end{equation}
In the laboratory frame, one should perform a Lorentz transformation to find the laboratory-frame energy of $a_1$ $E_{a_1}$:
\begin{equation}
    E_{a_1}=\gamma_{A_1}E_{a_1}^* (1+\beta_{A_1}\cos\theta_{a_1}^*)\,, \label{eq:Elab}
\end{equation}
where $\gamma_{A_1}=(1-\beta_{A_1}^2)^{-1/2}$ is the boost factor of $A_1$ in the laboratory frame and $\theta_{a_1}$ is the emission angle of $a_1$ in the $A_1$ rest frame with respect to $\vec{\beta}_{A_1}$. 
One can see that for any $\gamma_{A_1}$, $E_{a_1}^*$ is the (only) commonly included value in the distribution of $E_{a_1}$. Since $A_1$ is assumed unpolarized or scalar, $\cos\theta_{a_1}^*$ is a flat variable and so is $E_{a_1}$. Therefore, whatever distribution of $\gamma_{A_1}$ is given, the final $E_{a_1}$ distribution shows a peak at $E_{a_1}^*$~\cite{Agashe:2012bn}. 
This observation was made in the context of the cosmic $\pi^0$ decay~\cite{Carlson1950}, and then extended and generalized to the two-body decay of an unpolarized resonance at colliders~\cite{Agashe:2012bn,Kawabata:2013fta}. 

This energy-peak feature can be viewed in the logarithmic energy space. In Eq.~\eqref{eq:Elab}, the maximum and minimum laboratory-frame energy values $E_{a_1}^{\pm}$ arise at $\cos\theta^\ast_{a_1}=\pm1$, resulting in
\begin{equation}
    E_{a_1}^{\pm} =\gamma_{A_1}E_{a_1}^\ast(1\pm\beta_{A_1})\,,
\end{equation}
from which one can see that the rest-frame energy $E_{a_1}^\ast$ is the geometric mean of the maximum and minimum laboratory-frame energoies $E_{a_1}^\pm$ for any boost factor $\gamma_{A_1}$:
\begin{equation}
    \left( E_{a_1}^\ast \right)^2 = E_{a_1}^+ E_{a_1}^-\,.
\end{equation}
This further implies that $\ln E_{a_1}^\ast$ is the mean of $\ln E_{a_1}^+$ and $\ln E_{a_1}^-$ and the rectangular distribution in $E_a$ for a given $\gamma_{A_1}$ is log-symmetric with respect to $\ln E_{a_1}^\ast$.  
Once such log-symmetric rectangular distributions, which are weighted by the $\gamma_{A_1}$ distribution, are stacked up, the final $E_{a_1}$ distribution is automatically log-symmetric with respect to $\ln E_{a_1}^\ast$. 

\begin{figure}[t]
    \centering
    \includegraphics[width=0.95\linewidth]{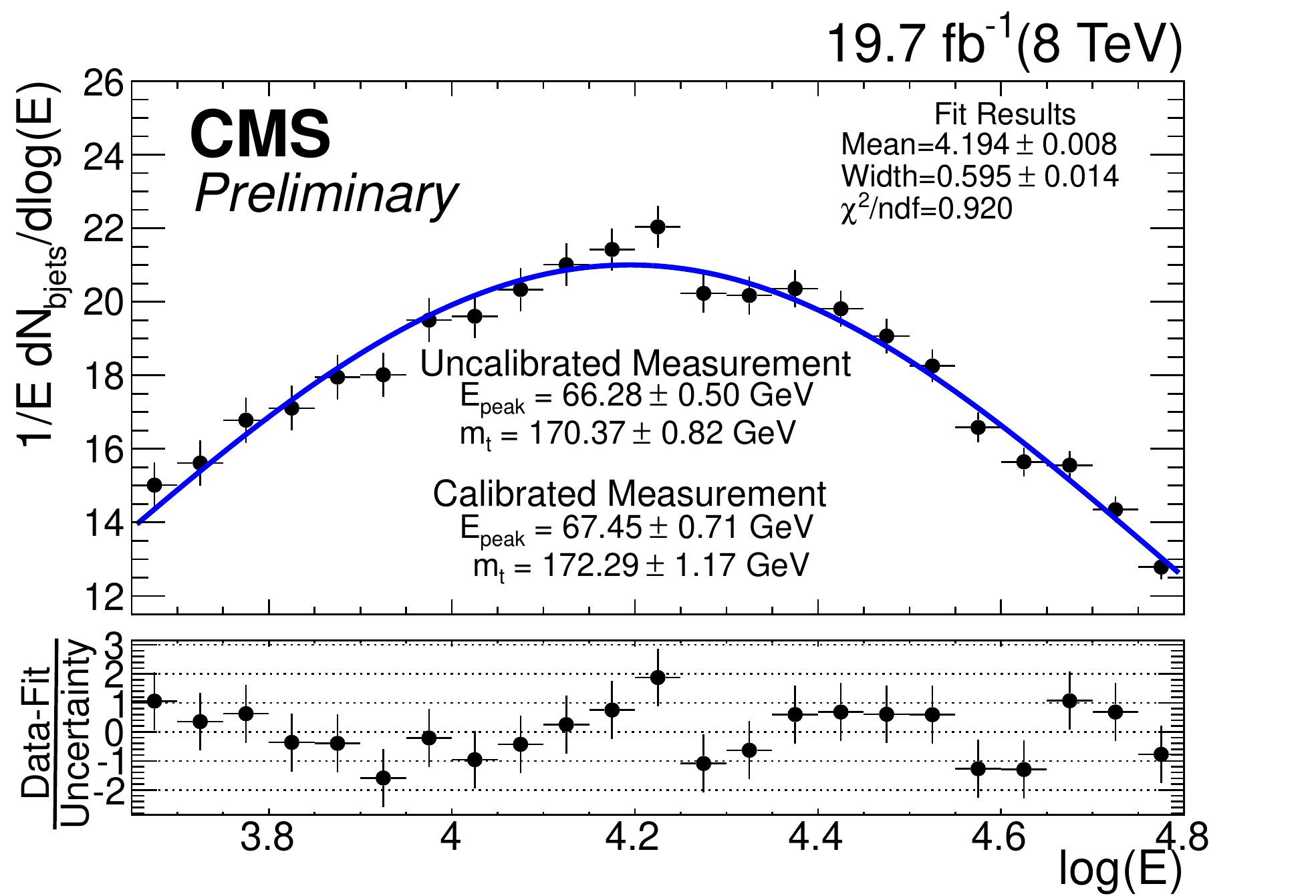}
    \caption{The CMS top quark mass measurement using the energy-peak method in the energy distribution of $b$-jets from the top quark decay, $t\to bW$. The distribution is log-symmetric with respect to $\ln E_b^\ast$, and the fit was performed with a Gaussian template. Taken from Ref.~\cite{CMS:2015jwa}. }
    \label{fig:cmsenergypeak}
\end{figure}

As suggested by Eq.~\eqref{eq:epeak}, the extraction of $E_{a_1}^*$ implies the measurement of a mass relation between $A_1$ and $a_0$. 
This kinematic feature is particularly useful in a hadron collider environment where the longitudinal boosts of individual events are {\it a priori} unknown, i.e., the $\gamma_{A_1}$ profile is unknown. 
In addition, since the method involves no combinatorial ambiguity, its applicability is nearly unaffected by high particle multiplicity. 
If the $a_0$ mass is known through independent measurements and if the peak in the $a_1$ energy distribution is extracted, the mass of $A_1$ can be readily determined using Eq.~\eqref{eq:epeak}. 
A well-motivated and practical physics application is the top quark mass measurement in the top quark decay, $t\to bW$, through the $b$-jet energy-peak method~\cite{Agashe:2012bn} and the weight function method~\cite{Kawabataa:2014osa}. 
The CMS Collaboration has measured the top quark mass by extracting the peak in the $b$-jet energy distribution in the $e\mu$ channel~\cite{CMS:2015jwa} as shown in Figure~\ref{fig:cmsenergypeak} where the aforementioned log-symmetric feature is evident. 
Another SM example is the $W$ mass determination, using the lepton energy spectrum in the case of associate production of leptonic $W$ along with other particles, i.e., $pp \to W X, W \to \ell \nu_\ell$~\cite{Bianchini:2019iey}. 
The method is not just restricted to SM processes but straightforwardly applicable to new particles mass measurements; examples include mass measurements of new resonances in models of supersymmetry~\cite{Low:2013aza,Agashe:2013eba,Agashe:2015wwa,Agashe:2015ike,Bianchini:2019iey} and in the context of potential cosmic $\gamma$-ray excesses~\cite{Kim:2015usa,Kim:2015gka,Boddy:2016fds,Boddy:2016hbp,Boddy:2018qur}. 

The crucial assumptions to retain the boost-invariant feature of the energy peak are that the visible decay product is {\it massless} and it comes from a {\it two-body} decay of an {\it unpolarized} (or {\it scalar}) heavy resonance. As some of them loosen, the validity of the method would be gradually degraded and a certain extent of prescriptions would be needed. 

First of all, if the visible decay product has a non-zero mass, the relation in Eq.~\eqref{eq:Elab} is modified to 
\begin{equation}
    E_{a_1}=\gamma_{A_1}(E_{a_1}^*+p_{a_1}^*\beta_{A_1}\cos\theta_{a_1}^*)\,,
\end{equation}
where $p_{a_1}^*$ is the magnitude of the $a_1$ momentum measured in the $A_1$ rest frame. Unlike the massless case, $E_{a_1}^*$ is no more commonly contained in the $E_{a_1}$ distribution for any boost $\gamma_{A_1}$. One can find that if $\gamma_{A_1}>\gamma_{A_1}^{\rm cr}\equiv 2\gamma_{a_1}^*-1$ with $\gamma_{a_1}^*$ being the boost factor of $a_1$ in the $A_1$ rest frame, the minimum $E_{a_1}$ occurring with $\cos\theta_{a_1}^*=-1$ becomes larger than $E_{a_1}^*$~\cite{Agashe:2012bn}. 
Therefore, for the $\gamma_{A_1}$ profile extending beyond $\gamma_{A_1}^{\rm cr}$, the peak in the overall $E_{a_1}$ distribution may be larger than $E_{a_1}^*$. 
In the example of top quark decay, $\gamma_b^*\approx 15$ and hence $\gamma_t^{\rm cr}\approx 450$. At the LHC, such a large boost factor of top quark is kinematically inaccessible, so the energy-peak method can safely go through for the top quark decay at the LHC. 
However, if $m_{a_1}$ is too large with respect to $m_{A_1}-m_{a_0}$, the shift of the energy peak is unavoidable. Nevertheless, one can still extract $E_{a_1}^*$, modeling the energy distribution by appropriately accommodating the shift. We refer to Ref.~\cite{Agashe:2015ike} for a more detailed discussion.  

Second, once the decay of interest accompanies additional decay products, $\cos\theta_{a_1}^*$ is no more a flat variable and hence the argument breaks down. It was analytically demonstrated that the resulting peak position is always less than $E_{a_1}^*$ predicted in the associated multi-body decay process of $A_1$~\cite{Agashe:2012fs}.  
In the top quark decay, the $b$ quark often comes with a hard gluon emission in the final state which is not captured as part of the $b$ jet, and thus the contamination from such $t\to bWg$ inevitably induces a systematic error in the $E_b^*$ extraction.\footnote{By contrast, initial state radiation from either incoming partons or top quark itself simply reshuffles the $\gamma_t$ profile, and hence it does not ruin the boost-invariant feature~\cite{Agashe:2012bn}.} 
The systematics was carefully assessed in the top quark decay at the next-to-leading order in Ref.~\cite{Agashe:2016bok} and it claimed that a $\lesssim 0.5\%$ level of the associated systematic error would be achievable for a 1\% jet energy scaling uncertainty. 
Another approach to treat the multi-body decays involving an invisible decay product is to interpret the $A_1$ decay as an effective two-body decay to $a_0$ and a composite visible system. Assuming that the visible particles are all massless, their invariant mass $M_{\rm vis}$ spans from 0 to $m_{A_1}-m_{a_0}$. One can then divide $M_{\rm vis}$ space into pieces, and for each such phase-space slice, the method for the massive visible particle that is briefly discussed earlier can be applied~\cite{Agashe:2015wwa}.  

Finally, polarized production of (non-scalar) $A_1$ induces a non-trivial angular dependence of $a_1$, i.e., $\cos\theta_{a_1}^*$ is non-flat.  Therefore, the peak position in the overall $E_{a_1}$ distribution can be larger or smaller than $E_{a_1}^*$ depending on the underlying decay dynamics~\cite{Agashe:2012bn}. 
At the LHC, the top quarks are predominantly produced via QCD, hence they are unpolarized. However, if certain new-physics dynamics produces polarized top quarks, an appreciable deviation from the $b$-jet energy peak can be interpreted as a sign of new physics. This further implies that the energy variable can be utilized as a cut to isolate the signal from SM backgrounds. 
This aspect of the energy peak was extensively investigated in Ref.~\cite{Low:2013aza} in the context of supersymmetric top quark decays. 

Energy peaks can also be applied in lepton colliders, where the predominance of electroweak interactions can potentially bring in more effects related to polarization. As one of the main top quark sources in $e^+e^-$ programs is the production close to the threshold, where particular care is needed to account for the production of bound state and slowly-moving top quarks. Reference~\cite{Bach:2017ggt} has studied   threshold effects on the energy spectrum of a $b$ quark at a 380~GeV $e^+e^-$ collider. As the production of top quarks at threshold strengthens the validity of the arguments behind the invariance of the energy peak, the correction to the energy distribution  are found to be very localized compared to the corrections to the transverse momentum.

The utility of the energy peak is not limited to particle mass measurements and cutting for signal versus background discrimination. A representative example is to distinguish the two-body decay topology from the $\geq3$-body ones. Imagine two scenarios where each of the pair-produced $A_1$'s follows either $A_1\to a_1 a_0$ (i.e., two-body topology) or $A_1 \to a_1 a_0 a_0'$ (i.e., $\geq 3$-body topology) with $A_0'$ representing additional invisible particle(s). 
It was demonstrated that half the $\mu$ parameter extracted from the $M_{T2}$ distribution is the same as (greater than) the peak position $E_{a_1}^{\rm peak}$ for the two-body (three-body) decay scenario~\cite{Agashe:2012fs}:
\begin{equation}
\frac{\mu}{2}=\frac{m_{A_1}^2-m_{a_0}^2}{2m_{A_1}}\left\{
\begin{array}{l c}
 = E_{a_1}^{\rm peak}    & \hbox{for the two-body decay,}  \\ [1em]
> E_{a_1}^{\rm peak}     & \hbox{for the three-body decay.} 
\end{array}\right. 
\end{equation}

\subsection{Timing  \label{sec:timing}}

The variables in the preceding sections involve quantities in energy-momentum space. Likewise, one may utilize the information in time-position space. These next two subsections are devoted to discussing variables designed with time and position information. 

In principle, the timing information is useful for the following situations including: i) the case where the signals of interest differ from the unwanted signals (or backgrounds) by the timing at which they hit the detector system; ii) the case where the subprocesses of the signal process come along in a time-ordered manner (e.g., sequential decays of heavy resonances).  
In practice, the timing information becomes useful when the resolution in the timing measurement is sufficiently good and the uncertainty stemming from the particle-beam pulse spread is small enough. 
In many of the collider experiments (more generally accelerator-based experiments) including the LHC, beam parameters are well under control and sufficiently narrow beam pulses can be generated. 
When it comes to the timing resolution, it is a few hundred pico-seconds as of Run II of the LHC and thus its utilization is somewhat limited especially for the (new) physics processes where all relevant hard interactions take place instantly.  

The use of timing information is receiving increasing attention, however, as higher-resolution timing information allows for improved pile-up-origin background suppression~\cite{CERN-LHCC-2018-023,Butler:2019rpu} and it provides a unique handle in the search for long-lived particles (LLPs)~\cite{Liu:2018wte,Kang:2019ukr,Dienes:2021cxr} which were targeted marginally in the earlier LHC operation as its detectors (e.g., ATLAS, CMS, and LHCb) were designed to be optimal to prompt processes. 
Therefore, the ATLAS~\cite{CERN-LHCC-2018-023}, CMS~\cite{Butler:2019rpu}, and LHCb~\cite{Perazzini:2022cdy} Collaborations are planning to install dedicated timing modules and develop appropriate trigger algorithms, expecting them to operate from Run IV. 

\begin{figure}[t]
    \centering
    \includegraphics[width=0.95\linewidth]{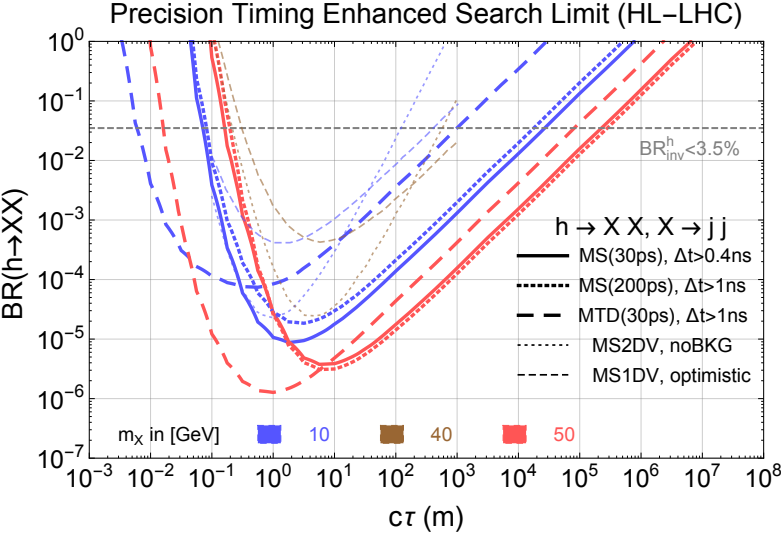}
    \caption{The 95\% CL limits on BR$(h \to XX)$ for signal
process $pp \to jh$ followed by subsequent decays $h \to XX$ and $X\to jj$ with $X$ being a new particle. Taken from Ref.~\cite{Liu:2018wte}.}
    \label{fig:timing}
\end{figure}

Indeed, it has been pointed out that the timing variable allows for a powerful separation between delayed new physics signal events and SM background events, given projected upgrades and implementations of high-capability timing modules and dedicated triggers at the LHC detectors~\cite{Liu:2018wte,Kang:2019ukr,Dienes:2021cxr}. 
For example, massive enough LLPs at the LHC can travel for a finite amount of time such that their decay products arrive at detectors with time delays around nanosecond scale, unlike the light SM particles. The strategies for utilizing this time delay feature can be applied in the search for LLPs, using the initial state radiation as a way of setting a reference timing and requiring at least one LLP to decay within the detector~\cite{Liu:2018wte}. It has been demonstrated that the strategies can improve the sensitivity to the lifetime of the LLPs by two orders of magnitude or more~\cite{Liu:2018wte}, in comparison to conventional search strategies for LLPs~\cite{CMS:2014wda,ATLAS:2015xit,Coccaro:2016lnz}, e.g., displaced vertex searches.  
Example sensitivity reaches for the scenario where the SM Higgs decays to a pair of LLPs $X$ are shown in Figure~\ref{fig:timing} in terms of limits on the branching fraction of $h\to XX$ as a function of the proper decay length of $X$~\cite{Liu:2018wte}. 

The timing information is useful not only for the discovery opportunities of LLPs but for the resonance mass reconstruction of the new particles involved in the associated decay process. 
The authors in Ref.~\cite{Kang:2019ukr} investigated the case where LLPs are pair-produced and each of them decays to an invisible particle and visible particle(s). They showed that the particle mass spectrum can be completely determined event by event, using the missing transverse momentum condition and the timing information, even in the case where the pair-produced LLPs are not identical. 
Another possibility that allows for the event-by-event mass measurement is the so-called ``tumbler'' scenario where a LLP decays to lighter LLPs sequentially. 
The authors in Ref.~\cite{Dienes:2021cxr} studied the simplest tumbler process, $A_2 \to a_2 A_1$ followed by $A_1 \to a_1 a_0$ with $A_2$ and $A_1$ being LLPs, $a_2$ and $a_1$ being visible particle systems, and $a_0$ being a collider-stable particle or dark-matter candidate. 
They demonstrated that the masses of $A_2$, $A_1$, and $a_0$ can be determined with the aid of timing information but without the recourse to the missing transverse momentum measurement. 

Timing is also considered as a discriminating variable to suppress the beam-induced background activities due to muon decays in muon beam machines~\cite{Ally:2022rgk,MuonCollider:2022ded,DiBenedetto:2018cpy,Bartosik:2019dzq}. Due to the unstable nature of the particles in the beam, the muon beams are a source of particles that go outside the control of the machine optics. The primary particles come from muon decay products: electron and neutrinos. They can interact with the accelerator apparatus and every material surrounding the beam, generating secondary radiation which can eventually reach the detectors. The tracking of these secondary particles is usually quite difficult, as they are a very large number of low-momentum particles. They can still be discriminated based on the fact that the production of secondary particles happens during the whole time of beam circulation, while interesting particles from the hard collisions appear only at beam crossing. The discrimination is further helped from the fact that only a limited length before the detector, hence  a specific time window, actually behaves as a source of secondaries that are potentially harmful for the physics analyses.

\subsection{Distance variables  \label{sec:distance} }

Mass measurements are inherently energy scale measurements.  However, excellent measurements of masses can, in principle, be conducted by accurately measuring the distances travelled by particles. The key relation is 
\begin{equation}
\left\langle d\right\rangle =c\beta\cdot \gamma\tau_{0} = \frac{\sqrt{E^2-m^2}}{m}\cdot c\tau_0\,,\label{eq:cbetagammatau}
\end{equation}
where $\left\langle d\right\rangle $ is the average distance travelled
before decaying, $\tau_{0}$ is the proper lifetime of the particle
at hand, and $\gamma=E/m=\left(1-\beta^{2}\right)^{-1/2}$ is the
usual Lorentz boost factor that governs the time dilation and length contraction.
In essence a mass measurement can be carried out from a sample of
particles all of which have identical boost factor -- or equivalently identical
energy -- by measuring the average decay length of the sample and measuring
elsewhere (or using theory predictions for) the proper lifetime $\tau_{0}$. 

A practical difficulty of pursuing this method is that in general
it is very hard to come across a sample of particles perfectly monochromatic in energy. 
A non-monochromatic particle sample can be used to carry
out a mass measurement if the energy distribution of these particles
is sufficiently well known.
The case of particles produced by parton collisions in $pp$ or $p\bar{p}$
colliders clearly shows how measurements of this sort need very
accurate knowledge of the source (i.e., PDFs in this case).

This idea can be exploited even for particles whose lifetime is 
short enough to make it impossible to measure $\left\langle d\right\rangle $.
If the prompt-decaying particle produces an unstable long-lived particle as a decay product, it can be viewed as the source of a new sample of particles as it inherits a certain extent of relevant information of the decaying particle. 
The average decay length of the unstable decay product can now be measured and turned into a mass measurement, provided that the proper lifetime and the energy distribution
of the measurable decay product are sufficiently well known.

Measurements of this sort have been proposed for SM particles such
as the top quark \cite{Hill:2005zy}. Concrete results for this strategy appeared from the CDF~\cite{Abulencia:2007cz} and CMS \cite{CMS-PAS-TOP-12-030}
experiments. A great deal of work in these measurements is devoted
in understanding the source of the unstable particles whose decay
length is measured and a number of issues having something to do with the formation
of hadrons and other aspects of QCD that impact the kinematics of
top quark decay products. 

In fact, in the case of the top quark mass measurement the accessible
unstable states are B-flavored hadrons (e.g., $B^{+}$, $B_{s}^{0}$,
$B^{0}$, $\Lambda_{b}^{0}$, ...) whose 
$c\tau_0$'s are in the
range of 100~$\mu$m hence can be measured in modern detectors,
especially when they enjoy typical Lorentz boost factor of $\mathcal{O}(10)$ as
expected from top quark decays. Relating the top quark mass to the
observed decay lengths of the $B$-hadrons has significant complications
with respect to the simple one-step monochromatic case sketched at
the beginning of this section. Indeed, one has to deal with multiple
species of $B$ hadrons, each of which has a different mass
and a different proper lifetime. As a consequence, the actual yield of each
type of $B$-hadrons does affect a measurement that is blind to the identification of each species. 
For this reason, the analyses of CDF \cite{Abulencia:2007cz}
and CMS \cite{CMS-PAS-TOP-12-030} are filled with details on the
treatment of hadronization effects, which become of primary interest
when a sub-percent measurement is attempted to be competitive with
other top quark mass determinations. We refer to the above studies for a
detailed discussion of these effects. 

Concerning the use of kinematic variables, we remark that so far the
CDF~\cite{Abulencia:2007cz} and CMS \cite{CMS-PAS-TOP-12-030} Collaborations
have managed to relate the top quark mass and the length distribution
of the $B$-hadrons only by producing templates of the length distribution
with a full chain of MC simulations. However, the insights on the
peak of the energy distribution can be translated into properties
of the length distribution, as entailed by the relation among energy, mass, and decay length as in Eq.~\eqref{eq:cbetagammatau}. Work is underway
to formulate new mass measurement strategies exploiting this insight~\cite{AgasheLength,Agashe:2022sxw}.

In addition to mass measurements, distance variables are useful discriminators for the identification of heavy flavor quarks and leptons. Particles with a measurably long proper lifetime, e.g., $B$-hadrons, result in a measurable impact parameter, defined as the transverse distance of closest approach of a track to the primary interaction vertex. This quantity can be shown \cite{Barger:1987nn} to be largely unaffected by the boost of the decaying particle, as time dilation contributes to displace further the decay point in the laboratory frame, but at the same time length contraction makes the direction of the decay product tend to align with the decaying particle momentum. As a consequence, the characteristic decay times of particles are translated to the characteristic impact parameters, which are very useful for particle identification. 

Modern experiments use impact parameter information, among many other inputs, to give a likelihood for particle identification. For a detailed explanation of the role played by the impact parameter in heavy flavor tagging, we refer to e.g., Ref.~\cite{rizzi2006track}. A modern incarnation that leverages the impact parameter in a neural network classifier is described in Ref.~\cite{note2020deep}. 

A major difficulty in analyzing collider events is the necessity to disentangle the useful particles potentially bearing information on interesting phenomena from the particles stemming out of ordinary collisions not carrying any useful information. Especially at hadron colliders the collision rate is so high that a number collisions can happen for each bunch crossing. A basic tool to discriminate particles from the collisions recorded at once is the position from which these particles momenta originate if extrapolated to the beam axis~\cite{Wells:1957370,1705.02211v1}. This simple observable keeps being a basic ingredient for current and future experiments and it is used in conjunction with the most theoretically sophisticated tools~\cite{1801.09721v2} to remove pile-up effects.

New physics models have  provided many examples of signatures involving LLPs and other exotic states (e.g., Refs.~\cite{2108.02204v1,Lee:2018aa,2103.08620v1,Schwaller:2015ek,Alimena:2019aa,Hewett:2004aa,Evans:2016il,Barnard:2015uq,Bomark:2013lh,Meade:2010kx,Dienes:2021cxr}) that can be analyzed with observables referring to length measurements. The most basic measurements involve the euclidian distance between the primary interaction point and the displaced vertex where the exotic particle decays. A summary of the power of this approach to search for exotic states is given, for example, in Ref.~\cite{Liu:2015eu} together with a comparison of the coverage of new physics models parameter space of the equivalent ``prompt'' searches not exploiting length measurements. The typical prompt search gets quickly ineffective when one considers distances greater than 100~$\mu$m $\div$ 1 cm for the lifetime of the exotic state, or the equivalent parameter that controls the appearance of displaced vertices in more complicated models. The threshold for the beginning of degradation of the prompt searches is process-dependent, but the general message that the displaced vertex searches can fully fill in this gap is very robust. As a matter of fact, when the experiments have looked for these exotic signals (e.g., Refs.~\cite{1810.10069v2,1806.07355v2,1905.09787v2,2012.01581v2,2107.04838v1}), the bounds from displaced vertices are stronger than the prompt counterparts.

\section{Other exclusive event variables \label{sec:otherexc}}

\subsection{Dimensionless variables  \label{sec:other} }

The variables in the preceding sections are dimensionful hence they allow to infer the scale information of the underlying physics processes. 
By contrast, dimensionless variables make it possible to extract scale-independent information. 
Here we briefly review a few dimensionless exclusive event variables most of which are developed for particular processes and/or event topologies. 

The first example is the $\alpha_T$ variable \cite{Randall:2008rw} which is introduced to efficiently reduce multijet events without a significant missing transverse momentum $\mptvec$. For dijet events, it can be defined as
\cite{CMS:2012rao}
\begin{equation}
\alpha_T = \frac{E_T^{j_2}}{m_T} \, , \label{eq:alphaT}
\end{equation}
where $E_T^{j_2}$ is the transverse energy of the second hardest jet and $m_T = \sqrt{(E_T^{j_1} + E_T^{j_2})^2 - (\vec p_T^{\, j_1} + \vec p_T^{\, j_2})^2}$ is the transverse mass of the dijet system. For a perfect dijet system where the two jets are back-to-back, $E_T^{j_1} = E_T^{j_2}$ and $\vec p_T^{\,j_1} = - \vec p_T^{\,j_2}$ which leads to $m_T = 2 E_T^{j_2}$ and $\alpha_T= 0.5$. $\alpha_T$ is significantly larger than 0.5, when the two jets are not back-to-back, recoiling against $\mptvec$.

In the case of events with three or more jets, one can form an equivalent dijet system by combining the jets in the event into two pseudo-jets. One chooses the combination such that the $E_T$ difference ($\Delta H_T$) between the two pseudo-jets is minimized. This simple clustering criterion provides a good separation between QCD multijet events and events with true $\mptvec$. In this case, the $\alpha_T$ is generalized as 
\begin{eqnarray}
\alpha_T &=& \frac{1}{2} \frac{H_T - \Delta H_T}{\sqrt{ H_T^2 - \mht^2}} \\
&=& \frac{1}{2} \frac{1 - \Delta H_T / H_T }{\sqrt{ 1 - (\mht / H_T)^2}} \, ,
\end{eqnarray}
where $H_T = \sum_{j=1}^{N_{\rm jets}} E_T^{j}$ and $\mht = |\sum_{j=1}^{N_{\rm jets}} \vec p_T^{j}|$.
Here $N_{\rm jets}$ is the number of jets with $E_T$ greater than a certain threshold, typically chosen to be 50 GeV \cite{CMS:2011bul}.

The second example is topness and higgsness for which the main idea is to use the value of $\chi^2$ as a cut. This method becomes more powerful, especially when one can define two (or more) independent $\chi^2$ values. 
Topness was originally introduced to reduce the $t\bar t$ background in the search for supersymmetric top quarks \cite{Graesser:2012qy} and later further fine-tuned in the search for double Higgs production \cite{Kim:2018cxf,Kim:2019wns,Huang:2022rne}.
Topness basically aims to check the consistency of a given event with $t\bar{t}$ production.
It is a minimized chi-square value constructed by using four on-shell constraints, $m_t$, $m_{\bar t}$, $m_{W^+}$ and $m_{W^-}$, and 6 unknowns (the three-momenta of the two neutrinos, $\vec p_{\nu}$ and $\vec p_{\bar\nu}$) 
\begin{eqnarray}
\chi^2_{ij} &\equiv& \min_{\tiny \mptvec = \vec p_{\nu T} + \vec p_{ \bar\nu T}}  \left [ 
\frac{\left ( m^2_{b_i \ell^+ \nu} - m^2_t \right )^2}{\sigma_t^4} \,   +
\frac{\left ( m^2_{\ell^+ \nu} - m^2_W \right )^2}{\sigma_W^4}  \, \right.  \label{eq:tt} \nonumber \\
&& \hspace*{0.5cm}\left . + \frac{\left ( m^2_{b_j \ell^- \bar \nu} - m^2_t \right )^2}{\sigma_t^4} \, +
\frac{\left ( m^2_{\ell^- \bar\nu} - m^2_W \right )^2}{\sigma_W^4}   \right ]  , \label{eq:Tness1}
\end{eqnarray}
subject to the constraint, $ \mptvec = \vec p_{\nu T} + \vec p_{ \bar\nu T}$.  
Here $\sigma_t$ and $\sigma_W$ determine the relative weight of the on-shell conditions, and should not be less than typical resolutions. 
Due to the twofold ambiguity in paring a $b$-jet (out of $b_1$ and $b_2$) and a lepton (out of $\ell^+$ and $\ell^-$), we define Topness as the smaller of the two possible chi-square values, $\chi_{12}^2$ and $\chi_{21}^2$:
\begin{eqnarray}
T &\equiv&  { \min} \left ( \chi^2_{12} \, , \, \chi^2_{21} \right ) \, .
\label{eq:Tness2}
\end{eqnarray}

Similarly, Higgsness aims to probe the consistency of a given event with double Higgs production. The challenge here is to find the sufficient number of constraints, as there are four unknowns, while there are only two intermediate on-shell particles.\footnote{See Ref. \cite{Alves:2022gnw} for the heavy Higgs decaying to two on-shell $W$ bosons.} The Higgsness is defined by
\begin{eqnarray}
H &\equiv&    {\rm min} \left [
 \frac{\left ( m^2_{\ell^+\ell^-\nu \bar\nu} - m^2_h \right )^2}{\sigma_{h_\ell}^4}    \right. 
 + \frac{ \left ( m_{\nu  \bar\nu}^2 -  m_{\nu\bar\nu, peak}^2 \right )^2}{ \sigma^4_{\nu}}
   \nonumber \\
 && \hspace*{-0.8cm}   + {\min} \left ( 
\frac{\left ( m^2_{\ell^+ \nu } - m^2_W \right )^2}{\sigma_W^4} + 
\frac{\left ( m^2_{\ell^- \bar \nu} - m^2_{W^*, peak} \right )^2}{\sigma_{W_*}^4}  \, , \right. \label{eq:Hness}  \\
 && \left.  \left .
\frac{\left ( m^2_{\ell^- \bar \nu} - m^2_W \right )^2}{\sigma_W^4} + 
\frac{\left ( m^2_{\ell^+ \nu} - m^2_{W^*, {\rm peak}} \right )^2}{\sigma_{W_*}^4}  
\right )  \right ]   \, , \nonumber
\end{eqnarray}
where $m_{W^*}$ is bounded from above, $m_{W^*} \leq m_h- m_W$, and its location of the peak can be estimated by
\begin{eqnarray} 
m_{W^*}^{\rm peak} &=& \frac{1}{\sqrt{3}} \sqrt{ 2 \left ( m_h^2 + m_W^2 \right ) - \sqrt{m_h^4 + 14 m_h^2 m_W^2 + m_W^4}} \nonumber \\
&\approx& 40 \text{ GeV} \, .
\end{eqnarray}

The $m_{\nu\bar\nu}^{\rm peak} = m_{\ell\ell}^{\rm peak} \approx 30$ GeV is the location of the peak in the invariant mass distribution of two neutrinos $\frac{d\sigma}{d m_{\nu\bar\nu}}$ (or $\frac{d\sigma}{d m_{\ell\ell}}$), which is bounded from above by $m_{\nu\bar\nu}^{\max} = m_{\ell\ell}^{\max} = \sqrt{m_h^2 - m_W^2}$.
The phase space distribution of $\frac{d\sigma}{dm_{\nu \bar\nu}}$ is given by
\begin{equation}
\frac{d\sigma}{dm_{\nu \bar\nu}} \propto \int dm_{W^*}^2 \lambda^{1/2}(m_h^2,m_W^2, m_{W^*}^2) f(m_{\nu \bar\nu}),
\end{equation}
where $\lambda(x,y,z)=x^2+y^2+z^2-2xy-2yz-2zx$ is the kinematic triangular function and $f(m_{\nu \bar\nu})$ is the invariant mass distribution of the antler topology (see also Section~\ref{sec:mass}) with $h\to W W^* \to \ell^+ \ell^- \nu \bar \nu$
\begin{eqnarray}
f(m_{\nu \bar\nu})\sim \left\{
\begin{array}{l l}
\eta\, m_{\nu \bar\nu} \, , & 0 \leq m_{\nu \bar\nu} \leq e^{-\eta}E, \\ [1mm]
m_{\nu \bar\nu} \ln(E/m_{\nu \bar\nu})  \, , & e^{-\eta}E \leq m_{\nu \bar\nu} \leq E,
\end{array}\right. \nonumber \\
\label{eq:antlerf}
\end{eqnarray}
where the endpoint $E$ and the parameter $\eta$ are defined in terms of the particle masses as~\cite{Cho:2012er}
\begin{eqnarray}
E &=&  \sqrt{ m_W m_{W^*} \, e^{\eta}}, \\
\cosh \eta &=&\left ( \frac{m_h^2- m_{W}^2 - m_{W^*}^2}{2 m_{W} m_{W^*} } \right ).
\end{eqnarray}
The actual peak of 30 GeV is slightly less than the result for pure phase space due to a helicity suppression in the $W$-$\ell$-$\nu$ vertex. 

The definitions of Topness and Higgsness involve $\sigma$ hyperparameters which represent experimental uncertainties and particle widths. However, in principle, they can be taken as free parameters. The precise values of these parameters are not crucial, as results are not sensitive to their numerical values.

\begin{figure}[t]
\centering
\includegraphics[width=4.3cm]{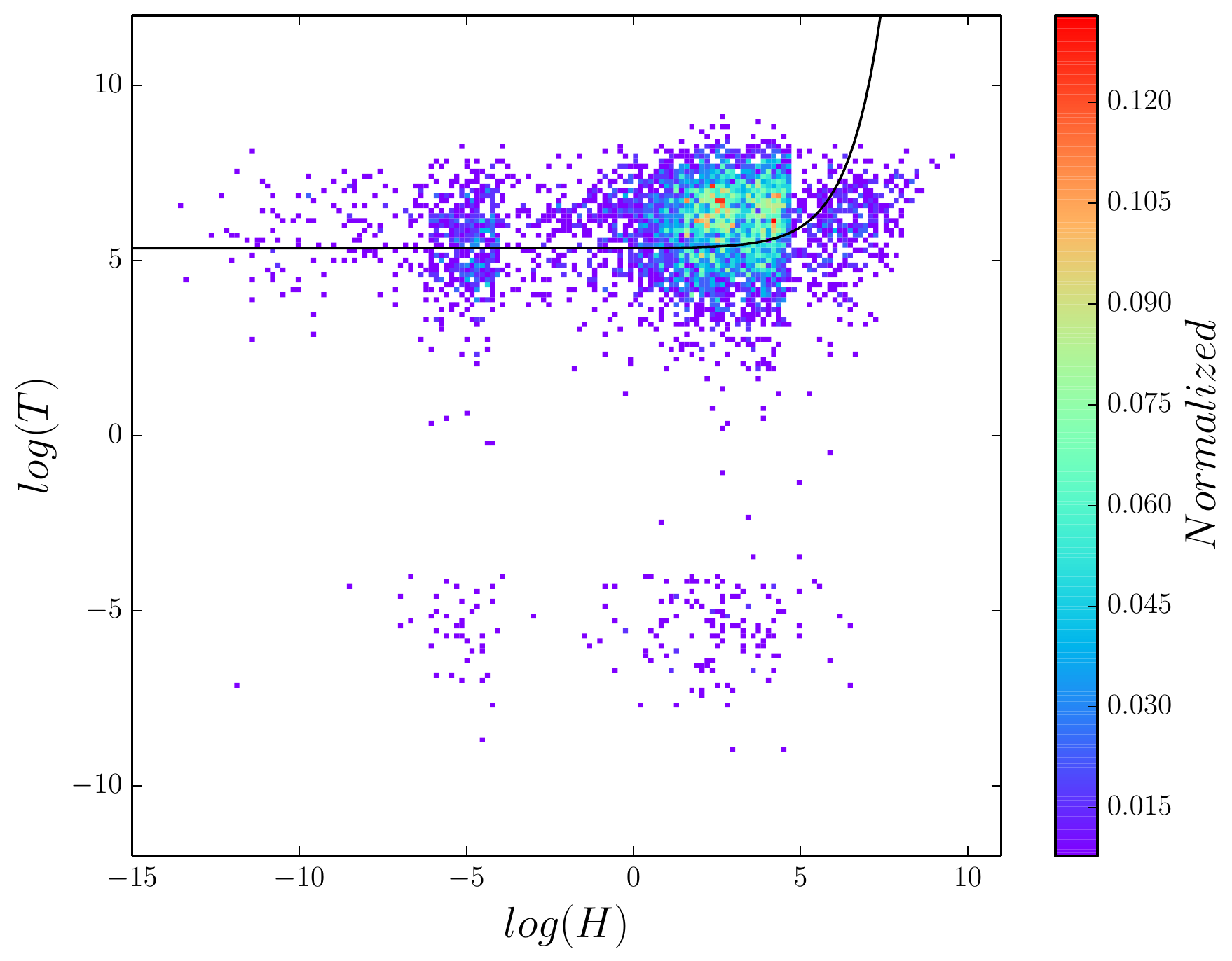}   \hspace*{-0.2cm}
\includegraphics[width=4.2cm]{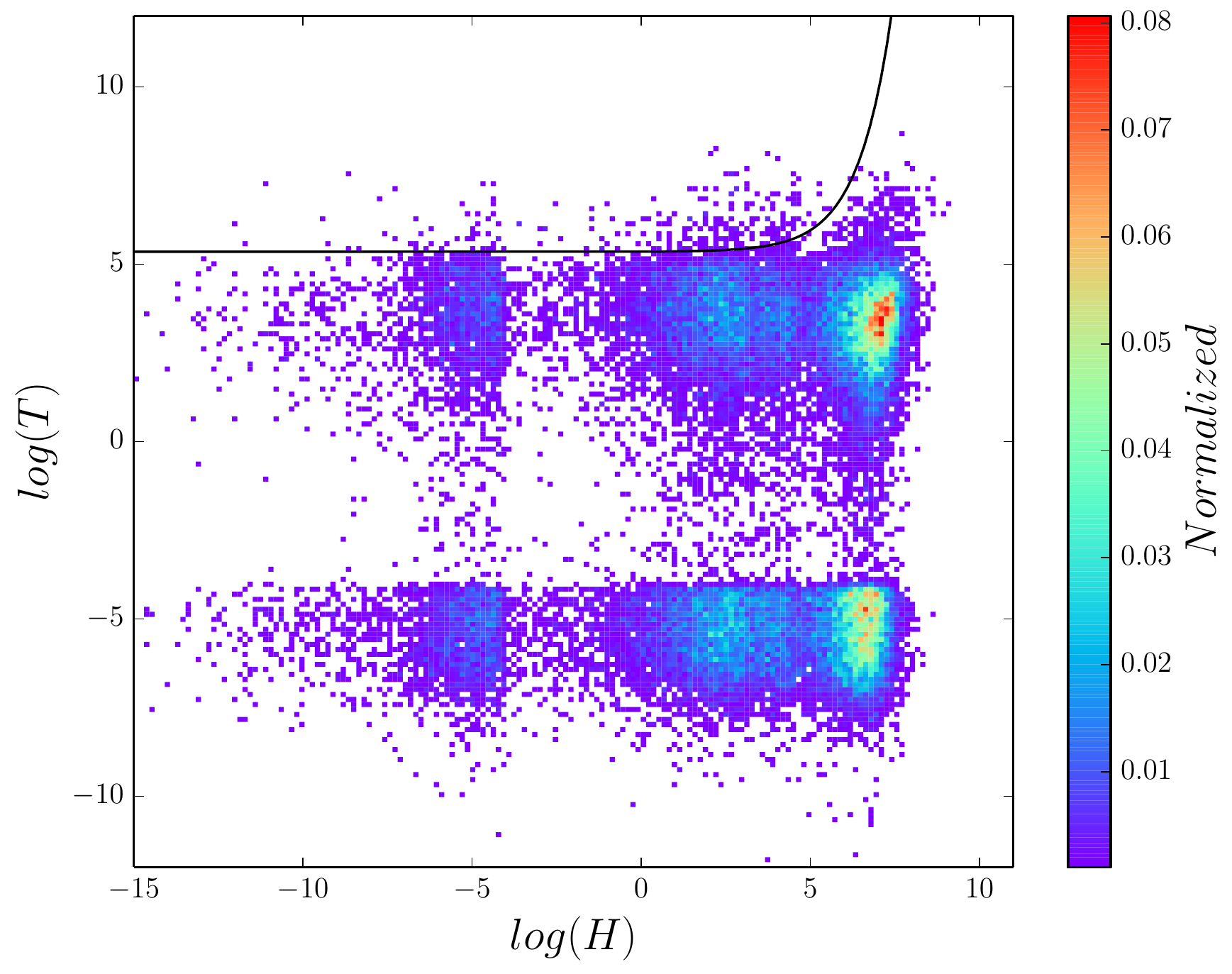} 
\caption{\label{fig:scatter} 
Scatter plot distribution of ($\ln H$, $\ln T$) for double Higgs production ($hh$) and backgrounds ($t \bar t$, $t\bar t h$, $t \bar t V$, $\ell\ell b j$ and $\tau\tau b b$) after loose baseline selection cuts. The black-solid curves in both panels are the same and represent the optimized cut. Taken from Ref. \cite{Kim:2018cxf}}
\end{figure}

As mentioned previously, the $\chi^2$ method becomes more useful when applied in more than one dimension. For example, Figure~\ref{fig:scatter} shows a scatter plot distribution of ($\ln H$, $\ln T$) for double Higgs production and backgrounds. The black-solid curves in both panels are the same and represent the optimized cut.

The third example is ratios of energy and transverse momenta which have been recognized as useful kinematic variables very early on, especially for problems in which part of the information on the kinematics is not accessible due to production of invisible particles. The ratio of energies and $p_{T}$ from cascade decays $A_{2}\to A_{1} a_{2}\to a_{1} a_{0} $ can provide extra information \cite{Nojiri:2000wq} on top of the ``classic'' invariant mass $M_{a_{1}a_{2}}$ of the visible decay products. Indeed, by properly combining information from the transverse energy ratio $\ln E_{1,T}/E_{2,T}$ and the invariant mass $M_{a_{1}a_{2}}$, it is possible to reconstruct the full information on the three masses involved in the cascade decay~\cite{Cheng:2011ya}.

Another interesting application is for the case where the numbers of final state invisible particles are different in the decays of pair-produced heavy resonances. For example, dark-matter ``partners'' charged under a $Z_2$ symmetry decay to a single dark-matter candidate, e.g., $A_1 \to a_1 a_0$ with $A_1$, $a_1$, and $a_0$ being dark-matter partner, visible particle(s), and dark-matter candidate, respectively. By contrast, those charged under a $Z_3$ symmetry can decay with one or two dark-matter candidates, namely, $A_1 \to a_1 a_0 a_0$ or $A_1 \to a_1 a_0$. This implies that if $Z_3$-charged $A_1$'s are pair-produced, each of their decays terminates with different numbers of $a_0$, the ratios of energy or $p_T$ of $a_1$ in both decay sides are likely to be unbalanced, whereas it is more likely to be balanced in the $Z_2$ case. 
The authors of Ref.~\cite{Agashe:2010tu} defined the ratio $R_{p_T}$ as
\begin{equation}
    R_{p_T} = \frac{\max\left(p_{T, a_1}^{(1)},p_{T, a_1}^{(2)}\right)}{\min\left(p_{T, a_1}^{(1)},p_{T, a_1}^{(2)}\right)},
\end{equation}
where $p_{T,a_1}^{(i)}$ denotes the $p_T$ of the visible particle in the $i$-th decay side ($i=1,2$). 
Therefore, if the underlying physics is $Z_3$ ($Z_2$), the ratios are typically larger than (close to) 1, and thus these models can be distinguished~\cite{Agashe:2010tu}.     

Finally, likelihood methods such as the MELA and other matrix-element based techniques can be categorized as dimensionless variables. Although fitting in the definition of dimensionless variables suitable for this section, they will be discussed separately in Section~\ref{sec:MLM}.

\subsection{ISR methods  
\label{sec:isrmethods}}

In this subsection, we discuss kinematic effects due to the presence of initial state radiation (ISR). 
We begin with a method which attempts to identify ISR. At hadron colliders, the production of heavy new particles is often accompanied by additional jets with a significant transverse momentum. These extra jets make the combinatorial problem worse and complicate the reconstruction of new particle masses. In Ref.~\cite{Alwall:2009zu}, a novel technique was discussed to reduce these effects and to reconstruct a clear kinematical endpoint, taking gluino pair production and decay at the LHC as an example. That analysis considered the three-body decay of the gluino, which leads to 4 jets plus missing transverse momentum, in which case the mass reconstruction is done via $M_{T2}$. To isolate the ISR jet, the authors introduced $M_{T2}^{\min} = \min\limits_{i=1,\cdots,5} M_{T2}(i)$, where $M_{T2}(i)$ is calculated from the five highest $p_T$ jets, excluding the $i$-th highest $p_T$ jet. Then the $i_{\min}$-th jet which satisfies $M_{T2}(i_{\min}) = M_{T2}^{\min}$ is tagged as the ISR jet. After all this, a strong correlation was found between the reconstructed ISR jet and true ISR jet. 

\begin{figure}[t]
    \centering
    \includegraphics[width=0.49\linewidth]{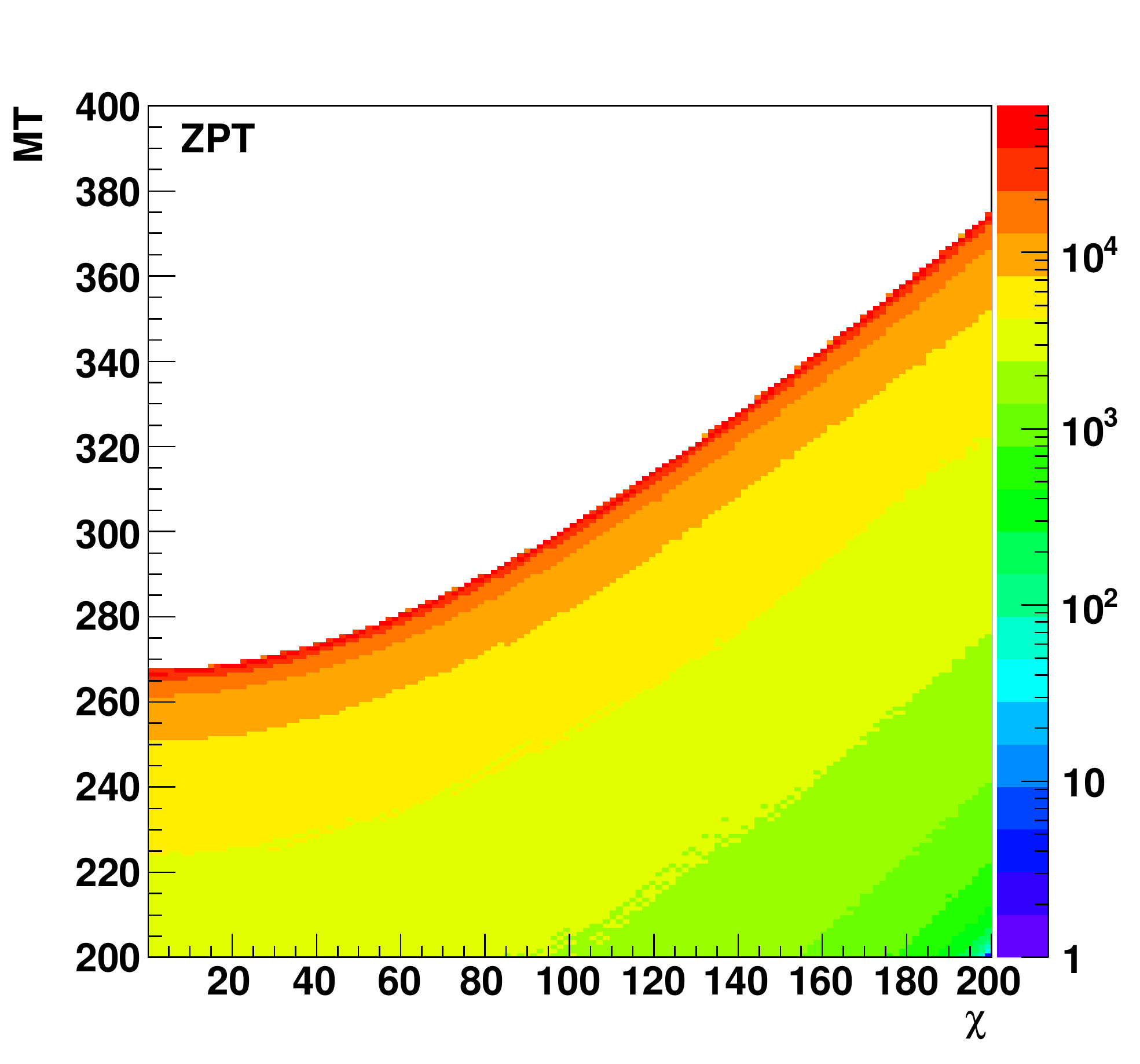}
    \includegraphics[width=0.49\linewidth]{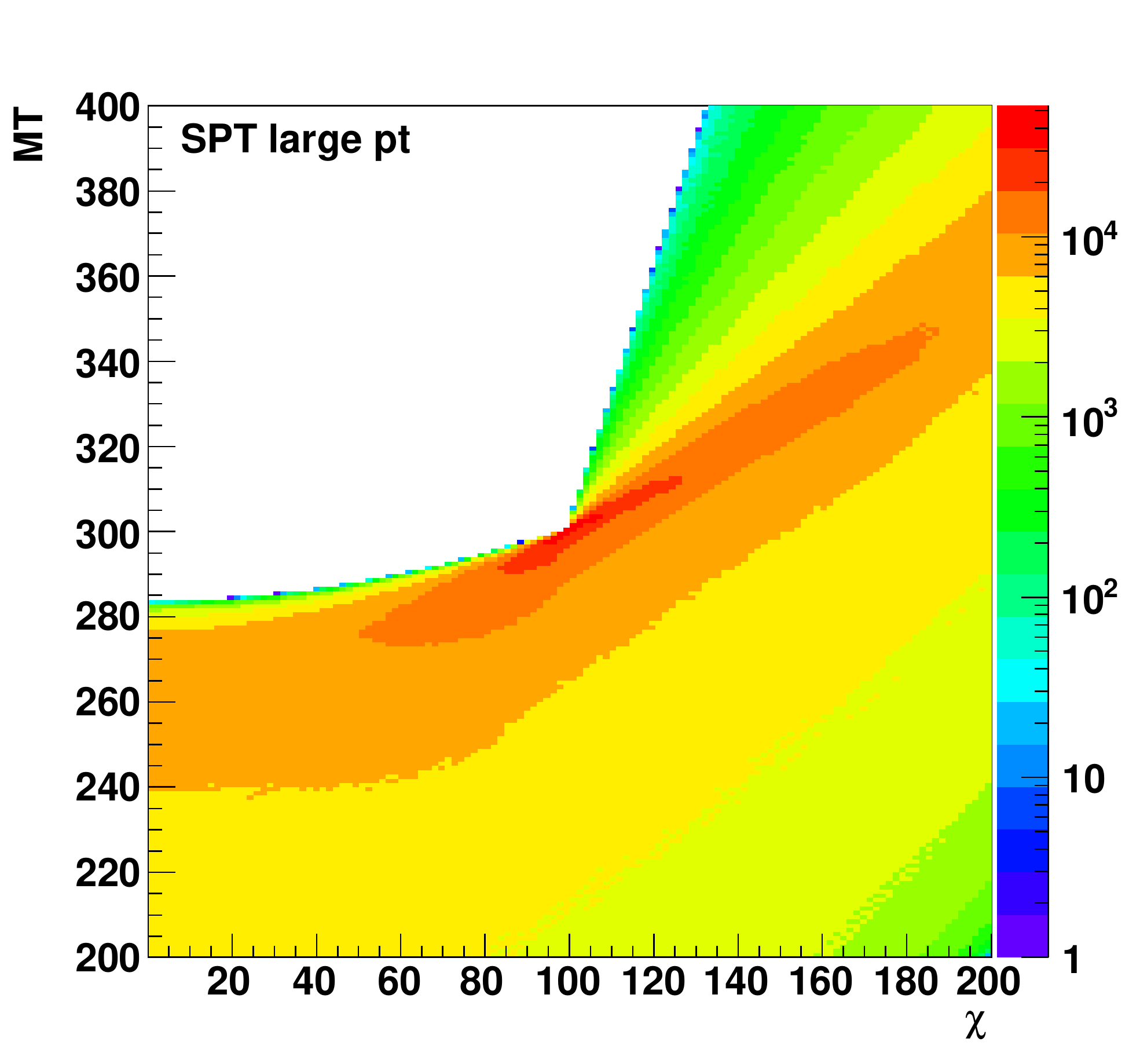}
    \caption{
    The transverse mass of the mother particle decaying semi-invisibly as a function of the daughter particle mass for the phase space Monte Carlo in which the mother has been constrained to be at rest in the laboratory frame with vanishing $p_T$ (left) and for the phase space Monte Carlo in which the mother can have large transverse momentum in the laboratory frame (right). Taken from Ref. \cite{Barr:2007hy}.}
    \label{fig:ISRkink}
\end{figure}

Occasionally, the ISR helps measurement of particle masses. The author of Ref.~\cite{Gripaios:2007is} considered the two-body decay of a particle at a hadron collider into a visible and an invisible particle, generalizing $W \to \ell \nu_\ell$, where the masses of the decaying particle and the invisible daughter particle are unknown. It was proved analytically that the transverse mass, when maximized over all possible kinematic configurations, can be used to determine both of the unknown masses. The authors of Ref. \cite{Barr:2007hy} generalized the idea for more complex decays of a singly-produced mother particle and for pair-produced particles. On the other hand, in the absence of ISR, one can in principle consider the upstream transverse momentum (UTM) playing the role of ISR, placing the system of interest under different momentum configurations \cite{Burns:2008va}.
In all cases, the mass variables ($M_T$, $M_{T2}$, or total invariant mass) optimized over all possible momentum configurations, which are given by either ISR or UTM, exhibit a kink structure at the true values of the mother and daughter particle masses, as illustrated in Figure \ref{fig:ISRkink}. 
An application of these ISR and kink methods is illustrated in Ref.~\cite{Matchev:2009fh} to determine the masses of the supersymmetric chargino and sneutrino in an inclusive manner, {\it i.e.} using the two well measured lepton momenta, while treating all other upstream objects in the event as a single entity of total transverse momentum $\mptvec$. This method takes full advantage of the large production rates of colored superpartners, but does not rely on the poorly measured hadronic jets, and avoids any jet combinatorics problems.

ISR plays a crucial role in studying dark-matter production at colliders as well. 
Under the hypothesis of classical weakly interacting massive particles (WIMPs), one can consider pair annihilation of dark matter ($\chi$) into a pair of SM particles, $\chi+\chi \to X_i + \bar{X}_i$, with $X_i=\ell,q,g,\cdots$.
The detailed balancing equation~\cite{Low:1958sn,Birkedal:2004xn} relates the pair-annihilation cross section and its inverse: 
\beq
\frac{\sigma(\chi+\chi\to X_i + \bar{X}_i)}
{\sigma(X_i+\bar{X}_i\to \chi+\chi)}\,=\,
2\,\frac{v_{X}^2 (2S_{X}+1)^2}{v_\chi^2 (2S_\chi+1)^2}\,, \label{eq:detailed}
\eeq
where $v_i$ and $S_i$ denote the velocity of initial-state species $i$ and the spin number of species $i$, respectively, and where the cross sections are averaged over spins but not other quantum numbers 
such as color. Then the WIMP production rate can be obtained as
\begin{eqnarray}
& &\sigma(X_i\bar{X}_i\to2\chi) = \nonumber \\ & &
2^{2(J_0-1)}\,\kappa_i \sigma_{\rm ann}\,
\frac{(2S_\chi+1)^2}{(2S_{\rm X}+1)^2}   
\left(1-\frac{4m_\chi^2}{s}\right)^{1/2+J_0}, \label{eq:prediction}
\end{eqnarray}
where the initial state particles are assumed to be relativistic 
($m_X\ll m_\chi$). Eq. (\ref{eq:prediction}) is written in terms of a small number of parameters with a clear physical meaning: the mass $m_\chi$ and the spin $S_\chi$ of the WIMP, the value of $J_0$ (either 0 or 1, depending on whether dark-matter annihilation is $s$-wave or $p$-wave annihilation), and the annihilation fraction $\kappa_i$ for the given initial state. Importantly, the overall scale for this prediction, the total annihilation cross section quantity $\sigma_{\rm ann}$ is provided by cosmology. 
This formula is only valid at center of mass energies 
slightly above the $2\chi$ threshold, $v=2v_\chi=2\sqrt{1-4m_\chi^2/s}\ll 1$, 
and receives corrections of order $v^2$. Taking $X_i=q$ or $g$ (or even $W$, $Z$)
for a hadron collider or $X_i=e$ for an electron-positron machine, Eq.~(\ref{eq:prediction}) provides a prediction of the WIMP production rate.

\begin{figure}[t]
    \centering
    \includegraphics[width=0.495\linewidth]{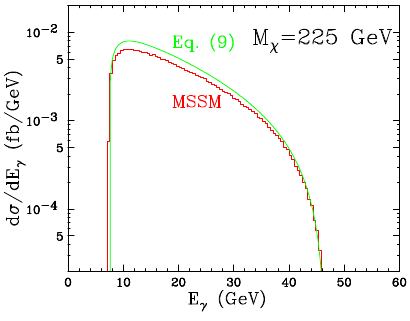}\hspace*{-0.1cm}
    \includegraphics[width=0.495\linewidth,height=0.4\linewidth]{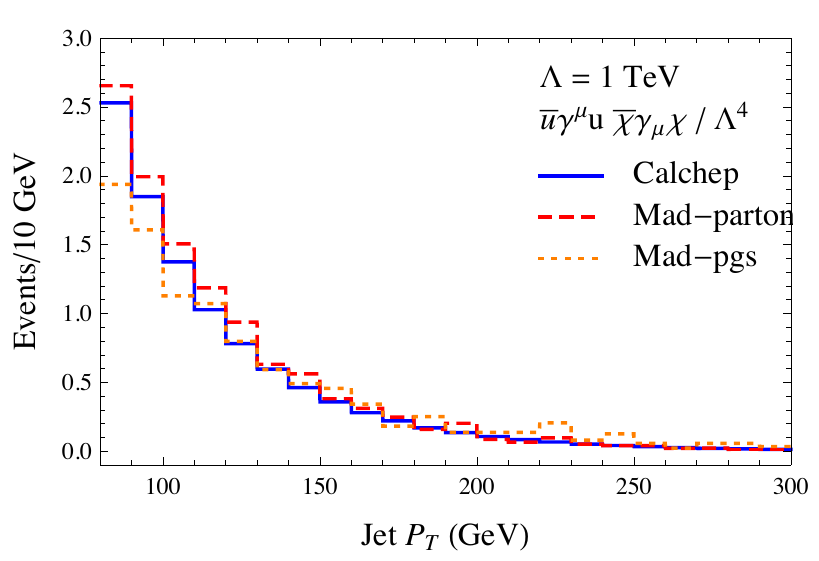}
    \caption{(left) Comparison between the photon spectra from the process $e^+e^- \to \met + \gamma$ in the explicit supersymmetric models (red) and the spectra predicted by Eq. (\ref{collinear}) (Eq. (9) in Ref. \cite{Birkedal:2004xn}) for a p-annihilator of the corresponding mass. 
    Taken from Ref. \cite{Birkedal:2004xn}.
    (right) Comparisons of simulated signal events from two different Monte-Carlo tools for the parton and the particle-level at Tevatron. Taken from Ref. \cite{Bai:2010hh}.
    }
    \label{fig:monophoton}
\end{figure}

Unfortunately, this dark-matter production process is not directly observalbe at colliders. At least one detectable particle is required for the event to pass the triggers and be recorded on tape. Therefore it is desirable to consider the production of two WIMPs in association with a photon or a gluon radiated from the known initial state. We consider a simple example given in Ref. \cite{Birkedal:2004xn}, $e^+e^-\to2\chi+\gamma$. 
If the emitted photon is either {\it soft} or {\it collinear} with 
the incoming electron or positron, soft/collinear factorization theorems 
provide a model-independent relation. The emission of collinear photons is given by 
\beq
\frac{d\sigma(e^+e^-\to 2\chi+\gamma)}{dx\, d\cos\theta} \approx
{\cal F}(x, \cos\theta)\,\hat{\sigma}(e^+e^-\to2\chi),
\label{collinear}
\eeq
where $x=2E_\gamma/\sqrt{s}$ ($E_\gamma$ is the photon energy), 
$\theta$ is the angle between the photon direction and the direction of the 
incoming electron beam, ${\cal F}$ denotes the collinear factor:
\beq
{\cal F}(x, \cos\theta) = \frac{\alpha}{\pi}\frac{1+(1-x)^2}{x} 
\frac{1}{\sin^2\theta}\,, \label{llog}
\eeq
and $\hat{\sigma}$ is the WIMP pair-production cross section evaluated at 
the reduced center of mass energy, $\hat{s}=(1-x)s$. Note that upon
integration over $\theta$, the above equation reproduces 
the familiar Weizsacker-Williams distribution function. The factor ${\cal F}$ 
is universal: it does not depend on the nature of the (electrically neutral) 
particles produced in association with the photon. 
Combining Eq. (\ref{eq:prediction}) and Eq. (\ref{collinear}), one can easily obtain the expression for $\dfrac{d\sigma(e^+e^-\to 2\chi+\gamma)}{dx\, d\cos\theta}$. The left panel in Figure \ref{fig:monophoton} shows the comparison between the photon spectra from the process $e^+e^- \to \met + \gamma$ in the explicit supersymmetric models (red) and the spectra predicted by above procedure \cite{Birkedal:2004xn}. At hadron colliders, the corresponding distributions show different shapes, as illustrated in the right panel, which shows the jet $p_T$ distribution for an EFT leading to mono-jet plus $\mptvec$ at the Tevatron \cite{Bai:2010hh}.

\begin{figure}[t!]
    \centering
    \includegraphics[width=0.5\linewidth]{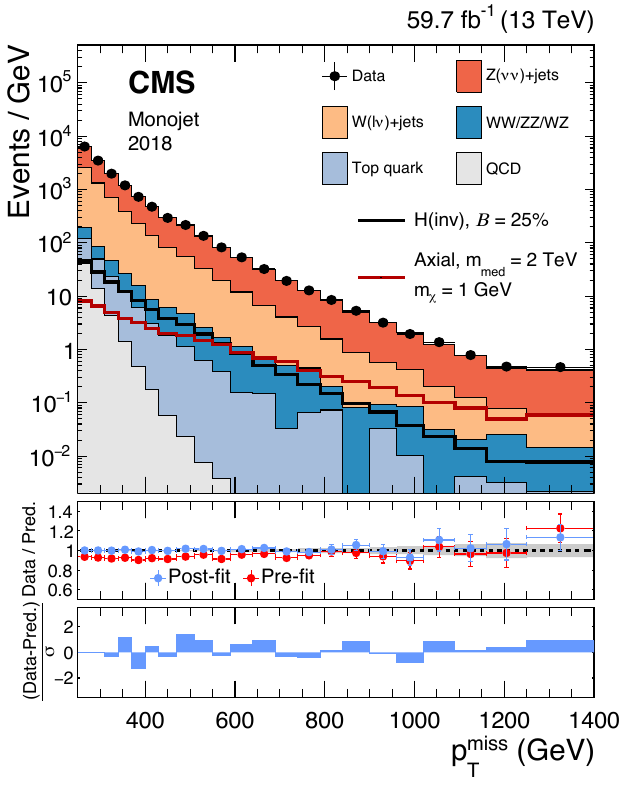}\hspace*{-0.1cm}
    \includegraphics[width=0.5\linewidth]{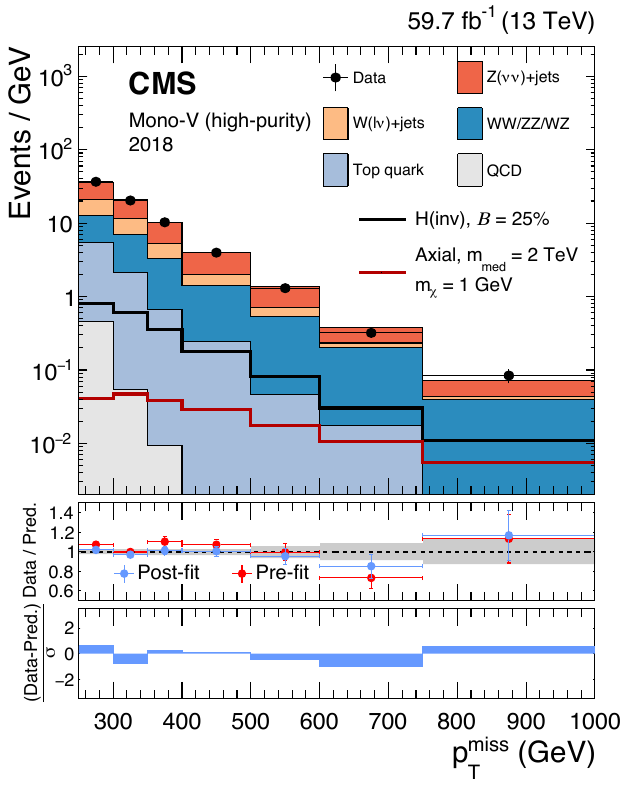}
    \caption{
    Comparison between data and the background prediction in the monojet (left) and monoV (right) signal region before and after the simultaneous fit. 
    Taken from Ref. \cite{CMS:2021far}.}
    \label{fig:monojet}
\end{figure}

Similar to $\gamma + \met$ at LC, one can consider dark-matter production at hadron colliders.
Dark-matter particles could lead to events with the large missing transverse momentum, if another visible object (e.g., an energetic jet) is produced at the same time. Such mono-jet process has been widely studied at the Tevatron, LHC and future colliders \cite{Bai:2010hh,Goodman:2010ku,Hubisz:2008gg,Bae:2017ebe}. The same idea has been extended to other standard model particles (such as $W/Z$, $b/t$, $h$, etc and a new particle such as $Z^\prime$) being produced together with dark-matter candidates. Such searches are often called mono-X searches \cite{Abercrombie:2015wmb} and are one of primary methods looking for dark-matter particles at the LHC. 
Figure \ref{fig:monojet} (taken from Ref. \cite{CMS:2021far}.) shows the missing transverse momentum comparison between data and the background prediction in the monojet (left) and monoV (right) signal region before and after the simultaneous fit. 
A similar result can be found from Ref. \cite{ATLAS:2021kxv}. (see Ref.~\cite{monoXLHC} for summary of recent results.)

ISR plays an important role in the search for new physics models, where the mass spectrum is generate ({\it i.e.,} see Refs. \cite{Martin:2007gf,LeCompte:2011cn} for compressed supersymmetry and Refs. \cite{Cheng:2002ab,Cheng:2002iz,Freitas:2017afm} for universal extra dimensions). For example, when the squark ($\tilde q$) and the lightest neutralino ($\tilde\chi_1^0$) are degenerate, the pair-produced squark leads to very soft decay products with little missing transverse momentum, which will be completely hidden under the QCD backgrounds. By requiring a hard ISR, one can boost the squark system with substantial $p_T$, and improve the signal sensitivity, especially in the region of $m_{\tilde\chi_1^0} \approx m_{\tilde q}$. 

 Another situation where the ISR is crucial is the low mass dijet resonance searches. The low mass dijet resonance is completely hidden under QCD backgrounds. Similarly to SUSY searches with degenerate mass spectrum, one requires a hard radiation ($\gamma$, $Z$, $j$ etc) from the initial state, which boosts the resonance to a high transverse momentum. Such boosted resonance will appear as unresolved fat-jet. The appropriate tagging algorithm and the hard radiation can overcome the huge QCD background \cite{Shimmin:2016vlc,An:2012ue}. Mono-jet searches can therefore be interpreted in several different ways --- as dark-matter production in association with a jet, squark production in SUSY with degenerate spectrum, or $Z^\prime$ production with invisible decay.

Finally, initial state radiation is also useful in the measurement of particle properties such as top quark mass measurement at the LHC \cite{Alioli:2013mxa,Fuster:2017rev} and ILC \cite{Boronat:2019cgt}.

\section{Variables and methods using ensembles of events}
\label{sec:ensemblesn}

\begin{figure}[t!]
    \centering
    \includegraphics[width=0.95\linewidth]{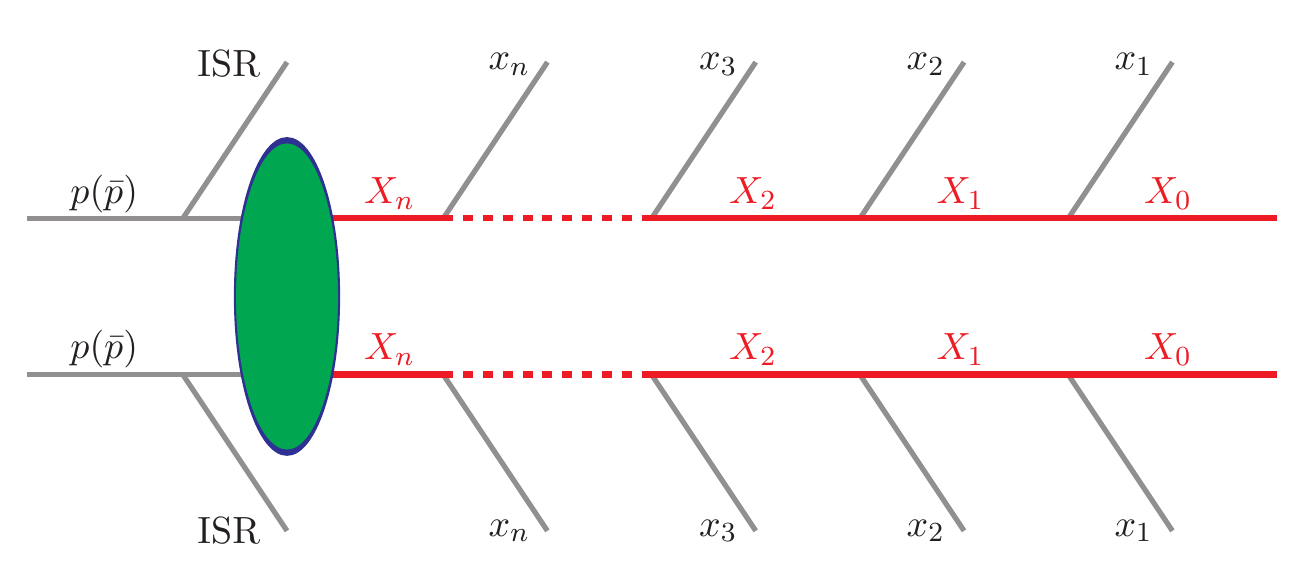}
    \caption{The generic event topology under consideration in Section~\ref{sec:ensemblesn}. The particles $X_i, 1\le i\le n$, are BSM particles which appear as promptly decaying, on-shell intermediate resonances. The particles $x_i$ are the corresponding SM decay products, which are all visible in the detector, i.e.,~we assume that there are no neutrinos among them. ISR stands for generic initial state radiation with total transverse momentum $\vec{p}_T$. $X_0$ is a BSM particle which is invisible in the detector. The integer $n$ counts the total number of {\em intermediate} BSM particles in each chain, so that the {\em total} number of BSM particles in each chain is $n+1$. For simplicity, in this review we shall only consider symmetric events, in which the two decay chains are identical. The generalization of the methods discussed here to asymmetric decay chains is straightforward \cite{Barr:2009jv,Konar:2009qr}. Taken from Ref. \cite{Burns:2008va}.}
    \label{fig:metevent}
\end{figure}

So far we have considered kinematic variables which can be calculated on an event-by-event basis. In this section, we introduce a variety of different types of kinematic methods, which use ensembles of events. For concreteness, let us consider the generic event topology in a collider analysis displayed in Figure~\ref{fig:metevent}. The particles $X_i, 1\le i\le n$, are BSM particles which appear as promptly decaying, on-shell intermediate resonances. The particles $x_i$ are the corresponding SM decay products, which are all visible in the detector. We begin with the so-called polynomial method \cite{Cheng:2007xv,Cheng:2008mg,Cheng:2008hk}. 

\subsection{Polynomial method  \label{sec:poly}}

We use the experimentally measured four-momenta $p^{\mu}_i$ (of all SM particles) as well as the missing transverse momentum $\mptvec$ in the event. We then impose the mass shell constraints for the intermediate BSM particles $X_i$ and attempt to solve the resulting system of equations for the 8 unknown components of the four-momenta $q_0^{\mu}$ of the two missing particles $X_0$. With the $(n+1)$ unknown BSM masses $m_i$ for $X_i$, the number of unknown parameters $N_p$ is given by
\begin{equation}
N_p = 8 + (n + 1)= n+9 \, .
\label{Np_pol1}
\end{equation}
The number of measurements (constraints) $N_m$ includes the two components of the missing transverse momentum condition and $2(n+1)$ mass-shell conditions (for each BSM particle $X_i$ belonging to one of the two decay chains shown in Figure~\ref{fig:metevent}): 
\begin{equation}
N_m = 2(n+1) + 2 = 2n + 4 \, .
\label{Nm_pol1}
\end{equation}
Then the number of undetermined parameters for any given event is 
readily obtained from Eqs.~\eqref{Np_pol1} and \eqref{Nm_pol1}
\begin{equation}
N_p - N_m = 5 - n.
\label{Ndif_pol1}
\end{equation}
Therefore if $n \geq 5$, one can in principle solve for the momenta of the invisible particles and reconstruct the entire final state (up to the combinatorial issue mentioned in Section~\ref{sec:challenges}). 

However, one might do better than this, by combining the information from
two or more events \cite{Nojiri:2003tu,Kawagoe:2004rz, Cheng:2008mg,Cheng:2009fw}. 
For example, consider another event of the same type.
Since the $(n+1)$ unknown masses were already counted in Eq.~(\ref{Np_pol1}),
the second event introduces only 8 new parameters 
(the four-momenta of the two $X_0$ particles in the second event),
bringing up the total number of unknowns in the two events to
\begin{equation}
N_p = 8 + 8 + (n + 1)= n+17 \ .
\label{Np_pol2}
\end{equation}
At the same time, all the constraints are still valid for the second event,
which results in $(2n+4)$ additional constraints. 
This brings the total number of constraints to
\begin{equation}
N_m = (2n+4) + (2n+4) = 4n + 8\ .
\label{Nm_pol2}
\end{equation}
Subtracting (\ref{Np_pol2}) and (\ref{Nm_pol2}), we get
\begin{equation}
N_p - N_m = 9 - 3n.
\label{Ndif_pol2}
\end{equation}
Comparing the previous result (\ref{Ndif_pol1}) 
with (\ref{Ndif_pol2}), we see that the latter decreases much faster with $n$. 
Therefore, when using the polynomial method, combining information from two different events is beneficial for large $n$ (in this example, for $n \ge 3$).

Following the same logic, one can generalize this parameter counting 
to the case where the polynomial method is applied for a group of $N_{\rm eve}$ 
different events of the same type at a time. The number of unknown parameters is
\begin{equation}
N_p = n + 1 + 8 N_{\rm eve} ,
\label{Np_polNev}
\end{equation}
the number of constraints is
\begin{equation}
N_m = (2n+4)N_{\rm eve} ,
\label{Nm_polNev}
\end{equation}
and therefore, the number of undetermined parameters is given by
\begin{equation}
N_p - N_m = n+1 -2(n-2)N_{\rm eve} .
\label{Ndif_polNev}
\end{equation}
For $N_{\rm eve}=1$ and $N_{\rm eve}=2$ this equation reduces to Eqs.~\eqref{Ndif_pol1} and \eqref{Ndif_pol2}, respectively. 
What is the optimal number of events $N_{\rm eve}$ for the polynomial method? 
The answer can be readily obtained 
from Eq.~(\ref{Ndif_polNev}), where $N_{\rm eve}$ enters the last term on
the right-hand side. If this term is negative, increasing $N_{\rm eve}$
would decrease the number of undetermined parameters, and therefore 
it would be beneficial to combine information from more and more 
different events. From Eq.~(\ref{Ndif_polNev}) we see that 
this would be the case if the decay chain is sufficiently long, i.e.,~$n \ge 3$. On the other hand, when $n = 1$, considering more than one event at a time is actually detrimental -- we are
adding more unknowns than constraints. In the case of $n=2$, the number of undetermined parameters $N_p-N_m$ is 
actually independent of $N_{\rm eve}$ and one might as well consider the simplest case of $N_{\rm eve}=1$. 
\begin{figure}[t!]
    \centering
    \includegraphics[width=0.8\linewidth]{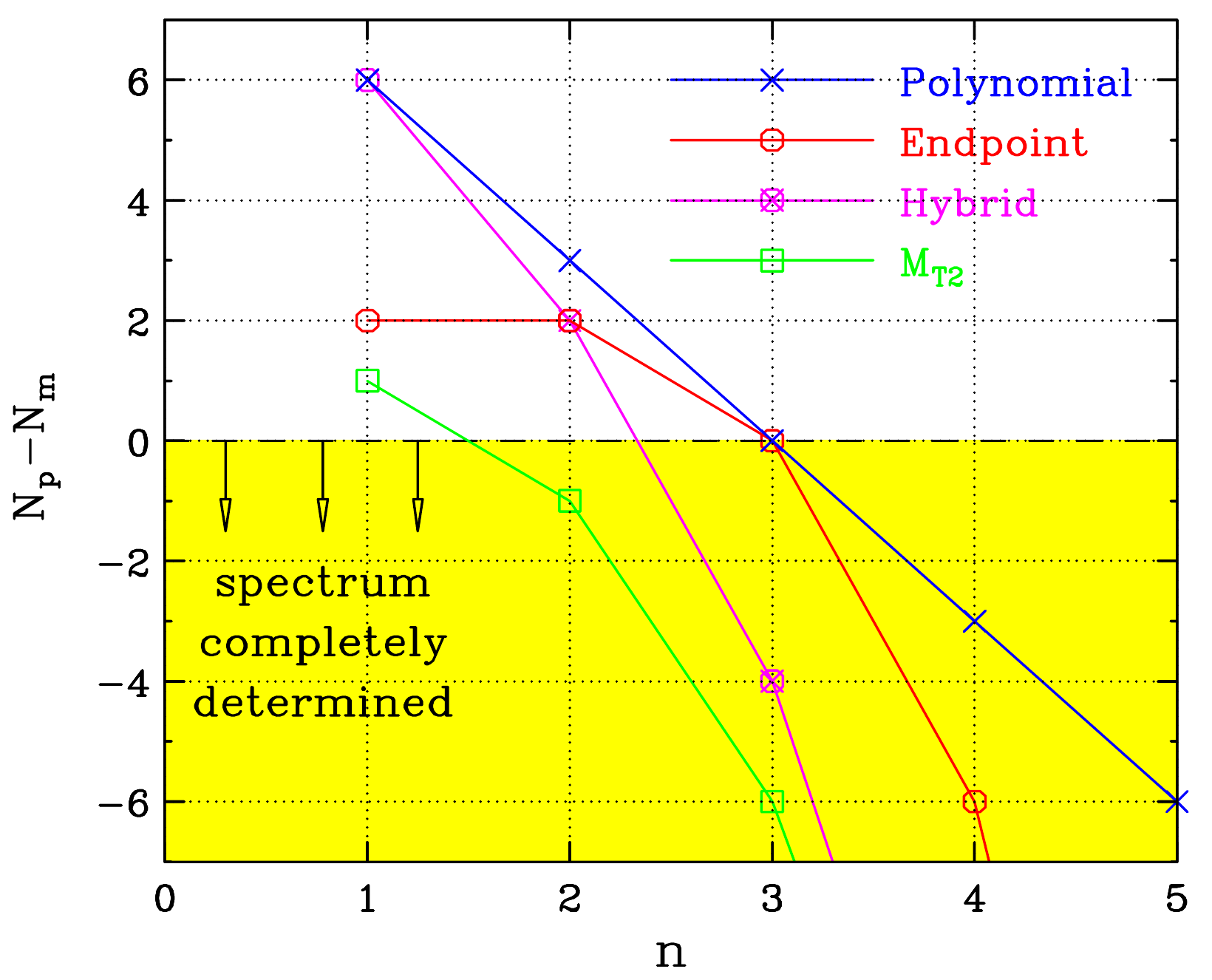}
    \caption{The dependence of the number of undetermined parameters $N_p-N_m$ as a function of the number $n$ of intermediate heavy resonances in the decay chains of Figure~\ref{fig:metevent}, for various mass determination methods: $M_{T2}$ method (green, open squares), endpoint method (red, open circles), polynomial method for $N_{\rm eve}=2$ (blue, $\times$ symbols), or a hybrid method which is a combination of the latter two methods (magenta, $\otimes$ symbols). Within the yellow-shaded region the number of unknowns $N_p$ does not exceed the number of measurements $N_m$ for the corresponding method, and the mass spectrum can be completely determined. Taken from Ref. \cite{Burns:2008va}.}
    \label{fig:param}
\end{figure}

Figure~\ref{fig:param} summarizes the dependence of the number of undetermined parameters $N_p-N_m$ as a function of the number $n$ of intermediate heavy resonances in the decay chains of Figure~\ref{fig:metevent}, for various mass determination methods: $M_{T2}$ method (green, open squares), endpoint method (red, open circles), polynomial method for $N_{\rm eve}=2$ (blue, $\times$ symbols), or a hybrid method which is a combination of the latter two methods (magenta, $\otimes$ symbols) \cite{Burns:2008va}. Within the yellow-shaded region the number of unknowns $N_p$ does not exceed the number of measurements $N_m$ for the corresponding method, and the mass spectrum can be completely determined. 
For readers who are interested in counting the number of undetermined
parameters of various methods, we refer to Ref. \cite{Burns:2008va} for more details including a hybrid method combining the techniques of the polynomial and endpoint methods \cite{Nojiri:2007pq}. A similar idea on  the mass determination in sequential particle decay chains is discussed in Ref. \cite{Webber:2009vm}.

The polynomial method described in this section relies on the presence of a sufficient number of kinematic constraints, so that the event kinematics becomes exactly solvable for the components of the invisible momenta. This typically requires complex event topologies, with several successive decays in each decay chain. Therefore, the method cannot be applied to simpler event topologies with fewer kinematics constraints, and new ideas are needed. Some of those alternative techniques are described in the following subsections.

\subsection{Focus point method  \label{sec:focus}}

\begin{figure}[t]
    \centering
    \includegraphics[height=0.45\columnwidth]{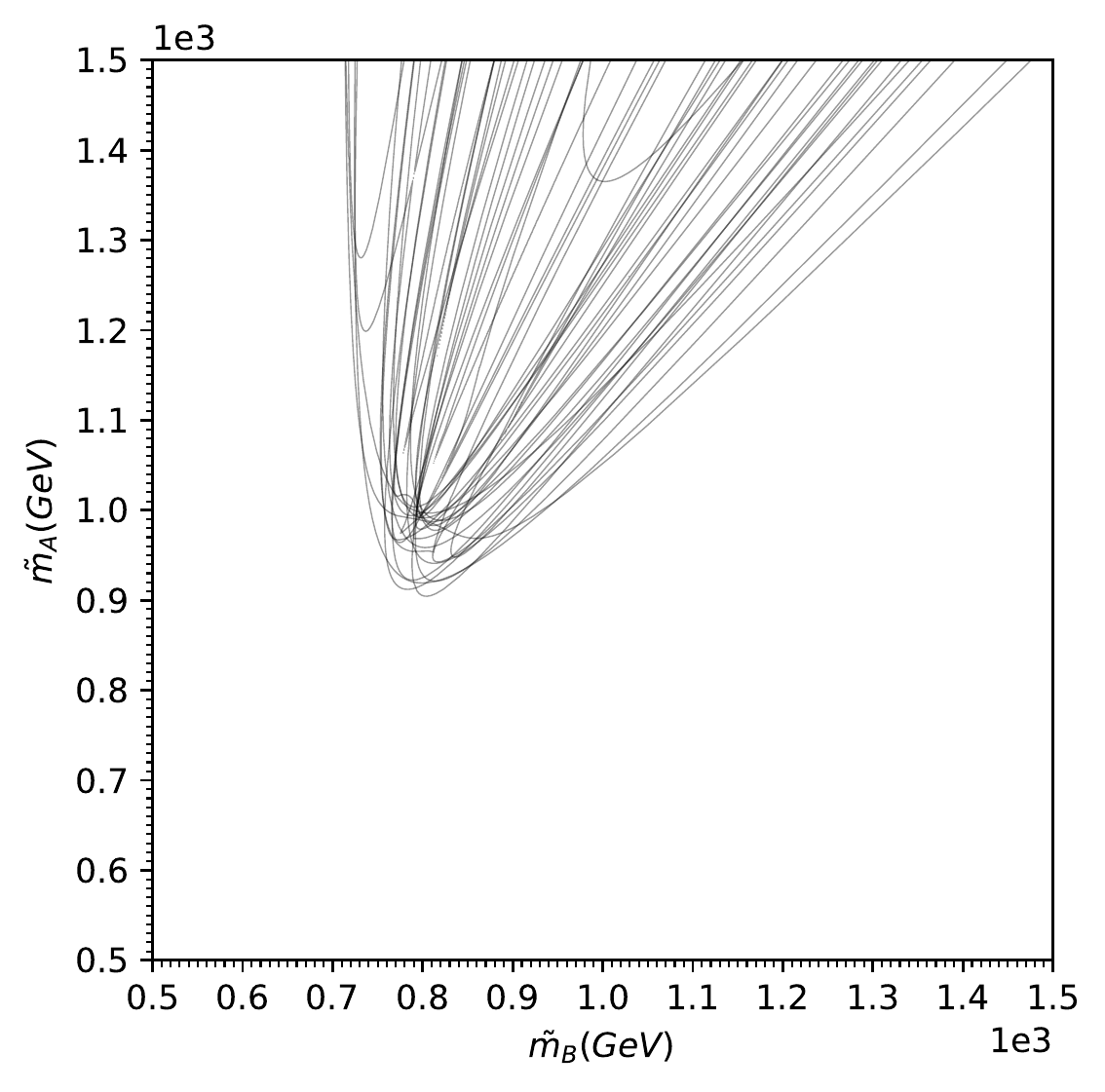}
    \includegraphics[height=0.45\columnwidth]{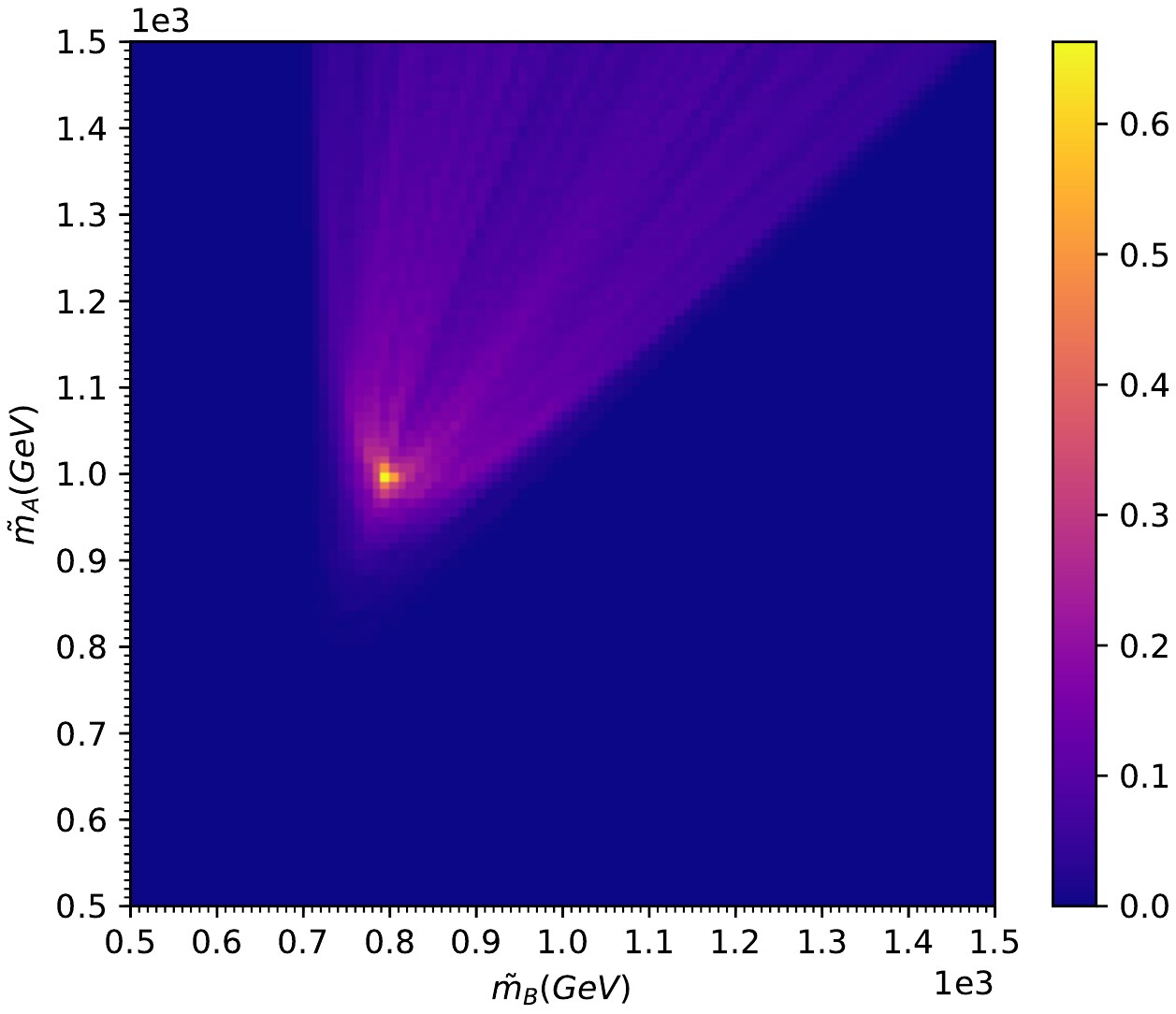}
    \caption{The left panel shows a plot of 20 extreme curves in the $(\tilde m_{A_1}, \tilde m_{A_2})$ plane for fixed $\tilde m_{a_0} = m_{a_0} = 700$ GeV, for the the $t\bar{t}$-like topology of Figure~\ref{fig:mt2diagram}. The right panel shows the fractional density of extreme curves, i.e., the fraction of events whose extreme curves pass through a given $10\times 10$ GeV pixel. (Taken from Ref.~\cite{Kim:2019prx}.) }
    \label{fig:focuspoints}
\end{figure}

The focus point method for mass measurement proposed in \cite{Kim:2019prx,Matchev:2019bon} can be applied to certain event topologies with underconstrained kinematics. The method relies on the fact that the projection onto the visible space will result in a relatively large number of events in the vicinity of a singularity, as illustrated in Figure~\ref{fig:projection}. Turning the argument around, one could ask, for any given event, what choice of the unknown mass parameters would place a singularity at that point. This condition delineates a hypersurface in mass parameter space, and we have one such ``extreme'' surface for each event in the data. As shown in \cite{Kim:2019prx}, the extreme surfaces for many of the events in the dataset pass close to the true values of the unknown masses. This leads to a technique for estimating the unknown masses simply as the ``focus-point'' of the extreme surfaces in the mass-parameter space. This is illustrated in Figure~\ref{fig:focuspoints} for the $t\bar{t}$-like topology of Figure~\ref{fig:mt2diagram}($b$), for $m_{A_2}=1000$ GeV, $m_{A_1}=800$ GeV and $m_{a_0}=700$ GeV. The left panel in the figure shows the kinematic boundaries of 100 events in the trial mass parameter space $(\tilde m_{A_1}, \tilde m_{A_2})$, with $\tilde m_{a_0}$ fixed to its true value $m_{a_0}=700$ GeV. Notice that the kinematic boundaries tend to focus at the true values of the masses of the parent particles $A_2$ and $A_1$ in this example. The right panel in Figure~\ref{fig:focuspoints} shows a heatmap of the density of extreme surfaces (curves in this case) per $10\times 10$ GeV bin. The bright spot in the figure clearly marks the true values of the masses. This technique can be readily generalized from $t\bar{t}$ events to more general event topologies in SUSY and beyond \cite{Kim:2019prx}. 

\subsection{Kinematic endpoint methods  \label{sec:endpoint}}

As we have discussed in preceding sections, the distributions of kinematic variables often allow us to infer the mass values of the (new) particles involved in the physics process of interest. For example, the endpoints of invariant mass, transverse mass, $M_{T2}$, and $M_2$ variables and energy peaks are determined purely by kinematics, regardless of detailed dynamics (see also Section~\ref{sec:exclusive}). Therefore, those observables have received great attention and a host of ideas have been proposed especially in the context of mass measurements of new particles because they do not require any prior knowledge about the exact details of the associated new physics. 

The classic supersymmetric ``$q\ell\ell$-chain'' introduced in Section~\ref{sec:mass} is one of the most extensively studied benchmark processes in this regard~\cite{Allanach:2000kt,Lester:2001zx,Gjelsten:2004ki,Gjelsten:2005aw,Miller:2005zp,Costanzo:2009mq,Burns:2009zi,Matchev:2009iw,Kim:2015bnd,Matchev:2019sqa}.
There are up to four unknown mass parameters, namely, $m_{\tilde{q}}$, $m_{\tilde{\chi}_2^0}$, $m_{\tilde{\ell}}$, and $m_{\tilde{\chi}_1^0}$, and four kinematic endpoints are readily available, e.g., $M_{q\ell_n}^{\max}$, $M_{q\ell_f}^{\max}$, $M_{\ell\ell}^{\max}$, and $M_{q\ell\ell}^{\max}$. To avoid the combinatorial ambiguity between $\ell_n$ and $\ell_f$, one uses instead the set $M_{q\ell}^{<,\max}$, $M_{q\ell}^{>,\max}$, $M_{\ell\ell}^{\max}$, and $M_{q\ell\ell}^{\max}$ (see Section~\ref{sec:mass}).   
Therefore, in principle, all the unknown mass parameters can be completely determined by ``inverting'' the four endpoint measurements. 
However, this mass determination can be sometimes ambiguous or underconstrained, and several degenerate solutions for the masses may arise~\cite{Gjelsten:2004ki,Arkani-Hamed:2005qjb,Gjelsten:2005sv,Gjelsten:2006as,Costanzo:2009mq}.

There are several reasons for these challenges in the mass determination. First of all, depending on the underlying mass spectrum, the four observables are not completely independent; in certain regions of parameter space, the following relation holds~\cite{Gjelsten:2004ki}:
\beq
\left(M_{q\ell\ell}^{\max} \right)^2 = \left(M_{q\ell}^{>,\max} \right)^2+\left( M_{\ell\ell}^{\max} \right)^2\,.
\eeq
In this case, another independent measurement is needed, and one such example is the lower kinematic endpoint $M_{q\ell\ell,{\rm const.}}^{\min}$~\cite{Allanach:2000kt}. Here the constrained variable $M_{q\ell\ell,{\rm const.}}$ is the usual $M_{q\ell\ell}$ subject to the condition $M_{\ell\ell}^{\max}/\sqrt{2}<M_{\ell\ell}<M_{\ell\ell}^{\max}$ which forces to choose events where the opening angle between the two leptons is greater than $\pi/2$ in the rest frame of $\tilde{\ell}$.
Second of all, the finite detector resolution smears each of the measured endpoint values away from the theoretically predicted ones, resulting in multiple and/or unphysical solutions~\cite{Gjelsten:2004ki,Gjelsten:2005aw,Gjelsten:2005sv}. Similarly, if an endpoint is identified as the (long) tail of the associated distribution, its measurement is highly sensitive to the data statistics and ``false'' solutions can emerge.  
However, even in the ideal case of a perfect experiment, it is possible that different mass spectra can result in the same set of endpoints so that an application of relevant inversion formulas (see Ref.~\cite{Burns:2009zi} for the full sets of formulas) would yield ``fake'' solutions~\cite{Burns:2009zi}. 
A possible way of resolving this ambiguity is to study the shapes of the boundaries of the bivariate distributions in $\left\{ M_{q\ell}^<, M_{q\ell}^>\right\}$ and $\left\{ M_{\ell\ell}, M_{q\ell\ell}\right\}$~\cite{Burns:2009zi}.

The above discussion requires one sufficiently long decay chain to determine the full mass spectrum. However, for the event topologies involving pair-produced heavy resonances, the mass determination can often be done in combination with transverse variables such as $M_{T2}$. A well-studied prototypical example is the fully-leptonic SM $t\bar{t}$ production. The $b\ell$ invariant mass endpoint encodes the mass relation
\beq
M_{b\ell}^{\max}=\frac{1}{m_W}\sqrt{(m_t^2-m_W^2)(m_W^2-m_\nu^2)}\,,
\eeq
and the two combinatorics-free subsystems, leptonic and bottom respectively, when applied to $M_{T2}$ or $M_{T2,\perp}$, allow one to extract two independent mass relations $\mu^{(\ell\ell)}$ and $\mu^{(bb)}$, respectively (see also Section~\ref{sec:mt2variables}):
\bea
\mu^{(\ell\ell)} &=& \frac{m_W}{2}\left(1-\frac{m_\nu^2}{m_W^2} \right)\,,\\
\mu^{(bb)} &=& \frac{m_t}{2}\left(1-\frac{m_W^2}{m_t^2} \right)\,.
\eea
The CMS Collaboration performed the top quark mass measurement using this idea at $\sqrt{s}=7$~TeV~\cite{CMS:2012eya,CMS:2013wbt} and $\sqrt{s}=8$~TeV~\cite{CMS:2016kgk}, and achieved a $\sim 2$~GeV-level and a $\lesssim1$~GeV-level systematics in the respective measurements by constraining the $W$ and $\nu$ masses from other independent measurements. 

In a similar fashion, the three mass parameters appearing in the two-step two-body cascade decay topology shown in Figure~\ref{fig:invmassdiagrams}($a$) can be determined in combination with the peak values in the $a_2$ and $a_1$ energy distributions, if both of $A_2$ and $A_1$ are either scalar or unpolarized. Again, the kinematic endpoint in the $M_{a_2a_1}$ distribution is
\beq
M_{a_2a_1}^{\max}=\frac{1}{m_{A_1}}\sqrt{(m_{A_2}^2-m_{A_1}^2)(m_{A_1}^2-m_{a_0}^2)}\,,
\eeq
and the two energy-peak values are given by
\bea
E_{a_2}^{\rm peak} &=& \frac{m_{A_2}^2-m_{A_1}^2}{2m_{A_2}}\,, \\
E_{a_1}^{\rm peak} &=& \frac{m_{A_1}^2-m_{a_0}^2}{2m_{A_1}}\,.
\eea 
The last three mass relations are completely independent, allowing one to determine the three mass parameters. 
The generic idea was first proposed in Ref.~\cite{Agashe:2013eba}, and applied to a supersymmetric gluino decay process, $\tilde{g}\to b \tilde{b}, \tilde{b} \to b \tilde{\chi}_1^0$, with combinatorial ambiguity in the $b$-jet energy distribution appropriately prescribed.  

\subsection{Matrix element and likelihood methods \label{sec:MLM} }

The Matrix Element Method (MEM) is one of the likelihood methods which utilizes the quantum amplitude of a process. 
The probability to observe visible particles $\{P_i^{\rm vis}\}$, $i=1, \cdots, N_{\rm vis}$ under the assumed process and parameters $\{\alpha\}$ for a single event is given by
\begin{eqnarray} \label{eq:likelihood}
&&{\cal P}(\{P_i^{\rm vis}\} |\alpha) = \frac{1}{\sigma_{\alpha}}
\biggr[ \prod_{i=1}^{N_{\rm vis}} \int \frac{d^3 p_i}{(2\pi)^3 2E_i} \biggr]
W(\{P_i^{\rm vis}\}, \{p_i\} ) \nonumber \\
&&\times  
\biggr[ \prod_{j=1}^{N_{\rm inv}} \int \frac{d^3 q_j}{(2\pi)^3 2E_j} \biggr]
\sum_{a,b}  \frac{f_a(x_1)f_b(x_2)}{2sx_1x_2} |{\cal M}_{\alpha}(\{p_i\},\{q_j\})|^2 \nonumber\\
&&\times (2 \pi)^4 \delta^4\left[p_a+p_b-\left(\sum_{i} p_i+\sum_{j} q_j^{\rm inv}\right)\right],
\end{eqnarray}
where $\{p_a, p_b\}$ are the four-momenta of the initial-state partons $a$ and $b$ and $\{f_a, f_b\}$ are their corresponding PDFs. 
Here one integrates out the unknown momenta $\{q_j\}$ of invisible particles $j=1, \cdots, N_{\rm inv}$ and considers various non-partonic effects including detector response and QCD activity with a transfer function 
$W(\{P_i^{\rm vis}\}, \{p_i\} )$ between the parton-level momentum $\{ p_i \}$ and the reconstructed momentum $\{ P_i^{\rm vis}\}$.  
Integrating the transfer functions often takes up most of the computing resources for a MEM analys. 
In the case where the visible particles consist of only light leptons (electron and muon), we can neglect the transfer functions, which leads to an important simplification for purely leptonic channels like Higgs to four leptons~\cite{Gao:2010qx,Bolognesi:2012mm,Avery:2012um}. We can construct the likelihood specific to $\{\alpha\}$, $\mathcal{L}_\alpha$, for a set of $N$ events with individual likelihoods for each event $n$ as
\beq
\mathcal{L}_\alpha \equiv \prod_n^N {\cal P}(\{P_i^{\rm vis}\}_n |\alpha).
\eeq
Thus one can expect to find the model parameters $\{\alpha\}$ by maximizing $\mathcal{L}_\alpha$.
By constructing a likelihood function, one can measure not only particle properties such as the mass and width of new particles but also coupling structures in the interaction vertices as demonstrated in~\cite{Betancur:2017kqe}. 

\subsection{Edge detection \label{sec:voronoi}}

The identification of kinematic endpoints in a certain {\it one}-dimensional distribution is often indicative of boundaries in the {\it high}-dimensional phase space that is being projected onto the one-dimensional subspace (see Figure~\ref{fig:projection}). At the same time, such kinematic endpoints can also be formed by signal events piling up on top of a smooth background distribution, and thus can be used for discovery \cite{Debnath:2015wra}. 

The traditional endpoint techniques (usually in the context of supersymmetry, see Ref.~\cite{Matchev:2019sqa} for a recent review) were typically applied to one-dimensional histograms, but that does not necessarily have to be the case --- while dimensional reduction increases the statistics near the kinematic endpoint, it may also result in the loss of useful information. This is why recent work advocated for edge detection in a two-dimensional space of observables~\cite{Karapostoli:2008rra,Huang:2008ae,Costanzo:2009mq,Burns:2009zi} and even for a (surface) boundary detection in three or more dimensions~\cite{Agrawal:2013uka,Debnath:2016gwz,Altunkaynak:2016bqe}.

The traditional approach to edge detection is to bin the data in a lower-dimensional observable space and identify a kinematic edge by comparing the counts in adjacent bins, looking for a significant variation~\cite{Agashe:2010tu,Curtin:2011ng}. Conventional edge-detection algorithms for machine vision have been developed mostly for two-dimensional image data and are not necessarily aligned with the goals of particle physics analyses, which need to account for smearing of the edges due to detector resolution, particle widths, etc.

When it comes to detecting kinematic features in the data,  an alternative approach to binning is offered by the {\em tessellation} of the data, where we can treat the full set of collider events as a point pattern in the observable multi-dimensional kinematic space. There exist different tessellation methods. For example, Ref.~\cite{Debnath:2015wra} proposed a phase-space edge detection method based on the {\em Voronoi} tessellation, which divides the original space into non-overlapping regions (Voronoi cells) so that the points within each region are closest to one of the original data points~\cite{Voronoi}. It was shown that
the value of the scaled standard deviation, 
\begin{equation}
    \frac{\sigma_a}{\bar{a}}=\frac{1}{\bar{a}}\sqrt{\sum_{n \in N_i}\frac{(a_n -\bar{a})^2}{|N_i|-1}},
\label{eq:scaleddev}    
\end{equation}
where $N_i$ is the set of neighbors of the $i$-th Voronoi cell and $\bar{a}(N_i)$ is their mean area, is indicative of whether the $i$-th cell is close to a boundary. This result can be easily understood intuitively by noting that for boundary cells, the neighbors on the dense side have small areas while the neighbors on the sparse side have large areas. 
Therefore, edge cells are expected to show relatively large scaled standard deviation.
The method was subsequently tested on the classic supersymmetric ``$q\ell\ell" chain$, $\tilde{q} \to \tilde{\chi}_2^0 j$, $\tilde{\chi}_2^0 \to \tilde{\ell}^\pm \ell^\mp$, $\tilde{\ell}^\pm \to \tilde{\chi}_1^0 \ell^\pm$: Reference~\cite{Debnath:2015wra} considered edge detection in the two-dimensional space $\left\{m_{\ell\ell}^2, (m_{j\ell\ell}^2-m_{\ell\ell}^2)/6 \right\}$, while Ref.~\cite{Debnath:2016mwb} demonstrated a surface boundary detection in the three-dimensional space $\left\{m_{j\ell_n}^2, m_{\ell\ell}^2, m_{j\ell_f}^2 \right\}$. The method can also be adapted to mass measurements of new particles~\cite{Debnath:2016gwz} and for enhancing the discovery opportunities in combination with the $\Delta_4$ variable~\cite{Debnath:2018azt}.  

The Delaunay triangulation is a tessellation that is the dual graph of a Voronoi tessellation. Therefore, Ref.~\cite{Matchev:2020vhr} proposed an alternative edge detection method which utilizes the Delaunay tessellation of the data instead. Since edge detection necessarily involves computing the gradient of the phase space density, the Delaunay cells, being formed by several neighboring data points, are the natural objects for computing local gradients.

\subsection{Interference effects  \label{sec:mgammagamma}}

When high precision is required in measurements of masses or other kinematic properties, subtle quantum effects can modify the theoretical predictions in ways that can be hard to predict or to interpret without a full theoretical understanding of kinematic variables.  

Even relatively simple observables like the invariant mass can be affected by subtle effects which may give apparently inconsistent results between measurements of the mass of a particle in two different channels. This is the case for instance for the Higgs boson whose most precise mass measurements are in the $\gamma\gamma$ and $4\ell$ channels. The key difference between these two channels is that the $4\ell$ channel is essentially free from background, whereas the $\gamma\gamma$ channel has substantial background from QED+QCD production of two photons. The presence of large background opens the possibility to have non-resonant features to redefine the expected signal shape even for perfect detectors. The studies in Refs.~\cite{Dixon:2003yb,Martin:2012xc,deFlorian:2013psa,Martin:2013ula,Coradeschi:2015tna,Cieri:2017kpq} have pointed out that the $\gamma\gamma$ peak in the $M_{\gamma\gamma}$ distribution is expected to be shifted and broadened by the interference between $gg\to\gamma\gamma$ and $gg\to h \to \gamma \gamma$. 
Therefore, the mass measured as the peak of the $M_{\gamma\gamma}$ distribution will differ from that measured in background-free channels such as $4\ell$. The subtraction of the peak from $4\ell$ events into the extraction of the peak in the $M_{\gamma\gamma}$  distribution may help to highlight this effects in a model-independent way. As the effect has to do with the Higgs boson width, it has also been pointed out that the mass shift can be used to constrain the Higgs boson width~\cite{Dixon:2013haa,Campbell:2017rke}. 

The shift also depends on experimental conditions such as the the diphoton mass resolution. Depending on how the measurement in the $\gamma\gamma$ channel is performed, the shift may range from a fraction of 100 MeV to a fraction of 1 GeV, which would be significantly large to be observed~\cite{Martin:2012xc}. The ATLAS Collaboration has evaluated the impact of this theoretical effect on a realistic mass measurement and has found a  non-vanishing effect around 35~MeV for the specific procedure used to extract the Higgs boson mass in the $\gamma\gamma$ channel~\cite{ATLAS:2016kvj}. 

The present Higgs boson mass appearing in the up-to-date Particle Data Group report is 
$125.25\pm 0.17$ \cite{ParticleDataGroup:2020ssz},
stemming from slightly disagreeing measurements from CMS~\cite{CMS:2020xrn} 
$m_h^{\gamma\gamma} = 125.78 \pm 0.26 $~GeV,
$m_h^{4\ell}        = 125.46 \pm 0.16 $~GeV and ATLAS \cite{ATLAS:2018tdk}
$m_h^{4\ell}        = 124.79 \pm 0.37 $~GeV,
$m_h^{\gamma\gamma} = 124.93 \pm 0.40 $~GeV.
Future HL-LHC measurements, exploiting a dataset about two orders of magnitude larger, could reach a measurement with a much smaller uncertainty than the present one. Systematic uncertainties will be relevant for such a large dataset, but a purely statistical rescaling of the present measurement would hint at the necessity to take into account these subtle effects from interference. For example, in Ref.~\cite{Cepeda:2019klc} a precision on $m_h$ in the $\gamma\gamma$ channel around 10-20~MeV is foreseen \cite{CMS-PAS-FTR-21-007,CMS-PAS-FTR-21-008}, thus calling for a careful evaluation of the interference effects.

\section{Kinematic variables in the machine learning era \label{sec:ML} 
}

\begin{figure}[t]
\begin{center}
\includegraphics[width=0.48\textwidth]{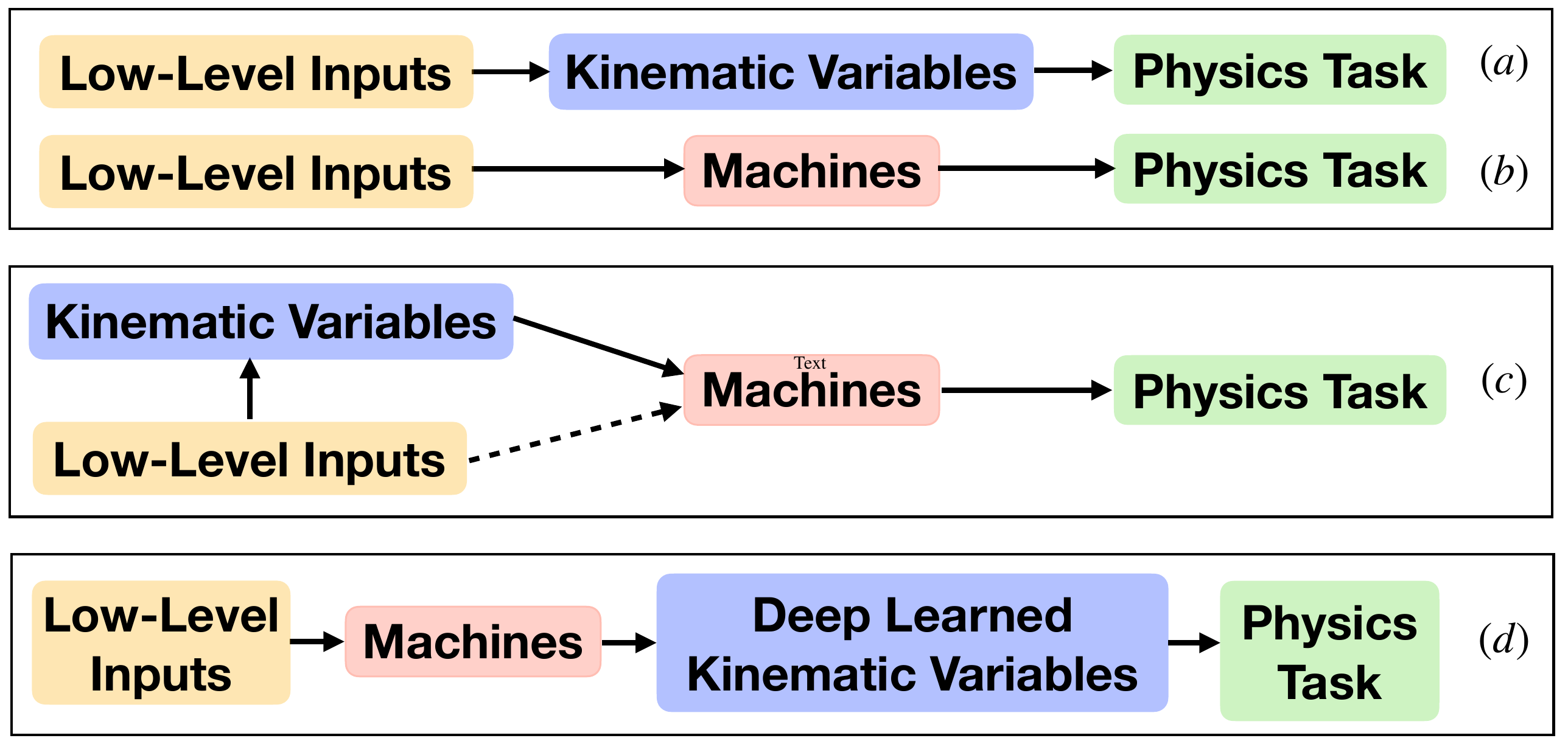}
\end{center}
\caption{\label{fig:mlchain} 
Various possible analysis chains for some physics-motivated task. 
($a$) The traditional (non-ML) analysis technique using kinematic variables; ($b$) ML-based analysis using only raw or low-level data as inputs; ($c$) ML-based analysis using in addition reconstructed objects and/or human-engineered high-level variables as inputs; ($d$) construction of sensitive analysis variables using machine learning techniques.}
\end{figure}

Recently, there has been an explosion of studies employing ML for various tasks in the analysis of high energy data. A collection of references is maintained at HEPML-LivingReview \cite{Feickert:2021ajf}. So far we have discussed traditional (i.e., non-ML) analysis techniques using kinematic variables [Figure~\ref{fig:mlchain}($a$)]. Here we will discuss the multifaceted synergy between ML and kinematic variables in particle physics.

\subsection{Feature engineering}
\noindent{\textbf{Feature Engineering for ML.}~~Kinematic variables are often used as input event-features in ML approaches.
While in principle, one can train machines using only raw or low-level data as inputs [Figure~\ref{fig:mlchain}($b$)], the dimensionality of such feature vectors will typically be large. 
When we feed low-level data, the machine could spend a lot of resources trying to extract useful information from it.
Furthermore, inaccuracies in the simulation models could lead to unknown and unquantified errors in the high-dimensional low-level simulated data, which could lead to unquantified errors in the subsequent ML-based analysis.
Both these issues can be ameliorated by using reconstructed objects and/or human-engineered high-level variables as inputs in ML applications [Figure~\ref{fig:mlchain}($c$)]. Optionally, one can also pass the low-level information to the machine, in addition to the high-level features. Carefully chosen high-level input features can efficiently retain the information from low-level data that is relevant to the task at hand, and facilitate efficient training of ML approaches. Furthermore, reducing the dimensionality of the input allows for easier and more meaningful validation of the simulation models (in the low-dimensional input space), for the purposes of the analysis at hand.

\noindent\textbf{Feature Engineering with ML.}~~An interesting development in the last few years is the construction of sensitive analysis variables using machine learning techniques [Figure~\ref{fig:mlchain}($d$)]. Any ML-approach, in which a machine (e.g., neural network, boosted decision tree) takes individual events as input and returns an output, can be thought of as constructing an analysis variable or observable. However discussing all such applications of machine learning, including classifiers \cite{Guest:2018yhq}, likelihood ratio estimators \cite{Cranmer:2015bka}, etc. is beyond the scope of this paper. Instead, we will focus on ML-approaches for constructing event observables, which are functionally similar to more traditional event observables.

At an abstract level, many ML-approaches for constructing collider observables share the following basic procedure. 1) Construct a trainable, machine-learning-based function which maps a high-dimensional event description to a low-dimensional observable. 2) Construct an evaluation metric to quantify the performance of the ML-based observable, for some task at hand. 3) Train the ML-based observable by optimizing the evaluation metric. Several novel ML-based observables have been proposed in recent years, each differing in the implementation details of the steps listed above~\cite{Datta:2019ndh,Kim:2021pcz}. 

For example, in Ref.~\cite{Datta:2019ndh}, jet observables are constructed as products of powers of $N$-subjettiness variables with unknown (trainable) exponents. The performance of the observable for distinguishing between different signatures ($H\rightarrow bb$ vs. $g\rightarrow bb$; $Z'$ vs. quarks and gluons) is used as the performance metric to be minimized in order to choose the exponents in the observable.

In Ref.~\cite{Kim:2021pcz}, kinematic variables for different event topologies are constructed as 
neural-network-based functions of reconstructed parton-level data. The variables are trained by maximizing the sensitivity of their distributions to the value of underlying parameters in the corresponding event topologies, as captured by the mutual information between the parameters and the variable. Variables trained using such an approach are sensitive over a range of unknown theory parameter values, and can subsequently be used for signal discovery or parameter measurement analyses involving the concerned topology.

\subsection{Domain-inspired machine learning}

Several HEP-inspired neural network architectures have been invented for use in ML for HEP. For example, Energy Flow Networks and Particle Flow Networks \cite{Komiske:2018cqr}, based on Deep Sets \cite{NIPS2017_f22e4747} and Energy Flow Polynomials \cite{Komiske:2017aww}, are neural network architectures designed for learning from collider events represented as unordered, variable-length sets of particles. As another example, Lorentz Boost Networks (LBN) \cite{Erdmann:2018shi, LBN, LBNtth} allow for the construction of composite particles and rest frames (both represented by combinations of particles) within the trainable layers of the network, using four-momenta of final state particles as input. The LBNs also Lorentz boost the composite particles into the constructed rest frames. The features thus constructed within the layers of the network can then be used to perform relevant physics tasks like classification or regression. Such domain-specific network architectures have been observed to outperform other domain-unspecific neural network architectures for collider event classification tasks \cite{Erdmann:2018shi}. Furthermore, they also allow for interpretation of the intermediate layers of the trained neural network.

\subsection{Interpretability and explainability}

Most machine learning approaches, including neural networks and boosted decision trees, act as blackbox systems with varying degrees of ``blackboxness'', depending on their architecture. This poses a challenge to the trustworthiness of ML-based analyses. This problem can be approached from three directions. The first approach is to interpret and explain the ML blackbox. 
The second approach is to try and make the machine less of a black box. 
The third approach is to design ML-based analyses techniques that are robust despite the blackbox nature of the machine. 
Kinematic variables can play a role in each of these approaches. For example, one can use kinematic variables to interpret and explain the decisions made by the machine learning algorithms~\cite{Chang:2017kvc,Faucett:2020vbu,Agarwal:2020fpt,Grojean:2020ech}. Using kinematic-variable-inspired neural architectures like energy flow and particle flow networks~\cite{Komiske:2018cqr} and lorentz boost networks~\cite{Erdmann:2018shi, LBN, LBNtth} can reduce the black-box nature of neural networks. Finally, using machine learning to construct \emph{low-dimensional event} observables that are not tuned to specific study points (i.e., are sensitive to the underlying physics over a range of unknown model parameters)~\cite{Kim:2021pcz}, akin to kinematic variables, allow for meaningful control-region validation of the simulations which in turn leads to robust analysis techniques. 

\subsection{Advantage of quantum computation for identifying event-topologies}

So far, most kinematic analyses are based on an assumed event topology which makes it possible to optimize a kinematic variable in each case. Due to the complicated structure of phase space and the limited information from invisible particles, various machine learning algorithms become useful, but their training data is also generated only for particular event topologies. 
\begin{table}[t]
\caption{\label{tab:counting}
The number of inequivalent event-topologies as a function of $1\le N_{v} \le 4$ and $1\le N_{\chi} \le 5$. Taken from Ref. \cite{Cho:2012er}.}
\begin{ruledtabular}
\begin{tabular}{c | c  c  c  c  c }
\multicolumn{1}{c|}{}                       & \multicolumn{5}{c}{$N_{\chi}$}   \\ 
\hline
$N_{v}$   &    1       &  2     &     3   &      4   &   5        \\   \hline\hline
1             &    1       &  2     &     4   &      8   &    16        \\
2             &    2       &  7    &     20  &    55   &  142   \\
3             &    4       &  20  &    78   &  270    & 860           \\
4             &    8       &  55  &    270 &  1138    & 4294   \\     
 \end{tabular}
\end{ruledtabular}
\end{table}

Given the lack of any clear signal of new physics so far at the LHC, we need to ask how one can perform an optimized analysis without any assumptions on the new physics model. This question is related to identifying the event-topology of the signal from data alone. Checking all possible event-topologies can be a very time-consuming task. For example, the number of possible event-topologies with $N_v =5$ visible particles and $1\le N_\chi\le 4$ ``assumed" invisible particles is $\mathcal{O}(5000)$ as shown in Table\,\,\ref{tab:counting}\,\cite{Cho:2012er}.
As the number of visible and invisible particles increases, the number of possible event-topologies grows exponentially, making the problem of identifying the event-topology an NP-hard problem. Since it is in the category of combinatorial optimization problems, where various quantum algorithms have been introduced \cite{djidjev2018efficient, farhi2014quantum} and shown to be successful, this suggests the use of quantum computers for this task.

When the produced particles are boosted, their decay products are organized into groups exhibiting characteristic structures and substructures. One can then utilize a shape variable used in clustering a jet with either a gate-type quantum computer or a quantum annealer~\cite{Wei:2019rqy, Pires:2020urc, Pires:2021fka, Delgado:2022snu}. 
Unlike QCD jet activity, the phase space of a hard process can exhibit a more complicated structure, in which case one can try to minimize some basic kinematic quantity, for example, the total invariant mass or sum/difference of invariant masses of the clusters.
If one restricts to a $2\to 2$ process, one can directly use a Quadratic Unconstrained Binary Optimization (QUBO) with an Ising model to cluster visible particles as in Figure~\ref{diagram_QA} \cite{Kim:2021wrr}. 
Once we identify the event-topology, we can proceed to optimize the analysis to measure the masses and spins of the new particles as usual. 

\begin{figure}[t]
\centering
\includegraphics[width=0.48\textwidth]{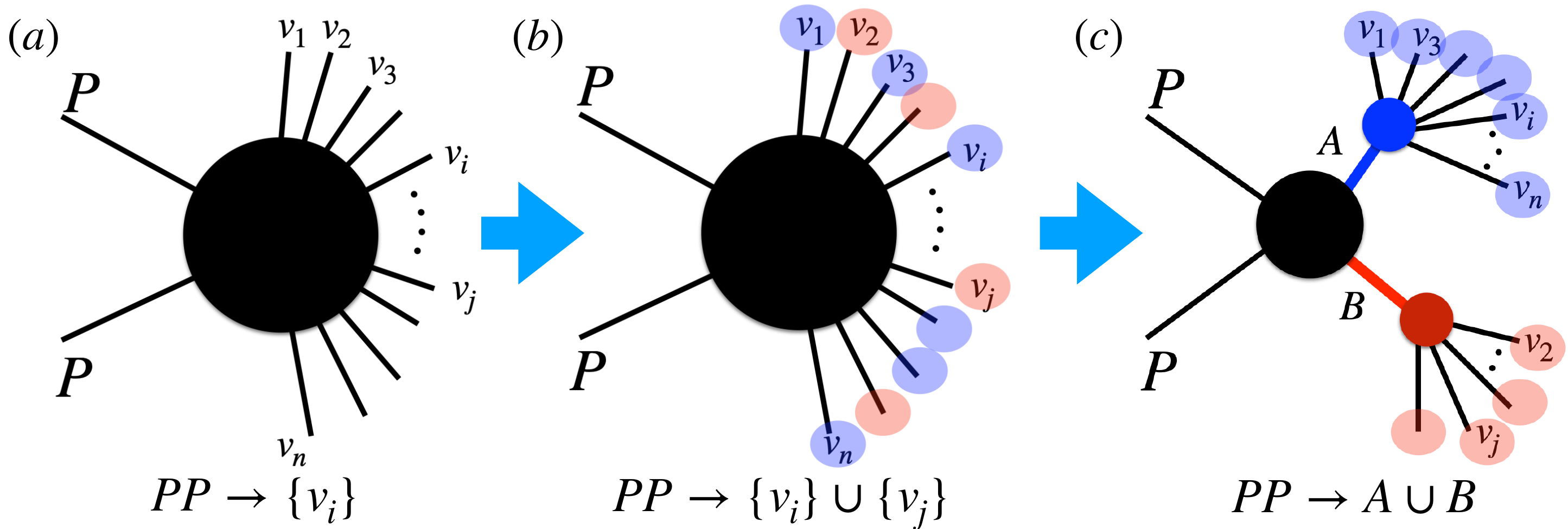}
\caption{(a) $n$-observed particles, (b) dividing $n$ particles into two groups for a $2\to2$ process, and (c) identified event-topology with $A$ and $B$. Taken from Ref. \cite{Kim:2021wrr}.
\label{diagram_QA}
}
\end{figure}

\section{Kinematic variables in different 
experiments}
\label{sec:otherexps}

While most of the recent developments in kinematic variables have been motivated by the phenomenology in collider experiments, especially hadron colliders, they are readily applicable to other experiments including accelerator-based experiments [e.g, fixed target or beam-dump type (neutrino) experiments], reactor-based (neutrino) experiments, dark-matter (in)direct detection experiments, and cosmic-particle telescopes. Here we briefly review existing usage of kinematic variables in non-collider experiments, and discuss future prospects of applications of kinematic variables to upcoming experiments. 

One of the crucial differences of non-collider experiments from collider-based experiments is that the transverse plane for a given event is usually ill-defined. In typical collider experiments, the initial-state particles have their momenta aligned with the beam axis, so that the transverse plane is literally transverse to the beam direction and most of the transverse variables are defined with respect to this plane. 
By contrast, this is not always the case for non-collider experiments. For example, in the beam-focused neutrino experiments, source particles of neutrinos (e.g., charged pions, kaons, and muons) are focused and aligned to the particle beam axis by the magnets in the horn system and then decay to neutrinos. Although the source particles are highly boosted in the forward direction, the neutrinos come with a non-zero angular spread with respect to the beam axis and as a consequence, we do not know the ``neutrino beam'' direction event-by-event. 
Likewise, in dark-matter or cosmic-ray detection experiments, the incoming direction of dark matter or cosmic particles is not known {\it a priori}. 
In these experiments, therefore, any variables defined on the beam-transverse plane do not allow for robust physical interpretations unless they are re-defined with appropriate prescriptions. Instead, basic quantities (e.g., energy, timing, etc), invariant quantities (e.g., invariant mass), or their combinations are more straightforwardly applicable.

\medskip

\noindent{\bf Energy}: Energy is one of the most widely used variables in particle physics experiments, as it is a basic physics quantity to measure at detectors. 
A few example uses follow. 
\begin{enumerate}[label=(\alph*),leftmargin=*,align=left] \itemsep1pt \parskip0pt \parsep0pt
\item In conventional dark-matter direct detection experiments targeting WIMP dark-matter candidates, the shape of the nuclear recoil energy spectrum carries information about the dark-matter properties. One can estimate the mass scale of dark matter~\cite{Jungman:1995df} or test whether the observed events are caused by inelastic dark matter~\cite{Tucker-Smith:2001myb}.
\item In stopped-pion neutrino experiments (e.g., COHERENT~\cite{COHERENT:2017ipa,COHERENT:2018gft,COHERENT:2019kwz} and CCM~\cite{CCM:2021leg}), the energy is used to eliminate the pion-induced muon neutrino events in the search for low-mass dark matter. Since the beam energy of these experiments is small, $\pi^+$'s produced in the beam target material lose their kinetic energy and stop before decaying to a $\mu^+$ and a $\nu_\mu$. The energy of $\nu_\mu$ is single-valued at $\sim30$~MeV so that the recoil energy of $\nu_\mu$ scattering events is bounded from above. By contrast, vector-portal dark matter coming from the $\pi^0$ decay through a dark photon is typically more energetic hence deposits more energy in the detector. An energy cut rejects $\nu_\mu$-induced events, leading to a dark-matter signal-rich region~\cite{deNiverville:2015mwa,Dutta:2019nbn,Dutta:2020vop,CCM:2021leg,Aguilar-Arevalo:2021sbh}.  
\item The energy peak is useful in the energy distribution of cosmic photons from neutral pion decays. In a cosmic shower, $\pi^0$'s are produced with various boost factors. Since $\pi^0$ is a scalar and its decay is a two-body process, the peak position is identified as half the $\pi^0$ mass~\cite{Carlson1950} as also discussed in Section~\ref{sec:energy} in a more general context.
\end{enumerate}

\medskip

\noindent{\bf Timing}: Timing is also a readily accessible quantity in many of the aforementioned experiments. As discussed in Section~\ref{sec:timing}, it is important to set the reference time (i.e., $t_0$) in order to render the timing values meaningful. This is deeply connected to the event triggering. In fixed target type experiments, the beam-on time is often set to be $t_0$. Three example applications are given below. 
\begin{enumerate}[label=(\alph*),leftmargin=*,align=left] \itemsep1pt \parskip0pt \parsep0pt
\item In the stopped pion neutrino experiments mentioned earlier whose proton beam energy is $\sim 1$~GeV, beam-related neutrons would give rise to an enormous amount of background. However, due to the scale of the beam energy, most of the produced neutrons are slowly moving so that their arrival timing at the detector is rather delayed, compared to the pion-induced neutrino events. Therefore, restricting to the prompt region, one can reject beam-related neutrons very efficiently (see, e.g., Refs.~\cite{CCM:2021leg,Aguilar-Arevalo:2021sbh}).
\item In a similar manner, neutrinos from muon decays can be vetoed using a timing cut in the stopped pion neutrino experiments. Since a muon is much longer-lived than a charged pion, the muon-induced neutrinos typically arrive at the detector much later than a pion-induced neutrino. In other words, the muon-induced neutrino events usually fall in delayed timing bins. The low-mass dark matter mentioned before is also prompt as it comes from the rare $\pi^0$ decay, and therefore, a timing cut can reduce significantly not only beam-related neutron backgrounds but also muon-induced ``delayed'' neutrino events, while keeping as many dark-matter events as possible~\cite{Dutta:2019nbn,Dutta:2020vop}. 
\item An inverted timing cut can be utilized in the beam-focused neutrino experiments. The MiniBooNE Collaboration sets the limit for WIMP dark-matter candidates, based on the fact that no significant number of events are observed in their delayed timing bins~\cite{MiniBooNE:2012jpi,MiniBooNEDM:2018cxm}. If a WIMP candidate with a sub-GeV or greater mass were produced in the MiniBooNE target, it would travel rather slowly because the beam energy is as small as $\sim 8$~GeV.  
\end{enumerate}

\medskip

\noindent{\bf Angle or directionality}: The angle variable is useful if the source point is well known, the momenta of visible particles can be measured, and the angular resolution is good enough. In typical accelerator-based experiments, the angle is usually defined with respect to the beam axis, visible particles are energetic enough to measure their three-momentum, and detectors (e.g., calorimeters, liquid argon time projection chamber detectors) are capable of measuring angles precisely.  

An angle can also be defined in the search for cosmogenic signals. For example, a class of boosted dark matter models built upon a non-minimal dark-sector framework~\cite{Belanger:2011ww} predict that a certain dark-matter component can be produced with a significant boost factor in the present universe by the pair-annihilation of the halo dark-matter component~\cite{Agashe:2014yua}.
Therefore, visible particles (e.g., recoil electrons) induced by the scattering of boosted dark matter can be not only energetic enough to allow for the measurement of their three-momenta but forward-directed enough to be aligned with the momentum of the incoming boosted dark matter. 
The dominant fraction of the flux of this so-called boosted dark matter comes from the regions where halo dark matter is densely populated, e.g., galactic center~\cite{Agashe:2014yua,Alhazmi:2016qcs,Kim:2016ixu}, dwarf galaxies~\cite{Necib:2016aez}, and the sun~\cite{Berger:2014sqa,Kong:2014mia,Alhazmi:2016qcs} in the solar-captured boosted dark matter scenarios. 
Therefore, the line extended between these source points and the detector location can be taken as a reference axis. 
If a detector features a good angular resolution, one may define a signal-rich region by selecting the events in which the angles of visible particles with respect to the reference axis are within a certain range. 
The Super-Kamiokande Collaboration used angle cuts to perform a search for galactic boosted dark matter and solar-captured boosted dark matter that is interacting with electrons,  and set the first limits for models of two-component boosted dark matter~\cite{Super-Kamiokande:2017dch}. 

\medskip

\noindent{\bf Complex variables}: Beyond the basic quantities that have been discussed thus far, complex kinematic variables constructed with basic quantities are being used in a wide range of non-collider experiments as they are equipped with high-capability detectors with good angular, spatial, and/or energy-momentum resolutions. A couple of examples are given below. 

\begin{enumerate}[label=(\alph*),leftmargin=*,align=left] \itemsep1pt \parskip0pt \parsep0pt
\item A wide range of dark-sector scenarios predict the production of low-mass dark matter (say, $\chi$) in various beam-induced neutrino experiments. 
An extensively investigated detection channel is the elastic scattering of dark matter off an electron inside the detector material: $\chi + e^- \to \chi+e^-$.  
One of the major backgrounds to this signal is the charged-current quasi-elastic (CCQE) scattering of electron neutrinos, i.e., $\nu_e + n \to e^- + p$ or $\bar{\nu}_e + p \to e^+ + n$, where the final-state nucleon is not energetic enough to be detected. 
It was demonstrated that this type of signal-faking events can be significantly rejected by an application of the $E_e\theta_e^2$ cut with $E_e$ and $\theta_e$ being the energy and the angle of the final-state electron. 
For example, a $E_e \theta_e^2 < 5$~MeV$\cdot$rad$^2$ cut can reduce about 99\% of CCQE events at the NO$\nu$A near detector~\cite{deNiverville:2018dbu} and a $E_e \theta_e^2 < 2$~MeV$\cdot$rad$^2$ cut can suppress the CCQE events by $\sim 99.9\%$ at the DUNE near detector~\cite{DeRomeri:2019kic}. 
\item A class of well-motivated new physics models that can be tested at neutrino experiments predict upscattering of the incoming particles: for example, upscattering of a SM neutrino to a heavier sterile neutrino (say, $N_R$) through  mixing~\cite{Bertuzzo:2018itn} and upscattering of dark matter to a heavier dark-sector state (say, $\chi^\prime$)~\cite{Tucker-Smith:2001myb,Izaguirre:2014dua,Kim:2016zjx}.
These upscattered states may decay into a set of visible particles in addition to the recoiling particle emerging from the primary scattering of the incoming neutrino or dark matter. 
Thus there can be multiple visible particles in the final state, allowing for constructing complex variables such as invariant masses~\cite{Kim:2016zjx}. 
\end{enumerate}

\section{Conclusions and outlook} 
\label{sec;conclusion}

In general, the outcome of any particle physics experiment (whether studying scattering or decay processes) is a measured probability distribution in the relevant phase space of the final state. Unfortunately, in typical situations, the phase space is high-dimensional, and the observed features are difficult to visualize. Furthermore, for many interesting signals, some information, e.g., related to the kinematics of invisible particles like neutrinos or dark-matter candidates, may be missing. In that case it makes sense to perform dimensional reduction to the lower-dimensional observable slice of the phase space. In doing so, a major goal is to use the proper kinematic variables which retain as much information as possible about the underlying physics, features, etc.

The higher-level variables which are derived from the measured particle kinematic information are generically referred to as ``kinematic variables''. Depending on the type of experimental signature and/or the goal of the analysis, many different kinematic variables have been introduced and discussed over the years in the particle phenomenology literature. The main purpose of this review was to collect and summarize all those recent developments in one place. We also provided the motivation for introducing each variable, its applicability and limitations, together with a guide to the relevant references.

Many of the traditional questions and approaches in particle kinematics are now being reevaluated using machine learning. The ability of ML to better capture the high-dimensional correlations in the data may lead to superior performance, at the expense of introducing unphysical hyperparameters and perhaps less transparency and interpretability. At the same time, kinematic variables can be incorporated into the ML approaches, thus boosting their performance and interpretability. The general methods which have guided particle phenomenologists in deriving these kinematic variables can be used in other fields of science and are therefore of interest outside the domain of particle physics.

\appendix
\section{Tools and codes for kinematic variables}
\label{sec:tools}

For the benefit of the users of kinematic variables, in Table~\ref{tab:codes} we list a few popular public codes for numerically computing some of the kinematic variables described in the main text. Note that the name of each code in the table is hyperlinked to the respective web-page or repository. We also provide the corresponding reference, language, and system requirements.

\begin{table*}[t!]
\centering
\begin{tabular}{c | c | c}
\hline \hline
Code  & Language/Requirements & Kinematic variables  \\
\hline
\href{https://pypi.org/project/mt2/}{$M_{T2}$} \cite{Lester:2014yga} & Python3 & $M_{T2}$ \\
\hline
\href{https://www.hep.phy.cam.ac.uk/~lester/mt2/}{Oxbridge Kinetics Library} \cite{OxbridgeKineticsLibrary} & Python, C++, Root & $\alpha_T$, $M_{T2}$, $M_{TGen}$ $M_{2C}$, $M_C$, $M_{CT}$, $M_{CT2}$ \\
\hline
\href{https://github.com/hepkosmos/OptiMass}{OPTIMASS} \cite{Cho:2015laa} & C++, Python2.7, Root & $M_2$ and friends\\
\hline
\href{https://github.com/cbpark/YAM2}{YAM2}~\cite{Park:2020bsu} & C++ & $M_2$\\
\hline
\href{https://github.com/KLFitter/KLFitter}{KLFitter}~\cite{Erdmann:2013rxa} & C++, Root & Top quark reconstruction \\
\hline
\href{https://http://restframes.com}{RestFrames}~\cite{Jackson:2017gcy} & C++, Root & Recursive Jigsaw Reconstruction  \\
\hline
\href{https://github.com/MoMEMta/MoMEMta}{MoMEMta}~\cite{Brochet:2018pqf} & C++, Root & Modular Matrix Element Method implementation.  \\
\hline
\href{https://root.cern.ch}{Root}~\cite{Antcheva:2011zz,Antcheva:2009zz} & Root & Basic kinematic variables and $M_T$ \\
\hline
\href{http://particle.physics.ucdavis.edu/hefti/projects/doku.php?id=wimpmass}{WIMPMASS}~\cite{Cheng:2009fw,Cheng:2008mg,Cheng:2008hk,Cheng:2007xv} & C++, Root & $M_{T2}$ using bisection method \\
\hline
\href{https://proj-clhep.web.cern.ch/proj-clhep/}{CLHEP}~\cite{Lonnblad:1994np} & C++ & Basic kinematic variables and $M_T$ \\
\hline
\href{http://fastjet.fr}{FASTJET}~\cite{Cacciari:2011ma} & C++ & Basic kinematic variables and $E_T$ \\
     \hline \hline
\end{tabular}
\caption{\label{tab:codes} A summary of public codes for numerically computing kinematic variables.}
\end{table*}

\section*{Acknowledgements}
This review was written at the request of the Theory Frontier Collider Phenomenology group (TF07) of the Snowmass 2021-22 Particle Physics Community Planning Exercise \cite{snowmass}. 
We would like to thank Roberto Di Nardo, Lance Dixon, Roy Forestano, Stephen Martin, Federico Meloni and Eyup Unlu for discussions and comments on the manuscript.
KK, KM, and PS would like to thank the Aspen Center for Physics for hospitality during the completion of this work, supported in part by National Science Foundation grant PHY-1607611.
DK is supported by the DOE Grant No. DE-SC0010813.  
KK is supported in part by DOE DE-SC0021447. KM is supported in part by DOE DE-SC0022148. MP is supported by NRF-2021R1A2C4002551.
PS is partially supported by the U.S. Department of Energy, Office of Science, Office of High Energy Physics QuantISED program under the grants ``HEP Machine Learning and Optimization Go Quantum'', Award Number 0000240323, and ``DOE QuantiSED Consortium QCCFP-QMLQCF'', Award Number DE-SC0019219. This manuscript has been authored by Fermi Research Alliance, LLC under Contract No. DEAC02-07CH11359 with the U.S. Department of Energy, Office of Science, Office of High Energy Physics.

\bibliographystyle{bib}
\bibliography{ref}

\providecommand{\href}[2]{#2}\begingroup\raggedright\begin{thebibliography}{100}

\bibitem{Albertsson:2018maf}
K.~Albertsson et~al., \emph{{Machine Learning in High Energy Physics Community
  White Paper}},
  \href{http://dx.doi.org/10.1088/1742-6596/1085/2/022008}{\emph{J. Phys. Conf.
  Ser.} {\bf 1085} (2018) 022008}, [\href{http://arxiv.org/abs/1807.02876}{{\tt
  1807.02876}}].

\bibitem{Ask:2012sm}
S.~Ask et~al., \emph{{From Lagrangians to Events: Computer Tutorial at the
  MC4BSM-2012 Workshop}},  \href{http://arxiv.org/abs/1209.0297}{{\tt
  1209.0297}}.

\bibitem{Barr:2010zj}
A.~J. Barr and C.~G. Lester, \emph{{A Review of the Mass Measurement Techniques
  proposed for the Large Hadron Collider}},
  \href{http://dx.doi.org/10.1088/0954-3899/37/12/123001}{\emph{J. Phys. G}
  {\bf 37} (2010) 123001}, [\href{http://arxiv.org/abs/1004.2732}{{\tt
  1004.2732}}].

\bibitem{Han:2005mu}
T.~Han, \emph{{Collider phenomenology: Basic knowledge and techniques}},  in
  \emph{{Theoretical Advanced Study Institute in Elementary Particle Physics}:
  {Physics in D $\geqq$ 4}}, pp.~407--454, 8, 2005.
\newblock \href{http://arxiv.org/abs/hep-ph/0508097}{{\tt hep-ph/0508097}}.
\newblock \href{http://dx.doi.org/10.1142/9789812773579_0008}{DOI}.

\bibitem{Perelstein:2010hh}
M.~Perelstein, \emph{{Introduction to Collider Physics}},  in
  \emph{{Theoretical Advanced Study Institute in Elementary Particle Physics}:
  {Physics of the Large and the Small}}, pp.~421--486, 2011.
\newblock \href{http://arxiv.org/abs/1002.0274}{{\tt 1002.0274}}.
\newblock \href{http://dx.doi.org/10.1142/9789814327183_0008}{DOI}.

\bibitem{Schwartz:2017hep}
M.~D. Schwartz, \emph{{TASI Lectures on Collider Physics}},  in
  \emph{{Proceedings, Theoretical Advanced Study Institute in Elementary
  Particle Physics : Anticipating the Next Discoveries in Particle Physics
  (TASI 2016)}: {Boulder, CO, USA, June 6-July 1, 2016}} (R.~Essig and I.~Low,
  eds.), pp.~65--100.
\newblock World Scientific, 2018.
\newblock \href{http://arxiv.org/abs/1709.04533}{{\tt 1709.04533}}.

\bibitem{Franceschini:2017dxe}
R.~Franceschini, \emph{{Energy peaks: A high energy physics outlook}},
  \href{http://dx.doi.org/10.1142/S0217732317300348}{\emph{Mod. Phys. Lett. A}
  {\bf 32} (2017) 1730034}, [\href{http://arxiv.org/abs/1711.02969}{{\tt
  1711.02969}}].

\bibitem{Barr:2011xt}
A.~Barr, T.~Khoo, P.~Konar, K.~Kong, C.~Lester, K.~Matchev et~al., \emph{{Guide
  to transverse projections and mass-constraining variables}},
  \href{http://dx.doi.org/10.1103/PhysRevD.84.095031}{\emph{Phys. Rev. D} {\bf
  84} (2011) 095031}, [\href{http://arxiv.org/abs/1105.2977}{{\tt 1105.2977}}].

\bibitem{Andrews:2018nwy}
M.~Andrews, M.~Paulini, S.~Gleyzer and B.~Poczos, \emph{{End-to-End Physics
  Event Classification with CMS Open Data: Applying Image-Based Deep Learning
  to Detector Data for the Direct Classification of Collision Events at the
  LHC}}, \href{http://dx.doi.org/10.1007/s41781-020-00038-8}{\emph{Comput.
  Softw. Big Sci.} {\bf 4} (2020) 6},
  [\href{http://arxiv.org/abs/1807.11916}{{\tt 1807.11916}}].

\bibitem{Andrews:2019faz}
M.~Andrews, J.~Alison, S.~An, P.~Bryant, B.~Burkle, S.~Gleyzer et~al.,
  \emph{{End-to-end jet classification of quarks and gluons with the CMS Open
  Data}}, \href{http://dx.doi.org/10.1016/j.nima.2020.164304}{\emph{Nucl.
  Instrum. Meth. A} {\bf 977} (2020) 164304},
  [\href{http://arxiv.org/abs/1902.08276}{{\tt 1902.08276}}].

\bibitem{Cogan:2014oua}
J.~Cogan, M.~Kagan, E.~Strauss and A.~Schwarztman, \emph{{Jet-Images: Computer
  Vision Inspired Techniques for Jet Tagging}},
  \href{http://dx.doi.org/10.1007/JHEP02(2015)118}{\emph{JHEP} {\bf 02} (2015)
  118}, [\href{http://arxiv.org/abs/1407.5675}{{\tt 1407.5675}}].

\bibitem{Kagan:2020yrm}
M.~Kagan, \emph{{Image-Based Jet Analysis}},
  \href{http://arxiv.org/abs/2012.09719}{{\tt 2012.09719}}.

\bibitem{CMS:2013kfa}
{\scshape CMS} collaboration, \emph{{Performance of quark/gluon discrimination
  in 8 TeV pp data}},  2013.

\bibitem{ATLAS:2014vax}
{\scshape ATLAS} collaboration, G.~Aad et~al., \emph{{Light-quark and gluon jet
  discrimination in $pp$ collisions at $\sqrt{s}=7\mathrm {\ TeV}$ with the
  ATLAS detector}},
  \href{http://dx.doi.org/10.1140/epjc/s10052-014-3023-z}{\emph{Eur. Phys. J.
  C} {\bf 74} (2014) 3023}, [\href{http://arxiv.org/abs/1405.6583}{{\tt
  1405.6583}}].

\bibitem{Komiske:2016rsd}
P.~T. Komiske, E.~M. Metodiev and M.~D. Schwartz, \emph{{Deep learning in
  color: towards automated quark/gluon jet discrimination}},
  \href{http://dx.doi.org/10.1007/JHEP01(2017)110}{\emph{JHEP} {\bf 01} (2017)
  110}, [\href{http://arxiv.org/abs/1612.01551}{{\tt 1612.01551}}].

\bibitem{Ball:2007zza}
{\scshape CMS} collaboration, G.~Bayatian et~al., \emph{{CMS technical design
  report, volume II: Physics performance}},
  \href{http://dx.doi.org/10.1088/0954-3899/34/6/S01}{\emph{J. Phys. G} {\bf
  34} (2007) 995--1579}.

\bibitem{Matsumoto:2006ws}
S.~Matsumoto, M.~M. Nojiri and D.~Nomura, \emph{{Hunting for the Top Partner in
  the Littlest Higgs Model with T-parity at the CERN LHC}},
  \href{http://dx.doi.org/10.1103/PhysRevD.75.055006}{\emph{Phys. Rev. D} {\bf
  75} (2007) 055006}, [\href{http://arxiv.org/abs/hep-ph/0612249}{{\tt
  hep-ph/0612249}}].

\bibitem{Lester:2007fq}
C.~Lester and A.~Barr, \emph{{MTGEN: Mass scale measurements in pair-production
  at colliders}},
  \href{http://dx.doi.org/10.1088/1126-6708/2007/12/102}{\emph{JHEP} {\bf 12}
  (2007) 102}, [\href{http://arxiv.org/abs/0708.1028}{{\tt 0708.1028}}].

\bibitem{Alwall:2009zu}
J.~Alwall, K.~Hiramatsu, M.~M. Nojiri and Y.~Shimizu, \emph{{Novel
  reconstruction technique for New Physics processes with initial state
  radiation}},
  \href{http://dx.doi.org/10.1103/PhysRevLett.103.151802}{\emph{Phys. Rev.
  Lett.} {\bf 103} (2009) 151802}, [\href{http://arxiv.org/abs/0905.1201}{{\tt
  0905.1201}}].

\bibitem{Albrow:1976jm}
M.~Albrow et~al., \emph{{A Search for Narrow Resonances in Proton Proton
  Collisions at 53-GeV Center-Of-Mass Energy}},
  \href{http://dx.doi.org/10.1016/0550-3213(76)90438-7}{\emph{Nucl. Phys. B}
  {\bf 114} (1976) 365--379}.

\bibitem{Hinchliffe:1996iu}
I.~Hinchliffe, F.~Paige, M.~Shapiro, J.~Soderqvist and W.~Yao, \emph{{Precision
  SUSY measurements at CERN LHC}},
  \href{http://dx.doi.org/10.1103/PhysRevD.55.5520}{\emph{Phys. Rev. D} {\bf
  55} (1997) 5520--5540}, [\href{http://arxiv.org/abs/hep-ph/9610544}{{\tt
  hep-ph/9610544}}].

\bibitem{Agashe:2015wwa}
K.~Agashe, R.~Franceschini, D.~Kim and K.~Wardlow, \emph{{Mass Measurement
  Using Energy Spectra in Three-body Decays}},
  \href{http://dx.doi.org/10.1007/JHEP05(2016)138}{\emph{JHEP} {\bf 05} (2016)
  138}, [\href{http://arxiv.org/abs/1503.03836}{{\tt 1503.03836}}].

\bibitem{Kim:2015bnd}
D.~Kim, K.~T. Matchev and M.~Park, \emph{{Using sorted invariant mass variables
  to evade combinatorial ambiguities in cascade decays}},
  \href{http://dx.doi.org/10.1007/JHEP02(2016)129}{\emph{JHEP} {\bf 02} (2016)
  129}, [\href{http://arxiv.org/abs/1512.02222}{{\tt 1512.02222}}].

\bibitem{Jackson:2017gcy}
P.~Jackson and C.~Rogan, \emph{{Recursive Jigsaw Reconstruction: HEP event
  analysis in the presence of kinematic and combinatoric ambiguities}},
  \href{http://dx.doi.org/10.1103/PhysRevD.96.112007}{\emph{Phys. Rev. D} {\bf
  96} (2017) 112007}, [\href{http://arxiv.org/abs/1705.10733}{{\tt
  1705.10733}}].

\bibitem{Baringer:2011nh}
P.~Baringer, K.~Kong, M.~McCaskey and D.~Noonan, \emph{{Revisiting
  Combinatorial Ambiguities at Hadron Colliders with $M_{T2}$}},
  \href{http://dx.doi.org/10.1007/JHEP10(2011)101}{\emph{JHEP} {\bf 10} (2011)
  101}, [\href{http://arxiv.org/abs/1109.1563}{{\tt 1109.1563}}].

\bibitem{Debnath:2017ktz}
D.~Debnath, D.~Kim, J.~H. Kim, K.~Kong and K.~T. Matchev, \emph{{Resolving
  Combinatorial Ambiguities in Dilepton $t\bar t$ Event Topologies with
  Constrained $M_2$ Variables}},
  \href{http://dx.doi.org/10.1103/PhysRevD.96.076005}{\emph{Phys. Rev. D} {\bf
  96} (2017) 076005}, [\href{http://arxiv.org/abs/1706.04995}{{\tt
  1706.04995}}].

\bibitem{Allanach:2000kt}
B.~Allanach, C.~Lester, M.~A. Parker and B.~Webber, \emph{{Measuring sparticle
  masses in nonuniversal string inspired models at the LHC}},
  \href{http://dx.doi.org/10.1088/1126-6708/2000/09/004}{\emph{JHEP} {\bf 09}
  (2000) 004}, [\href{http://arxiv.org/abs/hep-ph/0007009}{{\tt
  hep-ph/0007009}}].

\bibitem{Lester:1999tx}
C.~Lester and D.~Summers, \emph{{Measuring masses of semiinvisibly decaying
  particles pair produced at hadron colliders}},
  \href{http://dx.doi.org/10.1016/S0370-2693(99)00945-4}{\emph{Phys. Lett. B}
  {\bf 463} (1999) 99--103}, [\href{http://arxiv.org/abs/hep-ph/9906349}{{\tt
  hep-ph/9906349}}].

\bibitem{Cho:2008tj}
W.~S. Cho, K.~Choi, Y.~G. Kim and C.~B. Park, \emph{{M(T2)-assisted on-shell
  reconstruction of missing momenta and its application to spin measurement at
  the LHC}}, \href{http://dx.doi.org/10.1103/PhysRevD.79.031701}{\emph{Phys.
  Rev. D} {\bf 79} (2009) 031701}, [\href{http://arxiv.org/abs/0810.4853}{{\tt
  0810.4853}}].

\bibitem{Kim:2017awi}
D.~Kim, K.~T. Matchev, F.~Moortgat and L.~Pape, \emph{{Testing Invisible
  Momentum Ansatze in Missing Energy Events at the LHC}},
  \href{http://dx.doi.org/10.1007/JHEP08(2017)102}{\emph{JHEP} {\bf 08} (2017)
  102}, [\href{http://arxiv.org/abs/1703.06887}{{\tt 1703.06887}}].

\bibitem{Kim:2021pcz}
D.~Kim, K.~Kong, K.~T. Matchev, M.~Park and P.~Shyamsundar, \emph{{Deep-Learned
  Event Variables for Collider Phenomenology}},
  \href{http://arxiv.org/abs/2105.10126}{{\tt 2105.10126}}.

\bibitem{Cho:2007qv}
W.~S. Cho, K.~Choi, Y.~G. Kim and C.~B. Park, \emph{{Gluino Stransverse Mass}},
  \href{http://dx.doi.org/10.1103/PhysRevLett.100.171801}{\emph{Phys. Rev.
  Lett.} {\bf 100} (2008) 171801}, [\href{http://arxiv.org/abs/0709.0288}{{\tt
  0709.0288}}].

\bibitem{Burns:2008cp}
M.~Burns, K.~Kong, K.~T. Matchev and M.~Park, \emph{{A General Method for
  Model-Independent Measurements of Particle Spins, Couplings and Mixing Angles
  in Cascade Decays with Missing Energy at Hadron Colliders}},
  \href{http://dx.doi.org/10.1088/1126-6708/2008/10/081}{\emph{JHEP} {\bf 10}
  (2008) 081}, [\href{http://arxiv.org/abs/0808.2472}{{\tt 0808.2472}}].

\bibitem{Han:2009ss}
T.~Han, I.-W. Kim and J.~Song, \emph{{Kinematic Cusps: Determining the Missing
  Particle Mass at Colliders}},
  \href{http://dx.doi.org/10.1016/j.physletb.2010.09.010}{\emph{Phys. Lett. B}
  {\bf 693} (2010) 575--579}, [\href{http://arxiv.org/abs/0906.5009}{{\tt
  0906.5009}}].

\bibitem{deFavereau:2013fsa}
{\scshape DELPHES 3} collaboration, J.~de~Favereau, C.~Delaere, P.~Demin,
  A.~Giammanco, V.~Lema\^\i{}tre, A.~Mertens et~al., \emph{{DELPHES 3, A
  modular framework for fast simulation of a generic collider experiment}},
  \href{http://dx.doi.org/10.1007/JHEP02(2014)057}{\emph{JHEP} {\bf 02} (2014)
  057}, [\href{http://arxiv.org/abs/1307.6346}{{\tt 1307.6346}}].

\bibitem{CERN-LHCC-2017-027}
{\scshape CMS Collaboration} collaboration, \emph{{Technical proposal for a MIP
  timing detector in the CMS experiment Phase 2 upgrade}},  Dec, 2017.
\newblock 10.17181/CERN.2RSJ.UE8W.

\bibitem{Kogler:2018hem}
R.~Kogler et~al., \emph{{Jet Substructure at the Large Hadron Collider:
  Experimental Review}},
  \href{http://dx.doi.org/10.1103/RevModPhys.91.045003}{\emph{Rev. Mod. Phys.}
  {\bf 91} (2019) 045003}, [\href{http://arxiv.org/abs/1803.06991}{{\tt
  1803.06991}}].

\bibitem{Soyez:2018opl}
G.~Soyez, \emph{{Pileup mitigation at the LHC: A theorist\textquoteright{}s
  view}}, \href{http://dx.doi.org/10.1016/j.physrep.2019.01.007}{\emph{Phys.
  Rept.} {\bf 803} (2019) 1--158}, [\href{http://arxiv.org/abs/1801.09721}{{\tt
  1801.09721}}].

\bibitem{ATLAS:2012am}
{\scshape ATLAS} collaboration, G.~Aad et~al., \emph{{Jet mass and substructure
  of inclusive jets in $\sqrt{s}=7$ TeV $pp$ collisions with the ATLAS
  experiment}}, \href{http://dx.doi.org/10.1007/JHEP05(2012)128}{\emph{JHEP}
  {\bf 05} (2012) 128}, [\href{http://arxiv.org/abs/1203.4606}{{\tt
  1203.4606}}].

\bibitem{Bertolini:2014bba}
D.~Bertolini, P.~Harris, M.~Low and N.~Tran, \emph{{Pileup Per Particle
  Identification}},
  \href{http://dx.doi.org/10.1007/JHEP10(2014)059}{\emph{JHEP} {\bf 10} (2014)
  059}, [\href{http://arxiv.org/abs/1407.6013}{{\tt 1407.6013}}].

\bibitem{Komiske:2017ubm}
P.~T. Komiske, E.~M. Metodiev, B.~Nachman and M.~D. Schwartz, \emph{{Pileup
  Mitigation with Machine Learning (PUMML)}},
  \href{http://dx.doi.org/10.1007/JHEP12(2017)051}{\emph{JHEP} {\bf 12} (2017)
  051}, [\href{http://arxiv.org/abs/1707.08600}{{\tt 1707.08600}}].

\bibitem{ArjonaMartinez:2018eah}
J.~Arjona~Mart\'\i{}nez, O.~Cerri, M.~Pierini, M.~Spiropulu and J.-R. Vlimant,
  \emph{{Pileup mitigation at the Large Hadron Collider with graph neural
  networks}}, \href{http://dx.doi.org/10.1140/epjp/i2019-12710-3}{\emph{Eur.
  Phys. J. Plus} {\bf 134} (2019) 333},
  [\href{http://arxiv.org/abs/1810.07988}{{\tt 1810.07988}}].

\bibitem{Butterworth:2015oua}
J.~Butterworth et~al., \emph{{PDF4LHC recommendations for LHC Run II}},
  \href{http://dx.doi.org/10.1088/0954-3899/43/2/023001}{\emph{J. Phys. G} {\bf
  43} (2016) 023001}, [\href{http://arxiv.org/abs/1510.03865}{{\tt
  1510.03865}}].

\bibitem{Valassi:2020ueh}
{\scshape HSF Physics Event Generator WG} collaboration, S.~Amoroso et~al.,
  ``{Challenges in Monte Carlo event generator software for High-Luminosity
  LHC}.'' 4, 2020.

\bibitem{Matchev:2020jqz}
K.~T. Matchev and P.~Shyamsundar, \emph{{OASIS: Optimal Analysis-Specific
  Importance Sampling for event generation}},
  \href{http://dx.doi.org/10.21468/SciPostPhys.10.2.034}{\emph{SciPost Phys.}
  {\bf 10} (2021) 034}, [\href{http://arxiv.org/abs/2006.16972}{{\tt
  2006.16972}}].

\bibitem{PhysRevD.45.1531}
R.~H. Dalitz and G.~R. Goldstein, \emph{Decay and polarization properties of
  the top quark}, \href{http://dx.doi.org/10.1103/PhysRevD.45.1531}{\emph{Phys.
  Rev. D} {\bf 45} (Mar, 1992) 1531--1543}.

\bibitem{CMS:2013wfa}
{\scshape CMS} collaboration, \emph{{Projected improvement of the accuracy of
  top-quark mass measurements at the upgraded LHC}},  2013.

\bibitem{Frixione:2014ala}
S.~Frixione and A.~Mitov, \emph{{Determination of the top quark mass from
  leptonic observables}},
  \href{http://dx.doi.org/10.1007/JHEP09(2014)012}{\emph{JHEP} {\bf 09} (2014)
  012}, [\href{http://arxiv.org/abs/1407.2763}{{\tt 1407.2763}}].

\bibitem{FerrarioRavasio:2019vmq}
S.~Ferrario~Ravasio, T.~Je\v{z}o, P.~Nason and C.~Oleari, \emph{{A theoretical
  study of top-mass measurements at the LHC using NLO+PS generators of
  increasing accuracy}},
  \href{http://dx.doi.org/10.1140/epjc/s10052-019-7336-9}{\emph{Eur. Phys. J.
  C} {\bf 78} (2018) 458}, [\href{http://arxiv.org/abs/1906.09166}{{\tt
  1906.09166}}].

\bibitem{FerrarioRavasio:2019glk}
S.~Ferrario~Ravasio, \emph{{Top-mass observables: all-orders behaviour,
  renormalons and NLO + Parton Shower effects}}.
\newblock PhD thesis, Milan Bicocca U., 2018.
\newblock \href{http://arxiv.org/abs/1902.05035}{{\tt 1902.05035}}.

\bibitem{Corcella:2017rpt}
G.~Corcella, R.~Franceschini and D.~Kim, \emph{{Fragmentation Uncertainties in
  Hadronic Observables for Top-quark Mass Measurements}},
  \href{http://dx.doi.org/10.1016/j.nuclphysb.2018.02.012}{\emph{Nucl. Phys. B}
  {\bf 929} (2018) 485--526}, [\href{http://arxiv.org/abs/1712.05801}{{\tt
  1712.05801}}].

\bibitem{ATLAS:2019ezb}
{\scshape ATLAS} collaboration, \emph{{Measurement of the top quark mass using
  a leptonic invariant mass in pp collisions at $sqrt{s}$ = 13 TeV with the
  ATLAS detector}},  10, 2019.

\bibitem{Aaltonen:2009zi}
{\scshape CDF} collaboration, T.~Aaltonen et~al., \emph{{Measurement of the Top
  Quark Mass Using the Invariant Mass of Lepton Pairs in Soft Muon b-tagged
  Events}}, \href{http://dx.doi.org/10.1103/PhysRevD.80.051104}{\emph{Phys.
  Rev. D} {\bf 80} (2009) 051104}, [\href{http://arxiv.org/abs/0906.5371}{{\tt
  0906.5371}}].

\bibitem{Khachatryan:2016wqo}
{\scshape CMS} collaboration, V.~Khachatryan et~al., \emph{{Measurement of the
  top quark mass using charged particles in pp collisions at $\sqrt s =$ 8
  TeV}}, \href{http://dx.doi.org/10.1103/PhysRevD.93.092006}{\emph{Phys. Rev.
  D} {\bf 93} (2016) 092006}, [\href{http://arxiv.org/abs/1603.06536}{{\tt
  1603.06536}}].

\bibitem{Hill:2005zy}
C.~S. Hill, J.~R. Incandela and J.~M. Lamb, \emph{{A Method for measurement of
  the top quark mass using the mean decay length of $b$ hadrons in $t \bar{t}$
  events}}, \href{http://dx.doi.org/10.1103/PhysRevD.71.054029}{\emph{Phys.
  Rev. D} {\bf 71} (2005) 054029},
  [\href{http://arxiv.org/abs/hep-ex/0501043}{{\tt hep-ex/0501043}}].

\bibitem{CMS-PAS-TOP-12-030}
{CMS Collaboration}, \emph{{Measurement of the top quark mass using the
  B-hadron lifetime technique}}, {\emph{CMS-PAS-TOP-12-030} (10, 2013) }.

\bibitem{Czakon:2020qbd}
M.~Czakon, A.~Mitov and R.~Poncelet, \emph{{NNLO QCD corrections to leptonic
  observables in top-quark pair production and decay}},
  \href{http://arxiv.org/abs/2008.11133}{{\tt 2008.11133}}.

\bibitem{ATLAS:2017kdr}
{\scshape ATLAS} collaboration, \emph{{Measurement of lepton differential
  distributions and the top quark mass in $t\bar{t}$ production in $pp$
  collisions at $\sqrt{s}=8$ TeV with the ATLAS detector}},  7, 2017.

\bibitem{CMS-PAS-TOP-14-014}
{\scshape CMS Collaboration} collaboration, \emph{{Determination of the
  top-quark mass from the m(lb) distribution in dileptonic ttbar events at
  sqrt(s) = 8 TeV}},  2014.

\bibitem{Kharchilava:1999yj}
A.~Kharchilava, \emph{{Top mass determination in leptonic final states with
  $J/\psi$}},
  \href{http://dx.doi.org/10.1016/S0370-2693(00)00120-9}{\emph{Phys. Lett. B}
  {\bf 476} (2000) 73--78}, [\href{http://arxiv.org/abs/hep-ph/9912320}{{\tt
  hep-ph/9912320}}].

\bibitem{Czakon:2021ohs}
M.~L. Czakon, T.~Generet, A.~Mitov and R.~Poncelet, \emph{{B-hadron
  hadro-production in NNLO QCD: application to LHC $t\bar{t}$ events with
  leptonic decays}},  \href{http://arxiv.org/abs/2102.08267}{{\tt 2102.08267}}.

\bibitem{Alioli:2013mxa}
S.~Alioli, P.~Fernandez, J.~Fuster, A.~Irles, S.-O. Moch, P.~Uwer et~al.,
  \emph{{A new observable to measure the top-quark mass at hadron colliders}},
  \href{http://dx.doi.org/10.1140/epjc/s10052-013-2438-2}{\emph{Eur. Phys. J.
  C} {\bf 73} (2013) 2438}, [\href{http://arxiv.org/abs/1303.6415}{{\tt
  1303.6415}}].

\bibitem{Aad:2019mkw}
{\scshape ATLAS} collaboration, G.~Aad et~al., \emph{{Measurement of the
  top-quark mass in $t\bar{t}+1$-jet events collected with the ATLAS detector
  in $pp$ collisions at $\sqrt{s}=8$ TeV}},
  \href{http://dx.doi.org/10.1007/JHEP11(2019)150}{\emph{JHEP} {\bf 11} (2019)
  150}, [\href{http://arxiv.org/abs/1905.02302}{{\tt 1905.02302}}].

\bibitem{Agashe:2012bn}
K.~Agashe, R.~Franceschini and D.~Kim, \emph{{Simple
  \textquotedblleft{}invariance\textquotedblright{} of two-body decay
  kinematics}}, \href{http://dx.doi.org/10.1103/PhysRevD.88.057701}{\emph{Phys.
  Rev. D} {\bf 88} (2013) 057701}, [\href{http://arxiv.org/abs/1209.0772}{{\tt
  1209.0772}}].

\bibitem{CMS-PAS-TOP-15-002}
{\scshape CMS Collaboration} collaboration, \emph{{Measurement of the top-quark
  mass from the b jet energy spectrum}},  2015.

\bibitem{Corcella:2019tgt}
G.~Corcella, \emph{{The top-quark mass: challenges in definition and
  determination}},
  \href{http://dx.doi.org/10.3389/fphy.2019.00054}{\emph{Front. in Phys.} {\bf
  7} (2019) 54}, [\href{http://arxiv.org/abs/1903.06574}{{\tt 1903.06574}}].

\bibitem{Hoang:2020iah}
A.~H. Hoang, \emph{{What is the Top Quark Mass?}},
  \href{http://dx.doi.org/10.1146/annurev-nucl-101918-023530}{\emph{Ann. Rev.
  Nucl. Part. Sci.} {\bf 70} (2020) 225--255},
  [\href{http://arxiv.org/abs/2004.12915}{{\tt 2004.12915}}].

\bibitem{1702.07546v1}
A.~{Collaboration}, \emph{{Top-quark mass measurement in the all-hadronic
  $t\bar{t}$ decay channel at $\sqrt{s}$ = 8 TeV with the ATLAS detector}},
  \href{http://arxiv.org/abs/1702.07546v1}{{\tt 1702.07546v1}}.

\bibitem{CMS-PAS-TOP-14-001}
{\scshape CMS Collaboration} collaboration, \emph{{Measurement of the top-quark
  mass in t t-bar events with lepton+jets final states in pp collisions at
  sqrt(s)=8 TeV}},  2014.

\bibitem{Chatrchyan:2012kl}
S.~{Chatrchyan}, V.~{Khachatryan}, A.~M. {Sirunyan}, A.~{Tumasyan}, W.~{Adam},
  E.~{Aguilo} et~al., \emph{{Measurement of the top-quark mass in $t\bar{t}$
  events with lepton+jets final states in pp collisions at $\sqrt{s}$=7~TeV}},
  \href{http://dx.doi.org/10.1007/JHEP12(2012)105}{\emph{Journal of High Energy
  Physics} {\bf 12} (Dec., 2012) 105},
  [\href{http://arxiv.org/abs/1209.2319}{{\tt 1209.2319}}].

\bibitem{Hoang:2014la}
A.~H. {Hoang}, \emph{The top mass: Interpretation and theoretical
  uncertainties},  \href{http://arxiv.org/abs/1412.3649v1}{{\tt 1412.3649v1}}.

\bibitem{2004.12915v1}
A.~H. {Hoang}, \emph{{What is the Top Quark Mass?}},
  \href{http://arxiv.org/abs/2004.12915v1}{{\tt 2004.12915v1}}.

\bibitem{Nason:2017aa}
P.~{Nason}, \emph{{The Top Mass in Hadronic Collisions}}, {\emph{ArXiv
  e-prints} (Dec., 2017) }, [\href{http://arxiv.org/abs/1712.02796}{{\tt
  1712.02796}}].

\bibitem{Beneke:2016lr}
M.~{Beneke}, P.~{Marquard}, P.~{Nason} and M.~{Steinhauser}, \emph{{On the
  ultimate uncertainty of the top quark pole mass}}, {\emph{ArXiv e-prints}
  (May, 2016) }, [\href{http://arxiv.org/abs/1605.03609}{{\tt 1605.03609}}].

\bibitem{Ferrario-Ravasio:2019ab}
S.~{Ferrario Ravasio}, T.~{Jezo}, P.~{Nason} and C.~{Oleari}, \emph{{Addendum
  to: A Theoretical Study of Top-Mass Measurements at the LHC Using NLO+PS
  Generators of Increasing Accuracy}}, {\emph{arXiv e-prints} (June, 2019) },
  [\href{http://arxiv.org/abs/1906.09166}{{\tt 1906.09166}}].

\bibitem{Ferrario-Ravasio:2019aa}
S.~{Ferrario Ravasio}, \emph{{Top-mass observables: all-orders behaviour,
  renormalons and NLO + Parton Shower effects}}, {\emph{arXiv e-prints} (Feb.,
  2019) }, [\href{http://arxiv.org/abs/1902.05035}{{\tt 1902.05035}}].

\bibitem{Ferrario-Ravasio:2018aa}
S.~{Ferrario Ravasio}, T.~{Jezo}, P.~{Nason} and C.~{Oleari}, \emph{{A
  Theoretical Study of Top-Mass Measurements at the LHC Using NLO+PS Generators
  of Increasing Accuracy}}, {\emph{ArXiv e-prints} (Jan., 2018) },
  [\href{http://arxiv.org/abs/1801.03944}{{\tt 1801.03944}}].

\bibitem{Jezo:2016oj}
T.~{Je{\v z}o}, J.~M. {Lindert}, P.~{Nason}, C.~{Oleari} and S.~{Pozzorini},
  \emph{{An NLO+PS generator for $t\bar{t}$ and $Wt$ production and decay
  including non-resonant and interference effects}}, {\emph{ArXiv e-prints}
  (July, 2016) }, [\href{http://arxiv.org/abs/1607.04538}{{\tt 1607.04538}}].

\bibitem{CMS-PAS-TOP-11-027}
\emph{Mass determination in the t t-bar system with kinematic endpoints},
  2012.

\bibitem{Frixione:2014jk}
S.~{Frixione} and A.~{Mitov}, \emph{{Determination of the top quark mass from
  leptonic observables}},
  \href{http://dx.doi.org/10.1007/JHEP09(2014)012}{\emph{JHEP} {\bf 9} (Sept.,
  2014) 12}, [\href{http://arxiv.org/abs/1407.2763}{{\tt 1407.2763}}].

\bibitem{Kawabataa:2014osa}
S.~Kawabata, Y.~Shimizu, Y.~Sumino and H.~Yokoya, \emph{{Weight function method
  for precise determination of top quark mass at Large Hadron Collider}},
  \href{http://dx.doi.org/10.1016/j.physletb.2014.12.044}{\emph{Phys. Lett. B}
  {\bf 741} (2015) 232--238}, [\href{http://arxiv.org/abs/1405.2395}{{\tt
  1405.2395}}].

\bibitem{Aaltonen:2009zl}
{CDF Collaboration}, \emph{{Measurement of the top quark mass using the
  invariant mass of lepton pairs in soft muon b-tagged events}},
  \href{http://dx.doi.org/10.1103/PhysRevD.80.051104}{\emph{{Phys. Rev. D}}
  {\bf 80} (Sept., 2009) 051104}, [\href{http://arxiv.org/abs/0906.5371}{{\tt
  0906.5371}}].

\bibitem{Kharchilava:2000yk}
A.~{Kharchilava}, \emph{{Top mass determination in leptonic final states with
  J/{$\psi$}}},
  \href{http://dx.doi.org/10.1016/S0370-2693(00)00120-9}{\emph{Physics Letters
  B} {\bf 476} (Mar., 2000) 73--78},
  [\href{http://arxiv.org/abs/hep-ph/9912320}{{\tt hep-ph/9912320}}].

\bibitem{Corcella:2017aa}
G.~{Corcella}, R.~{Franceschini} and D.~{Kim}, \emph{{Fragmentation
  Uncertainties in Hadronic Observables for Top-quark Mass Measurements}},
  {\emph{ArXiv e-prints} (Dec., 2017) },
  [\href{http://arxiv.org/abs/1712.05801}{{\tt 1712.05801}}].

\bibitem{ATLAS-CONF-2020-050}
{The ATLAS collaboration}, \emph{{Measurements of $b$-jet moments sensitive to
  $b$-quark fragmentation in $t \bar{t}$ events at the LHC with the ATLAS
  detector}}, \href{http://dx.doi.org/oai:cds.cern.ch:2730444}{\emph{{CERN
  Note}} {\bf {}} ({}) {}},
  [\href{http://arxiv.org/abs/ATLAS-CONF-2020-050}{{\tt ATLAS-CONF-2020-050}}].

\bibitem{2108.11650v1}
A.~{Collaboration}, \emph{{Measurement of $b$-quark fragmentation properties in
  jets using the decay $B^{\pm} \to J/\psi K^{\pm}$ in $pp$ collisions at
  $\sqrt{s}=13$~TeV with the ATLAS detector}},
  \href{http://arxiv.org/abs/2108.11650v1}{{\tt 2108.11650v1}}.

\bibitem{ATLAS-Collaboration:2015ng}
{ATLAS Collaboration}, \emph{{Determination of the ratio of $b$-quark
  fragmentation fractions $f_s/f_d$ in $pp$ collisions at $\sqrt{s}=7$~TeV with
  the ATLAS detector}}, {\emph{ArXiv e-prints} (July, 2015) },
  [\href{http://arxiv.org/abs/1507.08925}{{\tt 1507.08925}}].

\bibitem{Agashe:2013sw}
K.~{Agashe}, R.~{Franceschini} and D.~{Kim}, \emph{{A simple, yet subtle
  'invariance' of two-body decay kinematics}},
  \href{http://dx.doi.org/10.1103/PhysRevD.88.057701}{\emph{Phys. Rev. D} {\bf
  88} (Sept., 2013) 057701}, [\href{http://arxiv.org/abs/1209.0772}{{\tt
  1209.0772}}].

\bibitem{Agashe:2016xq}
K.~{Agashe}, D.~{Kim}, R.~{Franceschini} and M.~{Schulze}, \emph{{Top quark
  mass determination from the energy peaks of b-jets and B-hadrons at NLO
  QCD}},  \href{http://arxiv.org/abs/1603.03445v1}{{\tt 1603.03445v1}}.

\bibitem{Hill:2005dq}
C.~S. {Hill}, J.~R. {Incandela} and J.~M. {Lamb}, \emph{{Method for measurement
  of the top quark mass using the mean decay length of b hadrons in $t\bar{t}$
  events}}, \href{http://dx.doi.org/10.1103/PhysRevD.71.054029}{\emph{{Phys.
  Rev. D}} {\bf 71} (Mar., 2005) 054029},
  [\href{http://arxiv.org/abs/hep-ex/0501043}{{\tt hep-ex/0501043}}].

\bibitem{peaklength}
{Agashe, Airen, Franceschini, Kim, Sathyan }, \emph{in preparation}, .

\bibitem{Kawabata:2016aya}
S.~Kawabata and H.~Yokoya, \emph{{Top-quark mass from the diphoton mass
  spectrum}},
  \href{http://dx.doi.org/10.1140/epjc/s10052-017-4884-8}{\emph{Eur. Phys. J.
  C} {\bf 77} (2017) 323}, [\href{http://arxiv.org/abs/1607.00990}{{\tt
  1607.00990}}].

\bibitem{Seidel:1498599}
K.~Seidel, F.~Simon and M.~Tesar, \emph{{Prospects for the Measurement of the
  Top Mass in a Threshold Scan at CLIC and ILC}}, .

\bibitem{Seidel:2013sqa}
K.~Seidel, F.~Simon, M.~Tesar and S.~Poss, \emph{{Top quark mass measurements
  at and above threshold at CLIC}},
  \href{http://dx.doi.org/10.1140/epjc/s10052-013-2530-7}{\emph{Eur. Phys. J.
  C} {\bf 73} (2013) 2530}, [\href{http://arxiv.org/abs/1303.3758}{{\tt
  1303.3758}}].

\bibitem{Maier:2019vll}
A.~Maier, \emph{{Top pair production and mass determination}},
  \href{http://dx.doi.org/10.23731/CYRM-2020-003.117}{\emph{CERN Yellow
  Reports: Monographs} {\bf 3} (2020) 117--122}.

\bibitem{Nowak:2021xmp}
K.~Nowak and A.~F. Zarnecki, \emph{{Optimising top-quark threshold scan at CLIC
  using genetic algorithm}},
  \href{http://dx.doi.org/10.1007/JHEP07(2021)070}{\emph{JHEP} {\bf 07} (2021)
  070}, [\href{http://arxiv.org/abs/2103.00522}{{\tt 2103.00522}}].

\bibitem{Beneke:2016kkb}
M.~Beneke, Y.~Kiyo, A.~Maier and J.~Piclum, \emph{{Near-threshold production of
  heavy quarks with $\tt{QQbar\_threshold}$}},
  \href{http://dx.doi.org/10.1016/j.cpc.2016.07.026}{\emph{Comput. Phys.
  Commun.} {\bf 209} (2016) 96--115},
  [\href{http://arxiv.org/abs/1605.03010}{{\tt 1605.03010}}].

\bibitem{Beneke:2015kwa}
M.~Beneke, Y.~Kiyo, P.~Marquard, A.~Penin, J.~Piclum and M.~Steinhauser,
  \emph{{Next-to-Next-to-Next-to-Leading Order QCD Prediction for the Top
  Antitop $S$-Wave Pair Production Cross Section Near Threshold in $e^+e^-$
  Annihilation}},
  \href{http://dx.doi.org/10.1103/PhysRevLett.115.192001}{\emph{Phys. Rev.
  Lett.} {\bf 115} (2015) 192001}, [\href{http://arxiv.org/abs/1506.06864}{{\tt
  1506.06864}}].

\bibitem{Beneke:2013kia}
M.~Beneke, J.~Piclum and T.~Rauh, \emph{{P-wave contribution to third-order
  top-quark pair production near threshold}},
  \href{http://dx.doi.org/10.1016/j.nuclphysb.2014.01.015}{\emph{Nucl. Phys. B}
  {\bf 880} (2014) 414--434}, [\href{http://arxiv.org/abs/1312.4792}{{\tt
  1312.4792}}].

\bibitem{Beneke:2010mp}
M.~Beneke, B.~Jantzen and P.~Ruiz-Femenia, \emph{{Electroweak non-resonant NLO
  corrections to e+ e- -\ensuremath{>} W+ W- b bbar in the t tbar resonance
  region}},
  \href{http://dx.doi.org/10.1016/j.nuclphysb.2010.07.006}{\emph{Nucl. Phys. B}
  {\bf 840} (2010) 186--213}, [\href{http://arxiv.org/abs/1004.2188}{{\tt
  1004.2188}}].

\bibitem{Hoang:1999zc}
A.~H. Hoang and T.~Teubner, \emph{{Top quark pair production close to
  threshold: Top mass, width and momentum distribution}},
  \href{http://dx.doi.org/10.1103/PhysRevD.60.114027}{\emph{Phys. Rev. D} {\bf
  60} (1999) 114027}, [\href{http://arxiv.org/abs/hep-ph/9904468}{{\tt
  hep-ph/9904468}}].

\bibitem{Hoang:1998xf}
A.~H. Hoang and T.~Teubner, \emph{{Top quark pair production at threshold:
  Complete next-to-next-to-leading order relativistic corrections}},
  \href{http://dx.doi.org/10.1103/PhysRevD.58.114023}{\emph{Phys. Rev. D} {\bf
  58} (1998) 114023}, [\href{http://arxiv.org/abs/hep-ph/9801397}{{\tt
  hep-ph/9801397}}].

\bibitem{Strassler:1990nw}
M.~J. Strassler and M.~E. Peskin, \emph{{The Heavy top quark threshold: QCD and
  the Higgs}}, \href{http://dx.doi.org/10.1103/PhysRevD.43.1500}{\emph{Phys.
  Rev. D} {\bf 43} (1991) 1500--1514}.

\bibitem{Nejad:2016bci}
B.~Chokouf\'e~Nejad, W.~Kilian, J.~M. Lindert, S.~Pozzorini, J.~Reuter and
  C.~Weiss, \emph{{NLO QCD predictions for off-shell $ t\overline{t} $ and $
  t\overline{t}H $ production and decay at a linear collider}},
  \href{http://dx.doi.org/10.1007/JHEP12(2016)075}{\emph{JHEP} {\bf 12} (2016)
  075}, [\href{http://arxiv.org/abs/1609.03390}{{\tt 1609.03390}}].

\bibitem{Bach:2017ggt}
F.~Bach, B.~C. Nejad, A.~Hoang, W.~Kilian, J.~Reuter, M.~Stahlhofen et~al.,
  \emph{{Fully-differential Top-Pair Production at a Lepton Collider: From
  Threshold to Continuum}},
  \href{http://dx.doi.org/10.1007/JHEP03(2018)184}{\emph{JHEP} {\bf 03} (2018)
  184}, [\href{http://arxiv.org/abs/1712.02220}{{\tt 1712.02220}}].

\bibitem{Hoang:2013uda}
A.~H. Hoang and M.~Stahlhofen, \emph{{The Top-Antitop Threshold at the ILC:
  NNLL QCD Uncertainties}},
  \href{http://dx.doi.org/10.1007/JHEP05(2014)121}{\emph{JHEP} {\bf 05} (2014)
  121}, [\href{http://arxiv.org/abs/1309.6323}{{\tt 1309.6323}}].

\bibitem{Boronat:2019cgt}
M.~Boronat, E.~Fullana, J.~Fuster, P.~Gomis, A.~Hoang, V.~Mateu et~al.,
  \emph{{Top quark mass measurement in radiative events at electron-positron
  colliders}},
  \href{http://dx.doi.org/10.1016/j.physletb.2020.135353}{\emph{Phys. Lett. B}
  {\bf 804} (2020) 135353}, [\href{http://arxiv.org/abs/1912.01275}{{\tt
  1912.01275}}].

\bibitem{Abramowicz:2018aa}
{Abramowicz et al.}, \emph{Top-quark physics at the clic electron-positron
  linear collider},  \href{http://arxiv.org/abs/1807.02441v1}{{\tt
  1807.02441v1}}.

\bibitem{ATLAS-Collaboration:2017ac}
{ATLAS Collaboration}, \emph{{Measurement of the $W$-boson mass in pp
  collisions at $\sqrt{s}=7$ TeV with the ATLAS detector}}, {\emph{ArXiv
  e-prints} (Jan., 2017) }, [\href{http://arxiv.org/abs/1701.07240}{{\tt
  1701.07240}}].

\bibitem{PhysRevLett.50.1738}
{Smith, J. and van Neerven, W. L. and Vermaseren, J. A. M.}, \emph{{Transverse
  Mass and Width of the $W$ Boson}},
  \href{http://dx.doi.org/10.1103/PhysRevLett.50.1738}{\emph{{Phys. Rev.
  Lett.}} {\bf 50} (May, 1983) 1738--1740}.

\bibitem{Bianchini:2019aa}
L.~{Bianchini} and G.~{Rolandi}, \emph{A critical point in the distribution of
  lepton energies from the decay of a spin-1 resonance},
  \href{http://arxiv.org/abs/1902.03028v1}{{\tt 1902.03028v1}}.

\bibitem{De-Rujula:2011lr}
A.~{De R{\'u}jula} and A.~{Galindo}, \emph{{Measuring the W-Boson mass at a
  hadron collider: a study of phase-space singularity methods}},
  \href{http://dx.doi.org/10.1007/JHEP08(2011)023}{\emph{Journal of High Energy
  Physics} {\bf 8} (Aug., 2011) 23},
  [\href{http://arxiv.org/abs/1106.0396}{{\tt 1106.0396}}].

\bibitem{The-ALEPH-Collaboration:2013kq}
{The ALEPH Collaboration}, {The DELPHI Collaboration}, {The L3 Collaboration},
  {The OPAL Collaboration} and {The LEP Electroweak Working Group},
  \emph{{Electroweak Measurements in Electron-Positron Collisions at
  W-Boson-Pair Energies at LEP}}, {\emph{ArXiv e-prints} (Feb., 2013) },
  [\href{http://arxiv.org/abs/1302.3415}{{\tt 1302.3415}}].

\bibitem{1907.02029v1}
M.~{Pili}, \emph{{Towards a W boson mass measurement with LHCb}},
  \href{http://arxiv.org/abs/1907.02029v1}{{\tt 1907.02029v1}}.

\bibitem{ATL-PHYS-PROC-2017-051}
{The ATLAS collaboration}, \emph{{W mass measurement}},
  \href{http://dx.doi.org/oai:cds.cern.ch:2264497}{\emph{{CERN Note}} {\bf {}}
  ({2017}) {71--78}}, [\href{http://arxiv.org/abs/ATL-PHYS-PROC-2017-051}{{\tt
  ATL-PHYS-PROC-2017-051}}].

\bibitem{1307.7627v2}
{CDF Collaboration and D0 Collaboration}, \emph{{Combination of CDF and D0
  W-Boson Mass Measurements}},  \href{http://arxiv.org/abs/1307.7627v2}{{\tt
  1307.7627v2}}.

\bibitem{ATL-PHYS-PUB-2018-026}
{The ATLAS collaboration}, \emph{{Prospects for the measurement of the W-boson
  mass at the HL- and HE-LHC}},
  \href{http://dx.doi.org/oai:cds.cern.ch:2645431}{\emph{{CERN Note}} {\bf {}}
  ({}) {}}, [\href{http://arxiv.org/abs/ATL-PHYS-PUB-2018-026}{{\tt
  ATL-PHYS-PUB-2018-026}}].

\bibitem{2107.04444v1}
P.~{Azzurri}, \emph{{The W mass and width measurement challenge at FCC-ee}},
  \href{http://arxiv.org/abs/2107.04444v1}{{\tt 2107.04444v1}}.

\bibitem{Baak:2014oq}
M.~{Baak}, J.~{Cuth}, J.~{Haller}, A.~{Hoecker}, R.~{Kogler}, K.~{Moenig}
  et~al., \emph{{The global electroweak fit at NNLO and prospects for the LHC
  and ILC}}, {\emph{ArXiv e-prints} (July, 2014) },
  [\href{http://arxiv.org/abs/1407.3792}{{\tt 1407.3792}}].

\bibitem{1501.05587v2}
G.~{Bozzi}, L.~{Citelli} and A.~{Vicini}, \emph{{PDF uncertainties on the W
  boson mass measurement from the lepton transverse momentum distribution}},
  \href{http://arxiv.org/abs/1501.05587v2}{{\tt 1501.05587v2}}.

\bibitem{1104.2056v1}
G.~{Bozzi}, J.~{Rojo} and A.~{Vicini}, \emph{{The impact of PDF uncertainties
  on the measurement of the W boson mass at the Tevatron and the LHC}},
  \href{http://arxiv.org/abs/1104.2056v1}{{\tt 1104.2056v1}}.

\bibitem{1910.04726v2}
E.~{Bagnaschi} and A.~{Vicini}, \emph{{Parton Density Uncertainties and the
  Determination of Electroweak Parameters at Hadron Colliders}},
  \href{http://arxiv.org/abs/1910.04726v2}{{\tt 1910.04726v2}}.

\bibitem{1508.06954v2}
G.~{Bozzi}, L.~{Citelli}, M.~{Vesterinen} and A.~{Vicini}, \emph{{Prospects for
  improving the LHC W boson mass measurement with forward muons}},
  \href{http://arxiv.org/abs/1508.06954v2}{{\tt 1508.06954v2}}.

\bibitem{2103.02671v1}
A.~{Behring}, F.~{Buccioni}, F.~{Caola}, M.~{Delto}, M.~{Jaquier},
  K.~{Melnikov} et~al., \emph{{Estimating the impact of mixed QCD-electroweak
  corrections on the $W$-mass determination at the LHC}},
  \href{http://arxiv.org/abs/2103.02671v1}{{\tt 2103.02671v1}}.

\bibitem{Kim:2009si}
I.-W. Kim, \emph{{Algebraic Singularity Method for Mass Measurement with
  Missing Energy}},
  \href{http://dx.doi.org/10.1103/PhysRevLett.104.081601}{\emph{Phys. Rev.
  Lett.} {\bf 104} (2010) 081601}, [\href{http://arxiv.org/abs/0910.1149}{{\tt
  0910.1149}}].

\bibitem{Bianchini:2019iey}
L.~Bianchini and G.~Rolandi, \emph{{A critical point in the distribution of
  lepton energies from the decay of a spin-1 resonance}},
  \href{http://dx.doi.org/10.1007/JHEP05(2019)044}{\emph{JHEP} {\bf 05} (2019)
  044}, [\href{http://arxiv.org/abs/1902.03028}{{\tt 1902.03028}}].

\bibitem{Konar:2008ei}
P.~Konar, K.~Kong and K.~T. Matchev, \emph{{$\sqrt{\hat{s}}_{min}$ : A Global
  inclusive variable for determining the mass scale of new physics in events
  with missing energy at hadron colliders}},
  \href{http://dx.doi.org/10.1088/1126-6708/2009/03/085}{\emph{JHEP} {\bf 03}
  (2009) 085}, [\href{http://arxiv.org/abs/0812.1042}{{\tt 0812.1042}}].

\bibitem{Banfi:2010xy}
A.~Banfi, G.~P. Salam and G.~Zanderighi, \emph{{Phenomenology of event shapes
  at hadron colliders}},
  \href{http://dx.doi.org/10.1007/JHEP06(2010)038}{\emph{JHEP} {\bf 06} (2010)
  038}, [\href{http://arxiv.org/abs/1001.4082}{{\tt 1001.4082}}].

\bibitem{Bjorken:1969wi}
J.~Bjorken and S.~J. Brodsky, \emph{{Statistical Model for electron-Positron
  Annihilation Into Hadrons}},
  \href{http://dx.doi.org/10.1103/PhysRevD.1.1416}{\emph{Phys. Rev. D} {\bf 1}
  (1970) 1416--1420}.

\bibitem{Brandt:1964sa}
S.~Brandt, C.~Peyrou, R.~Sosnowski and A.~Wroblewski, \emph{{The Principal axis
  of jets. An Attempt to analyze high-energy collisions as two-body
  processes}},
  \href{http://dx.doi.org/10.1016/0031-9163(64)91176-X}{\emph{Phys. Lett.} {\bf
  12} (1964) 57--61}.

\bibitem{Farhi:1977sg}
E.~Farhi, \emph{{A QCD Test for Jets}},
  \href{http://dx.doi.org/10.1103/PhysRevLett.39.1587}{\emph{Phys. Rev. Lett.}
  {\bf 39} (1977) 1587--1588}.

\bibitem{Fox:1978vw}
G.~C. Fox and S.~Wolfram, \emph{{Event Shapes in e+ e- Annihilation}},
  \href{http://dx.doi.org/10.1016/0550-3213(79)90120-2}{\emph{Nucl. Phys. B}
  {\bf 149} (1979) 413}.

\bibitem{Stewart:2010tn}
I.~W. Stewart, F.~J. Tackmann and W.~J. Waalewijn, \emph{{N-Jettiness: An
  Inclusive Event Shape to Veto Jets}},
  \href{http://dx.doi.org/10.1103/PhysRevLett.105.092002}{\emph{Phys. Rev.
  Lett.} {\bf 105} (2010) 092002}, [\href{http://arxiv.org/abs/1004.2489}{{\tt
  1004.2489}}].

\bibitem{Thaler:2010tr}
J.~Thaler and K.~Van~Tilburg, \emph{{Identifying Boosted Objects with
  N-subjettiness}},
  \href{http://dx.doi.org/10.1007/JHEP03(2011)015}{\emph{JHEP} {\bf 03} (2011)
  015}, [\href{http://arxiv.org/abs/1011.2268}{{\tt 1011.2268}}].

\bibitem{Basham:1978bw}
C.~L. Basham, L.~S. Brown, S.~D. Ellis and S.~T. Love, \emph{{Energy
  Correlations in electron - Positron Annihilation: Testing QCD}},
  \href{http://dx.doi.org/10.1103/PhysRevLett.41.1585}{\emph{Phys. Rev. Lett.}
  {\bf 41} (1978) 1585}.

\bibitem{Basham:1978zq}
C.~L. Basham, L.~S. Brown, S.~D. Ellis and S.~T. Love, \emph{{Energy
  Correlations in electron-Positron Annihilation in Quantum Chromodynamics:
  Asymptotically Free Perturbation Theory}},
  \href{http://dx.doi.org/10.1103/PhysRevD.19.2018}{\emph{Phys. Rev. D} {\bf
  19} (1979) 2018}.

\bibitem{ShapeVariableTable:Maltoni}
F.~Maltoni, \emph{{Basics of QCD For the LHC}}, .

\bibitem{Mulders:1484921}
M.~Mulders, ed., \emph{{2013 CERN - Latin-American School of High-Energy
  Physics: Arequipa, Peru 6 - 19 Mar 2013. Proceedings, 7th
  CERN–Latin-American School of High-Energy Physics (CLASHEP2013). Arequipa,
  Peru, March 6-19, 2013. 7th CERN - Latin-American School of High-Energy
  Physics}}, (Geneva), CERN, CERN, May, 2016.
\newblock 10.5170/CERN-2015-001.

\bibitem{Ford:2004dp}
M.~T. Ford, \emph{{Studies of event shape observables with the OPAL detector at
  LEP}}.
\newblock PhD thesis, Cambridge U., 2004.
\newblock \href{http://arxiv.org/abs/hep-ex/0405054}{{\tt hep-ex/0405054}}.

\bibitem{DELPHI:2003yqh}
{\scshape DELPHI} collaboration, J.~Abdallah et~al., \emph{{A Study of the
  energy evolution of event shape distributions and their means with the DELPHI
  detector at LEP}},
  \href{http://dx.doi.org/10.1140/epjc/s2003-01198-0}{\emph{Eur. Phys. J. C}
  {\bf 29} (2003) 285--312}, [\href{http://arxiv.org/abs/hep-ex/0307048}{{\tt
  hep-ex/0307048}}].

\bibitem{Weber:2011kor}
M.~Weber, \emph{{Measurement of hadronic event shapes with the CMS detector in
  7 TeV pp collisions at the LHC}}.
\newblock PhD thesis, Zurich, ETH, 2011.
\newblock 10.3929/ethz-a-006717052.

\bibitem{Weber:2009bhh}
{\scshape CMS} collaboration, M.~A. Weber, \emph{{Hadronic Event Shapes at
  CMS}},  in \emph{{17th International Workshop on Deep-Inelastic Scattering
  and Related Subjects}}, (Berlin, Germany), p.~82, Science Wise Publ., 2009.

\bibitem{ATLAS:2020vup}
{\scshape ATLAS} collaboration, G.~Aad et~al., \emph{{Measurement of hadronic
  event shapes in high-p$_{T}$ multijet final states at $ \sqrt{s} $ = 13 TeV
  with the ATLAS detector}},
  \href{http://dx.doi.org/10.1007/JHEP01(2021)188}{\emph{JHEP} {\bf 01} (2021)
  188}, [\href{http://arxiv.org/abs/2007.12600}{{\tt 2007.12600}}].

\bibitem{Banfi:2010zz}
A.~Banfi, \emph{{Event-shape variables at hadron colliders}},  in
  \emph{{Physics at the LHC 2010}}, pp.~186--189, 12, 2010.
\newblock \href{http://dx.doi.org/10.3204/DESY-PROC-2010-01/248}{DOI}.

\bibitem{Lenz:2017lqo}
A.~Lenz, M.~Spannowsky and G.~Tetlalmatzi-Xolocotzi, \emph{{Double-charming
  Higgs boson identification using machine-learning assisted jet shapes}},
  \href{http://dx.doi.org/10.1103/PhysRevD.97.016001}{\emph{Phys. Rev. D} {\bf
  97} (2018) 016001}, [\href{http://arxiv.org/abs/1708.03517}{{\tt
  1708.03517}}].

\bibitem{Fox:1978vu}
G.~C. Fox and S.~Wolfram, \emph{{Observables for the Analysis of Event Shapes
  in e+ e- Annihilation and Other Processes}},
  \href{http://dx.doi.org/10.1103/PhysRevLett.41.1581}{\emph{Phys. Rev. Lett.}
  {\bf 41} (1978) 1581}.

\bibitem{Bernaciak:2012nh}
C.~Bernaciak, M.~S.~A. Buschmann, A.~Butter and T.~Plehn, \emph{{Fox-Wolfram
  Moments in Higgs Physics}},
  \href{http://dx.doi.org/10.1103/PhysRevD.87.073014}{\emph{Phys. Rev. D} {\bf
  87} (2013) 073014}, [\href{http://arxiv.org/abs/1212.4436}{{\tt 1212.4436}}].

\bibitem{Chen:2011ah}
C.~Chen, \emph{{New approach to identifying boosted hadronically-decaying
  particle using jet substructure in its center-of-mass frame}},
  \href{http://dx.doi.org/10.1103/PhysRevD.85.034007}{\emph{Phys. Rev. D} {\bf
  85} (2012) 034007}, [\href{http://arxiv.org/abs/1112.2567}{{\tt 1112.2567}}].

\bibitem{Englert:2012ct}
C.~Englert, M.~Spannowsky and M.~Takeuchi, \emph{{Measuring Higgs CP and
  couplings with hadronic event shapes}},
  \href{http://dx.doi.org/10.1007/JHEP06(2012)108}{\emph{JHEP} {\bf 06} (2012)
  108}, [\href{http://arxiv.org/abs/1203.5788}{{\tt 1203.5788}}].

\bibitem{Dixon:2018qgp}
L.~J. Dixon, M.-X. Luo, V.~Shtabovenko, T.-Z. Yang and H.~X. Zhu,
  \emph{{Analytical Computation of Energy-Energy Correlation at Next-to-Leading
  Order in QCD}},
  \href{http://dx.doi.org/10.1103/PhysRevLett.120.102001}{\emph{Phys. Rev.
  Lett.} {\bf 120} (2018) 102001}, [\href{http://arxiv.org/abs/1801.03219}{{\tt
  1801.03219}}].

\bibitem{Akrawy:1989rg}
{\scshape OPAL} collaboration, M.~Z. Akrawy et~al., \emph{{A Study of Jet
  Production Rates and a Test of QCD on the Z0 Resonance}},
  \href{http://dx.doi.org/10.1016/0370-2693(90)91983-I}{\emph{Phys. Lett. B}
  {\bf 235} (1990) 389--398}.

\bibitem{Catani:1991hj}
S.~Catani, Y.~L. Dokshitzer, M.~Olsson, G.~Turnock and B.~R. Webber, \emph{{New
  clustering algorithm for multi - jet cross-sections in e+ e- annihilation}},
  \href{http://dx.doi.org/10.1016/0370-2693(91)90196-W}{\emph{Phys. Lett. B}
  {\bf 269} (1991) 432--438}.

\bibitem{Moult:2016cvt}
I.~Moult, L.~Necib and J.~Thaler, \emph{{New Angles on Energy Correlation
  Functions}}, \href{http://dx.doi.org/10.1007/JHEP12(2016)153}{\emph{JHEP}
  {\bf 12} (2016) 153}, [\href{http://arxiv.org/abs/1609.07483}{{\tt
  1609.07483}}].

\bibitem{Cesarotti:2020hwb}
C.~Cesarotti and J.~Thaler, \emph{{A Robust Measure of Event Isotropy at
  Colliders}}, \href{http://dx.doi.org/10.1007/JHEP08(2020)084}{\emph{JHEP}
  {\bf 08} (2020) 084}, [\href{http://arxiv.org/abs/2004.06125}{{\tt
  2004.06125}}].

\bibitem{Cesarotti:2020ngq}
C.~Cesarotti, M.~Reece and M.~J. Strassler, \emph{{The Efficacy of Event
  Isotropy as an Event Shape Observable}},
  \href{http://arxiv.org/abs/2011.06599}{{\tt 2011.06599}}.

\bibitem{CMS:2017jdm}
{\scshape CMS} collaboration, A.~M. Sirunyan et~al., \emph{{Search for dark
  matter produced with an energetic jet or a hadronically decaying W or Z boson
  at $ \sqrt{s}=13 $ TeV}},
  \href{http://dx.doi.org/10.1007/JHEP07(2017)014}{\emph{JHEP} {\bf 07} (2017)
  014}, [\href{http://arxiv.org/abs/1703.01651}{{\tt 1703.01651}}].

\bibitem{ATLAS:2017bfj}
{\scshape ATLAS} collaboration, M.~Aaboud et~al., \emph{{Search for dark matter
  and other new phenomena in events with an energetic jet and large missing
  transverse momentum using the ATLAS detector}},
  \href{http://dx.doi.org/10.1007/JHEP01(2018)126}{\emph{JHEP} {\bf 01} (2018)
  126}, [\href{http://arxiv.org/abs/1711.03301}{{\tt 1711.03301}}].

\bibitem{ATLAS:2017nga}
{\scshape ATLAS} collaboration, M.~Aaboud et~al., \emph{{Search for dark matter
  at $\sqrt{s}=13$ TeV in final states containing an energetic photon and large
  missing transverse momentum with the ATLAS detector}},
  \href{http://dx.doi.org/10.1140/epjc/s10052-017-4965-8}{\emph{Eur. Phys. J.
  C} {\bf 77} (2017) 393}, [\href{http://arxiv.org/abs/1704.03848}{{\tt
  1704.03848}}].

\bibitem{CMS:2017qyo}
{\scshape CMS} collaboration, A.~M. Sirunyan et~al., \emph{{Search for new
  physics in the monophoton final state in proton-proton collisions at $
  \sqrt{s}=13 $ TeV}},
  \href{http://dx.doi.org/10.1007/JHEP10(2017)073}{\emph{JHEP} {\bf 10} (2017)
  073}, [\href{http://arxiv.org/abs/1706.03794}{{\tt 1706.03794}}].

\bibitem{CMS:2017ret}
{\scshape CMS} collaboration, A.~M. Sirunyan et~al., \emph{{Search for dark
  matter and unparticles in events with a Z boson and missing transverse
  momentum in proton-proton collisions at $ \sqrt{s}=13 $ TeV}},
  \href{http://dx.doi.org/10.1007/JHEP03(2017)061}{\emph{JHEP} {\bf 03} (2017)
  061}, [\href{http://arxiv.org/abs/1701.02042}{{\tt 1701.02042}}].

\bibitem{ATLAS:2017nyv}
{\scshape ATLAS} collaboration, M.~Aaboud et~al., \emph{{Search for an
  invisibly decaying Higgs boson or dark matter candidates produced in
  association with a $Z$ boson in $pp$ collisions at $\sqrt{s} =$ 13 TeV with
  the ATLAS detector}},
  \href{http://dx.doi.org/10.1016/j.physletb.2017.11.049}{\emph{Phys. Lett. B}
  {\bf 776} (2018) 318--337}, [\href{http://arxiv.org/abs/1708.09624}{{\tt
  1708.09624}}].

\bibitem{ATLAS:2018nda}
{\scshape ATLAS} collaboration, M.~Aaboud et~al., \emph{{Search for dark matter
  in events with a hadronically decaying vector boson and missing transverse
  momentum in $pp$ collisions at $\sqrt{s} = 13$ TeV with the ATLAS detector}},
  \href{http://dx.doi.org/10.1007/JHEP10(2018)180}{\emph{JHEP} {\bf 10} (2018)
  180}, [\href{http://arxiv.org/abs/1807.11471}{{\tt 1807.11471}}].

\bibitem{CMS:2020ulv}
{\scshape CMS} collaboration, A.~M. Sirunyan et~al., \emph{{Search for dark
  matter produced in association with a leptonically decaying Z boson in
  proton-proton collisions at $\sqrt{s} =$ 13 TeV}},
  \href{http://dx.doi.org/10.1140/epjc/s10052-020-08739-5}{\emph{Eur. Phys. J.
  C} {\bf 81} (2021) 13}, [\href{http://arxiv.org/abs/2008.04735}{{\tt
  2008.04735}}].

\bibitem{ATLAS:2017pzz}
{\scshape ATLAS} collaboration, M.~Aaboud et~al., \emph{{Search for dark matter
  in association with a Higgs boson decaying to two photons at $\sqrt{s}$ = 13
  TeV with the ATLAS detector}},
  \href{http://dx.doi.org/10.1103/PhysRevD.96.112004}{\emph{Phys. Rev. D} {\bf
  96} (2017) 112004}, [\href{http://arxiv.org/abs/1706.03948}{{\tt
  1706.03948}}].

\bibitem{ATLAS:2017uis}
{\scshape ATLAS} collaboration, M.~Aaboud et~al., \emph{{Search for Dark Matter
  Produced in Association with a Higgs Boson Decaying to $b\bar b$ using 36
  fb$^{-1}$ of $pp$ collisions at $\sqrt s=13$ TeV with the ATLAS Detector}},
  \href{http://dx.doi.org/10.1103/PhysRevLett.119.181804}{\emph{Phys. Rev.
  Lett.} {\bf 119} (2017) 181804}, [\href{http://arxiv.org/abs/1707.01302}{{\tt
  1707.01302}}].

\bibitem{CMS:2019ykj}
{\scshape CMS} collaboration, A.~M. Sirunyan et~al., \emph{{Search for dark
  matter particles produced in association with a Higgs boson in proton-proton
  collisions at $ \sqrt{\mathrm{s}} $ = 13 TeV}},
  \href{http://dx.doi.org/10.1007/JHEP03(2020)025}{\emph{JHEP} {\bf 03} (2020)
  025}, [\href{http://arxiv.org/abs/1908.01713}{{\tt 1908.01713}}].

\bibitem{CMS:2011bgj}
{\scshape CMS} collaboration, S.~Chatrchyan et~al., \emph{{Missing transverse
  energy performance of the CMS detector}},
  \href{http://dx.doi.org/10.1088/1748-0221/6/09/P09001}{\emph{JINST} {\bf 6}
  (2011) P09001}, [\href{http://arxiv.org/abs/1106.5048}{{\tt 1106.5048}}].

\bibitem{ATLAS:2018txj}
{\scshape ATLAS} collaboration, M.~Aaboud et~al., \emph{{Performance of missing
  transverse momentum reconstruction with the ATLAS detector using
  proton-proton collisions at $\sqrt{s}$ = 13 TeV}},
  \href{http://dx.doi.org/10.1140/epjc/s10052-018-6288-9}{\emph{Eur. Phys. J.
  C} {\bf 78} (2018) 903}, [\href{http://arxiv.org/abs/1802.08168}{{\tt
  1802.08168}}].

\bibitem{Tovey:2000wk}
D.~Tovey, \emph{{Measuring the SUSY mass scale at the LHC}},
  \href{http://dx.doi.org/10.1016/S0370-2693(00)01363-0}{\emph{Phys. Lett. B}
  {\bf 498} (2001) 1--10}, [\href{http://arxiv.org/abs/hep-ph/0006276}{{\tt
  hep-ph/0006276}}].

\bibitem{Papaefstathiou:2009hp}
A.~Papaefstathiou and B.~Webber, \emph{{Effects of QCD radiation on inclusive
  variables for determining the scale of new physics at hadron colliders}},
  \href{http://dx.doi.org/10.1088/1126-6708/2009/06/069}{\emph{JHEP} {\bf 06}
  (2009) 069}, [\href{http://arxiv.org/abs/0903.2013}{{\tt 0903.2013}}].

\bibitem{Konar:2010ma}
P.~Konar, K.~Kong, K.~T. Matchev and M.~Park, \emph{{RECO level
  $\sqrt{s}_{min}$ and subsystem $\sqrt{s}_{min}$: Improved global inclusive
  variables for measuring the new physics mass scale in $E_T$ events at hadron
  colliders}}, \href{http://dx.doi.org/10.1007/JHEP06(2011)041}{\emph{JHEP}
  {\bf 06} (2011) 041}, [\href{http://arxiv.org/abs/1006.0653}{{\tt
  1006.0653}}].

\bibitem{Robens:2011zm}
T.~Robens, \emph{{$\sqrt{\hat{s}}_{\rm min}$ resurrected}},
  \href{http://dx.doi.org/10.1007/JHEP02(2012)051}{\emph{JHEP} {\bf 02} (2012)
  051}, [\href{http://arxiv.org/abs/1109.1018}{{\tt 1109.1018}}].

\bibitem{Swain:2014dha}
A.~K. Swain and P.~Konar, \emph{{Constrained $\sqrt{\hat{S}_{min}}$ and
  reconstructing with semi-invisible production at hadron colliders}},
  \href{http://dx.doi.org/10.1007/JHEP03(2015)142}{\emph{JHEP} {\bf 03} (2015)
  142}, [\href{http://arxiv.org/abs/1412.6624}{{\tt 1412.6624}}].

\bibitem{Bhardwaj:2016lcu}
A.~Bhardwaj, P.~Konar, P.~Sharma and A.~K. Swain, \emph{{Exploring CP phase in
  $\tau$-lepton Yukawa coupling in Higgs decays at the LHC}},
  \href{http://dx.doi.org/10.1088/1361-6471/ab2ee5}{\emph{J. Phys. G} {\bf 46}
  (2019) 105001}, [\href{http://arxiv.org/abs/1612.01417}{{\tt 1612.01417}}].

\bibitem{Konar:2016wbh}
P.~Konar and A.~K. Swain, \emph{{Reconstructing semi-invisible events in
  resonant tau pair production from Higgs}},
  \href{http://dx.doi.org/10.1016/j.physletb.2016.03.070}{\emph{Phys. Lett. B}
  {\bf 757} (2016) 211--215}, [\href{http://arxiv.org/abs/1602.00552}{{\tt
  1602.00552}}].

\bibitem{Cho:2014naa}
W.~S. Cho, J.~S. Gainer, D.~Kim, K.~T. Matchev, F.~Moortgat, L.~Pape et~al.,
  \emph{{On-shell constrained $M_2$ variables with applications to mass
  measurements and topology disambiguation}},
  \href{http://dx.doi.org/10.1007/JHEP08(2014)070}{\emph{JHEP} {\bf 08} (2014)
  070}, [\href{http://arxiv.org/abs/1401.1449}{{\tt 1401.1449}}].

\bibitem{Rogan:2010kb}
C.~Rogan, \emph{{Kinematical variables towards new dynamics at the LHC}},
  \href{http://arxiv.org/abs/1006.2727}{{\tt 1006.2727}}.

\bibitem{Byers:1964ryc}
N.~Byers and C.~N. Yang, \emph{{Physical Regions in Invariant Variables for n
  Particles and the Phase-Space Volume Element}},
  \href{http://dx.doi.org/10.1103/RevModPhys.36.595}{\emph{Rev. Mod. Phys.}
  {\bf 36} (1964) 595--609}.

\bibitem{UA1:1983mne}
{\scshape UA1} collaboration, G.~Arnison et~al., \emph{{Experimental
  Observation of Lepton Pairs of Invariant Mass Around 95-GeV/c**2 at the CERN
  SPS Collider}},
  \href{http://dx.doi.org/10.1016/0370-2693(83)90188-0}{\emph{Phys. Lett. B}
  {\bf 126} (1983) 398--410}.

\bibitem{UA2:1983mlz}
{\scshape UA2} collaboration, P.~Bagnaia et~al., \emph{{Evidence for $Z^{0} \to
  e^+ e^-$ at the CERN $\bar{p} p$ Collider}},
  \href{http://dx.doi.org/10.1016/0370-2693(83)90744-X}{\emph{Phys. Lett. B}
  {\bf 129} (1983) 130--140}.

\bibitem{E598:1974sol}
{\scshape E598} collaboration, J.~J. Aubert et~al., \emph{{Experimental
  Observation of a Heavy Particle $J$}},
  \href{http://dx.doi.org/10.1103/PhysRevLett.33.1404}{\emph{Phys. Rev. Lett.}
  {\bf 33} (1974) 1404--1406}.

\bibitem{SLAC-SP-017:1974ind}
{\scshape SLAC-SP-017} collaboration, J.~E. Augustin et~al., \emph{{Discovery
  of a Narrow Resonance in $e^+ e^-$ Annihilation}},
  \href{http://dx.doi.org/10.1103/PhysRevLett.33.1406}{\emph{Phys. Rev. Lett.}
  {\bf 33} (1974) 1406--1408}.

\bibitem{Herb:1977ek}
S.~W. Herb et~al., \emph{{Observation of a Dimuon Resonance at 9.5-GeV in
  400-GeV Proton-Nucleus Collisions}},
  \href{http://dx.doi.org/10.1103/PhysRevLett.39.252}{\emph{Phys. Rev. Lett.}
  {\bf 39} (1977) 252--255}.

\bibitem{ATLAS:2012yve}
{\scshape ATLAS} collaboration, G.~Aad et~al., \emph{{Observation of a new
  particle in the search for the Standard Model Higgs boson with the ATLAS
  detector at the LHC}},
  \href{http://dx.doi.org/10.1016/j.physletb.2012.08.020}{\emph{Phys. Lett. B}
  {\bf 716} (2012) 1--29}, [\href{http://arxiv.org/abs/1207.7214}{{\tt
  1207.7214}}].

\bibitem{CMS:2012qbp}
{\scshape CMS} collaboration, S.~Chatrchyan et~al., \emph{{Observation of a New
  Boson at a Mass of 125 GeV with the CMS Experiment at the LHC}},
  \href{http://dx.doi.org/10.1016/j.physletb.2012.08.021}{\emph{Phys. Lett. B}
  {\bf 716} (2012) 30--61}, [\href{http://arxiv.org/abs/1207.7235}{{\tt
  1207.7235}}].

\bibitem{Cheng:2002ab}
H.-C. Cheng, K.~T. Matchev and M.~Schmaltz, \emph{{Bosonic supersymmetry?
  Getting fooled at the CERN LHC}},
  \href{http://dx.doi.org/10.1103/PhysRevD.66.056006}{\emph{Phys. Rev. D} {\bf
  66} (2002) 056006}, [\href{http://arxiv.org/abs/hep-ph/0205314}{{\tt
  hep-ph/0205314}}].

\bibitem{Barr:2004ze}
A.~J. Barr, \emph{{Determining the spin of supersymmetric particles at the LHC
  using lepton charge asymmetry}},
  \href{http://dx.doi.org/10.1016/j.physletb.2004.06.074}{\emph{Phys. Lett. B}
  {\bf 596} (2004) 205--212}, [\href{http://arxiv.org/abs/hep-ph/0405052}{{\tt
  hep-ph/0405052}}].

\bibitem{Smillie:2005ar}
J.~M. Smillie and B.~R. Webber, \emph{{Distinguishing spins in supersymmetric
  and universal extra dimension models at the large hadron collider}},
  \href{http://dx.doi.org/10.1088/1126-6708/2005/10/069}{\emph{JHEP} {\bf 10}
  (2005) 069}, [\href{http://arxiv.org/abs/hep-ph/0507170}{{\tt
  hep-ph/0507170}}].

\bibitem{Datta:2005zs}
A.~Datta, K.~Kong and K.~T. Matchev, \emph{{Discrimination of supersymmetry and
  universal extra dimensions at hadron colliders}},
  \href{http://dx.doi.org/10.1103/PhysRevD.72.119901}{\emph{Phys. Rev. D} {\bf
  72} (2005) 096006}, [\href{http://arxiv.org/abs/hep-ph/0509246}{{\tt
  hep-ph/0509246}}].

\bibitem{Athanasiou:2006ef}
C.~Athanasiou, C.~G. Lester, J.~M. Smillie and B.~R. Webber,
  \emph{{Distinguishing Spins in Decay Chains at the Large Hadron Collider}},
  \href{http://dx.doi.org/10.1088/1126-6708/2006/08/055}{\emph{JHEP} {\bf 08}
  (2006) 055}, [\href{http://arxiv.org/abs/hep-ph/0605286}{{\tt
  hep-ph/0605286}}].

\bibitem{Alves:2006df}
A.~Alves, O.~Eboli and T.~Plehn, \emph{{It's a gluino}},
  \href{http://dx.doi.org/10.1103/PhysRevD.74.095010}{\emph{Phys. Rev. D} {\bf
  74} (2006) 095010}, [\href{http://arxiv.org/abs/hep-ph/0605067}{{\tt
  hep-ph/0605067}}].

\bibitem{Wang:2006hk}
L.-T. Wang and I.~Yavin, \emph{{Spin measurements in cascade decays at the
  LHC}}, \href{http://dx.doi.org/10.1088/1126-6708/2007/04/032}{\emph{JHEP}
  {\bf 04} (2007) 032}, [\href{http://arxiv.org/abs/hep-ph/0605296}{{\tt
  hep-ph/0605296}}].

\bibitem{Kilic:2007zk}
C.~Kilic, L.-T. Wang and I.~Yavin, \emph{{On the existence of angular
  correlations in decays with heavy matter partners}},
  \href{http://dx.doi.org/10.1088/1126-6708/2007/05/052}{\emph{JHEP} {\bf 05}
  (2007) 052}, [\href{http://arxiv.org/abs/hep-ph/0703085}{{\tt
  hep-ph/0703085}}].

\bibitem{Csaki:2007xm}
C.~Csaki, J.~Heinonen and M.~Perelstein, \emph{{Testing gluino spin with
  three-body decays}},
  \href{http://dx.doi.org/10.1088/1126-6708/2007/10/107}{\emph{JHEP} {\bf 10}
  (2007) 107}, [\href{http://arxiv.org/abs/0707.0014}{{\tt 0707.0014}}].

\bibitem{Grossman:2011nh}
Y.~Grossman, M.~Martone and D.~J. Robinson, \emph{{Kinematic Edges with Flavor
  Oscillation and Non-Zero Widths}},
  \href{http://dx.doi.org/10.1007/JHEP10(2011)127}{\emph{JHEP} {\bf 10} (2011)
  127}, [\href{http://arxiv.org/abs/1108.5381}{{\tt 1108.5381}}].

\bibitem{Kim:2017qdi}
D.~Kim and K.~T. Matchev, \emph{{How to prove that a $E_T$ excess at the LHC is
  not due to dark matter}},
  \href{http://dx.doi.org/10.1103/PhysRevD.98.055018}{\emph{Phys. Rev. D} {\bf
  98} (2018) 055018}, [\href{http://arxiv.org/abs/1712.07620}{{\tt
  1712.07620}}].

\bibitem{Han:2012nm}
T.~Han, I.-W. Kim and J.~Song, \emph{{Kinematic Cusps With Two Missing
  Particles I: Antler Decay Topology}},
  \href{http://dx.doi.org/10.1103/PhysRevD.87.035003}{\emph{Phys. Rev. D} {\bf
  87} (2013) 035003}, [\href{http://arxiv.org/abs/1206.5633}{{\tt 1206.5633}}].

\bibitem{Edelhauser:2012xb}
L.~Edelhauser, K.~T. Matchev and M.~Park, \emph{{Spin effects in the antler
  event topology at hadron colliders}},
  \href{http://dx.doi.org/10.1007/JHEP11(2012)006}{\emph{JHEP} {\bf 11} (2012)
  006}, [\href{http://arxiv.org/abs/1205.2054}{{\tt 1205.2054}}].

\bibitem{Cho:2012er}
W.~S. Cho, D.~Kim, K.~T. Matchev and M.~Park, \emph{{Probing Resonance Decays
  to Two Visible and Multiple Invisible Particles}},
  \href{http://dx.doi.org/10.1103/PhysRevLett.112.211801}{\emph{Phys. Rev.
  Lett.} {\bf 112} (2014) 211801}, [\href{http://arxiv.org/abs/1206.1546}{{\tt
  1206.1546}}].

\bibitem{Agashe:2010gt}
K.~Agashe, D.~Kim, M.~Toharia and D.~G. Walker, \emph{{Distinguishing Dark
  Matter Stabilization Symmetries Using Multiple Kinematic Edges and Cusps}},
  \href{http://dx.doi.org/10.1103/PhysRevD.82.015007}{\emph{Phys. Rev. D} {\bf
  82} (2010) 015007}, [\href{http://arxiv.org/abs/1003.0899}{{\tt 1003.0899}}].

\bibitem{Cho:2015nxy}
W.~S. Cho, D.~Kim, K.~Kong, S.~H. Lim, K.~T. Matchev, J.-C. Park et~al.,
  \emph{{750 GeV Diphoton Excess May Not Imply a 750 GeV Resonance}},
  \href{http://dx.doi.org/10.1103/PhysRevLett.116.151805}{\emph{Phys. Rev.
  Lett.} {\bf 116} (2016) 151805}, [\href{http://arxiv.org/abs/1512.06824}{{\tt
  1512.06824}}].

\bibitem{Lester:2001zx}
C.~G. Lester, \emph{{Model independent sparticle mass measurements at ATLAS}}.
\newblock PhD thesis, Cambridge U., 2001.

\bibitem{Gjelsten:2004ki}
B.~Gjelsten, D.~Miller and P.~Osland, \emph{{Measurement of SUSY masses via
  cascade decays for SPS 1a}},
  \href{http://dx.doi.org/10.1088/1126-6708/2004/12/003}{\emph{JHEP} {\bf 12}
  (2004) 003}, [\href{http://arxiv.org/abs/hep-ph/0410303}{{\tt
  hep-ph/0410303}}].

\bibitem{Gjelsten:2005aw}
B.~Gjelsten, D.~Miller and P.~Osland, \emph{{Measurement of the gluino mass via
  cascade decays for SPS 1a}},
  \href{http://dx.doi.org/10.1088/1126-6708/2005/06/015}{\emph{JHEP} {\bf 06}
  (2005) 015}, [\href{http://arxiv.org/abs/hep-ph/0501033}{{\tt
  hep-ph/0501033}}].

\bibitem{Miller:2005zp}
D.~J. Miller, P.~Osland and A.~R. Raklev, \emph{{Invariant mass distributions
  in cascade decays}},
  \href{http://dx.doi.org/10.1088/1126-6708/2006/03/034}{\emph{JHEP} {\bf 03}
  (2006) 034}, [\href{http://arxiv.org/abs/hep-ph/0510356}{{\tt
  hep-ph/0510356}}].

\bibitem{Costanzo:2009mq}
D.~Costanzo and D.~R. Tovey, \emph{{Supersymmetric particle mass measurement
  with invariant mass correlations}},
  \href{http://dx.doi.org/10.1088/1126-6708/2009/04/084}{\emph{JHEP} {\bf 04}
  (2009) 084}, [\href{http://arxiv.org/abs/0902.2331}{{\tt 0902.2331}}].

\bibitem{Burns:2009zi}
M.~Burns, K.~T. Matchev and M.~Park, \emph{{Using kinematic boundary lines for
  particle mass measurements and disambiguation in SUSY-like events with
  missing energy}},
  \href{http://dx.doi.org/10.1088/1126-6708/2009/05/094}{\emph{JHEP} {\bf 05}
  (2009) 094}, [\href{http://arxiv.org/abs/0903.4371}{{\tt 0903.4371}}].

\bibitem{Matchev:2009iw}
K.~T. Matchev, F.~Moortgat, L.~Pape and M.~Park, \emph{{Precise reconstruction
  of sparticle masses without ambiguities}},
  \href{http://dx.doi.org/10.1088/1126-6708/2009/08/104}{\emph{JHEP} {\bf 08}
  (2009) 104}, [\href{http://arxiv.org/abs/0906.2417}{{\tt 0906.2417}}].

\bibitem{Matchev:2019sqa}
K.~T. Matchev, F.~Moortgat and L.~Pape, \emph{{Dreaming Awake: Disentangling
  the Underlying Physics in Case of a SUSY-like Discovery at the LHC}},
  \href{http://dx.doi.org/10.1088/1361-6471/ab3bb8}{\emph{J. Phys. G} {\bf 46}
  (2019) 115002}, [\href{http://arxiv.org/abs/1902.11267}{{\tt 1902.11267}}].

\bibitem{Bisset:2008hm}
M.~Bisset, R.~Lu and N.~Kersting, \emph{{Improving SUSY Spectrum Determinations
  at the LHC with Wedgebox Technique}},
  \href{http://dx.doi.org/10.1007/JHEP05(2011)095}{\emph{JHEP} {\bf 05} (2011)
  095}, [\href{http://arxiv.org/abs/0806.2492}{{\tt 0806.2492}}].

\bibitem{Dev:2015kca}
P.~S.~B. Dev, D.~Kim and R.~N. Mohapatra, \emph{{Disambiguating Seesaw Models
  using Invariant Mass Variables at Hadron Colliders}},
  \href{http://dx.doi.org/10.1007/JHEP01(2016)118}{\emph{JHEP} {\bf 01} (2016)
  118}, [\href{http://arxiv.org/abs/1510.04328}{{\tt 1510.04328}}].

\bibitem{UA1:1983crd}
{\scshape UA1} collaboration, G.~Arnison et~al., \emph{{Experimental
  Observation of Isolated Large Transverse Energy Electrons with Associated
  Missing Energy at $\sqrt{s}= 540$ GeV}},
  \href{http://dx.doi.org/10.1016/0370-2693(83)91177-2}{\emph{Phys. Lett. B}
  {\bf 122} (1983) 103--116}.

\bibitem{UA2:1983tsx}
{\scshape UA2} collaboration, M.~Banner et~al., \emph{{Observation of Single
  Isolated Electrons of High Transverse Momentum in Events with Missing
  Transverse Energy at the CERN anti-p p Collider}},
  \href{http://dx.doi.org/10.1016/0370-2693(83)91605-2}{\emph{Phys. Lett. B}
  {\bf 122} (1983) 476--485}.

\bibitem{CDF:2022hxs}
{\scshape CDF} collaboration, T.~Aaltonen et~al., \emph{{High-precision
  measurement of the W boson mass with the CDF II detector}},
  \href{http://dx.doi.org/10.1126/science.abk1781}{\emph{Science} {\bf 376}
  (2022) 170--176}.

\bibitem{Barr:2003rg}
A.~Barr, C.~Lester and P.~Stephens, \emph{{m(T2): The Truth behind the
  glamour}}, \href{http://dx.doi.org/10.1088/0954-3899/29/10/304}{\emph{J.
  Phys. G} {\bf 29} (2003) 2343--2363},
  [\href{http://arxiv.org/abs/hep-ph/0304226}{{\tt hep-ph/0304226}}].

\bibitem{Barr:2009jv}
A.~J. Barr, B.~Gripaios and C.~G. Lester, \emph{{Transverse masses and
  kinematic constraints: from the boundary to the crease}},
  \href{http://dx.doi.org/10.1088/1126-6708/2009/11/096}{\emph{JHEP} {\bf 11}
  (2009) 096}, [\href{http://arxiv.org/abs/0908.3779}{{\tt 0908.3779}}].

\bibitem{Konar:2009qr}
P.~Konar, K.~Kong, K.~T. Matchev and M.~Park, \emph{{Dark Matter Particle
  Spectroscopy at the LHC: Generalizing M(T2) to Asymmetric Event Topologies}},
  \href{http://dx.doi.org/10.1007/JHEP04(2010)086}{\emph{JHEP} {\bf 04} (2010)
  086}, [\href{http://arxiv.org/abs/0911.4126}{{\tt 0911.4126}}].

\bibitem{Kawagoe:2004rz}
K.~Kawagoe, M.~M. Nojiri and G.~Polesello, \emph{{A New SUSY mass
  reconstruction method at the CERN LHC}},
  \href{http://dx.doi.org/10.1103/PhysRevD.71.035008}{\emph{Phys. Rev. D} {\bf
  71} (2005) 035008}, [\href{http://arxiv.org/abs/hep-ph/0410160}{{\tt
  hep-ph/0410160}}].

\bibitem{Burns:2008va}
M.~Burns, K.~Kong, K.~T. Matchev and M.~Park, \emph{{Using Subsystem MT2 for
  Complete Mass Determinations in Decay Chains with Missing Energy at Hadron
  Colliders}},
  \href{http://dx.doi.org/10.1088/1126-6708/2009/03/143}{\emph{JHEP} {\bf 03}
  (2009) 143}, [\href{http://arxiv.org/abs/0810.5576}{{\tt 0810.5576}}].

\bibitem{Lester:2011nj}
C.~G. Lester, \emph{{The stransverse mass, MT2, in special cases}},
  \href{http://dx.doi.org/10.1007/JHEP05(2011)076}{\emph{JHEP} {\bf 05} (2011)
  076}, [\href{http://arxiv.org/abs/1103.5682}{{\tt 1103.5682}}].

\bibitem{Lally:2012uj}
C.~H. Lally and C.~G. Lester, \emph{{Properties of MT2 in the massless limit}},
   \href{http://arxiv.org/abs/1211.1542}{{\tt 1211.1542}}.

\bibitem{Cho:2007dh}
W.~S. Cho, K.~Choi, Y.~G. Kim and C.~B. Park, \emph{{Measuring superparticle
  masses at hadron collider using the transverse mass kink}},
  \href{http://dx.doi.org/10.1088/1126-6708/2008/02/035}{\emph{JHEP} {\bf 02}
  (2008) 035}, [\href{http://arxiv.org/abs/0711.4526}{{\tt 0711.4526}}].

\bibitem{Gripaios:2007is}
B.~Gripaios, \emph{{Transverse observables and mass determination at hadron
  colliders}},
  \href{http://dx.doi.org/10.1088/1126-6708/2008/02/053}{\emph{JHEP} {\bf 02}
  (2008) 053}, [\href{http://arxiv.org/abs/0709.2740}{{\tt 0709.2740}}].

\bibitem{Barr:2007hy}
A.~J. Barr, B.~Gripaios and C.~G. Lester, \emph{{Weighing Wimps with Kinks at
  Colliders: Invisible Particle Mass Measurements from Endpoints}},
  \href{http://dx.doi.org/10.1088/1126-6708/2008/02/014}{\emph{JHEP} {\bf 02}
  (2008) 014}, [\href{http://arxiv.org/abs/0711.4008}{{\tt 0711.4008}}].

\bibitem{Choi:2009hn}
K.~Choi, S.~Choi, J.~S. Lee and C.~B. Park, \emph{{Reconstructing the Higgs
  boson in dileptonic W decays at hadron collider}},
  \href{http://dx.doi.org/10.1103/PhysRevD.80.073010}{\emph{Phys. Rev. D} {\bf
  80} (2009) 073010}, [\href{http://arxiv.org/abs/0908.0079}{{\tt 0908.0079}}].

\bibitem{Choi:2010dw}
K.~Choi, J.~S. Lee and C.~B. Park, \emph{{Measuring the Higgs boson mass with
  transverse mass variables}},
  \href{http://dx.doi.org/10.1103/PhysRevD.82.113017}{\emph{Phys. Rev. D} {\bf
  82} (2010) 113017}, [\href{http://arxiv.org/abs/1008.2690}{{\tt 1008.2690}}].

\bibitem{Barr:2011he}
A.~J. Barr, S.~T. French, J.~A. Frost and C.~G. Lester, \emph{{Speedy Higgs
  boson discovery in decays to tau lepton pairs : h-\ensuremath{>}tau,tau}},
  \href{http://dx.doi.org/10.1007/JHEP10(2011)080}{\emph{JHEP} {\bf 10} (2011)
  080}, [\href{http://arxiv.org/abs/1106.2322}{{\tt 1106.2322}}].

\bibitem{Park:2011uz}
C.~B. Park, \emph{{Reconstructing the heavy resonance at hadron colliders}},
  \href{http://dx.doi.org/10.1103/PhysRevD.84.096001}{\emph{Phys. Rev. D} {\bf
  84} (2011) 096001}, [\href{http://arxiv.org/abs/1106.6087}{{\tt 1106.6087}}].

\bibitem{Guadagnoli:2013xia}
D.~Guadagnoli and C.~B. Park, \emph{{$M_{T2}$-reconstructed invisible momenta
  as spin analizers, and an application to top polarization}},
  \href{http://dx.doi.org/10.1007/JHEP01(2014)030}{\emph{JHEP} {\bf 01} (2014)
  030}, [\href{http://arxiv.org/abs/1308.2226}{{\tt 1308.2226}}].

\bibitem{Konar:2009wn}
P.~Konar, K.~Kong, K.~T. Matchev and M.~Park, \emph{{Superpartner Mass
  Measurement Technique using 1D Orthogonal Decompositions of the Cambridge
  Transverse Mass Variable $M_{T2}$}},
  \href{http://dx.doi.org/10.1103/PhysRevLett.105.051802}{\emph{Phys. Rev.
  Lett.} {\bf 105} (2010) 051802}, [\href{http://arxiv.org/abs/0910.3679}{{\tt
  0910.3679}}].

\bibitem{Matchev:2009ad}
K.~T. Matchev and M.~Park, \emph{{A General method for determining the masses
  of semi-invisibly decaying particles at hadron colliders}},
  \href{http://dx.doi.org/10.1103/PhysRevLett.107.061801}{\emph{Phys. Rev.
  Lett.} {\bf 107} (2011) 061801}, [\href{http://arxiv.org/abs/0910.1584}{{\tt
  0910.1584}}].

\bibitem{Agashe:2010tu}
K.~Agashe, D.~Kim, D.~G. Walker and L.~Zhu, \emph{{Using $M_{T2}$ to
  Distinguish Dark Matter Stabilization Symmetries}},
  \href{http://dx.doi.org/10.1103/PhysRevD.84.055020}{\emph{Phys. Rev. D} {\bf
  84} (2011) 055020}, [\href{http://arxiv.org/abs/1012.4460}{{\tt 1012.4460}}].

\bibitem{Mahbubani:2012kx}
R.~Mahbubani, K.~T. Matchev and M.~Park, \emph{{Re-interpreting the Oxbridge
  stransverse mass variable MT2 in general cases}},
  \href{http://dx.doi.org/10.1007/JHEP03(2013)134}{\emph{JHEP} {\bf 03} (2013)
  134}, [\href{http://arxiv.org/abs/1212.1720}{{\tt 1212.1720}}].

\bibitem{Kim:2016ixu}
D.~Kim, \emph{{Distinguishing dark matter stabilization symmetries at hadron
  colliders}}, \href{http://dx.doi.org/10.1063/1.4953274}{\emph{AIP Conf.
  Proc.} {\bf 1743} (2016) 020007}.

\bibitem{CDF:2009zjw}
{\scshape CDF} collaboration, T.~Aaltonen et~al., \emph{{Measurement of the Top
  Quark Mass in the Dilepton Channel Using $m_{T2}$ at CDF}},
  \href{http://dx.doi.org/10.1103/PhysRevD.81.031102}{\emph{Phys. Rev. D} {\bf
  81} (2010) 031102}, [\href{http://arxiv.org/abs/0911.2956}{{\tt 0911.2956}}].

\bibitem{CMS:2012eya}
{\scshape CMS} collaboration, \emph{{Mass determination in the t t-bar system
  with kinematic endpoints}},  2012.

\bibitem{CMS:2016kgk}
{\scshape CMS} collaboration, \emph{{Measurement of the top quark mass in the
  dileptonic ttbar decay channel using the Mbl, MT2, and MAOS Mblv
  observables}},  8, 2016.

\bibitem{Kim:2015uea}
D.~Kim and K.~Kong, \emph{{Kinematic discrimination of tW and $t \overline{t}$
  productions using initial state radiation}},
  \href{http://dx.doi.org/10.1016/j.physletb.2015.11.010}{\emph{Phys. Lett. B}
  {\bf 751} (2015) 512--524}, [\href{http://arxiv.org/abs/1503.03872}{{\tt
  1503.03872}}].

\bibitem{Cho:2015laa}
W.~S. Cho, J.~S. Gainer, D.~Kim, S.~H. Lim, K.~T. Matchev, F.~Moortgat et~al.,
  \emph{{OPTIMASS: A Package for the Minimization of Kinematic Mass Functions
  with Constraints}},
  \href{http://dx.doi.org/10.1007/JHEP01(2016)026}{\emph{JHEP} {\bf 01} (2016)
  026}, [\href{http://arxiv.org/abs/1508.00589}{{\tt 1508.00589}}].

\bibitem{Park:2020bsu}
C.~B. Park, \emph{{YAM2: Yet another library for the $M_2$ variables using
  sequential quadratic programming}},
  \href{http://arxiv.org/abs/2007.15537}{{\tt 2007.15537}}.

\bibitem{Cheng:2008hk}
H.-C. Cheng and Z.~Han, \emph{{Minimal Kinematic Constraints and m(T2)}},
  \href{http://dx.doi.org/10.1088/1126-6708/2008/12/063}{\emph{JHEP} {\bf 12}
  (2008) 063}, [\href{http://arxiv.org/abs/0810.5178}{{\tt 0810.5178}}].

\bibitem{Lester:2014yga}
C.~G. Lester and B.~Nachman, \emph{{Bisection-based asymmetric M$_{T2}$
  computation: a higher precision calculator than existing symmetric methods}},
  \href{http://dx.doi.org/10.1007/JHEP03(2015)100}{\emph{JHEP} {\bf 03} (2015)
  100}, [\href{http://arxiv.org/abs/1411.4312}{{\tt 1411.4312}}].

\bibitem{Cho:2014yma}
W.~S. Cho, J.~S. Gainer, D.~Kim, K.~T. Matchev, F.~Moortgat, L.~Pape et~al.,
  \emph{{Improving the sensitivity of stop searches with on-shell constrained
  invariant mass variables}},
  \href{http://dx.doi.org/10.1007/JHEP05(2015)040}{\emph{JHEP} {\bf 05} (2015)
  040}, [\href{http://arxiv.org/abs/1411.0664}{{\tt 1411.0664}}].

\bibitem{Ross:2007rm}
G.~G. Ross and M.~Serna, \emph{{Mass determination of new states at hadron
  colliders}},
  \href{http://dx.doi.org/10.1016/j.physletb.2008.06.003}{\emph{Phys. Lett. B}
  {\bf 665} (2008) 212--218}, [\href{http://arxiv.org/abs/0712.0943}{{\tt
  0712.0943}}].

\bibitem{Lim:2016ymd}
S.~H. Lim, \emph{{Identifying the production process of new physics at
  colliders; symmetric or asymmetric?}},
  \href{http://dx.doi.org/10.1007/JHEP06(2016)105}{\emph{JHEP} {\bf 06} (2016)
  105}, [\href{http://arxiv.org/abs/1603.01981}{{\tt 1603.01981}}].

\bibitem{Baer:2016wkz}
H.~Baer, V.~Barger, J.~S. Gainer, P.~Huang, M.~Savoy, D.~Sengupta et~al.,
  \emph{{Gluino reach and mass extraction at the LHC in radiatively-driven
  natural SUSY}},
  \href{http://dx.doi.org/10.1140/epjc/s10052-017-5067-3}{\emph{Eur. Phys. J.
  C} {\bf 77} (2017) 499}, [\href{http://arxiv.org/abs/1612.00795}{{\tt
  1612.00795}}].

\bibitem{Alhazmi:2022qbf}
H.~Alhazmi, Z.~Dong, L.~Huang, J.~H. Kim, K.~Kong and D.~Shih, \emph{{Resolving
  combinatorial ambiguities in dilepton tt\textasciimacron{} event topologies
  with neural networks}},
  \href{http://dx.doi.org/10.1103/PhysRevD.105.115011}{\emph{Phys. Rev. D} {\bf
  105} (2022) 115011}, [\href{http://arxiv.org/abs/2202.05849}{{\tt
  2202.05849}}].

\bibitem{Goncalves:2018agy}
D.~Gon\c{c}alves, K.~Kong and J.~H. Kim, \emph{{Probing the top-Higgs Yukawa CP
  structure in dileptonic $ t\overline{t}h $ with M$_{2}$-assisted
  reconstruction}},
  \href{http://dx.doi.org/10.1007/JHEP06(2018)079}{\emph{JHEP} {\bf 06} (2018)
  079}, [\href{http://arxiv.org/abs/1804.05874}{{\tt 1804.05874}}].

\bibitem{Goncalves:2021dcu}
D.~Gon\c{c}alves, J.~H. Kim, K.~Kong and Y.~Wu, \emph{{Direct Higgs-top
  CP-phase measurement with $ t\overline{t}h $ at the 14 TeV LHC and 100 TeV
  FCC}}, \href{http://dx.doi.org/10.1007/JHEP01(2022)158}{\emph{JHEP} {\bf 01}
  (2022) 158}, [\href{http://arxiv.org/abs/2108.01083}{{\tt 2108.01083}}].

\bibitem{Kim:2014ana}
D.~Kim, H.-S. Lee and M.~Park, \emph{{Invisible dark gauge boson search in top
  decays using a kinematic method}},
  \href{http://dx.doi.org/10.1007/JHEP03(2015)134}{\emph{JHEP} {\bf 03} (2015)
  134}, [\href{http://arxiv.org/abs/1411.0668}{{\tt 1411.0668}}].

\bibitem{Konar:2015hea}
P.~Konar and A.~K. Swain, \emph{{Mass reconstruction with $M_2$ under
  constraint in semi-invisible production at a hadron collider}},
  \href{http://dx.doi.org/10.1103/PhysRevD.93.015021}{\emph{Phys. Rev. D} {\bf
  93} (2016) 015021}, [\href{http://arxiv.org/abs/1509.00298}{{\tt
  1509.00298}}].

\bibitem{Konar:2017kgm}
P.~Konar and A.~K. Swain, \emph{{Mass restricting variables in semi-invisible
  production at the LHC}},
  \href{http://dx.doi.org/10.1007/s12043-017-1453-5}{\emph{Pramana} {\bf 89}
  (2017) 56}.

\bibitem{Tovey:2008ui}
D.~R. Tovey, \emph{{On measuring the masses of pair-produced semi-invisibly
  decaying particles at hadron colliders}},
  \href{http://dx.doi.org/10.1088/1126-6708/2008/04/034}{\emph{JHEP} {\bf 04}
  (2008) 034}, [\href{http://arxiv.org/abs/0802.2879}{{\tt 0802.2879}}].

\bibitem{Cho:2009ve}
W.~S. Cho, J.~E. Kim and J.-H. Kim, \emph{{Amplification of endpoint structure
  for new particle mass measurement at the LHC}},
  \href{http://dx.doi.org/10.1103/PhysRevD.81.095010}{\emph{Phys. Rev. D} {\bf
  81} (2010) 095010}, [\href{http://arxiv.org/abs/0912.2354}{{\tt 0912.2354}}].

\bibitem{Cho:2010vz}
W.~S. Cho, W.~Klemm and M.~M. Nojiri, \emph{{Mass measurement in boosted decay
  systems at hadron colliders}},
  \href{http://dx.doi.org/10.1103/PhysRevD.84.035018}{\emph{Phys. Rev. D} {\bf
  84} (2011) 035018}, [\href{http://arxiv.org/abs/1008.0391}{{\tt 1008.0391}}].

\bibitem{Barr:2010ii}
A.~J. Barr, C.~Gwenlan, C.~G. Lester and C.~J.~S. Young, \emph{{A comment on
  `Amplification of endpoint structure for new particle mass measurement at the
  LHC'}}, \href{http://dx.doi.org/10.1103/PhysRevD.83.118701}{\emph{Phys. Rev.
  D} {\bf 83} (2011) 118701}, [\href{http://arxiv.org/abs/1006.2568}{{\tt
  1006.2568}}].

\bibitem{CMS:2018rst}
{\scshape CMS} collaboration, A.~M. Sirunyan et~al., \emph{{Inclusive search
  for supersymmetry in pp collisions at $ \sqrt{s}=13 $ TeV using razor
  variables and boosted object identification in zero and one lepton final
  states}}, \href{http://dx.doi.org/10.1007/JHEP03(2019)031}{\emph{JHEP} {\bf
  03} (2019) 031}, [\href{http://arxiv.org/abs/1812.06302}{{\tt 1812.06302}}].

\bibitem{CMS:2015adc}
{\scshape CMS} collaboration, V.~Khachatryan et~al., \emph{{Search for
  Supersymmetry Using Razor Variables in Events with $b$-Tagged Jets in $pp$
  Collisions at $\sqrt{s} =$ 8 TeV}},
  \href{http://dx.doi.org/10.1103/PhysRevD.91.052018}{\emph{Phys. Rev. D} {\bf
  91} (2015) 052018}, [\href{http://arxiv.org/abs/1502.00300}{{\tt
  1502.00300}}].

\bibitem{CMS:2012yua}
{\scshape CMS} collaboration, \emph{{Search for supersymmetry with the razor
  variables at CMS}},  2012.

\bibitem{Rujula:2011qn}
A.~Rujula and A.~Galindo, \emph{{Measuring the W-Boson mass at a hadron
  collider: a study of phase-space singularity methods}},
  \href{http://dx.doi.org/10.1007/JHEP08(2011)023}{\emph{JHEP} {\bf 08} (2011)
  023}, [\href{http://arxiv.org/abs/1106.0396}{{\tt 1106.0396}}].

\bibitem{DeRujula:2012ns}
A.~De~Rujula and A.~Galindo, \emph{{Singular ways to search for the Higgs
  boson}}, \href{http://dx.doi.org/10.1007/JHEP06(2012)091}{\emph{JHEP} {\bf
  06} (2012) 091}, [\href{http://arxiv.org/abs/1202.2552}{{\tt 1202.2552}}].

\bibitem{Matchev:2019bon}
K.~T. Matchev and P.~Shyamsundar, \emph{{Singularity Variables for Missing
  Energy Event Kinematics}},
  \href{http://dx.doi.org/10.1007/JHEP04(2020)027}{\emph{JHEP} {\bf 04} (2020)
  027}, [\href{http://arxiv.org/abs/1911.01913}{{\tt 1911.01913}}].

\bibitem{Park:2021lwa}
C.~B. Park, \emph{{Could M$_{T2}$ be a singularity variable?}},
  \href{http://dx.doi.org/10.1007/JHEP11(2021)042}{\emph{JHEP} {\bf 11} (2021)
  042}, [\href{http://arxiv.org/abs/2108.13820}{{\tt 2108.13820}}].

\bibitem{Agrawal:2013uka}
P.~Agrawal, C.~Kilic, C.~White and J.-H. Yu, \emph{{Improved Mass Measurement
  Using the Boundary of Many-Body Phase Space}},
  \href{http://dx.doi.org/10.1103/PhysRevD.89.015021}{\emph{Phys. Rev. D} {\bf
  89} (2014) 015021}, [\href{http://arxiv.org/abs/1308.6560}{{\tt 1308.6560}}].

\bibitem{Debnath:2018azt}
D.~Debnath, J.~S. Gainer, C.~Kilic, D.~Kim, K.~T. Matchev and Y.-P. Yang,
  \emph{{Enhancing the discovery prospects for SUSY-like decays with a
  forgotten kinematic variable}},
  \href{http://dx.doi.org/10.1007/JHEP05(2019)008}{\emph{JHEP} {\bf 05} (2019)
  008}, [\href{http://arxiv.org/abs/1809.04517}{{\tt 1809.04517}}].

\bibitem{Park:2020rol}
C.~B. Park, \emph{{A singular way to search for heavy resonances in missing
  energy events}}, \href{http://dx.doi.org/10.1007/JHEP07(2020)089}{\emph{JHEP}
  {\bf 07} (2020) 089}, [\href{http://arxiv.org/abs/2005.12297}{{\tt
  2005.12297}}].

\bibitem{Kim:2019prx}
D.~Kim, K.~T. Matchev and P.~Shyamsundar, \emph{{Kinematic Focus Point Method
  for Particle Mass Measurements in Missing Energy Events}},
  \href{http://dx.doi.org/10.1007/JHEP10(2019)154}{\emph{JHEP} {\bf 10} (2019)
  154}, [\href{http://arxiv.org/abs/1906.02821}{{\tt 1906.02821}}].

\bibitem{Carlson1950}
A.~Carlson, J.~Hooper and D.~King, \emph{Lxiii. nuclear transmutations produced
  by cosmic-ray particles of great energy.---part v. the neutral mesons},
  \href{http://dx.doi.org/10.1080/14786445008561001}{\emph{The London,
  Edinburgh, and Dublin Philosophical Magazine and Journal of Science} {\bf 41}
  (1950) 701--724},
  [\href{http://arxiv.org/abs/https://doi.org/10.1080/14786445008561001}{{\tt
  https://doi.org/10.1080/14786445008561001}}].

\bibitem{Kawabata:2013fta}
S.~Kawabata, Y.~Shimizu, Y.~Sumino and H.~Yokoya, \emph{{Measurement of
  physical parameters with a weight function method and its application to the
  Higgs boson mass reconstruction}},
  \href{http://dx.doi.org/10.1007/JHEP08(2013)129}{\emph{JHEP} {\bf 08} (2013)
  129}, [\href{http://arxiv.org/abs/1305.6150}{{\tt 1305.6150}}].

\bibitem{CMS:2015jwa}
{\scshape CMS} collaboration, \emph{{Measurement of the top-quark mass from the
  b jet energy spectrum}},  9, 2015.

\bibitem{Low:2013aza}
I.~Low, \emph{{Polarized charginos (and top quarks) in scalar top quark
  decays}}, \href{http://dx.doi.org/10.1103/PhysRevD.88.095018}{\emph{Phys.
  Rev. D} {\bf 88} (2013) 095018}, [\href{http://arxiv.org/abs/1304.0491}{{\tt
  1304.0491}}].

\bibitem{Agashe:2013eba}
K.~Agashe, R.~Franceschini and D.~Kim, \emph{{Using Energy Peaks to Measure New
  Particle Masses}},
  \href{http://dx.doi.org/10.1007/JHEP11(2014)059}{\emph{JHEP} {\bf 11} (2014)
  059}, [\href{http://arxiv.org/abs/1309.4776}{{\tt 1309.4776}}].

\bibitem{Agashe:2015ike}
K.~Agashe, R.~Franceschini, S.~Hong and D.~Kim, \emph{{Energy spectra of
  massive two-body decay products and mass measurement}},
  \href{http://dx.doi.org/10.1007/JHEP04(2016)151}{\emph{JHEP} {\bf 04} (2016)
  151}, [\href{http://arxiv.org/abs/1512.02265}{{\tt 1512.02265}}].

\bibitem{Kim:2015usa}
D.~Kim and J.-C. Park, \emph{{Energy peak: back to the Galactic Center GeV
  gamma-ray excess}},
  \href{http://dx.doi.org/10.1016/j.dark.2016.01.001}{\emph{Phys. Dark Univ.}
  {\bf 11} (2016) 74--78}, [\href{http://arxiv.org/abs/1507.07922}{{\tt
  1507.07922}}].

\bibitem{Kim:2015gka}
D.~Kim and J.-C. Park, \emph{{An alternative interpretation for cosmic ray
  peaks}}, \href{http://dx.doi.org/10.1016/j.physletb.2015.09.070}{\emph{Phys.
  Lett. B} {\bf 750} (2015) 552--558},
  [\href{http://arxiv.org/abs/1508.06640}{{\tt 1508.06640}}].

\bibitem{Boddy:2016fds}
K.~K. Boddy, K.~R. Dienes, D.~Kim, J.~Kumar, J.-C. Park and B.~Thomas,
  \emph{{Lines and Boxes: Unmasking Dynamical Dark Matter through Correlations
  in the MeV Gamma-Ray Spectrum}},
  \href{http://dx.doi.org/10.1103/PhysRevD.94.095027}{\emph{Phys. Rev. D} {\bf
  94} (2016) 095027}, [\href{http://arxiv.org/abs/1606.07440}{{\tt
  1606.07440}}].

\bibitem{Boddy:2016hbp}
K.~K. Boddy, K.~R. Dienes, D.~Kim, J.~Kumar, J.-C. Park and B.~Thomas,
  \emph{{Boxes, Boosts, and Energy Duality: Understanding the Galactic-Center
  Gamma-Ray Excess through Dynamical Dark Matter}},
  \href{http://dx.doi.org/10.1103/PhysRevD.95.055024}{\emph{Phys. Rev. D} {\bf
  95} (2017) 055024}, [\href{http://arxiv.org/abs/1609.09104}{{\tt
  1609.09104}}].

\bibitem{Boddy:2018qur}
K.~Boddy, J.~Kumar, D.~Marfatia and P.~Sandick, \emph{{Model-independent
  constraints on dark matter annihilation in dwarf spheroidal galaxies}},
  \href{http://dx.doi.org/10.1103/PhysRevD.97.095031}{\emph{Phys. Rev. D} {\bf
  97} (2018) 095031}, [\href{http://arxiv.org/abs/1802.03826}{{\tt
  1802.03826}}].

\bibitem{Agashe:2012fs}
K.~Agashe, R.~Franceschini, D.~Kim and K.~Wardlow, \emph{{Using Energy Peaks to
  Count Dark Matter Particles in Decays}},
  \href{http://dx.doi.org/10.1016/j.dark.2013.03.003}{\emph{Phys. Dark Univ.}
  {\bf 2} (2013) 72--82}, [\href{http://arxiv.org/abs/1212.5230}{{\tt
  1212.5230}}].

\bibitem{Agashe:2016bok}
K.~Agashe, R.~Franceschini, D.~Kim and M.~Schulze, \emph{{Top quark mass
  determination from the energy peaks of b-jets and B-hadrons at NLO QCD}},
  \href{http://dx.doi.org/10.1140/epjc/s10052-016-4494-x}{\emph{Eur. Phys. J.
  C} {\bf 76} (2016) 636}, [\href{http://arxiv.org/abs/1603.03445}{{\tt
  1603.03445}}].

\bibitem{CERN-LHCC-2018-023}
{\scshape ATLAS Collaboration} collaboration, \emph{{Technical Proposal: A
  High-Granularity Timing Detector for the ATLAS Phase-II Upgrade}},  Jun,
  2018.

\bibitem{Butler:2019rpu}
{\scshape CMS} collaboration, J.~N. Butler and T.~Tabarelli~de Fatis, \emph{{A
  MIP Timing Detector for the CMS Phase-2 Upgrade}}, .

\bibitem{Liu:2018wte}
J.~Liu, Z.~Liu and L.-T. Wang, \emph{{Enhancing Long-Lived Particles Searches
  at the LHC with Precision Timing Information}},
  \href{http://dx.doi.org/10.1103/PhysRevLett.122.131801}{\emph{Phys. Rev.
  Lett.} {\bf 122} (2019) 131801}, [\href{http://arxiv.org/abs/1805.05957}{{\tt
  1805.05957}}].

\bibitem{Kang:2019ukr}
Z.~Flowers, Q.~Meier, C.~Rogan, D.~W. Kang and S.~C. Park, \emph{{Timing
  information at HL-LHC: Complete determination of masses of Dark Matter and
  Long lived particle}},
  \href{http://dx.doi.org/10.1007/JHEP03(2020)132}{\emph{JHEP} {\bf 03} (2020)
  132}, [\href{http://arxiv.org/abs/1903.05825}{{\tt 1903.05825}}].

\bibitem{Dienes:2021cxr}
K.~R. Dienes, D.~Kim, T.~Leininger and B.~Thomas, \emph{{Tumblers: A Novel
  Collider Signature for Long-Lived Particles}},
  \href{http://arxiv.org/abs/2108.02204}{{\tt 2108.02204}}.

\bibitem{Perazzini:2022cdy}
{\scshape LHCb ECAL Upgrade-2 R\&D Group} collaboration, S.~Perazzini,
  F.~Ferrari and V.~M. Vagnoni, \emph{{Development of an MCP-Based Timing Layer
  for the LHCb ECAL Upgrade-2}},
  \href{http://dx.doi.org/10.3390/instruments6010007}{\emph{Instruments} {\bf
  6} (2022) 7}.

\bibitem{CMS:2014wda}
{\scshape CMS} collaboration, V.~Khachatryan et~al., \emph{{Search for
  Long-Lived Neutral Particles Decaying to Quark-Antiquark Pairs in
  Proton-Proton Collisions at $\sqrt{s} =$ 8 TeV}},
  \href{http://dx.doi.org/10.1103/PhysRevD.91.012007}{\emph{Phys. Rev. D} {\bf
  91} (2015) 012007}, [\href{http://arxiv.org/abs/1411.6530}{{\tt 1411.6530}}].

\bibitem{ATLAS:2015xit}
{\scshape ATLAS} collaboration, G.~Aad et~al., \emph{{Search for long-lived,
  weakly interacting particles that decay to displaced hadronic jets in
  proton-proton collisions at $\sqrt{s}=8$ TeV with the ATLAS detector}},
  \href{http://dx.doi.org/10.1103/PhysRevD.92.012010}{\emph{Phys. Rev. D} {\bf
  92} (2015) 012010}, [\href{http://arxiv.org/abs/1504.03634}{{\tt
  1504.03634}}].

\bibitem{Coccaro:2016lnz}
A.~Coccaro, D.~Curtin, H.~J. Lubatti, H.~Russell and J.~Shelton,
  \emph{{Data-driven Model-independent Searches for Long-lived Particles at the
  LHC}}, \href{http://dx.doi.org/10.1103/PhysRevD.94.113003}{\emph{Phys. Rev.
  D} {\bf 94} (2016) 113003}, [\href{http://arxiv.org/abs/1605.02742}{{\tt
  1605.02742}}].

\bibitem{Ally:2022rgk}
D.~Ally, L.~Carpenter, T.~Holmes, L.~Lee and P.~Wagenknecht, \emph{{Strategies
  for Beam-Induced Background Reduction at Muon Colliders}},
  \href{http://arxiv.org/abs/2203.06773}{{\tt 2203.06773}}.

\bibitem{MuonCollider:2022ded}
{\scshape Muon Collider} collaboration, N.~Bartosik et~al., \emph{{Simulated
  Detector Performance at the Muon Collider}},  in \emph{{2022 Snowmass Summer
  Study}}, 3, 2022.
\newblock \href{http://arxiv.org/abs/2203.07964}{{\tt 2203.07964}}.

\bibitem{DiBenedetto:2018cpy}
V.~Di~Benedetto, C.~Gatto, A.~Mazzacane, N.~V. Mokhov, S.~I. Striganov and
  N.~K. Terentiev, \emph{{A Study of Muon Collider Background Rejection
  Criteria in Silicon Vertex and Tracker Detectors}},
  \href{http://dx.doi.org/10.1088/1748-0221/13/09/P09004}{\emph{JINST} {\bf 13}
  (2018) P09004}, [\href{http://arxiv.org/abs/1807.00074}{{\tt 1807.00074}}].

\bibitem{Bartosik:2019dzq}
N.~Bartosik et~al., \emph{{Preliminary Report on the Study of Beam-Induced
  Background Effects at a Muon Collider}},
  \href{http://arxiv.org/abs/1905.03725}{{\tt 1905.03725}}.

\bibitem{Abulencia:2007cz}
A.~{Abulencia}, J.~{Adelman}, T.~{Affolder}, T.~{Akimoto}, M.~G. {Albrow},
  D.~{Ambrose} et~al., \emph{{Measurement of the top quark mass in $p\bar{p}$
  collisions at s=1.96TeV using the decay length technique}},
  \href{http://dx.doi.org/10.1103/PhysRevD.75.071102}{\emph{\prd} {\bf 75}
  (Apr., 2007) 071102}, [\href{http://arxiv.org/abs/hep-ex/0612061}{{\tt
  hep-ex/0612061}}].

\bibitem{AgasheLength}
{K.~Agashe}, \emph{\href{https://indico.fnal.gov/event/43738/}{New ideas for
  top quark mass measurements}}, .

\bibitem{Agashe:2022sxw}
K.~Agashe, S.~Airen, R.~Franceschini, D.~Kim and D.~Sathyan,
  \emph{{Snowmass2021 - White Paper, Implications of Energy Peak for Collider
  Phenomenology: Top Quark Mass Determination and Beyond}},  in \emph{{2022
  Snowmass Summer Study}}, 4, 2022.
\newblock \href{http://arxiv.org/abs/2204.02928}{{\tt 2204.02928}}.

\bibitem{Barger:1987nn}
V.~D. Barger and R.~J.~N. Phillips, \emph{{COLLIDER PHYSICS}}.
\newblock Addison-Wesley Publishing Company, 1987.

\bibitem{rizzi2006track}
A.~Rizzi, F.~Palla and G.~Segneri, \emph{Track impact parameter based b-tagging
  with cms},  tech. rep., CERN-CMS-NOTE-2006-019, 2006.

\bibitem{note2020deep}
A.~P. Note, \emph{Deep sets based neural networks for impact parameter flavour
  tagging in atlas}, .

\bibitem{Wells:1957370}
P.~Wells, \emph{{Pileup Mitigation at the HL-LHC}}, .

\bibitem{1705.02211v1}
A.~{Collaboration}, \emph{{Identification and rejection of pile-up jets at high
  pseudorapidity with the ATLAS detector}},
  \href{http://arxiv.org/abs/1705.02211v1}{{\tt 1705.02211v1}}.

\bibitem{1801.09721v2}
G.~{Soyez}, \emph{{Pileup mitigation at the LHC: a theorist's view}},
  \href{http://arxiv.org/abs/1801.09721v2}{{\tt 1801.09721v2}}.

\bibitem{2108.02204v1}
K.~R. {Dienes}, D.~{Kim}, T.~{Leininger} and B.~{Thomas}, \emph{{Tumblers: A
  Novel Collider Signature for Long-Lived Particles}},
  \href{http://arxiv.org/abs/2108.02204v1}{{\tt 2108.02204v1}}.

\bibitem{Lee:2018aa}
L.~{Lee}, C.~{Ohm}, A.~{Soffer} and T.-T. {Yu}, \emph{{Collider Searches for
  Long-Lived Particles Beyond the Standard Model}}, {\emph{ArXiv e-prints}
  (Oct., 2018) }, [\href{http://arxiv.org/abs/1810.12602}{{\tt 1810.12602}}].

\bibitem{2103.08620v1}
D.~{Linthorne} and D.~{Stolarski}, \emph{{Triggering on Emerging Jets}},
  \href{http://arxiv.org/abs/2103.08620v1}{{\tt 2103.08620v1}}.

\bibitem{Schwaller:2015ek}
P.~{Schwaller}, D.~{Stolarski} and A.~{Weiler}, \emph{{Emerging Jets}},
  {\emph{ArXiv e-prints} (Feb., 2015) },
  [\href{http://arxiv.org/abs/1502.05409}{{\tt 1502.05409}}].

\bibitem{Alimena:2019aa}
{Alimena et al.}, \emph{{Searching for long-lived particles beyond the Standard
  Model at the Large Hadron Collider}}, {\emph{arXiv e-prints} (Mar., 2019) },
  [\href{http://arxiv.org/abs/1903.04497}{{\tt 1903.04497}}].

\bibitem{Hewett:2004aa}
J.~A.~L. {Hewett}, B.~{Lillie}, M.~{Masip} and T.~G. {Rizzo}, \emph{{Signatures
  of long-lived gluinos in split supersymmetry}},
  \href{http://dx.doi.org/10.1088/1126-6708/2004/09/070}{\emph{Journal of High
  Energy Physics} {\bf 9} (Sept., 2004) 070},
  [\href{http://arxiv.org/abs/hep-ph/0408248}{{\tt hep-ph/0408248}}].

\bibitem{Evans:2016il}
J.~A. {Evans} and J.~{Shelton}, \emph{{Long-lived staus and displaced leptons
  at the LHC}}, \href{http://dx.doi.org/10.1007/JHEP04(2016)056}{\emph{Journal
  of High Energy Physics} {\bf 4} (Apr., 2016) No 56},
  [\href{http://arxiv.org/abs/1601.01326}{{\tt 1601.01326}}].

\bibitem{Barnard:2015uq}
J.~{Barnard}, P.~{Cox}, T.~{Gherghetta} and A.~{Spray}, \emph{{Long-Lived,
  Colour-Triplet Scalars from Unnaturalness}}, {\emph{ArXiv e-prints} (Oct.,
  2015) }, [\href{http://arxiv.org/abs/1510.06405}{{\tt 1510.06405}}].

\bibitem{Bomark:2013lh}
N.-E. {Bomark}, A.~{Kvellestad}, S.~{Lola}, P.~{Osland} and A.~R. {Raklev},
  \emph{{Long lived charginos in Natural SUSY?}}, {\emph{ArXiv e-prints} (Oct.,
  2013) }, [\href{http://arxiv.org/abs/1310.2788}{{\tt 1310.2788}}].

\bibitem{Meade:2010kx}
P.~{Meade}, M.~{Reece} and D.~{Shih}, \emph{{Long-lived neutralino NLSPs}},
  \href{http://dx.doi.org/10.1007/JHEP10(2010)067}{\emph{Journal of High Energy
  Physics} {\bf 10} (Oct., 2010) 67},
  [\href{http://arxiv.org/abs/1006.4575}{{\tt 1006.4575}}].

\bibitem{Liu:2015eu}
Z.~{Liu} and B.~{Tweedie}, \emph{The fate of long-lived superparticles with
  hadronic decays after lhc run 1},
  \href{http://arxiv.org/abs/1503.05923v1}{{\tt 1503.05923v1}}.

\bibitem{1810.10069v2}
C.~{Collaboration}, \emph{{Search for new particles decaying to a jet and an
  emerging jet}},  \href{http://arxiv.org/abs/1810.10069v2}{{\tt
  1810.10069v2}}.

\bibitem{1806.07355v2}
ATLAS, \emph{{Search for the Higgs boson produced in association with a vector
  boson and decaying into two spin-0 particles in the $H \to aa \to 4b$ channel
  in $pp$ collisions at $\sqrt{s} = 13$ TeV with the ATLAS detector}},
  \href{http://arxiv.org/abs/1806.07355v2}{{\tt 1806.07355v2}}.

\bibitem{1905.09787v2}
A.~{Collaboration}, \emph{{Search for heavy neutral leptons in decays of $W$
  bosons produced in 13 TeV $pp$ collisions using prompt and displaced
  signatures with the ATLAS detector}},
  \href{http://arxiv.org/abs/1905.09787v2}{{\tt 1905.09787v2}}.

\bibitem{2012.01581v2}
C.~{Collaboration}, \emph{{Search for long-lived particles using displaced jets
  in proton-proton collisions at $\sqrt{s} = $ 13 TeV}},
  \href{http://arxiv.org/abs/2012.01581v2}{{\tt 2012.01581v2}}.

\bibitem{2107.04838v1}
C.~{Collaboration}, \emph{{Search for long-lived particles decaying in the CMS
  endcap muon detectors in proton-proton collisions at $\sqrt{s} = $ 13 TeV}},
  \href{http://arxiv.org/abs/2107.04838v1}{{\tt 2107.04838v1}}.

\bibitem{Randall:2008rw}
L.~Randall and D.~Tucker-Smith, \emph{{Dijet Searches for Supersymmetry at the
  LHC}}, \href{http://dx.doi.org/10.1103/PhysRevLett.101.221803}{\emph{Phys.
  Rev. Lett.} {\bf 101} (2008) 221803},
  [\href{http://arxiv.org/abs/0806.1049}{{\tt 0806.1049}}].

\bibitem{CMS:2012rao}
{\scshape CMS} collaboration, S.~Chatrchyan et~al., \emph{{Search for
  Supersymmetry in Final States with Missing Transverse Energy and 0, 1, 2, or
  at Least 3 b-Quark Jets in 7 TeV pp Collisions using the Variable
  $\alpha_T$}}, \href{http://dx.doi.org/10.1007/JHEP01(2013)077}{\emph{JHEP}
  {\bf 01} (2013) 077}, [\href{http://arxiv.org/abs/1210.8115}{{\tt
  1210.8115}}].

\bibitem{CMS:2011bul}
{\scshape CMS} collaboration, S.~Chatrchyan et~al., \emph{{Search for
  Supersymmetry at the LHC in Events with Jets and Missing Transverse Energy}},
  \href{http://dx.doi.org/10.1103/PhysRevLett.107.221804}{\emph{Phys. Rev.
  Lett.} {\bf 107} (2011) 221804}, [\href{http://arxiv.org/abs/1109.2352}{{\tt
  1109.2352}}].

\bibitem{Graesser:2012qy}
M.~L. Graesser and J.~Shelton, \emph{{Hunting Mixed Top Squark Decays}},
  \href{http://dx.doi.org/10.1103/PhysRevLett.111.121802}{\emph{Phys. Rev.
  Lett.} {\bf 111} (2013) 121802}, [\href{http://arxiv.org/abs/1212.4495}{{\tt
  1212.4495}}].

\bibitem{Kim:2018cxf}
J.~H. Kim, K.~Kong, K.~T. Matchev and M.~Park, \emph{{Probing the Triple Higgs
  Self-Interaction at the Large Hadron Collider}},
  \href{http://dx.doi.org/10.1103/PhysRevLett.122.091801}{\emph{Phys. Rev.
  Lett.} {\bf 122} (2019) 091801}, [\href{http://arxiv.org/abs/1807.11498}{{\tt
  1807.11498}}].

\bibitem{Kim:2019wns}
J.~H. Kim, M.~Kim, K.~Kong, K.~T. Matchev and M.~Park, \emph{{Portraying Double
  Higgs at the Large Hadron Collider}},
  \href{http://dx.doi.org/10.1007/JHEP09(2019)047}{\emph{JHEP} {\bf 09} (2019)
  047}, [\href{http://arxiv.org/abs/1904.08549}{{\tt 1904.08549}}].

\bibitem{Huang:2022rne}
L.~Huang, S.-b. Kang, J.~H. Kim, K.~Kong and J.~S. Pi, \emph{{Portraying Double
  Higgs at the Large Hadron Collider II}},
  \href{http://arxiv.org/abs/2203.11951}{{\tt 2203.11951}}.

\bibitem{Alves:2022gnw}
A.~Alves and C.~H. Yamaguchi, \emph{{Reconstruction of Missing Resonances
  Combining Nearest Neighbors Regressors and Neural Network Classifiers}},
  \href{http://arxiv.org/abs/2203.03662}{{\tt 2203.03662}}.

\bibitem{Nojiri:2000wq}
M.~M. Nojiri, D.~Toya and T.~Kobayashi, \emph{{Lepton energy asymmetry and
  precision SUSY study at hadron colliders}},
  \href{http://dx.doi.org/10.1103/PhysRevD.62.075009}{\emph{Phys. Rev. D} {\bf
  62} (2000) 075009}, [\href{http://arxiv.org/abs/hep-ph/0001267}{{\tt
  hep-ph/0001267}}].

\bibitem{Cheng:2011ya}
H.-C. Cheng and J.~Gu, \emph{{Measuring Invisible Particle Masses Using a
  Single Short Decay Chain}},
  \href{http://dx.doi.org/10.1007/JHEP10(2011)094}{\emph{JHEP} {\bf 10} (2011)
  094}, [\href{http://arxiv.org/abs/1109.3471}{{\tt 1109.3471}}].

\bibitem{Matchev:2009fh}
K.~T. Matchev, F.~Moortgat, L.~Pape and M.~Park, \emph{{Precision sparticle
  spectroscopy in the inclusive same-sign dilepton channel at LHC}},
  \href{http://dx.doi.org/10.1103/PhysRevD.82.077701}{\emph{Phys. Rev. D} {\bf
  82} (2010) 077701}, [\href{http://arxiv.org/abs/0909.4300}{{\tt 0909.4300}}].

\bibitem{Low:1958sn}
F.~E. Low, \emph{{Bremsstrahlung of very low-energy quanta in elementary
  particle collisions}},
  \href{http://dx.doi.org/10.1103/PhysRev.110.974}{\emph{Phys. Rev.} {\bf 110}
  (1958) 974--977}.

\bibitem{Birkedal:2004xn}
A.~Birkedal, K.~Matchev and M.~Perelstein, \emph{{Dark matter at colliders: A
  Model independent approach}},
  \href{http://dx.doi.org/10.1103/PhysRevD.70.077701}{\emph{Phys. Rev. D} {\bf
  70} (2004) 077701}, [\href{http://arxiv.org/abs/hep-ph/0403004}{{\tt
  hep-ph/0403004}}].

\bibitem{Bai:2010hh}
Y.~Bai, P.~J. Fox and R.~Harnik, \emph{{The Tevatron at the Frontier of Dark
  Matter Direct Detection}},
  \href{http://dx.doi.org/10.1007/JHEP12(2010)048}{\emph{JHEP} {\bf 12} (2010)
  048}, [\href{http://arxiv.org/abs/1005.3797}{{\tt 1005.3797}}].

\bibitem{Goodman:2010ku}
J.~Goodman, M.~Ibe, A.~Rajaraman, W.~Shepherd, T.~M.~P. Tait and H.-B. Yu,
  \emph{{Constraints on Dark Matter from Colliders}},
  \href{http://dx.doi.org/10.1103/PhysRevD.82.116010}{\emph{Phys. Rev. D} {\bf
  82} (2010) 116010}, [\href{http://arxiv.org/abs/1008.1783}{{\tt 1008.1783}}].

\bibitem{Hubisz:2008gg}
J.~Hubisz, J.~Lykken, M.~Pierini and M.~Spiropulu, \emph{{Missing energy
  look-alikes with 100 pb$^{-1}$ at the LHC}},
  \href{http://dx.doi.org/10.1103/PhysRevD.78.075008}{\emph{Phys. Rev. D} {\bf
  78} (2008) 075008}, [\href{http://arxiv.org/abs/0805.2398}{{\tt 0805.2398}}].

\bibitem{Bae:2017ebe}
K.~J. Bae, T.~H. Jung and M.~Park, \emph{{Spectral Decomposition of Missing
  Transverse Energy at Hadron Colliders}},
  \href{http://dx.doi.org/10.1103/PhysRevLett.119.261801}{\emph{Phys. Rev.
  Lett.} {\bf 119} (2017) 261801}, [\href{http://arxiv.org/abs/1706.04512}{{\tt
  1706.04512}}].

\bibitem{Abercrombie:2015wmb}
D.~Abercrombie et~al., \emph{{Dark Matter benchmark models for early LHC Run-2
  Searches: Report of the ATLAS/CMS Dark Matter Forum}},
  \href{http://dx.doi.org/10.1016/j.dark.2019.100371}{\emph{Phys. Dark Univ.}
  {\bf 27} (2020) 100371}, [\href{http://arxiv.org/abs/1507.00966}{{\tt
  1507.00966}}].

\bibitem{CMS:2021far}
{\scshape CMS} collaboration, A.~Tumasyan et~al., \emph{{Search for new
  particles in events with energetic jets and large missing transverse momentum
  in proton-proton collisions at $ \sqrt{s} $ = 13 TeV}},
  \href{http://dx.doi.org/10.1007/JHEP11(2021)153}{\emph{JHEP} {\bf 11} (2021)
  153}, [\href{http://arxiv.org/abs/2107.13021}{{\tt 2107.13021}}].

\bibitem{ATLAS:2021kxv}
{\scshape ATLAS} collaboration, G.~Aad et~al., \emph{{Search for new phenomena
  in events with an energetic jet and missing transverse momentum in $pp$
  collisions at $\sqrt {s}$ =13 TeV with the ATLAS detector}},
  \href{http://dx.doi.org/10.1103/PhysRevD.103.112006}{\emph{Phys. Rev. D} {\bf
  103} (2021) 112006}, [\href{http://arxiv.org/abs/2102.10874}{{\tt
  2102.10874}}].

\bibitem{monoXLHC}
Y.~Abulaiti, \emph{{Status of searches for dark matter at the LHC}},
  {\emph{ATL-PHYS-PROC-2022-003} (2022) }.

\bibitem{Martin:2007gf}
S.~P. Martin, \emph{{Compressed supersymmetry and natural neutralino dark
  matter from top squark-mediated annihilation to top quarks}},
  \href{http://dx.doi.org/10.1103/PhysRevD.75.115005}{\emph{Phys. Rev. D} {\bf
  75} (2007) 115005}, [\href{http://arxiv.org/abs/hep-ph/0703097}{{\tt
  hep-ph/0703097}}].

\bibitem{LeCompte:2011cn}
T.~J. LeCompte and S.~P. Martin, \emph{{Large Hadron Collider reach for
  supersymmetric models with compressed mass spectra}},
  \href{http://dx.doi.org/10.1103/PhysRevD.84.015004}{\emph{Phys. Rev. D} {\bf
  84} (2011) 015004}, [\href{http://arxiv.org/abs/1105.4304}{{\tt 1105.4304}}].

\bibitem{Cheng:2002iz}
H.-C. Cheng, K.~T. Matchev and M.~Schmaltz, \emph{{Radiative corrections to
  Kaluza-Klein masses}},
  \href{http://dx.doi.org/10.1103/PhysRevD.66.036005}{\emph{Phys. Rev. D} {\bf
  66} (2002) 036005}, [\href{http://arxiv.org/abs/hep-ph/0204342}{{\tt
  hep-ph/0204342}}].

\bibitem{Freitas:2017afm}
A.~Freitas, K.~Kong and D.~Wiegand, \emph{{Radiative corrections to masses and
  couplings in Universal Extra Dimensions}},
  \href{http://dx.doi.org/10.1007/JHEP03(2018)093}{\emph{JHEP} {\bf 03} (2018)
  093}, [\href{http://arxiv.org/abs/1711.07526}{{\tt 1711.07526}}].

\bibitem{Shimmin:2016vlc}
C.~Shimmin and D.~Whiteson, \emph{{Boosting low-mass hadronic resonances}},
  \href{http://dx.doi.org/10.1103/PhysRevD.94.055001}{\emph{Phys. Rev. D} {\bf
  94} (2016) 055001}, [\href{http://arxiv.org/abs/1602.07727}{{\tt
  1602.07727}}].

\bibitem{An:2012ue}
H.~An, R.~Huo and L.-T. Wang, \emph{{Searching for Low Mass Dark Portal at the
  LHC}}, \href{http://dx.doi.org/10.1016/j.dark.2013.03.002}{\emph{Phys. Dark
  Univ.} {\bf 2} (2013) 50--57}, [\href{http://arxiv.org/abs/1212.2221}{{\tt
  1212.2221}}].

\bibitem{Fuster:2017rev}
J.~Fuster, A.~Irles, D.~Melini, P.~Uwer and M.~Vos, \emph{{Extracting the
  top-quark running mass using $t\bar{t} + \hbox {1-jet}$ events produced at
  the Large Hadron Collider}},
  \href{http://dx.doi.org/10.1140/epjc/s10052-017-5354-z}{\emph{Eur. Phys. J.
  C} {\bf 77} (2017) 794}, [\href{http://arxiv.org/abs/1704.00540}{{\tt
  1704.00540}}].

\bibitem{Cheng:2007xv}
H.-C. Cheng, J.~F. Gunion, Z.~Han, G.~Marandella and B.~McElrath, \emph{{Mass
  determination in SUSY-like events with missing energy}},
  \href{http://dx.doi.org/10.1088/1126-6708/2007/12/076}{\emph{JHEP} {\bf 12}
  (2007) 076}, [\href{http://arxiv.org/abs/0707.0030}{{\tt 0707.0030}}].

\bibitem{Cheng:2008mg}
H.-C. Cheng, D.~Engelhardt, J.~F. Gunion, Z.~Han and B.~McElrath,
  \emph{{Accurate Mass Determinations in Decay Chains with Missing Energy}},
  \href{http://dx.doi.org/10.1103/PhysRevLett.100.252001}{\emph{Phys. Rev.
  Lett.} {\bf 100} (2008) 252001}, [\href{http://arxiv.org/abs/0802.4290}{{\tt
  0802.4290}}].

\bibitem{Nojiri:2003tu}
M.~M. Nojiri, G.~Polesello and D.~R. Tovey, \emph{{Proposal for a new
  reconstruction technique for SUSY processes at the LHC}},  in \emph{{3rd Les
  Houches Workshop on Physics at TeV Colliders}}, 12, 2003.
\newblock \href{http://arxiv.org/abs/hep-ph/0312317}{{\tt hep-ph/0312317}}.

\bibitem{Cheng:2009fw}
H.-C. Cheng, J.~F. Gunion, Z.~Han and B.~McElrath, \emph{{Accurate Mass
  Determinations in Decay Chains with Missing Energy. II}},
  \href{http://dx.doi.org/10.1103/PhysRevD.80.035020}{\emph{Phys. Rev. D} {\bf
  80} (2009) 035020}, [\href{http://arxiv.org/abs/0905.1344}{{\tt 0905.1344}}].

\bibitem{Nojiri:2007pq}
M.~M. Nojiri, G.~Polesello and D.~R. Tovey, \emph{{A Hybrid method for
  determining SUSY particle masses at the LHC with fully identified cascade
  decays}}, \href{http://dx.doi.org/10.1088/1126-6708/2008/05/014}{\emph{JHEP}
  {\bf 05} (2008) 014}, [\href{http://arxiv.org/abs/0712.2718}{{\tt
  0712.2718}}].

\bibitem{Webber:2009vm}
B.~Webber, \emph{{Mass determination in sequential particle decay chains}},
  \href{http://dx.doi.org/10.1088/1126-6708/2009/09/124}{\emph{JHEP} {\bf 09}
  (2009) 124}, [\href{http://arxiv.org/abs/0907.5307}{{\tt 0907.5307}}].

\bibitem{Arkani-Hamed:2005qjb}
N.~Arkani-Hamed, G.~L. Kane, J.~Thaler and L.-T. Wang, \emph{{Supersymmetry and
  the LHC inverse problem}},
  \href{http://dx.doi.org/10.1088/1126-6708/2006/08/070}{\emph{JHEP} {\bf 08}
  (2006) 070}, [\href{http://arxiv.org/abs/hep-ph/0512190}{{\tt
  hep-ph/0512190}}].

\bibitem{Gjelsten:2005sv}
B.~K. Gjelsten, D.~J. Miller and P.~Osland, \emph{{Resolving ambiguities in
  mass determinations at future colliders}}, {\emph{eConf} {\bf C050318} (2005)
  0211}, [\href{http://arxiv.org/abs/hep-ph/0507232}{{\tt hep-ph/0507232}}].

\bibitem{Gjelsten:2006as}
B.~K. Gjelsten, D.~J. Miller, P.~Osland and A.~R. Raklev, \emph{{Mass
  ambiguities in cascade decays}}, {\emph{Conf. Proc. C} {\bf 060726} (2006)
  1171--1174}, [\href{http://arxiv.org/abs/hep-ph/0611080}{{\tt
  hep-ph/0611080}}].

\bibitem{CMS:2013wbt}
{\scshape CMS} collaboration, S.~Chatrchyan et~al., \emph{{Measurement of
  Masses in the $t \bar{t}$ System by Kinematic Endpoints in pp Collisions at
  $\sqrt{s}$ = 7 TeV}},
  \href{http://dx.doi.org/10.1140/epjc/s10052-013-2494-7}{\emph{Eur. Phys. J.
  C} {\bf 73} (2013) 2494}, [\href{http://arxiv.org/abs/1304.5783}{{\tt
  1304.5783}}].

\bibitem{Gao:2010qx}
Y.~Gao, A.~V. Gritsan, Z.~Guo, K.~Melnikov, M.~Schulze and N.~V. Tran,
  \emph{{Spin Determination of Single-Produced Resonances at Hadron
  Colliders}}, \href{http://dx.doi.org/10.1103/PhysRevD.81.075022}{\emph{Phys.
  Rev. D} {\bf 81} (2010) 075022}, [\href{http://arxiv.org/abs/1001.3396}{{\tt
  1001.3396}}].

\bibitem{Bolognesi:2012mm}
S.~Bolognesi, Y.~Gao, A.~V. Gritsan, K.~Melnikov, M.~Schulze, N.~V. Tran
  et~al., \emph{{On the spin and parity of a single-produced resonance at the
  LHC}}, \href{http://dx.doi.org/10.1103/PhysRevD.86.095031}{\emph{Phys. Rev.
  D} {\bf 86} (2012) 095031}, [\href{http://arxiv.org/abs/1208.4018}{{\tt
  1208.4018}}].

\bibitem{Avery:2012um}
P.~Avery et~al., \emph{{Precision studies of the Higgs boson decay channel
  $H\to ZZ\to 4 \ell$ with MEKD}},
  \href{http://dx.doi.org/10.1103/PhysRevD.87.055006}{\emph{Phys. Rev. D} {\bf
  87} (2013) 055006}, [\href{http://arxiv.org/abs/1210.0896}{{\tt 1210.0896}}].

\bibitem{Betancur:2017kqe}
A.~Betancur, D.~Debnath, J.~S. Gainer, K.~T. Matchev and P.~Shyamsundar,
  \emph{{Measuring the mass, width, and couplings of semi-invisible resonances
  with the Matrix Element Method}},
  \href{http://dx.doi.org/10.1103/PhysRevD.99.116007}{\emph{Phys. Rev. D} {\bf
  99} (2019) 116007}, [\href{http://arxiv.org/abs/1708.07641}{{\tt
  1708.07641}}].

\bibitem{Debnath:2015wra}
D.~Debnath, J.~S. Gainer, D.~Kim and K.~T. Matchev, \emph{{Edge Detecting New
  Physics the Voronoi Way}},
  \href{http://dx.doi.org/10.1209/0295-5075/114/41001}{\emph{EPL} {\bf 114}
  (2016) 41001}, [\href{http://arxiv.org/abs/1506.04141}{{\tt 1506.04141}}].

\bibitem{Karapostoli:2008rra}
G.~Karapostoli, \emph{{Observation and measurement of the supersymmetric
  process $\tilde{\chi}_{2}^{0} \rightarrow \tilde{\chi}_{1}^{0} \ell \ell$
  with the CMS experiment at LHC}}.
\newblock PhD thesis, Athens U., 2008.

\bibitem{Huang:2008ae}
P.~Huang, N.~Kersting and H.~H. Yang, \emph{{Hidden Thresholds: A Technique for
  Reconstructing New Physics Masses at Hadron Colliders}},
  \href{http://arxiv.org/abs/0802.0022}{{\tt 0802.0022}}.

\bibitem{Debnath:2016gwz}
D.~Debnath, J.~S. Gainer, C.~Kilic, D.~Kim, K.~T. Matchev and Y.-P. Yang,
  \emph{{Detecting kinematic boundary surfaces in phase space: particle mass
  measurements in SUSY-like events}},
  \href{http://dx.doi.org/10.1007/JHEP06(2017)092}{\emph{JHEP} {\bf 06} (2017)
  092}, [\href{http://arxiv.org/abs/1611.04487}{{\tt 1611.04487}}].

\bibitem{Altunkaynak:2016bqe}
B.~Altunkaynak, C.~Kilic and M.~D. Klimek, \emph{{Multidimensional phase space
  methods for mass measurements and decay topology determination}},
  \href{http://dx.doi.org/10.1140/epjc/s10052-017-4631-1}{\emph{Eur. Phys. J.
  C} {\bf 77} (2017) 61}, [\href{http://arxiv.org/abs/1611.09764}{{\tt
  1611.09764}}].

\bibitem{Curtin:2011ng}
D.~Curtin, \emph{{Mixing It Up With MT2: Unbiased Mass Measurements at Hadron
  Colliders}}, \href{http://dx.doi.org/10.1103/PhysRevD.85.075004}{\emph{Phys.
  Rev. D} {\bf 85} (2012) 075004}, [\href{http://arxiv.org/abs/1112.1095}{{\tt
  1112.1095}}].

\bibitem{Voronoi}
A.~Okabe, B.~Boots and K.~Sugihara, \emph{Spatial Tessellations: Concepts and
  Applications of Voronoi Diagrams}.
\newblock John Wiley \& Sons, Inc., USA, 1992.

\bibitem{Debnath:2016mwb}
D.~Debnath, J.~S. Gainer, C.~Kilic, D.~Kim, K.~T. Matchev and Y.-P. Yang,
  \emph{{Identifying Phase Space Boundaries with Voronoi Tessellations}},
  \href{http://dx.doi.org/10.1140/epjc/s10052-016-4431-z}{\emph{Eur. Phys. J.
  C} {\bf 76} (2016) 645}, [\href{http://arxiv.org/abs/1606.02721}{{\tt
  1606.02721}}].

\bibitem{Matchev:2020vhr}
K.~T. Matchev, A.~Roman and P.~Shyamsundar, \emph{{Finding Wombling Boundaries
  in LHC Data with Voronoi and Delaunay Tessellations}},
  \href{http://arxiv.org/abs/2006.06582}{{\tt 2006.06582}}.

\bibitem{Dixon:2003yb}
L.~J. Dixon and M.~S. Siu, \emph{{Resonance continuum interference in the
  diphoton Higgs signal at the LHC}},
  \href{http://dx.doi.org/10.1103/PhysRevLett.90.252001}{\emph{Phys. Rev.
  Lett.} {\bf 90} (2003) 252001},
  [\href{http://arxiv.org/abs/hep-ph/0302233}{{\tt hep-ph/0302233}}].

\bibitem{Martin:2012xc}
S.~P. Martin, \emph{{Shift in the LHC Higgs Diphoton Mass Peak from
  Interference with Background}},
  \href{http://dx.doi.org/10.1103/PhysRevD.86.073016}{\emph{Phys. Rev. D} {\bf
  86} (2012) 073016}, [\href{http://arxiv.org/abs/1208.1533}{{\tt 1208.1533}}].

\bibitem{deFlorian:2013psa}
D.~de~Florian, N.~Fidanza, R.~J. Hern\'andez-Pinto, J.~Mazzitelli,
  Y.~Rotstein~Habarnau and G.~F.~R. Sborlini, \emph{{A complete $O(\alpha_S^2)$
  calculation of the signal-background interference for the Higgs diphoton
  decay channel}},
  \href{http://dx.doi.org/10.1140/epjc/s10052-013-2387-9}{\emph{Eur. Phys. J.
  C} {\bf 73} (2013) 2387}, [\href{http://arxiv.org/abs/1303.1397}{{\tt
  1303.1397}}].

\bibitem{Martin:2013ula}
S.~P. Martin, \emph{{Interference of Higgs Diphoton Signal and Background in
  Production with a Jet at the LHC}},
  \href{http://dx.doi.org/10.1103/PhysRevD.88.013004}{\emph{Phys. Rev. D} {\bf
  88} (2013) 013004}, [\href{http://arxiv.org/abs/1303.3342}{{\tt 1303.3342}}].

\bibitem{Coradeschi:2015tna}
F.~Coradeschi, D.~de~Florian, L.~J. Dixon, N.~Fidanza, S.~H\"oche, H.~Ita
  et~al., \emph{{Interference effects in the $H(\rightarrow \gamma\gamma) + 2$
  jets channel at the LHC}},
  \href{http://dx.doi.org/10.1103/PhysRevD.92.013004}{\emph{Phys. Rev. D} {\bf
  92} (2015) 013004}, [\href{http://arxiv.org/abs/1504.05215}{{\tt
  1504.05215}}].

\bibitem{Cieri:2017kpq}
L.~Cieri, F.~Coradeschi, D.~de~Florian and N.~Fidanza,
  \emph{{Transverse-momentum resummation for the signal-background interference
  in the H\textrightarrow{}\ensuremath{\gamma}\ensuremath{\gamma} channel at
  the LHC}}, \href{http://dx.doi.org/10.1103/PhysRevD.96.054003}{\emph{Phys.
  Rev. D} {\bf 96} (2017) 054003}, [\href{http://arxiv.org/abs/1706.07331}{{\tt
  1706.07331}}].

\bibitem{Dixon:2013haa}
L.~J. Dixon and Y.~Li, \emph{{Bounding the Higgs Boson Width Through
  Interferometry}},
  \href{http://dx.doi.org/10.1103/PhysRevLett.111.111802}{\emph{Phys. Rev.
  Lett.} {\bf 111} (2013) 111802}, [\href{http://arxiv.org/abs/1305.3854}{{\tt
  1305.3854}}].

\bibitem{Campbell:2017rke}
J.~Campbell, M.~Carena, R.~Harnik and Z.~Liu, \emph{{Interference in the
  $gg\rightarrow h \rightarrow \gamma\gamma$ On-Shell Rate and the Higgs Boson
  Total Width}},
  \href{http://dx.doi.org/10.1103/PhysRevLett.119.181801}{\emph{Phys. Rev.
  Lett.} {\bf 119} (2017) 181801}, [\href{http://arxiv.org/abs/1704.08259}{{\tt
  1704.08259}}].

\bibitem{ATLAS:2016kvj}
{\scshape ATLAS} collaboration, \emph{{Estimate of the $m_H$ shift due to
  interference between signal and background processes in the $H \rightarrow
  \gamma\gamma$ channel, for the $\sqrt{s} = 8$ TeV dataset recorded by
  ATLAS}},  4, 2016.

\bibitem{ParticleDataGroup:2020ssz}
{\scshape Particle Data Group} collaboration, P.~A. Zyla et~al., \emph{{Review
  of Particle Physics}},
  \href{http://dx.doi.org/10.1093/ptep/ptaa104}{\emph{PTEP} {\bf 2020} (2020)
  083C01}.

\bibitem{CMS:2020xrn}
{\scshape CMS} collaboration, A.~M. Sirunyan et~al., \emph{{A measurement of
  the Higgs boson mass in the diphoton decay channel}},
  \href{http://dx.doi.org/10.1016/j.physletb.2020.135425}{\emph{Phys. Lett. B}
  {\bf 805} (2020) 135425}, [\href{http://arxiv.org/abs/2002.06398}{{\tt
  2002.06398}}].

\bibitem{ATLAS:2018tdk}
{\scshape ATLAS} collaboration, M.~Aaboud et~al., \emph{{Measurement of the
  Higgs boson mass in the $H\rightarrow ZZ^* \rightarrow 4\ell$ and $H
  \rightarrow \gamma\gamma$ channels with $\sqrt{s}=13$ TeV $pp$ collisions
  using the ATLAS detector}},
  \href{http://dx.doi.org/10.1016/j.physletb.2018.07.050}{\emph{Phys. Lett. B}
  {\bf 784} (2018) 345--366}, [\href{http://arxiv.org/abs/1806.00242}{{\tt
  1806.00242}}].

\bibitem{Cepeda:2019klc}
M.~Cepeda et~al., \emph{{Report from Working Group 2}: {Higgs Physics at the
  HL-LHC and HE-LHC}},
  \href{http://dx.doi.org/10.23731/CYRM-2019-007.221}{\emph{CERN Yellow Rep.
  Monogr.} {\bf 7} (2019) 221--584},
  [\href{http://arxiv.org/abs/1902.00134}{{\tt 1902.00134}}].

\bibitem{CMS-PAS-FTR-21-007}
{CMS Collaboration}, \emph{{Projection of the Higgs boson mass and on-shell
  width measurements in $H \rightarrow ZZ \rightarrow 4\ell$ decay channel at
  the HL-LHC}}, \href{http://dx.doi.org/oai:cds.cern.ch:2804004}{\emph{{CERN
  Note}} {\bf {}} ({}) {}},
  [\href{http://arxiv.org/abs/CMS-PAS-FTR-21-007}{{\tt CMS-PAS-FTR-21-007}}].

\bibitem{CMS-PAS-FTR-21-008}
{CMS Collaboration}, \emph{{A projection of the precision of the Higgs boson
  mass measurement in the diphoton decay channel at the High Luminosity LHC}},
  \href{http://dx.doi.org/oai:cds.cern.ch:2804042}{\emph{{CERN Note}} {\bf {}}
  ({}) {}}, [\href{http://arxiv.org/abs/CMS-PAS-FTR-21-008}{{\tt
  CMS-PAS-FTR-21-008}}].

\bibitem{Feickert:2021ajf}
M.~Feickert and B.~Nachman, \emph{{A Living Review of Machine Learning for
  Particle Physics}},  \href{http://arxiv.org/abs/2102.02770}{{\tt
  2102.02770}}.

\bibitem{Guest:2018yhq}
D.~Guest, K.~Cranmer and D.~Whiteson, \emph{{Deep Learning and its Application
  to LHC Physics}},
  \href{http://dx.doi.org/10.1146/annurev-nucl-101917-021019}{\emph{Ann. Rev.
  Nucl. Part. Sci.} {\bf 68} (2018) 161--181},
  [\href{http://arxiv.org/abs/1806.11484}{{\tt 1806.11484}}].

\bibitem{Cranmer:2015bka}
K.~Cranmer, J.~Pavez and G.~Louppe, \emph{{Approximating Likelihood Ratios with
  Calibrated Discriminative Classifiers}},
  \href{http://arxiv.org/abs/1506.02169}{{\tt 1506.02169}}.

\bibitem{Datta:2019ndh}
K.~Datta, A.~Larkoski and B.~Nachman, \emph{{Automating the Construction of Jet
  Observables with Machine Learning}},
  \href{http://dx.doi.org/10.1103/PhysRevD.100.095016}{\emph{Phys. Rev. D} {\bf
  100} (2019) 095016}, [\href{http://arxiv.org/abs/1902.07180}{{\tt
  1902.07180}}].

\bibitem{Komiske:2018cqr}
P.~T. Komiske, E.~M. Metodiev and J.~Thaler, \emph{{Energy Flow Networks: Deep
  Sets for Particle Jets}},
  \href{http://dx.doi.org/10.1007/JHEP01(2019)121}{\emph{JHEP} {\bf 01} (2019)
  121}, [\href{http://arxiv.org/abs/1810.05165}{{\tt 1810.05165}}].

\bibitem{NIPS2017_f22e4747}
M.~Zaheer, S.~Kottur, S.~Ravanbakhsh, B.~Poczos, R.~R. Salakhutdinov and A.~J.
  Smola, \emph{Deep sets},  in \emph{Advances in Neural Information Processing
  Systems} (I.~Guyon, U.~V. Luxburg, S.~Bengio, H.~Wallach, R.~Fergus,
  S.~Vishwanathan et~al., eds.), vol.~30, Curran Associates, Inc., 2017.

\bibitem{Komiske:2017aww}
P.~T. Komiske, E.~M. Metodiev and J.~Thaler, \emph{{Energy flow polynomials: A
  complete linear basis for jet substructure}},
  \href{http://dx.doi.org/10.1007/JHEP04(2018)013}{\emph{JHEP} {\bf 04} (2018)
  013}, [\href{http://arxiv.org/abs/1712.07124}{{\tt 1712.07124}}].

\bibitem{Erdmann:2018shi}
M.~Erdmann, E.~Geiser, Y.~Rath and M.~Rieger, \emph{{Lorentz Boost Networks:
  Autonomous Physics-Inspired Feature Engineering}},
  \href{http://dx.doi.org/10.1088/1748-0221/14/06/P06006}{\emph{JINST} {\bf 14}
  (2019) P06006}, [\href{http://arxiv.org/abs/1812.09722}{{\tt 1812.09722}}].

\bibitem{LBN}
P.~Jung, \emph{{A Deep Learning Based Reconstruction of Two Neutrinos in the
  Di-lepton Decays of Top Quark Pairs with the CMS Experiment, bachelor thesis,
  RWTH Aachen, 2019}}, .

\bibitem{LBNtth}
B.~Idaszek, \emph{{RECONSTRUCTION OF THE NEUTRINO MOMENTUM IN TOP PAIR
  ASSOCIATED HIGGS BOSON PRODUCTION USING DEEP LEARNING, bachelor thesis, RWTH
  Aachen, 2019}}, .

\bibitem{Chang:2017kvc}
S.~Chang, T.~Cohen and B.~Ostdiek, \emph{{What is the Machine Learning?}},
  \href{http://dx.doi.org/10.1103/PhysRevD.97.056009}{\emph{Phys. Rev. D} {\bf
  97} (2018) 056009}, [\href{http://arxiv.org/abs/1709.10106}{{\tt
  1709.10106}}].

\bibitem{Faucett:2020vbu}
T.~Faucett, J.~Thaler and D.~Whiteson, \emph{{Mapping Machine-Learned Physics
  into a Human-Readable Space}},
  \href{http://dx.doi.org/10.1103/PhysRevD.103.036020}{\emph{Phys. Rev. D} {\bf
  103} (2021) 036020}, [\href{http://arxiv.org/abs/2010.11998}{{\tt
  2010.11998}}].

\bibitem{Agarwal:2020fpt}
G.~Agarwal, L.~Hay, I.~Iashvili, B.~Mannix, C.~McLean, M.~Morris et~al.,
  \emph{{Explainable AI for ML jet taggers using expert variables and layerwise
  relevance propagation}},
  \href{http://dx.doi.org/10.1007/JHEP05(2021)208}{\emph{JHEP} {\bf 05} (2021)
  208}, [\href{http://arxiv.org/abs/2011.13466}{{\tt 2011.13466}}].

\bibitem{Grojean:2020ech}
C.~Grojean, A.~Paul and Z.~Qian, \emph{{Resurrecting $ b\overline{b}h $ with
  kinematic shapes}},
  \href{http://dx.doi.org/10.1007/JHEP04(2021)139}{\emph{JHEP} {\bf 04} (2021)
  139}, [\href{http://arxiv.org/abs/2011.13945}{{\tt 2011.13945}}].

\bibitem{djidjev2018efficient}
H.~N. Djidjev, G.~Chapuis, G.~Hahn and G.~Rizk, \emph{Efficient combinatorial
  optimization using quantum annealing}, {\emph{arXiv preprint
  arXiv:1801.08653} (2018) }.

\bibitem{farhi2014quantum}
E.~Farhi, J.~Goldstone and S.~Gutmann, \emph{A quantum approximate optimization
  algorithm}, {\emph{arXiv preprint arXiv:1411.4028} (2014) }.

\bibitem{Wei:2019rqy}
A.~Y. Wei, P.~Naik, A.~W. Harrow and J.~Thaler, \emph{{Quantum Algorithms for
  Jet Clustering}},
  \href{http://dx.doi.org/10.1103/PhysRevD.101.094015}{\emph{Phys. Rev. D} {\bf
  101} (2020) 094015}, [\href{http://arxiv.org/abs/1908.08949}{{\tt
  1908.08949}}].

\bibitem{Pires:2020urc}
D.~Pires, Y.~Omar and J.~a. Seixas, \emph{{Adiabatic Quantum Algorithm for
  Multijet Clustering in High Energy Physics}},
  \href{http://arxiv.org/abs/2012.14514}{{\tt 2012.14514}}.

\bibitem{Pires:2021fka}
D.~Pires, P.~Bargassa, J.~a. Seixas and Y.~Omar, \emph{{A Digital Quantum
  Algorithm for Jet Clustering in High-Energy Physics}},
  \href{http://arxiv.org/abs/2101.05618}{{\tt 2101.05618}}.

\bibitem{Delgado:2022snu}
A.~Delgado and J.~Thaler, \emph{{Quantum Annealing for Jet Clustering with
  Thrust}},  \href{http://arxiv.org/abs/2205.02814}{{\tt 2205.02814}}.

\bibitem{Kim:2021wrr}
M.~Kim, P.~Ko, J.-h. Park and M.~Park, \emph{{Leveraging Quantum Annealer to
  identify an Event-topology at High Energy Colliders}},
  \href{http://arxiv.org/abs/2111.07806}{{\tt 2111.07806}}.

\bibitem{Jungman:1995df}
G.~Jungman, M.~Kamionkowski and K.~Griest, \emph{{Supersymmetric dark matter}},
  \href{http://dx.doi.org/10.1016/0370-1573(95)00058-5}{\emph{Phys. Rept.} {\bf
  267} (1996) 195--373}, [\href{http://arxiv.org/abs/hep-ph/9506380}{{\tt
  hep-ph/9506380}}].

\bibitem{Tucker-Smith:2001myb}
D.~Tucker-Smith and N.~Weiner, \emph{{Inelastic dark matter}},
  \href{http://dx.doi.org/10.1103/PhysRevD.64.043502}{\emph{Phys. Rev. D} {\bf
  64} (2001) 043502}, [\href{http://arxiv.org/abs/hep-ph/0101138}{{\tt
  hep-ph/0101138}}].

\bibitem{COHERENT:2017ipa}
{\scshape COHERENT} collaboration, D.~Akimov et~al., \emph{{Observation of
  Coherent Elastic Neutrino-Nucleus Scattering}},
  \href{http://dx.doi.org/10.1126/science.aao0990}{\emph{Science} {\bf 357}
  (2017) 1123--1126}, [\href{http://arxiv.org/abs/1708.01294}{{\tt
  1708.01294}}].

\bibitem{COHERENT:2018gft}
{\scshape COHERENT} collaboration, D.~Akimov et~al., \emph{{COHERENT 2018 at
  the Spallation Neutron Source}},  \href{http://arxiv.org/abs/1803.09183}{{\tt
  1803.09183}}.

\bibitem{COHERENT:2019kwz}
{\scshape COHERENT} collaboration, D.~Akimov et~al., \emph{{Sensitivity of the
  COHERENT Experiment to Accelerator-Produced Dark Matter}},
  \href{http://dx.doi.org/10.1103/PhysRevD.102.052007}{\emph{Phys. Rev. D} {\bf
  102} (2020) 052007}, [\href{http://arxiv.org/abs/1911.06422}{{\tt
  1911.06422}}].

\bibitem{CCM:2021leg}
{\scshape CCM} collaboration, A.~A. Aguilar-Arevalo et~al., \emph{{First Dark
  Matter Search Results From Coherent CAPTAIN-Mills}},
  \href{http://arxiv.org/abs/2105.14020}{{\tt 2105.14020}}.

\bibitem{deNiverville:2015mwa}
P.~deNiverville, M.~Pospelov and A.~Ritz, \emph{{Light new physics in coherent
  neutrino-nucleus scattering experiments}},
  \href{http://dx.doi.org/10.1103/PhysRevD.92.095005}{\emph{Phys. Rev. D} {\bf
  92} (2015) 095005}, [\href{http://arxiv.org/abs/1505.07805}{{\tt
  1505.07805}}].

\bibitem{Dutta:2019nbn}
B.~Dutta, D.~Kim, S.~Liao, J.-C. Park, S.~Shin and L.~E. Strigari, \emph{{Dark
  matter signals from timing spectra at neutrino experiments}},
  \href{http://dx.doi.org/10.1103/PhysRevLett.124.121802}{\emph{Phys. Rev.
  Lett.} {\bf 124} (2020) 121802}, [\href{http://arxiv.org/abs/1906.10745}{{\tt
  1906.10745}}].

\bibitem{Dutta:2020vop}
B.~Dutta, D.~Kim, S.~Liao, J.-C. Park, S.~Shin, L.~E. Strigari et~al.,
  \emph{{Searching for Dark Matter Signals in Timing Spectra at Neutrino
  Experiments}},  \href{http://arxiv.org/abs/2006.09386}{{\tt 2006.09386}}.

\bibitem{Aguilar-Arevalo:2021sbh}
A.~A. Aguilar-Arevalo et~al., \emph{{First Leptophobic Dark Matter Search from
  Coherent CAPTAIN-Mills}},  \href{http://arxiv.org/abs/2109.14146}{{\tt
  2109.14146}}.

\bibitem{MiniBooNE:2012jpi}
{\scshape MiniBooNE} collaboration, R.~Dharmapalan et~al., \emph{{Low Mass WIMP
  Searches with a Neutrino Experiment: A Proposal for Further MiniBooNE
  Running}},  \href{http://arxiv.org/abs/1211.2258}{{\tt 1211.2258}}.

\bibitem{MiniBooNEDM:2018cxm}
{\scshape MiniBooNE DM} collaboration, A.~A. Aguilar-Arevalo et~al.,
  \emph{{Dark Matter Search in Nucleon, Pion, and Electron Channels from a
  Proton Beam Dump with MiniBooNE}},
  \href{http://dx.doi.org/10.1103/PhysRevD.98.112004}{\emph{Phys. Rev. D} {\bf
  98} (2018) 112004}, [\href{http://arxiv.org/abs/1807.06137}{{\tt
  1807.06137}}].

\bibitem{Belanger:2011ww}
G.~Belanger and J.-C. Park, \emph{{Assisted freeze-out}},
  \href{http://dx.doi.org/10.1088/1475-7516/2012/03/038}{\emph{JCAP} {\bf 03}
  (2012) 038}, [\href{http://arxiv.org/abs/1112.4491}{{\tt 1112.4491}}].

\bibitem{Agashe:2014yua}
K.~Agashe, Y.~Cui, L.~Necib and J.~Thaler, \emph{{(In)direct Detection of
  Boosted Dark Matter}},
  \href{http://dx.doi.org/10.1088/1475-7516/2014/10/062}{\emph{JCAP} {\bf 10}
  (2014) 062}, [\href{http://arxiv.org/abs/1405.7370}{{\tt 1405.7370}}].

\bibitem{Alhazmi:2016qcs}
H.~Alhazmi, K.~Kong, G.~Mohlabeng and J.-C. Park, \emph{{Boosted Dark Matter at
  the Deep Underground Neutrino Experiment}},
  \href{http://dx.doi.org/10.1007/JHEP04(2017)158}{\emph{JHEP} {\bf 04} (2017)
  158}, [\href{http://arxiv.org/abs/1611.09866}{{\tt 1611.09866}}].

\bibitem{Necib:2016aez}
L.~Necib, J.~Moon, T.~Wongjirad and J.~M. Conrad, \emph{{Boosted Dark Matter at
  Neutrino Experiments}},
  \href{http://dx.doi.org/10.1103/PhysRevD.95.075018}{\emph{Phys. Rev. D} {\bf
  95} (2017) 075018}, [\href{http://arxiv.org/abs/1610.03486}{{\tt
  1610.03486}}].

\bibitem{Berger:2014sqa}
J.~Berger, Y.~Cui and Y.~Zhao, \emph{{Detecting Boosted Dark Matter from the
  Sun with Large Volume Neutrino Detectors}},
  \href{http://dx.doi.org/10.1088/1475-7516/2015/02/005}{\emph{JCAP} {\bf 02}
  (2015) 005}, [\href{http://arxiv.org/abs/1410.2246}{{\tt 1410.2246}}].

\bibitem{Kong:2014mia}
K.~Kong, G.~Mohlabeng and J.-C. Park, \emph{{Boosted dark matter signals
  uplifted with self-interaction}},
  \href{http://dx.doi.org/10.1016/j.physletb.2015.02.057}{\emph{Phys. Lett. B}
  {\bf 743} (2015) 256--266}, [\href{http://arxiv.org/abs/1411.6632}{{\tt
  1411.6632}}].

\bibitem{Super-Kamiokande:2017dch}
{\scshape Super-Kamiokande} collaboration, C.~Kachulis et~al., \emph{{Search
  for Boosted Dark Matter Interacting With Electrons in Super-Kamiokande}},
  \href{http://dx.doi.org/10.1103/PhysRevLett.120.221301}{\emph{Phys. Rev.
  Lett.} {\bf 120} (2018) 221301}, [\href{http://arxiv.org/abs/1711.05278}{{\tt
  1711.05278}}].

\bibitem{deNiverville:2018dbu}
P.~deNiverville and C.~Frugiuele, \emph{{Hunting sub-GeV dark matter with the
  NO$\nu$A near detector}},
  \href{http://dx.doi.org/10.1103/PhysRevD.99.051701}{\emph{Phys. Rev. D} {\bf
  99} (2019) 051701}, [\href{http://arxiv.org/abs/1807.06501}{{\tt
  1807.06501}}].

\bibitem{DeRomeri:2019kic}
V.~De~Romeri, K.~J. Kelly and P.~A.~N. Machado, \emph{{DUNE-PRISM Sensitivity
  to Light Dark Matter}},
  \href{http://dx.doi.org/10.1103/PhysRevD.100.095010}{\emph{Phys. Rev. D} {\bf
  100} (2019) 095010}, [\href{http://arxiv.org/abs/1903.10505}{{\tt
  1903.10505}}].

\bibitem{Bertuzzo:2018itn}
E.~Bertuzzo, S.~Jana, P.~A.~N. Machado and R.~Zukanovich~Funchal, \emph{{Dark
  Neutrino Portal to Explain MiniBooNE excess}},
  \href{http://dx.doi.org/10.1103/PhysRevLett.121.241801}{\emph{Phys. Rev.
  Lett.} {\bf 121} (2018) 241801}, [\href{http://arxiv.org/abs/1807.09877}{{\tt
  1807.09877}}].

\bibitem{Izaguirre:2014dua}
E.~Izaguirre, G.~Krnjaic, P.~Schuster and N.~Toro, \emph{{Physics motivation
  for a pilot dark matter search at Jefferson Laboratory}},
  \href{http://dx.doi.org/10.1103/PhysRevD.90.014052}{\emph{Phys. Rev. D} {\bf
  90} (2014) 014052}, [\href{http://arxiv.org/abs/1403.6826}{{\tt 1403.6826}}].

\bibitem{Kim:2016zjx}
D.~Kim, J.-C. Park and S.~Shin, \emph{{Dark Matter
  \textquotedblleft{}Collider\textquotedblright{} from Inelastic Boosted Dark
  Matter}}, \href{http://dx.doi.org/10.1103/PhysRevLett.119.161801}{\emph{Phys.
  Rev. Lett.} {\bf 119} (2017) 161801},
  [\href{http://arxiv.org/abs/1612.06867}{{\tt 1612.06867}}].

\bibitem{OxbridgeKineticsLibrary}
C.~Lester, \emph{{Oxbridge Kinetics Library}}, .

\bibitem{Erdmann:2013rxa}
J.~Erdmann, S.~Guindon, K.~Kroeninger, B.~Lemmer, O.~Nackenhorst, A.~Quadt
  et~al., \emph{{A likelihood-based reconstruction algorithm for top-quark
  pairs and the KLFitter framework}},
  \href{http://dx.doi.org/10.1016/j.nima.2014.02.029}{\emph{Nucl. Instrum.
  Meth. A} {\bf 748} (2014) 18--25},
  [\href{http://arxiv.org/abs/1312.5595}{{\tt 1312.5595}}].

\bibitem{Brochet:2018pqf}
S.~Brochet, C.~Delaere, B.~Fran\c{c}ois, V.~Lema\^\i{}tre, A.~Mertens,
  A.~Saggio et~al., \emph{{MoMEMta, a modular toolkit for the Matrix Element
  Method at the LHC}},
  \href{http://dx.doi.org/10.1140/epjc/s10052-019-6635-5}{\emph{Eur. Phys. J.
  C} {\bf 79} (2019) 126}, [\href{http://arxiv.org/abs/1805.08555}{{\tt
  1805.08555}}].

\bibitem{Antcheva:2011zz}
I.~Antcheva et~al., \emph{{ROOT: A C++ framework for petabyte data storage,
  statistical analysis and visualization}},
  \href{http://dx.doi.org/10.1016/j.cpc.2011.02.008}{\emph{Comput. Phys.
  Commun.} {\bf 182} (2011) 1384--1385}.

\bibitem{Antcheva:2009zz}
I.~Antcheva et~al., \emph{{ROOT: A C++ framework for petabyte data storage,
  statistical analysis and visualization}},
  \href{http://dx.doi.org/10.1016/j.cpc.2009.08.005}{\emph{Comput. Phys.
  Commun.} {\bf 180} (2009) 2499--2512},
  [\href{http://arxiv.org/abs/1508.07749}{{\tt 1508.07749}}].

\bibitem{Lonnblad:1994np}
L.~Lonnblad, \emph{{CLHEP: A project for designing a C++ class library for
  high-energy physics}},
  \href{http://dx.doi.org/10.1016/0010-4655(94)90217-8}{\emph{Comput. Phys.
  Commun.} {\bf 84} (1994) 307--316}.

\bibitem{Cacciari:2011ma}
M.~Cacciari, G.~P. Salam and G.~Soyez, \emph{{FastJet User Manual}},
  \href{http://dx.doi.org/10.1140/epjc/s10052-012-1896-2}{\emph{Eur. Phys. J.
  C} {\bf 72} (2012) 1896}, [\href{http://arxiv.org/abs/1111.6097}{{\tt
  1111.6097}}].

\bibitem{snowmass}
``Snowmass 2021 particle physics community planning exercise: Theory frontier:
  Topical group 07: Collider phenomenology.''
  \url{https://snowmass21.org/theory/phenomenology}.

\end{thebibliography}\endgroup

\end{document}